\documentclass[11pt,twoside]{article}
\usepackage{subeqnarray,psfig}

\newcommand{\sw}{s_w}
\newcommand{\cw}{c_w}

\newcommand{\SLASH}[2]{\makebox[#2ex][l]{$#1$}/}

\newcommand{\kslash}{\SLASH{k}{.15}}
\newcommand{\pslash}{\SLASH{p}{.2}}

\newcommand{\MA}{M_{A^0}}

\newcommand{\id}{{\rm 1\kern-.12em
\rule{0.3pt}{1.5ex}\raisebox{0.0ex}{\rule{0.1em}{0.3pt}}}}
\newcommand{\lsim}
{\;\raisebox{-.3em}{$\stackrel{\displaystyle <}{\sim}$}\;}

\newcommand{\OL}[1]{\overline{#1}}
\newcommand{\D}{\displaystyle}
\newcommand{\BS}{\begin{samepage}}
\newcommand{\ES}{\end{samepage}}

\newcommand{\VL}{\left( \begin{array}{c}}
\newcommand{\VR}{\end{array} \right)}
\newcommand{\ML}{\left( \begin{array}{cc}}
\newcommand{\MLv}{\left( \begin{array}{cccc}}
\newcommand{\MR}{\end{array} \right)}
\newcommand{\dd}{\partial}
\newcommand{\dmu}{\partial_{\mu}}

\newcommand{\re}{\Re\mbox{e}}

\newcommand{\Tb}{\tan \beta\hspace{1mm}}

\newcommand{\Sb}{\sin \beta\hspace{1mm}}
\newcommand{\SQb}{\sin^2\beta\hspace{1mm}}
\newcommand{\Cb}{\cos \beta\hspace{1mm}}
\newcommand{\CQb}{\cos^2\beta\hspace{1mm}}
\newcommand{\Sa}{\sin \alpha\hspace{1mm}}

\newcommand{\Ca}{\cos \alpha\hspace{1mm}}

\newcommand{\Sba}{\sin (\beta - \alpha)\hspace{1mm}}
\newcommand{\Cba}{\cos (\beta - \alpha)\hspace{1mm}}

\newcommand{\SZb}{\sin 2\beta\hspace{1mm}}

\newcommand{\CZb}{\cos 2\beta\hspace{1mm}}
\newcommand{\CQZb}{\cos^2 2\beta\hspace{1mm}}

\newcommand{\CQba}{\cos^2 (\beta - \alpha)\hspace{1mm}}
\newcommand{\SQba}{\sin^2 (\beta - \alpha)\hspace{1mm}}

\newcommand{\CtT}{c_{\tilde{\theta}}}
\newcommand{\StT}{s_{\tilde{\theta}}}
\newcommand{\CtTp}{c_{\tilde{\theta'}}}
\newcommand{\StTp}{s_{\tilde{\theta'}}}
\newcommand{\vorn}{\hspace{-0.6cm}}
\newcommand{\BC}{\begin{center}}
\newcommand{\EC}{\end{center}}
\newcommand{\BE}{\begin{equation}}
\newcommand{\EE}{\end{equation}}
\newcommand{\BEA}{\begin{eqnarray}}
\newcommand{\EEA}{\end{eqnarray}}
\newcommand{\BSUB}{\begin{subeqnarray}}
\newcommand{\ESUB}{\end{subeqnarray}}

\newcommand{\GeV}{\hspace{1mm}{\mbox {GeV}}}

\newcounter{dummy}
\pssilent
\begin{document}


\thispagestyle{empty}

\hfill KA-TP-3-1998

\hfill hep-ph/9807427

\vspace*{2cm}

\begin{center}

{\LARGE \bf
  Supersymmetric one-loop corrections \\ 
  to the process $e^+e^- \to f\bar{f}$}

\vspace{2cm}

\renewcommand{\thefootnote}{\fnsymbol{footnote}}
{\Large \sc W. Hollik\footnote[1]{E-mail: 
hollik@particle.physik.uni-karlsruhe.de} and 
C. Schappacher\footnote[2]{E-mail: cs@particle.physik.uni-karlsruhe.de}}
\renewcommand{\thefootnote}{\arabic{footnote}}

\vspace{2cm}

{\sl Institut f{\"u}r Theoretische Physik}

{\sl Universit{\"a}t Karlsruhe}

{\sl Kaiserstra{\ss}e 12}

{\sl D-76128 Karlsruhe, Germany}

\end{center}

\vspace{2cm}

\begin{abstract}
\noindent 
Radiative one-loop corrections to fermion pair production in $e^+e^-$ 
annihilation are calculated in the Minimal Supersymmetric Standard 
Model (MSSM). The size of the non-standard corrections is discussed for
the process $e^+e^- \to \mu^+\mu^-$ and $e^+e^- \to$ hadrons at LEP2 
energies and for $e^+e^- \to t\bar{t}$ at a high energy $e^+e^-$ 
collider. The relative difference between the predictions of the 
MSSM and the Standard Model is typically below $10\%$.
\end{abstract}

\newpage
\clearpage


\section{Introduction}
\label{chap1}
\setcounter{page}{1}
\parskip1ex

In the light of electroweak precision experiments the Standard Model 
(SM) performance is almost perfect \cite{LEPEWWG}. From a more
theoretical point of view, however, supersymmetric versions of the SM
seem more appropriate to overcome some of the conceptual problems of 
the SM, like the hierarchy problem or the non-occurrence of gauge 
coupling unification at high energies.

Tests of supersymmetric extensions of the SM hence are a central theme 
at present and future colliders. Supersymmetry (SUSY) \cite{nilles} as
an additional supporting symmetry is realized in its minimal version
in terms of the Minimal Supersymmetric Standard Model (MSSM) 
\cite{habkan,gunhab,hhg}.
Besides the possibility of direct production of SUSY particles at
sufficiently high energies, the process of standard fermion pair 
production in $e^+e^-$ annihilation offers the indirect search for
virtual SUSY particles through quantum effects in terms of loop 
corrections.

In previous studies, complete 1-loop calculations have been performed
for electroweak precision observables at the $Z$ resonance and for the
$M_W$-$M_Z$ mass interdependence \cite{precision}, with recent 
improvements by the 2-loop SUSY-QCD corrections to the
$\rho$-parameter \cite{rho}.
Investigation of $e^+e^- \to f\bar{f}$ above the $Z$ resonance, at
LEP2 or a future linear collider, requires the extension of the
previous calculations to the continuum region, which is the content of
this paper. As a process of special interest, $t\bar{t}$ production is
considered which will become experimentally accessible with high
accuracy at a possible future $e^+e^-$ collider.

In this paper we present a complete MSSM 1-loop calculation of the 
electroweak corrections and the non-standard part of the QCD corrections
(SUSY-QCD corrections) to $e^+e^- \to f\bar{f}$ with $f=\mu,\tau, q$ and
discuss the size of the possible virtual effects. Approximate results
based on the SUSY 1-loop terms of ${\cal O}(\alpha\, m_t^2/M_W^2)$ have
already been considered in \cite{Chang_Li}, and SUSY-QCD corrections in
\cite{Djouadi}. For the decay of the top quark $t \to b\,W^+$, the MSSM 
1-loop corrections are also available \cite{topdecay}.

This paper is organized as follows:
Section \ref{chap2}  defines conventions and notations and gives the
structure of the helicity amplitudes in terms of basic matrix
elements, including the higher order terms. The results are discussed
in Section \ref{chap3} for the light fermion case and, with special 
emphasis, for top pair production. The appendix collects all the
analytic expressions, required at the 1-loop level, in order to make
the paper complete and self-contained.

\newpage
\clearpage


\section{The  process $e^+ e^- \to f\bar{f}$}
\label{chap2}
\setcounter{equation}{0}
\setcounter{figure}{0}
\setcounter{table}{0}

\subsection{Cross section and helicity amplitudes}
\label{Bezeichnungen}

First we give the notations for the process $e^+ e^- \to f\bar{f}$ 
($f=\mu,\tau,q$). The momenta and helicities of the incoming 
electron and positron are denoted by $p$, $\kappa$ and $\bar{p}$, 
$\bar{\kappa}$, respectively. Correspondingly, $k$, $\eta$ and 
$\bar{k}$, $\bar{\eta}$ are used for the outgoing fermion $f$ and 
its antiparticle $\bar{f}$. The signs '+' and '$-$' of the 
variables $\kappa$, $\bar{\kappa}$ and $\eta$, $\bar{\eta}$ refer to 
helicities $+\frac{1}{2}$ and $-\frac{1}{2}$, respectively. The 
Mandelstam variables are defined by
\BEA
s &=& (p + \bar{p})^2 \; = \; (k + \bar{k})^2~, \nonumber \\
t &=& (p - k)^2 \; = \; (\bar{p} - \bar{k})^2~, \nonumber \\
u &=& (p - \bar{k})^2 \; = \; (\bar{p} - k)^2~.
\label{mandel}
\EEA
The mass of the electron is neglected whenever possible. Since the 
helicity amplitudes for $e^+ e^- \to f\bar{f}$ ($f=\mu,\tau,q$) vanish
for $\bar{\kappa}=\kappa$ in the limit $m_e \to 0$, we can write for the 
non-vanishing amplitudes
\BE
{\cal M}(\kappa,\bar{\kappa}=-\kappa,\eta,\bar{\eta};s,t) = 
{\cal M}(\kappa,\eta,\bar{\eta};s,t)~. 
\EE
Summing over the final helicities but keeping $\kappa$ fixed, we
obtain the differential cross section for polarized initial states as
follows:
\BE
\frac{d\sigma}{d\Omega}(\kappa,s,\theta) = 
\frac{1}{64\,\pi^2\,s}\, \beta_f\, N_C^f\, 
\sum_{\eta,\bar{\eta}} |{\cal M}(\kappa,\eta,\bar{\eta};s,t)|^2
\EE
with 
\BE
\beta_f=\sqrt{1-\frac{4m_f^2}{s}} \qquad ; \qquad
N_C^f = 3\; (1)\quad \mbox{for quarks (leptons)}~.
\EE
The scattering angle $\theta$ between $e^-$ and $f$ in the 
center-of-mass system is related to $t$, $u$ via
\BE
t=m_f^2 - \frac{s}{2}(1-\beta_f \cos\theta)~, \qquad
u=m_f^2 - \frac{s}{2}(1+\beta_f \cos\theta)~.
\EE
The cross section for unpolarized $e^{\pm}$ in the initial state is
given by
\BE
\frac{d\sigma}{d\Omega}(s,t) = \frac{1}{4}\, \left[
   \frac{d\sigma}{d\Omega}(+;s,t) + 
   \frac{d\sigma}{d\Omega}(-;s,t) \right]~.
\EE
Besides the integrated unpolarized cross section 
\BE
\sigma=\int\! d\Omega\; \frac{d\sigma}{d\Omega}(s,t) 
\label{sigtot}
\EE
it is of interest to consider the forward-backward asymmetry
\BE
A_{FB}=\frac{\sigma^F-\sigma^B}{\sigma^F+\sigma^B}~, \qquad
\sigma^{F(B)}=\int_{\theta<\frac{\pi}{2}(\theta>\frac{\pi}{2})} 
d\Omega\; \frac{d\sigma}{d\Omega}(s,t)
\label{afb}
\EE
and the left-right asymmetry for polarized beams
\BE
A_{LR}=\frac{\sigma_L-\sigma_R}{\sigma_L+\sigma_R}~, \qquad
\sigma_{R,L}=\int\! d\Omega\;\frac{d\sigma}{d\Omega}(\pm;s,t)~.
\label{alr}
\EE

For practical calculations at the 1-loop level, it is convenient to
decompose the helicity amplitude ${\cal M}$ into a set of basic matrix 
elements ${\cal M}_i^{\kappa\lambda}$ and corresponding invariant
functions $L_i^{\kappa\lambda}$ according to
\BE
{\cal M}(\kappa,\eta,\bar{\eta};s,t) = 
\sum_{i,\lambda} {\cal M}_i^{\kappa\lambda} L_i^{\kappa\lambda}
\EE
with ($\bar{\kappa}=-\kappa$):
\BSUB
\label{alle}
{\cal M}_1^{\kappa\lambda} &=& \OL{v}(\bar{p},\bar{\kappa})\, 
   \gamma_\mu\, \omega^\kappa\, u(p,\kappa)\,
   \OL{u}(k,\eta)\, \gamma^\mu\, \omega^\lambda\, 
    v(\bar{k},\bar{\eta})~,  
\slabel{bornmat1} \\
{\cal M}_2^{\kappa\lambda} &=& \OL{v}(\bar{p},\bar{\kappa}) 
   (\kslash-\bar{\kslash})\, \omega^\kappa\, u(p,\kappa)\, 
   \OL{u}(k,\eta) (\bar{\pslash}-\pslash)\, \omega^\lambda\, 
   v(\bar{k},\bar{\eta})~,
\label{matel2} \\
{\cal M}_3^{\kappa\lambda} &=& \OL{v}(\bar{p},\bar{\kappa}) 
   (\kslash-\bar{\kslash})\, \omega^\kappa\, u(p,\kappa)\,
   \OL{u}(k,\eta)\, \omega^\lambda\, v(\bar{k},\bar{\eta})~,  
\slabel{matl3} \\
{\cal M}_4^{\kappa\lambda} &=& \OL{v}(\bar{p},\bar{\kappa})\, 
   \gamma_\mu\, \omega^\kappa\, u(p,\kappa)\,
   \OL{u}(k,\eta) (\bar{\pslash} \gamma^\mu + \gamma^\mu \pslash)\, 
   \omega^\lambda\, v(\bar{k},\bar{\eta})~,
\slabel{matel4} \\
{\cal M}_5^{\kappa\lambda} &=& -\OL{v}(\bar{p},\bar{\kappa})\, 
   \gamma_\mu\, \omega^\lambda\, v(\bar{k},\bar{\eta})\,
   \OL{u}(k,\eta)\, \gamma^\mu\, \omega^\kappa\, u(p,\kappa)~,
\slabel{matel5} \\
{\cal M}_6^{\kappa\lambda} &=& -\OL{v}(\bar{p},\bar{\kappa})\, 
   \kslash\, \omega^\lambda\, v(\bar{k},\bar{\eta})\,
   \OL{u}(k,\eta)\, \bar{\kslash}\, \omega^\kappa\, u(p,\kappa)~,
\slabel{matel6} \\
{\cal M}_7^{\kappa\lambda} &=& -\OL{v}(\bar{p},\bar{\kappa})\, 
   \omega^\lambda\, v(\bar{k},\bar{\eta})\, 
   \OL{u}(k,\eta)\, \omega^\kappa\, u(p,\kappa)~,
\slabel{matel7} \\
{\cal M}_8^{\kappa\lambda} &=& -\OL{v}(\bar{p},\bar{\kappa})\, 
   \kslash\, \omega^\lambda\, v(\bar{k},\bar{\eta})\,
   \OL{u}(k,\eta)\, \omega^\kappa\, u(p,\kappa)~,  
\slabel{matel8} \\
{\cal M}_9^{\kappa\lambda} &=& -\OL{v}(\bar{p},\bar{\kappa})\, 
   \omega^\lambda\, v(\bar{k},\bar{\eta})\,
   \OL{u}(k,\eta)\, \bar{\kslash}\, \omega^\kappa\, u(p,\kappa)~, 
\slabel{matel9} \\
{\cal M}_{10}^{\kappa\lambda} &=& \OL{v}(\bar{p},\bar{\kappa})\, 
    \gamma_\mu\, \omega^\lambda\, v(k,\eta)\,
    \OL{u}(\bar{k},\bar{\eta})\, \gamma^\mu\, \omega^\kappa\, 
    u(p,\kappa)~,
\slabel{matel10} \\
{\cal M}_{11}^{\kappa\lambda} &=& \OL{v}(\bar{p},\bar{\kappa})\, 
    \bar{\kslash}\, \omega^\lambda\, v(k,\eta)\,
    \OL{u}(\bar{k},\bar{\eta})\, \kslash\, \omega^\kappa\, u(p,\kappa)~,
\slabel{matel11} \\
{\cal M}_{12}^{\kappa\lambda} &=& \OL{v}(\bar{p},\bar{\kappa})\, 
    \omega^\lambda\, v(k,\eta)\, 
    \OL{u}(\bar{k},\bar{\eta})\, \omega^\kappa\, u(p,\kappa)~,
\slabel{matel12} \\
{\cal M}_{13}^{\kappa\lambda} &=& \OL{v}(\bar{p},\bar{\kappa})\, 
    \bar{\kslash}\, \omega^\lambda\, v(k,\eta)\,
    \OL{u}(\bar{k},\bar{\eta})\, \omega^\kappa\, u(p,\kappa)~,
\slabel{matel13} \\
{\cal M}_{14}^{\kappa\lambda} &=& \OL{v}(\bar{p},\bar{\kappa})\, 
    \omega^\lambda\, v(k,\eta)\,
    \OL{u}(\bar{k},\bar{\eta})\, \kslash\, \omega^\kappa\, u(p,\kappa)~.
\slabel{matel14}
\ESUB 
The chirality projectors $\omega^\pm$ are defined by
\BE
\omega^\pm = \frac{1}{2} (\id \pm \gamma_5)~.
\EE

\begin{figure}[t]
\centerline{\psfig{figure=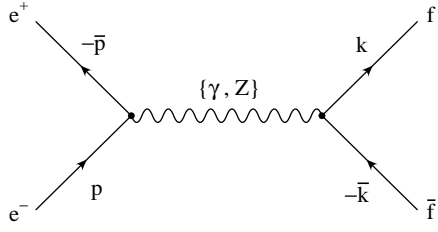,width=6cm}}
\caption[]{Born diagrams for the process $e^+ e^- \to f\bar{f}$.}
\label{fig:Borngraphen}
\end{figure}

In the Born approximation only the two diagrams of Fig. 
\ref{fig:Borngraphen} are relevant since the electron-Higgs coupling 
is negligible. Therefore only (\ref{bornmat1}) contributes to the 
lowest-order amplitude ${\cal M}^{(0)}$: 
\BE
{\cal M}^{(0)}(\kappa,\eta,\bar{\eta};s,t) = 
\sum_\lambda {\cal M}_1^{\kappa\lambda} L_1^{(0)\kappa\lambda}
\label{M0}
\EE
with
\BE
L_1^{(0)\kappa\lambda} = i e^2 \left( \frac{Q_e Q_f}{s} 
                     + \frac{g_e^\kappa g_f^\lambda}{s-M_Z^2} \right)~,
\label{bornform}
\EE
$Q_e$ $(Q_f)$ denotes the charge of the electron (fermion) and the 
couplings $g_e^\kappa$ and $g_f^\lambda$ can be found in the last line
of Table \ref{tabsecoup}.

Denoting the 1-loop contribution to the helicity amplitude by
\BE
{\cal M}^{(1)}(\kappa,\eta,\bar{\eta};s,t) = 
\sum_{i,\lambda} {\cal M}_i^{\kappa\lambda} L_i^{(1)\kappa\lambda}~,
\label{one-loop contribution}
\EE
the differential cross section including ${\cal O}(\alpha)$
corrections is given by
\BEA
\frac{d\sigma}{d\Omega}(\kappa,s,\theta)
&=& \frac{1}{64\,\pi^2\,s}\, \beta_f\, N_C^f\, 
    \sum_{\eta,\bar{\eta}} \bigg\{ 
    |{\cal M}(\kappa,\eta,\bar{\eta};s,t)|^2 
    \nonumber \\
& & + 2\, \re \left[
    {\cal M}^{(1)}(\kappa,\eta,\bar{\eta};s,t)
    {\cal M}^{(0)*}(\kappa,\eta,\bar{\eta};s,t) \right]
    \bigg\}~.
\EEA

Since the invariant functions $L_i^{\kappa\lambda}$ are independent
of the polarizations of the final-state fermions, the summation over
these polarizations can be carried out directly for the products of
the basic matrix elements:
\BSUB
\label{matsum}
\sum_{\eta,\bar{\eta}} {\cal M}_1^{\kappa\lambda} 
 \left( {\cal M}_1^{\kappa\lambda^\prime} \right)^* &=& 
 4 \left\{ \delta_{\lambda^\prime}^\lambda
           \left[ \delta_\kappa^\lambda (u-m_f^2)^2
     + \delta_{-\kappa}^\lambda (t-m_f^2)^2 \right]
     + \delta_{-\lambda^\prime}^\lambda m_f^2 s \right\}~,
\slabel{bornmatsum} \\
\sum_{\eta,\bar{\eta}} {\cal M}_2^{\kappa\lambda} 
 \left( {\cal M}_1^{\kappa\lambda^\prime} \right)^* &=& 
 8(ut-m_f^4)\, \delta_{\lambda^\prime}^\lambda
     \left[ \delta_\kappa^\lambda (u-m_f^2)
     - \delta_{-\kappa}^\lambda (t-m_f^2) \right]~,  \\
\sum_{\eta,\bar{\eta}} {\cal M}_3^{\kappa\lambda} 
 \left( {\cal M}_1^{\kappa\lambda^\prime} \right)^* &=& 
 -4 \, m_f \, (ut-m_f^4)~,
\slabel{matsum2} \\
\sum_{\eta,\bar{\eta}} {\cal M}_4^{\kappa\lambda} 
 \left( {\cal M}_1^{\kappa\lambda^\prime} \right)^* &=& 
 4\, m_f \, s \left( \delta_{\lambda^\prime}^\lambda -
              \delta_{-\lambda^\prime}^\lambda \right)
        \left[ \delta_\kappa^{\lambda^\prime} (u-m_f^2)
        - \delta_{-\kappa}^{\lambda^\prime} (t-m_f^2) \right]~, \\
\sum_{\eta,\bar{\eta}} {\cal M}_5^{\kappa\lambda} 
 \left( {\cal M}_1^{\kappa\lambda^\prime} \right)^* &=& 
 4\, \delta_{\lambda}^{\kappa}
   \left[ \delta_\lambda^{\lambda^\prime} (u-m_f^2)^2
   + \delta_{-\lambda}^{\lambda^\prime} \, s\, m_f^2 \right]~, \\
\sum_{\eta,\bar{\eta}} {\cal M}_6^{\kappa\lambda} 
 \left( {\cal M}_1^{\kappa\lambda^\prime} \right)^* &=& 
 2\, \delta_{\lambda}^{\kappa}
   \bigg\{ \delta_\lambda^{\lambda^\prime} \Big[ (u-m_f^2)^2
   (s-2m_f^2) - 2m_f^2(u-m_f^2)(t-m_f^2) \nonumber \\
& & + s\, m_f^4 \Big] 
    + \delta_{-\lambda}^{\lambda^\prime}\, m_f^2(u-m_f^2)^2 \bigg\}~, \\
\sum_{\eta,\bar{\eta}} {\cal M}_7^{\kappa\lambda} 
 \left( {\cal M}_1^{\kappa\lambda^\prime} \right)^* &=& 
 -2\, \delta_{-\lambda}^{\kappa}
   \left[ \delta_\lambda^{\lambda^\prime} (m_f^2-t)^2
   + \delta_{-\lambda}^{\lambda^\prime} \, s\, m_f^2 \right]~, \\
\sum_{\eta,\bar{\eta}} {\cal M}_8^{\kappa\lambda} 
 \left( {\cal M}_1^{\kappa\lambda^\prime} \right)^* &=& 
 m_f \, \delta_{\lambda}^{\kappa}
   \bigg\{ \delta_\lambda^{\lambda^\prime} \left[ (u-m_f^2)^2
         + s (s-2m_f^2) - (t-m_f^2)^2 \right] \nonumber \\
& & + \delta_{-\lambda}^{\lambda^\prime} \, 
    2(m_f^2-t)(m_f^2-u)  \bigg\}~, \\
\sum_{\eta,\bar{\eta}} {\cal M}_9^{\kappa\lambda} 
 \left( {\cal M}_1^{\kappa\lambda^\prime} \right)^* &=& 
 -m_f\, \delta_{-\lambda}^{\kappa}
    \bigg\{ \delta_\lambda^{\lambda^\prime} 2(m_f^2-t)(m_f^2-u)
    \nonumber \\ 
& & + \delta_{-\lambda}^{\lambda^\prime} 
      \left[ s (s-2m_f^2)+(u-m_f^2)^2 -(t-m_f^2)^2 \right] \bigg\}~, \\ 
\sum_{\eta,\bar{\eta}} {\cal M}_{10}^{\kappa\lambda} 
 \left( {\cal M}_1^{\kappa\lambda^\prime} \right)^* &=& 
 -4\, \delta_{\lambda}^{\kappa}
   \left[ \delta_\lambda^{-\lambda^\prime} (t-m_f^2)^2
   + \delta_\lambda^{\lambda^\prime} \, s\, m_f^2 \right]~, \\
\sum_{\eta,\bar{\eta}} {\cal M}_{11}^{\kappa\lambda} 
 \left( {\cal M}_1^{\kappa\lambda^\prime} \right)^* &=& 
 -2\, \delta_{\lambda}^{\kappa}
   \bigg\{ \delta_\lambda^{-\lambda^\prime} \Big[ (t-m_f^2)^2
   (s-2m_f^2) - 2m_f^2(t-m_f^2)(u-m_f^2) \nonumber \\
& & + s\, m_f^4 \Big]
    + \delta_\lambda^{\lambda^\prime}\, m_f^2(t-m_f^2)^2 \bigg\}~, \\
\sum_{\eta,\bar{\eta}} {\cal M}_{12}^{\kappa\lambda} 
 \left( {\cal M}_1^{\kappa\lambda^\prime} \right)^* &=& 
 2\, \delta_{-\lambda}^{\kappa}
   \left[ \delta_\lambda^{-\lambda^\prime} (m_f^2-u)^2
   + \delta_\lambda^{\lambda^\prime} \, s\, m_f^2 \right]~, \\
\sum_{\eta,\bar{\eta}} {\cal M}_{13}^{\kappa\lambda} 
 \left( {\cal M}_1^{\kappa\lambda^\prime} \right)^* &=& 
 -m_f \, \delta_{\lambda}^{\kappa}
   \bigg\{ \delta_\lambda^{-\lambda^\prime} \left[ (t-m_f^2)^2
         + s (s-2m_f^2) - (u-m_f^2)^2 \right] \nonumber \\
& & + \delta_\lambda^{\lambda^\prime} \, 
    2(m_f^2-t)(m_f^2-u)  \bigg\}~, \\
\sum_{\eta,\bar{\eta}} {\cal M}_{14}^{\kappa\lambda} 
 \left( {\cal M}_1^{\kappa\lambda^\prime} \right)^* &=& 
 m_f \, \delta_{-\lambda}^{\kappa}
    \bigg\{ \delta_\lambda^{-\lambda^\prime} 2(m_f^2-t)(m_f^2-u)
    \nonumber \\ 
& & + \delta_\lambda^{\lambda^\prime} 
      \left[ s (s-2m_f^2)+(t-m_f^2)^2 -(u-m_f^2)^2 \right] \bigg\}~.  
\ESUB

\subsection{One-loop corrections}

The 1-loop contribution ${\cal M}^{(1)}$ (\ref{one-loop contribution}) 
to the helicity amplitude contains the 
$\gamma$ and $Z$ self-energies, the $\gamma$ and $Z$ vertex corrections
together with the external wave function renormalization, and the box 
diagrams:
\BE
{\cal M}^{(1)}={\cal M}_S+{\cal M}_{eV}+{\cal M}_{fV}+{\cal M}_B~.
\label{Heliamp1}
\EE
The complete set of vertex corrections comprises the QED corrections 
with virtual photons and the QCD corrections with virtual gluons for 
quark final states. They need real photon and gluon bremsstrahlung for
a infrared finite result. The gauge invariant subclasses of 
``standard QED'' and ``standard QCD'' corrections are identical to
those in the Standard Model and are available in the literature 
\cite{hollik91,Zerwas}. We therefore concentrate our discussion on the
residual set of model dependent and IR-finite virtual corrections, 
which correspond to (\ref{Heliamp1}) without the diagrams involving 
virtual photons and gluons.

The supersymmetric part of the QCD corrections, arising from virtual 
gluinos, is included in (\ref{Heliamp1}) as part of the final state 
vertex correction ${\cal M}_{fV}$.

The concrete calculations are performed in the 't~Hooft-Feynman gauge.
Since the lowest order amplitudes contain only standard particles, we
can apply the {\sc on-shell} renormalization scheme for the Standard 
Model \cite{hollik95} also for the MSSM case at 1-loop. 
The formal relations between 1-loop vertex functions and counter
terms can be literally taken from the Standard Model; their explicit 
evaluation, however, requires the inclusion of the supersymmetric 
particles and the replacement of the standard Higgs contributions by 
those of the MSSM two Higgs doublets.

\newpage
\clearpage

\subsubsection{Vector boson propagator corrections}

\begin{figure}[ht]
\centerline{\psfig{figure=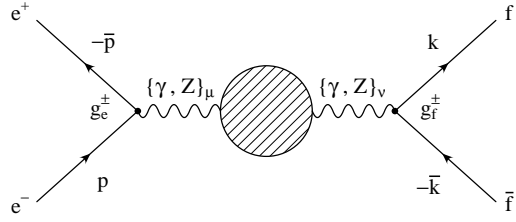,width=7cm}}
\caption[]{The self-energy corrections to $e^+e^- \to f\bar{f}$.}
\label{fig:sekorr}
\end{figure}

\vspace{4mm}

The 1-loop propagator correction appears in the sum (\ref{Heliamp1})
as the term (Fig. \ref{fig:sekorr}) 
\BE
{\cal M}_S = -i\, e^2 \sum_\lambda \sum_{a,b=\gamma,Z} 
             \frac{\hat{\Sigma}^{ab}(s)}{(s-M_a^2)(s-M_b^2)}\,  
             g_e^\kappa\, g_f^\lambda\, {\cal M}_1^{\kappa\lambda}
\label{M_S}
\EE
with ${\cal M}_1^{\kappa\lambda}$ from (\ref{bornmat1}), $M_\gamma=0$,
and the renormalized self-energies $\hat{\Sigma}^{ab}(s)$ of the 
neutral vector bosons:
\BEA
\hat{\Sigma}^{\gamma \gamma}(s) &=& 
  \Sigma^{\gamma \gamma}(s) - s\, \Pi^{\gamma \gamma}(0)~, \qquad 
  \Pi^{\gamma \gamma}(0) \equiv 
  \frac{\dd \Sigma^{\gamma\gamma}(q^2)}{\dd q^2}\Big|_{q^2=0}~, 
  \nonumber \\ 
\hat{\Sigma}^{ZZ}(s) &=& 
    \Sigma^{ZZ}(s) - \re\, \Sigma^{ZZ}(M_Z^2) 
    + (s-M_Z^2) \bigg\{ -\Pi^{\gamma\gamma}(0) 
    - 2\, \frac{\cw^2-\sw^2}{\sw\cw}\, 
    \frac{\Sigma^{\gamma Z}(0)}{M_Z^2} \nonumber \\
& & + \frac{\cw^2-\sw^2}{\sw^2}\, \re\, \Big[ 
    \frac{\Sigma^{ZZ}(M_Z^2)}{M_Z^2} 
    - \frac{\Sigma^{WW}(M_W^2)}{M_W^2} \Big] \bigg\}~, \nonumber \\
\hat{\Sigma}^{\gamma Z}(s) &=& \Sigma^{\gamma Z}(s) 
    + \Big( \frac{2s}{M_Z^2} - 1 \Big) \Sigma^{\gamma Z}(0) 
    - s\, \frac{\cw}{\sw}\, \re\, \Big[ \frac{\Sigma^{ZZ}(M_Z^2)}{M_Z^2} 
    - \frac{\Sigma^{WW}(M_W^2)}{M_W^2} \Big]~.  
\EEA
In the {\sc on-shell} scheme one has
\BE
\cw \equiv \cos{\theta_w} = \frac{M_W}{M_Z}~, \qquad
\sw \equiv \sin{\theta_w}~,
\label{Theta}
\EE
and the couplings in (\ref{M_S}) and Fig. \ref{fig:sekorr} are listed 
in Table \ref{tabsecoup} with the following abbreviations:
\BE
g_{Z,R}^{f(e)} = -\frac{\sw}{\cw}\, Q_{f(e)}~, \qquad
g_{Z,L}^{f(e)} = \frac{I_3^{f(e)}-\sw^2 Q_{f(e)}}{\sw\cw}~.
\EE
The quantities $\Sigma^{ab}(q^2)\; (a,b=\gamma,Z,W)$ denote the 
unrenormalized self-energies as the transverse coefficients in the
expansion
\BE
\Sigma^{ab}_{\mu \nu}(q) = -g_{\mu\nu} \Sigma^{ab}(q^2)
    +\frac{q_\mu q_\nu}{q^2} \left[ \Sigma^{ab}(q^2)
    -\Sigma^{ab}_L(q^2) \right]~.
\EE
They are explicitly given in Appendix \ref{appb1}. The 
$q_\mu q_\nu$-terms yield only contributions $\propto m_e^2$ in the 
{\sc on-shell} amplitudes and hence vanish in the limit $m_e \to 0$.

Since the energy domain above the $Z$ resonance is under consideration
in this paper, the subtleties of the higher order contributions to
(\ref{M_S}), which are important for $Z$ physics (see, e.g., Ref. 
\cite{Yellow Book}), can be omitted in view of the experimental 
accuracy. Therefore we can restrict ourselves to the simplified 
${\cal O}(\alpha)$ treatment according to (\ref{M_S}), but we include 
the leading $\log$ resummation of the terms involving the light fermions.

\begin{table}[ht]
\begin{center}$
\renewcommand{\arraystretch}{1.4}
\arraycolsep6mm
\begin{array}{|c|c||*{4}{c|}}
\hline
a & b & g_e^+ & g_e^- & g_f^+ & g_f^- \\
\hline\hline 
\gamma & \gamma & -Q_e & -Q_e & -Q_f & -Q_f \\
\hline 
\gamma & Z & -Q_e & -Q_e & g_{Z,R}^f & g_{Z,L}^f \\
\hline 
Z & \gamma & g_{Z,R}^e & g_{Z,L}^e & -Q_f & -Q_f \\
\hline 
Z & Z & g_{Z,R}^e & g_{Z,L}^e & g_{Z,R}^f & g_{Z,L}^f \\
\hline
\end{array}$
\end{center}
\caption[]{\label{tabsecoup} Couplings.}
\end{table}

\subsubsection{The vertex corrections}

\begin{figure}[h]
\centerline{\psfig{figure=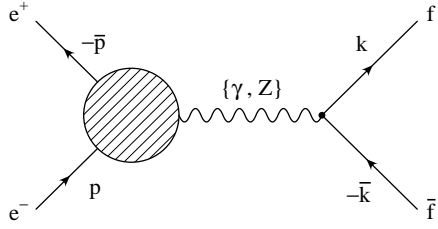,width=6cm}}
\caption[]{The electron vertex correction.}
\label{fig:Everkorr}
\end{figure}

\begin{figure}[h]
\centerline{\psfig{figure=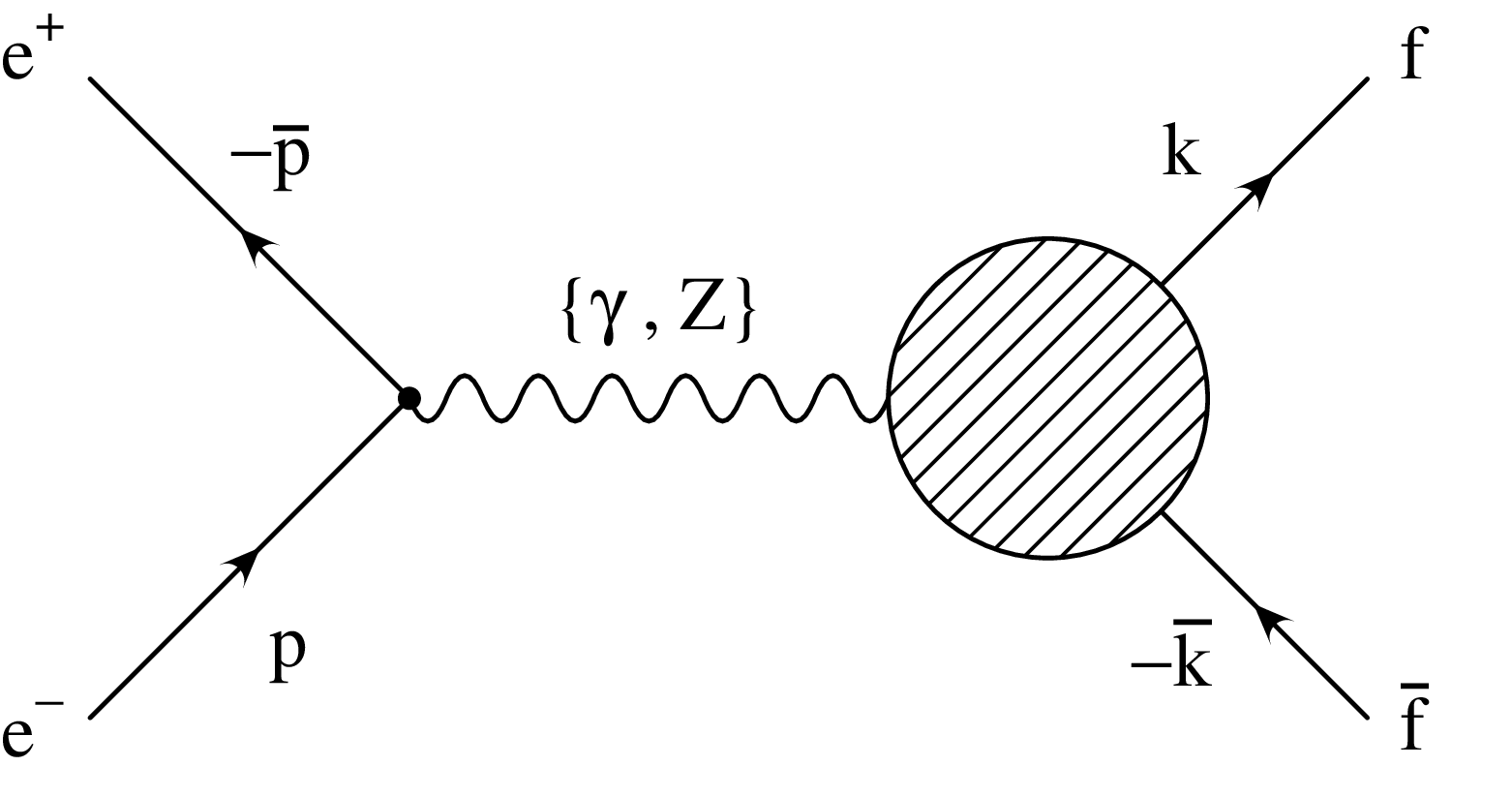,width=6cm}}
\caption[]{The vertex correction for outgoing fermions.}
\label{fig:Topverkorr}
\end{figure}

The renormalized initial and final state vertex corrections together
with the external wave function renormalizations are summarized by the
contributions ${\cal M}_{eV}$ and ${\cal M}_{fV}$ to
${\cal M}^{(1)}(\kappa,\eta,\bar{\eta};s,t)$ in (\ref{Heliamp1}) 
(see Figs. \ref{fig:Everkorr} and \ref{fig:Topverkorr}):
\BE
{\cal M}_{eV} = i\alpha^2 \sum_\lambda \Big(
                  -\frac{1}{s}\, Q_f\, A^\kappa_{e\gamma}\,
                  +\frac{1}{s-M_Z^2}\, g^\lambda_f\, A^\kappa_{eZ}\,
                  \Big) {\cal M}_1^{\kappa\lambda}
\label{inver}
\EE

\vspace{8mm}

\BEA
{\cal M}_{fV} &=& i\alpha^2 \sum_\lambda \bigg\{ \Big(
    -\frac{1}{s}\, Q_e \, A^\lambda_{f\gamma}\,
    +\frac{1}{s-M_Z^2}\, g^\kappa_e\, A^\lambda_{fZ}\,
    \Big) {\cal M}_1^{\kappa\lambda} \nonumber \\
& & \hspace{1.4cm} +\Big( -\frac{1}{s}\, Q_e \, B^\lambda_{f\gamma}\,
    +\frac{1}{s-M_Z^2}\, g^\kappa_e\, B^\lambda_{fZ}\,
    \Big) {\cal M}_3^{\kappa\lambda} \bigg\}
\label{outver}
\EEA

The coefficients $A^\lambda$ and $B^\lambda$ are given by the following
combinations $(a=\gamma,Z)$
\BSUB
A^\kappa_{ea} &=& \left[ A_I^\kappa + A_{I\!I}^\kappa + A_{V}^\kappa 
                  +A_{V\!I}^\kappa \right]_{ea}+\delta C_{ea}^\kappa~, \\
A^\lambda_{fa} &=& \left[ A_I^\lambda + A_{I\!I}^\lambda
                   + A_{I\!I\!I}^\lambda + A_{I\!V}^\lambda
                   + A_{V}^\lambda + A_{V\!I}^\lambda \right]_{fa}
                   + \delta C_{fa}^\lambda~, \\
B^\lambda_{fa} &=& \left[ B_I^\lambda + B_{I\!I}^\lambda
                   + B_{I\!I\!I}^\lambda + B_{I\!V}^\lambda
                   + B_{V}^\lambda + B_{V\!I}^\lambda \right]_{fa}~.
\ESUB
The quantities in the brackets are listed in Appendix \ref{appc}.
They are obtained as the form factors resulting from the various
vertex classes $I,\ldots ,V\!I$, summed over all individual 
contributions inside each class.

The counterterms for $F=e,f$ ($f=\mu,\tau,q$) read
\BEA
\frac{\alpha}{4 \pi}\, \delta C_{F \gamma}^- 
&=& \frac{I_3^F}{\sw\cw}\, \frac{\Sigma^{\gamma Z}(0)}{M_Z^2}
    + Q_F\, \delta Z_F^{(-)}~,  \nonumber \\ 
\frac{\alpha}{4 \pi}\, \delta C_{F \gamma}^+ 
&=& Q_F\, \delta Z_F^{(+)}~.
\label{Photon-CT}
\EEA
In the case of the $Z$-fermion vertex the counterterms are given by
\BEA
\frac{\alpha}{4 \pi}\, \delta C_{FZ}^- 
&=& - \frac{I_3^F\, \Sigma^{\gamma Z}(0)}{\sw^2\, M_Z^2}
    - \frac{I_3^F-\sw^2 Q_F}{\sw \cw}\, \delta Z_F^{(-)}~,
    \nonumber \\ 
\frac{\alpha}{4 \pi}\, \delta C_{FZ}^+ 
&=& \frac{\sw}{\cw}\, Q_F\, \delta Z_F^{(+)}~,
\label{Z-CT}
\EEA
with the fermion field renormalization constants 
\BE
\delta Z_F^{(\pm)} = \re\, \Sigma_V^F(m_F^2) + 2 m_F^2\, \re\, \Big[ 
    \Sigma_V^{\prime F}(m_F^2) + \Sigma_S^{\prime F}(m_F^2) \Big]
    \pm \re\, \Sigma_A^F(m_F^2)~.
\EE
The fermion self-energies $\Sigma_{S,V,A}^F$ are listed explicitly in 
Appendix \ref{appb2}.

\subsubsection{The box diagrams}
\label{Die Boxdiagramme}

For the process $e^+e^- \to f\bar{f}$ ($f=\mu,\tau,q$) there are two 
different topologies of box diagrams, which we denote as 'direct' 
(Fig. \ref{fig:Boxtopd}) and 'crossed' (Fig. \ref{fig:Boxtopc}) box
diagrams. The corresponding contribution to the 1-loop matrix element 
are labeled by $D$ and $C$, accordingly.

Since Higgs boson exchanges are negligible, we only have to consider
the standard box graphs with $Z$ and $W^\pm$ exchange and the SUSY box
graphs with neutralino and chargino exchange. As far as the charged 
current box diagrams are considered, for $I_3=-1/2$ only the direct
box diagrams with $W^\pm$ and chargino exchange contribute, and
for $I_3=+1/2$ only the crossed box diagrams with $W^\pm$ and chargino 
exchange.

The box contribution to the helicity amplitude (\ref{Heliamp1}) thus 
can be decomposed as follows:
\BE
{\cal M}_B = \sum_{V=Z,W} \left( {\cal M}_D^V + {\cal M}_C^V \right) 
  + \sum_{\tilde{\chi}=\tilde{\chi}^0,\tilde{\chi}^+} \left( 
    {\cal M}_D^{\tilde{\chi}} + {\cal M}_C^{\tilde{\chi}} \right)~.
\EE
The summation has to be understood as extended over all neutralino and
chargino configurations, and the corresponding 
$\tilde{e}_{1,2}, \tilde{\nu}_e$ and $\tilde{f}_{1,2}$ states.

We use the following shorthand notations for combinations of 4-point
tensor integral coefficients $D_{kl}$ and the scalar integrals 
$C_0$ and $D_0$:

\BEA
P_1 &=& 4 i\alpha^2 D_{27}~, \nonumber \\
P_2 &=& 2 i\alpha^2\left[ 2C_0 +\left\{{t \atop u}\right\}
          (D_{11}+D_{12}-D_{13}) - m_f^2 (D_{11}-D_{12}+D_{13}) \right] 
          \; \left\{{ \mbox{direct box} \atop \mbox{crossed box}}
          \right.~, \nonumber \\
P_3 &=& i \alpha^2 D_{11}~, \nonumber \\
P_4 &=& i \alpha^2 (D_{11} - 2 D_{13})~, \nonumber \\
P_5 &=& i \alpha^2 (D_{11} - 2 D_{12})~, \nonumber \\
P_6 &=& i \alpha^2 (D_{11}+D_{24}-D_{25})~, \nonumber \\
P_7 &=& i \alpha^2 \Big[ D_{11} + 2(D_{23}-D_{13}-D_{26}) 
                       + D_{24} - D_{25} \Big]~, \nonumber \\
P_8 &=& i \alpha^2 \Big[ D_{11} + 2(D_{26}-D_{12}-D_{22}) 
                       + D_{24} - D_{25} \Big]~, \nonumber \\
P_9 &=& i\alpha^2 (D_{25}-D_{26})~, \nonumber \\
P_{10} &=& i\alpha^2 (D_{24}-D_{26}) ~,\nonumber \\
P_{11} &=& i\alpha^2 D_{12}~, \nonumber \\
P_{12} &=& i\alpha^2 D_0~, \nonumber \\
P_{13} &=& i\alpha^2 D_{13}~, \nonumber \\
P_{14} &=& i\alpha^2 (D_{25}-D_{23})~, \nonumber \\
P_{15} &=& i\alpha^2 (D_{24}-D_{22})~.
\EEA
Further specifications are given in Appendix \ref{appd}.

With these abbreviations and with the matrix elements (\ref{alle}),
the result for a direct vector boson box diagram is given by
\BEA
{\cal M}_D^V &=& {\cal M}_1^{\kappa\kappa} \left[ 
                 P_1 \, \lambda_\kappa^+
                 - m_f m_3 (P_5 \, \bar{\lambda}_\kappa^-
                 - P_4 \, \bar{\lambda}_\kappa^+) \right]
                 + {\cal M}_1^{\kappa-\kappa} \left[ P_2 \, 
                 \lambda_\kappa^-
                 + m_f m_3 (P_5 \, \bar{\lambda}_\kappa^+
                 - P_4 \, \bar{\lambda}_\kappa^-) \right] \nonumber \\
             & & + {\cal M}_2^{\kappa\kappa} \, P_6 \, \lambda_\kappa^+ 
                 \nonumber \\
             & & + {\cal M}_3^{\kappa\kappa} \left[ m_f P_7 \, 
                 \lambda_\kappa^+ - m_3 P_4 \, 
                 \bar{\lambda}_\kappa^+ \right] 
                 - {\cal M}_3^{\kappa-\kappa} \left[ m_f P_8 \, 
                 \lambda_\kappa^+ - m_3 P_5 \, 
                 \bar{\lambda}_\kappa^-\right] \nonumber \\
             & & + {\cal M}_4^{\kappa\kappa}\, m_3\,  
                 P_3 \, \bar{\lambda}_\kappa^+ 
                 - {\cal M}_4^{\kappa-\kappa}\, m_3\, 
                 P_3 \, \bar{\lambda}_\kappa^-~,
\label{MDV}
\EEA
and for a crossed vector boson box diagram by
\BEA
{\cal M}_C^V &=& {\cal M}_1^{\kappa\kappa} \left[ 
                 - P_2 \, \lambda_\kappa^+
                 - m_f m_3 (P_5 \, \bar{\lambda}_\kappa^-
                 - P_4 \, \bar{\lambda}_\kappa^+) \right] 
                 + {\cal M}_1^{\kappa-\kappa} \left[ 
                 - P_1 \, \lambda_\kappa^-
                 + m_f m_3 (P_5 \, \bar{\lambda}_\kappa^+
                 - P_4 \, \bar{\lambda}_\kappa^-) \right] \nonumber \\
             & & + {\cal M}_2^{\kappa-\kappa} \, 
                 P_6 \, \lambda_\kappa^- \nonumber \\
             & & - {\cal M}_3^{\kappa\kappa} \left[ 
                 m_f P_7 \, \lambda_\kappa^- - m_3 
                 P_4 \, \bar{\lambda}_\kappa^- \right]
                 + {\cal M}_3^{\kappa-\kappa} \left[ 
                 m_f\, P_8 \, \lambda_\kappa^- - m_3 
                 P_5 \, \bar{\lambda}_\kappa^+ \right] \nonumber \\
             & & - {\cal M}_4^{\kappa\kappa}\, m_3\, 
                 P_3 \, \bar{\lambda}_\kappa^- 
                 + {\cal M}_4^{\kappa-\kappa}\, m_3\, 
                 P_3 \, \bar{\lambda}_\kappa^+~.
\EEA
$m_3$ can be read off from the Tables \ref{tabdVb} and \ref{tabcVb} in
Appendix \ref{appd}.

The expression for a direct SUSY box diagram reads 
\BEA
{\cal M}_D^{\tilde{\chi}} &=& \frac{1}{4}
      {\cal M}_5^{\kappa\kappa}\, P_1\, \bar{\eta}_\kappa^+ 
    - {\cal M}_6^{\kappa\kappa}\, P_9\, \bar{\eta}_\kappa^+ 
    \nonumber \\
& & + {\cal M}_7^{\kappa -\kappa} \Big[ 
      m_f^2\, P_{10}\, \bar{\eta}_\kappa^+ 
    - m_f\, m_2\, (P_{3}-P_{13})\, \eta_\kappa^- 
    - m_f\, m_4\, P_{11}\, \eta_\kappa^+
    + m_2\, m_4\, 
      P_{12}\, \bar{\eta}_\kappa^-  \Big] \nonumber \\
& & + {\cal M}_8^{\kappa\kappa} \Big[ 
      m_4\, P_{13}\, \eta_\kappa^+ 
    - m_f\, P_{14}\, \bar{\eta}_\kappa^+ \Big] \nonumber \\
& & + {\cal M}_9^{\kappa -\kappa} \Big[ 
      m_f\, P_{15}\, \bar{\eta}_\kappa^+ 
    - m_2\, (P_3-P_{11})\, \eta_\kappa^- \Big]~. 
\EEA
For a crossed SUSY box diagram we have
\BEA
{\cal M}_C^{\tilde{\chi}} &=& \frac{1}{4}
      {\cal M}_{10}^{\kappa\kappa}\, P_1\, \bar{\eta}_\kappa^+ 
    - {\cal M}_{11}^{\kappa\kappa}\, P_9\, \bar{\eta}_\kappa^+ 
    \nonumber \\
& & + {\cal M}_{12}^{\kappa -\kappa} \Big[ 
      m_f^2\, P_{10}\, \bar{\eta}_\kappa^+ 
    - m_f\, m_2 \, (P_{3}-P_{13})\, \eta_\kappa^-
    - m_f\, m_4 \, P_{11}\, \eta_\kappa^+ 
    + m_2\, m_4\, 
      P_{12}\, \bar{\eta}_\kappa^- \Big] \nonumber \\
& & + {\cal M}_{13}^{\kappa\kappa} \Big[ 
      m_4\, P_{13}\, \eta_\kappa^+ 
    - m_f\, P_{14}\, \bar{\eta}_\kappa^+ \Big] \nonumber \\
& & + {\cal M}_{14}^{\kappa -\kappa} \Big[ 
      m_f\, P_{15}\, \bar{\eta}_\kappa^+ 
    - m_2\, (P_3-P_{11})\, \eta_\kappa^- \Big]~. 
\label{MCS}
\EEA
$m_2$, $m_4$ denotes charginos or neutralinos, respectively, and 
can be read off from the Tables \ref{tabdNb}, \ref{tabcNb}, \ref{tabdCb}
and \ref{tabcCb} in Appendix \ref{appd}.

The following abbreviations for products of the couplings
$g^\kappa_{1,2,3,4}$ specified in Appendix \ref{appd} have been used
in the expressions (\ref{MDV}) -- (\ref{MCS}):
\BE
\begin{array}{*{7}{l}}
\lambda_\kappa^\pm &=& g_1^\kappa\, g_2^\kappa\, g_3^{\pm \kappa}\,
                       g_4^{\pm \kappa}~, & \hspace{1.8cm} &
\bar{\lambda}_\kappa^\pm &=& g_1^\kappa\, g_2^\kappa\, g_3^{\pm \kappa}\,
                             g_4^{\mp \kappa}~, \\[1.2ex]
\eta_\kappa^\pm &=& g_1^{-\kappa}\, g_2^\kappa\, g_3^{\pm \kappa}\,
                    g_4^{\pm \kappa}~,  &  &
\bar{\eta}_\kappa^\pm &=& g_1^{-\kappa}\, g_2^\kappa\, 
                          g_3^{\pm \kappa}\, g_4^{\mp \kappa}~.
\end{array}
\EE

\newpage
\clearpage


\section{Numerical analysis and discussion}
\label{chap3}
\setcounter{equation}{0}
\setcounter{figure}{0}
\setcounter{table}{0}

In this section we discuss the numerical effects of the weak and 
of the SUSY-QCD corrections in the reaction $e^+e^- \to f\bar{f}$ 
($f=\mu,\tau,q$).

For cross-checking our results\footnote{We used the program library 
{\bf AAFF} \cite{Oldenborgh} for the calculation of the tensor 
integrals.}, numerical comparisons have been performed with the 
already available subset of the SM and 2-Higgs doublet contributions 
\cite{arnd} and with the MSSM gauge boson self-energies as given in 
\cite{ds} by utilizing the computer codes of Ref. \cite{arnd,ds}. The 
comparisons show perfect agreement in all cases.

In order to exhibit the deviations from the SM induced by
supersymmetry, we introduce the quantity
\BE
\Delta (s) = \frac{\sigma^{\rm MSSM}(s)
            -\sigma^{\rm SM}(s)}{\sigma^{(0)}(s)}~,
\label{Delta}
\EE
for the unpolarized cross section, and the differences
\BEA
\delta A_{FB}(s) &=& A_{FB}^{\rm MSSM}(s)-A_{FB}^{\rm SM}(s)~,
                     \nonumber \\
\delta A_{LR}(s) &=& A_{LR}^{\rm MSSM}(s)-A_{LR}^{\rm SM}(s)~,
\EEA
for the forward-backward and left-right asymmetries.
$\sigma^{(0)}(s)$ is the total cross section in the Born 
approximation corresponding to Eqs. (\ref{M0}) and (\ref{bornform}).

For the numerical analysis, we used the MSSM parameters as specified
in Appendix \ref{appa}; they are chosen to be consistent with the 
experimental constraints \cite{LEPEWWG,PDG}. We further used the top 
quark mass of $m_t=175\GeV$ in the whole paper. Note that without the 
SUSY-GUT relation (\ref{M-GUT}) for $M_3$, the gluino mass 
$m_{\tilde{g}} \equiv |M_3|$ is another input parameter.
In the case of mixed sfermions (see Appendix \ref{appa}) we chose
for further simplicity $A_u=A_d=A$.

We split the discussion into two parts: the light fermion case, which
is experimentally accessible at LEP, and the case of $t\bar{t}$ 
production.

\subsection{Results for $e^+e^- \to \mu^+\mu^-$}
\label{section_mu}

In the absence of direct discoveries of SUSY particles at LEP, the
question for indirect effects in the standard cross sections,
resulting from virtual SUSY contributions, is of special interest.

In view of the claimed accuracy in the cross section measurement of 1.3\%
for $\mu^+\mu^-$ and 0.7\% for hadron final states \cite{LEP2}, it is
of basic importance to study the size of the MSSM 1-loop contributions
obtained from varying the model parameters in a wide range consistent
with experimental and theoretical constraints. Since for heavy 
non-standard particles one recovers the Standard Model with a light
Higgs boson ($m_{h^0}\lsim 130\GeV$), the largest deviations will occur
for light SUSY particles close to their detection limits.

\begin{figure}[htbp]
\centerline{\psfig{bbllx=20pt,bblly=90pt,bburx=600pt,bbury=420pt,%
figure=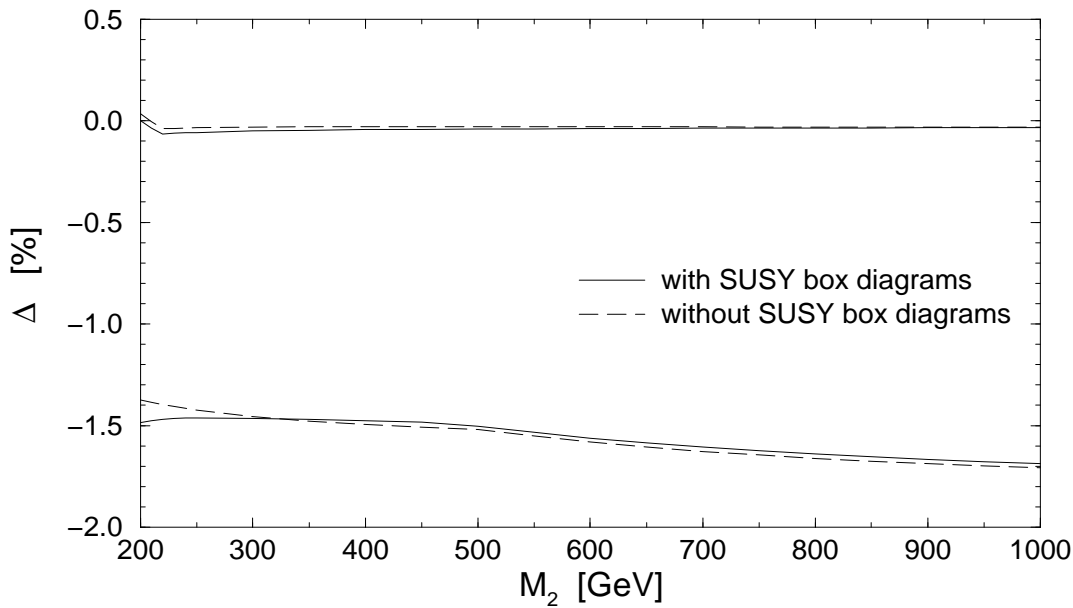,width=15cm}}
\caption[]{$e^+e^- \to \mu^+\mu^-$. Predicted range of $\Delta$ for 
  $\sqrt{s}=192$ GeV, $\Tb=1.6$, $M_A=500$ GeV, 
  $100\GeV\leq m_{\tilde{t}_1}\leq 500\GeV$, 
  $-400\GeV\leq A\leq 400\GeV$ and $-500\GeV\leq\mu\leq 500\GeV$ (but 
  $-100\GeV<\mu<100\GeV$ is excluded). The predicted range is between 
  the two upper and the two lower lines.}
\label{mue1.bat}
\end{figure}

In Fig. \ref{mue1.bat} we display the range for the quantity $\Delta$,
Eq. (\ref{Delta}), from scanning the SUSY parameters, varying
$m_{\tilde{t}_1}$ between 100 and 500 GeV, $\mu$ between $-500$ and 
500 GeV (with the region $-100\GeV<\mu<100\GeV$ excluded) in steps of 
50 GeV, and $M_2$ between 200 and 1000 GeV in steps of 10 GeV in a low
$\Tb$ scenario\footnote{In addition, we chose steps of $10\GeV$
  (instead of $50\GeV$) for the case $A=m_{\tilde{t}_1}$ without SUSY
  box diagrams (which is not depicted here) and the result differs
  less than $\pm 0.1\%$ from both results in Fig. \ref{mue1.bat}.} 
($\Tb=1.6$). 
The deviations from the SM are always negative, and $\Delta$ can reach
the maximum of $-1.7\%$ at $M_2=1000\GeV$ which is close to the 
experimental precision. 
There is (almost) no difference between the choice of a fixed 
$A=m_{\tilde{t}_1}$ (which is not depicted here) and for 
$-400\GeV\leq A\leq 400\GeV$. The large negative values are due to the
light sfermion masses which can occur for our set of parameters.

In the large $\Tb$ scenario ($\Tb=40$) the predicted range for the
same parameter variations is smaller than in Fig. \ref{mue1.bat} 
($-1\% \lsim \Delta \lsim 0\%$). This is essentially due to the fact 
that for large $\Tb$ the lightest {\it allowed} sfermion masses are
in general larger than in the low $\Tb$ case. Since the largest values
of $\Delta$ occur for the lightest possible sfermion masses this leads
to the smaller result.

\subsection{Results for $e^+e^- \to \sum_{q \neq t} q\bar{q}$}

At LEP energies, the hadron production process from primary quarks
$\neq t$ is experimentally accessible, with a large cross section and
a final accuracy of 0.7\% \cite{LEP2}. 

\begin{figure}[htbp]
\centerline{\psfig{bbllx=20pt,bblly=90pt,bburx=600pt,bbury=420pt,%
figure=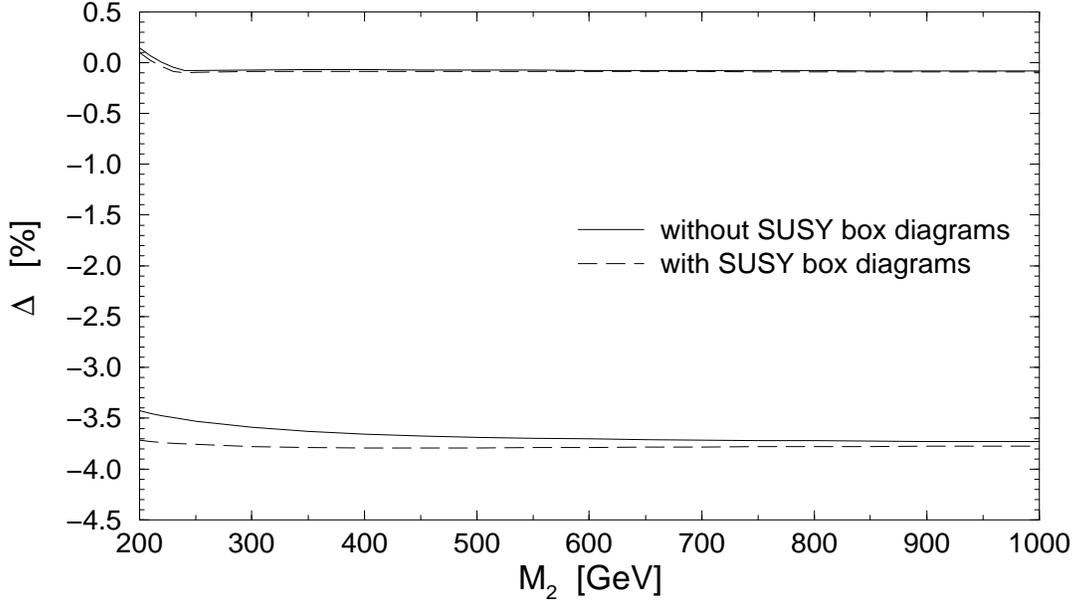,width=15cm}}
\caption[]{$e^+e^- \to \sum_{q \neq t} q\bar{q}$.
  Predicted range of $\Delta$ for $\sqrt{s}=192$ GeV, 
  $\Tb=1.6$, $M_A=500$ GeV, $100\GeV\leq m_{\tilde{t}_1}\leq 500\GeV$, 
  $-400\GeV\leq A\leq 400\GeV$ and $-500\GeV\leq\mu\leq 500\GeV$ (but 
  $-100\GeV<\mu<100\GeV$ is excluded). The predicted range is between 
  the two upper and the two lower lines.}
\label{quark1.bat}
\end{figure}

In Fig. \ref{quark1.bat} we display the results for $\Delta$ from SUSY
parameter scans, in which we vary $m_{\tilde{t}_1}$ between 100 and
500 GeV, $\mu$ between $-500$ and 500 GeV (but $-100\GeV<\mu<100\GeV$ 
is excluded) in steps of 50 GeV, and $M_2$ between 200 and 1000 GeV,
in steps of 10 GeV in the low $\Tb$ scenario. 

The deviations from the SM are always negative and $\Delta$ can reach
up to $-3.7\%$. The difference to the lepton case (Fig. \ref{mue1.bat}) 
is essentially caused by the presence of the extra gluino exchange in 
the SUSY-QCD corrections. There is (almost) no difference between the 
choice of a fixed $A=m_{\tilde{t}_1}$ (which is not depicted here) or for 
$-400\GeV\leq A\leq 400\GeV$.

In the large $\Tb$ scenario the predicted range for the same parameter
variations is smaller than in Fig. \ref{quark1.bat}
($-1.8\% \lsim \Delta \lsim 0\%$), for the same reasons as in Section
\ref{section_mu}.

\subsection{Results for $e^+e^- \to t\bar{t}$}

The production of $t\bar{t}$-pairs is a process of basic interest
at a future high energy $e^+e^-$ collider and will provide precision 
studies of the top quark properties (see, e.g., Ref. \cite{LINAC}). 
From the large top quark mass one expects a special sensitivity to the
structure of the Yukawa couplings.

1-loop electroweak corrections for $t\bar{t}$ production have been
calculated in the SM both for the threshold region \cite{kuhn} and the
continuum \cite{hollik91,beenakker}. Also for Two-Higgs-Doublet Models
(2HDM) the electroweak radiative corrections are available 
\cite{arnd,Guth}. When the SUSY constraints on the 2HDM Higgs sector
are imposed, the potentially large Higgs-induced corrections are kept 
typically below $\sim 2\%$ \cite{arnd}.

Within the MSSM the SUSY-QCD corrections to $t\bar{t}$ production
have been calculated in \cite{Djouadi}, and the subclass of
electroweak 1-loop contributions $\propto m_t^2/M_W^2$ resulting
from the top Yukawa couplings has been discussed in \cite{Chang_Li}. 
Our results are in qualitative agreement with Ref. \cite{Chang_Li}. 
Quantitatively they differ by a few percent, which, however, is not 
surprising in view of the approximations in Ref. \cite{Chang_Li}. It
is also known from the standard model case that the restriction to the
terms $\propto m_t^2/M_W^2$ only yields a poor approximation of the 
complete result. Our discussion in this section is complete at the 
1-loop level, including also the SUSY-QCD corrections.

\begin{figure}[htbp]
\centerline{\psfig{bbllx=20pt,bblly=90pt,bburx=600pt,bbury=420pt,%
figure=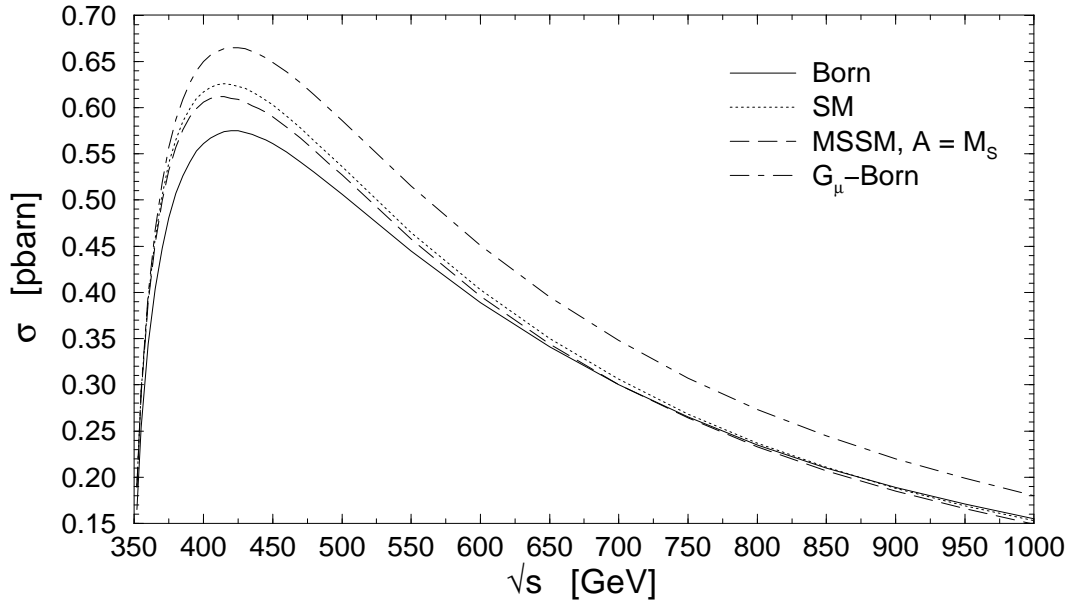,width=15cm}}
\caption[]{$e^+e^- \to t\bar{t}$. Energy dependence of $\sigma$ for 
  $\Tb=40$, $M_A=200$ GeV, $A=M_S=200$ GeV, $\mu=-150$ GeV and 
  $M_2=250$ GeV.}
\label{born.bat}
\end{figure}

\begin{figure}[htbp]
\centerline{\psfig{bbllx=20pt,bblly=90pt,bburx=600pt,bbury=420pt,%
figure=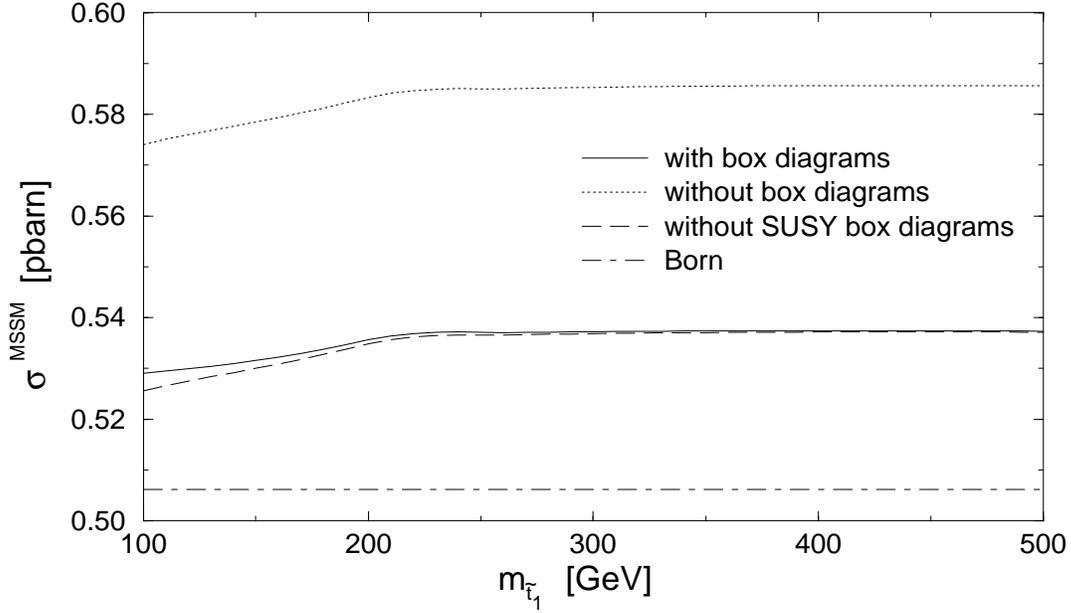,width=15cm}}
\caption[]{$e^+e^- \to t\bar{t}$. The $m_{\tilde{t}_1}$ dependence of 
  $\sigma$ for $\sqrt{s}=500$ GeV, $\Tb=1.6$, $M_A=200$ GeV, 
  $\mu=-150$ GeV, $M_2=200$ GeV and $A=m_{\tilde{t}_1}$.}
\label{box1.bat}
\end{figure}

In order to give an idea of the absolute size of the total cross 
section, we plot $\sigma^{(0)}(s)$ and at the 1-loop level 
$\sigma^{\rm SM}(s)$ and $\sigma^{\rm MSSM}(s)$ in Fig. 
\ref{born.bat} for a typical choice of parameters as a function of 
the center-of-mass energy. The strong deviation of the Born cross
section, parameterized by $G_\mu$, from the 1-loop result, both for
SM and MSSM, is essentially due to the box diagram contributions.  
In addition, Fig. \ref{box1.bat} displays the Born cross section and the
total cross section $\sigma^{\rm MSSM}(s)$ with and without 
box diagram contributions in dependence of the light top squark mass 
$m_{\tilde{t}_1}$, also for a typical set of parameters. 
This figure shows that the box diagram contributions to $\Delta$ can 
reach about 10\% of the MSSM 1-loop result and are therefore 
non-negligible. 

\begin{figure}[htbp]
\centerline{\psfig{bbllx=20pt,bblly=90pt,bburx=600pt,bbury=420pt,%
figure=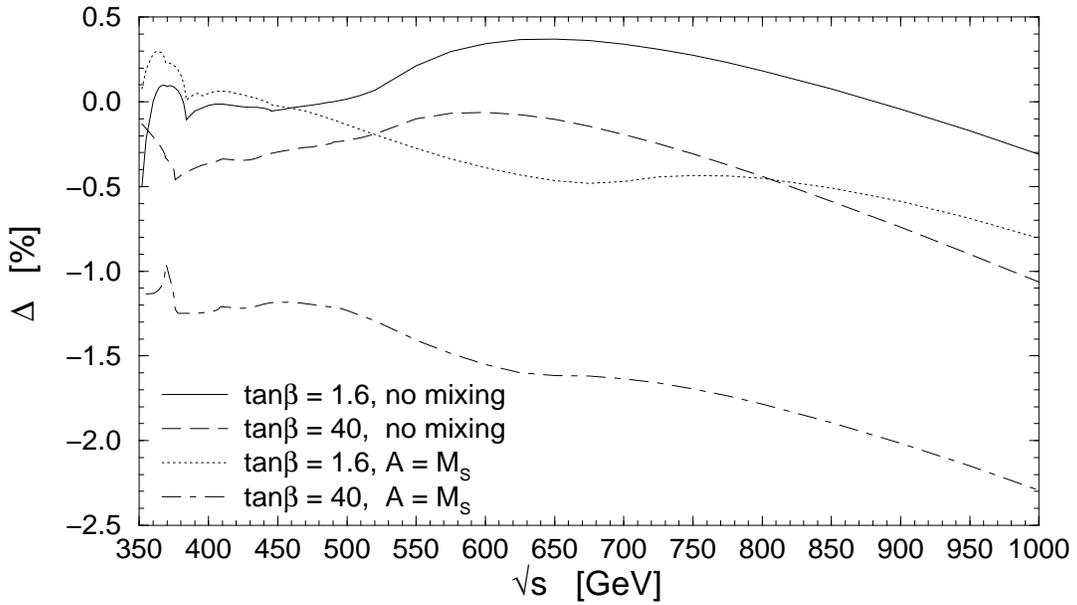,width=15cm}}
\caption[]{$e^+e^- \to t\bar{t}$. The $\sqrt{s}$ dependence of
  $\Delta$ for $M_A=200$ GeV, $M_S=200$ GeV, $\mu=-150$ GeV and 
  $M_2=200$ GeV. ``no mixing'' denotes unmixed sfermions.}
\label{s1.bat}
\end{figure}

The energy dependence of $\Delta$ is illustrated in Fig. \ref{s1.bat}
for low and high $\Tb$ with and without sfermion mixing for a fixed
set of SUSY parameters. With increasing $s$, the values of $\Delta$
become negative in all cases.

\begin{figure}[htbp]
\centerline{\psfig{bbllx=20pt,bblly=90pt,bburx=600pt,bbury=420pt,%
figure=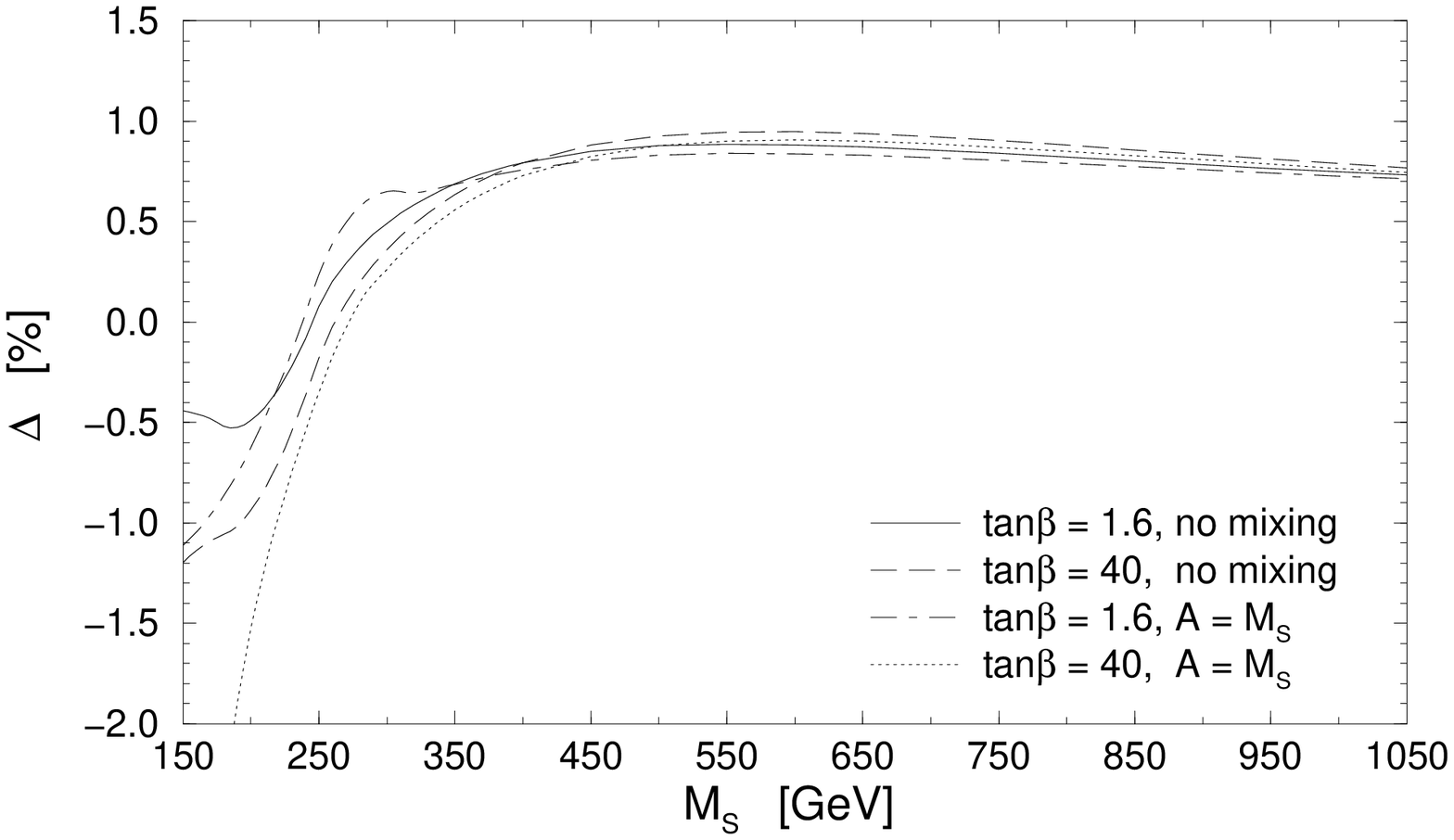,width=15cm}}
\caption[]{$e^+e^- \to t\bar{t}$. The $M_S$ dependence of $\Delta$ for
  $\sqrt{s}=500$ GeV, $M_A=200$ GeV, $\mu=-100$ GeV and $M_2=250$ GeV. 
  ``no mixing'' denotes unmixed sfermions.}
\label{susy11.bat}
\end{figure}

The dependence of $\Delta$ on the common squark mass parameter $M_S$,
Eq. (\ref{Msusy}), is shown in Fig. \ref{susy11.bat} at 
$\sqrt{s}=500\GeV$. For large $M_S$, $\Delta$ is very small in 
accordance with the decoupling properties of heavy sfermions \cite{AC}. 
The steep behavior of the dotted curve for $\Tb=40$ at low 
$M_S$ is related to low chargino and sbottom masses, close to their 
present exclusion limits. It enters through the top wave function
renormalization with a threshold singularity at 
$m_t=m_{\tilde{\chi}_1^+}+m_{\tilde{b}_2}$. Releasing the GUT
constraint (\ref{M-GUT}) for the gluino mass and treating
$m_{\tilde{g}}$ as an independent parameter leads to a variation of
the results in Fig. \ref{susy11.bat} within $\pm 1\%$.
The asymmetries $A_{LR}$, $A_{FB}$ (which are not depicted here) are
not very sensitive to the SUSY contributions\footnote{$A_{LR}$ and 
  $A_{FB}$ are of the order ${\cal O}(5\times 10^{-1})$}: 
their variations are in the ranges 
$-9\times 10^{-3} < \delta A_{LR} < 2\times 10^{-3}$ and 
$-7\times 10^{-3} < \delta A_{FB} < 1\times 10^{-3}$ for the same
parameters as in Fig. \ref{susy11.bat}.

\begin{figure}[htbp]
\centerline{\psfig{bbllx=20pt,bblly=90pt,bburx=600pt,bbury=420pt,%
figure=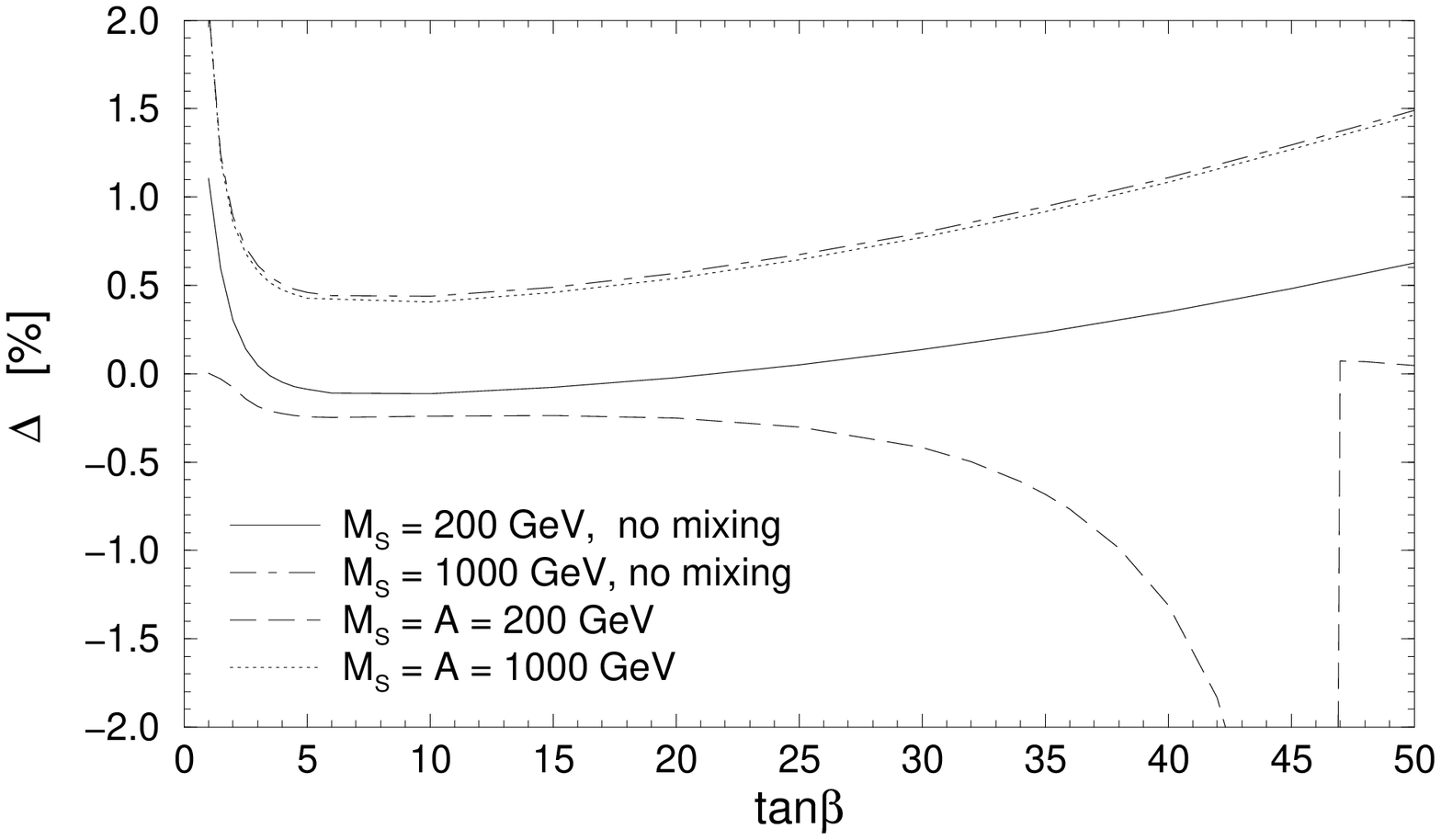,width=15cm}}
\caption[]{$e^+e^- \to t\bar{t}$. The $\Tb$ dependence of $\Delta$ for
  $\sqrt{s}=500$ GeV, $M_A=200$ GeV, $\mu=-150$ GeV and $M_2=100$ GeV. 
  ``no mixing'' denotes unmixed sfermions.}
\label{tan1.bat}
\end{figure}

Fig. \ref{tan1.bat} illustrates the $\Tb$ dependence of $\Delta$ for
various values of $M_S$ with and without sfermion mixing. Again, the
singularity for high $\Tb$ is related to the threshold 
$m_t=m_{\tilde{\chi}_1^+}+m_{\tilde{b}_2}$ in the top wave function
renormalization. 
For the parameters of Fig. \ref{tan1.bat} the contributions to the 
asymmetries cover the ranges 
$-9\times 10^{-3} < \delta A_{LR} < 5\times 10^{-3}$ and
$-6 \times 10^{-3} < \delta A_{FB} < 1\times 10^{-3}$.

\begin{figure}[htbp]
\centerline{\psfig{bbllx=20pt,bblly=90pt,bburx=600pt,bbury=420pt,%
figure=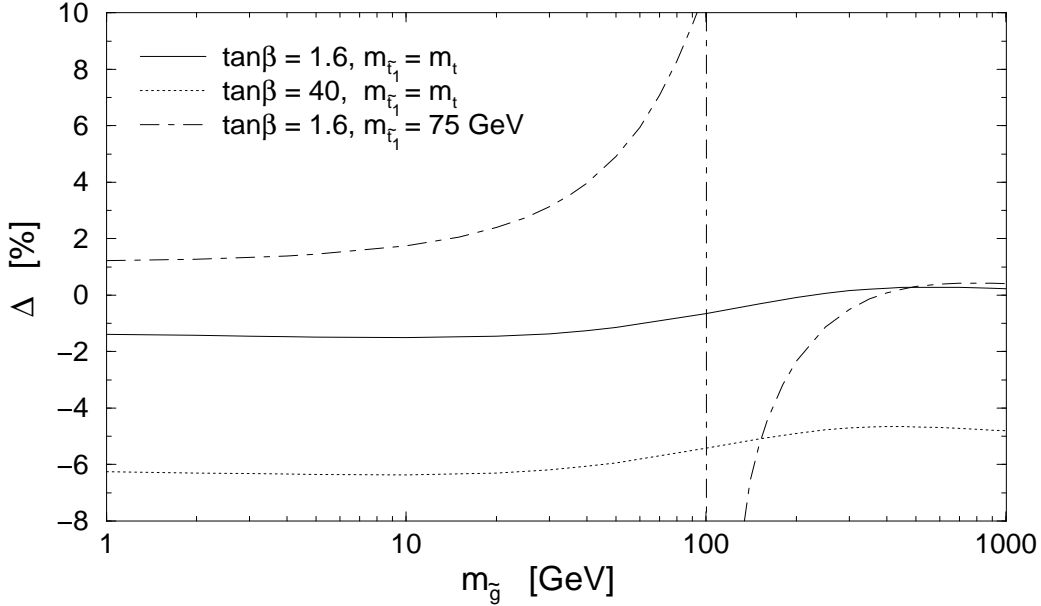,width=15cm}}
\caption[]{$e^+e^- \to t\bar{t}$. The $m_{\tilde{g}}$ dependence of 
  $\Delta$ for $\sqrt{s}=500$ GeV, $M_A=150$ GeV, $\mu=-150$ GeV, 
  $A=m_{\tilde{t}_1}$ and $M_2=150$ GeV.}
\label{mgluino3.bat}
\end{figure}

The dependence of $\Delta$ on the gluino mass $m_{\tilde{g}}$ is shown
in Fig. \ref{mgluino3.bat}. Now $m_{\tilde{g}}$ is treated as an 
independent parameter. For the purpose of illustration we also keep
the range of small gluino masses below $\sim 150\GeV$.
As it can be seen outside the singularity\footnote{The singularity in
Fig. \ref{mgluino3.bat} corresponds to the case 
$m_t=m_{\tilde{g}}+m_{\tilde{t}_1}$.} the $m_{\tilde{g}}$-dependence
is smooth and even very light gluinos do not give rise to significantly
different predictions. For large $\Tb$ and for the specific
constellation of masses $m_{\tilde{t}_1}=m_t$, corrections of the
order $-6\%$ are obtained. But this is due to the fact that for these
specific parameters we are near a threshold singularity at  
$m_t=m_{\tilde{\chi}^+_1}+m_{\tilde{b}_2}$, which is an electroweak 
effect.
Our results for the SUSY-QCD part of the 1-loop corrections are in 
agreement with the results of Ref. \cite{Djouadi}.


\begin{figure}[htbp]
\centerline{\psfig{bbllx=20pt,bblly=90pt,bburx=600pt,bbury=420pt,%
figure=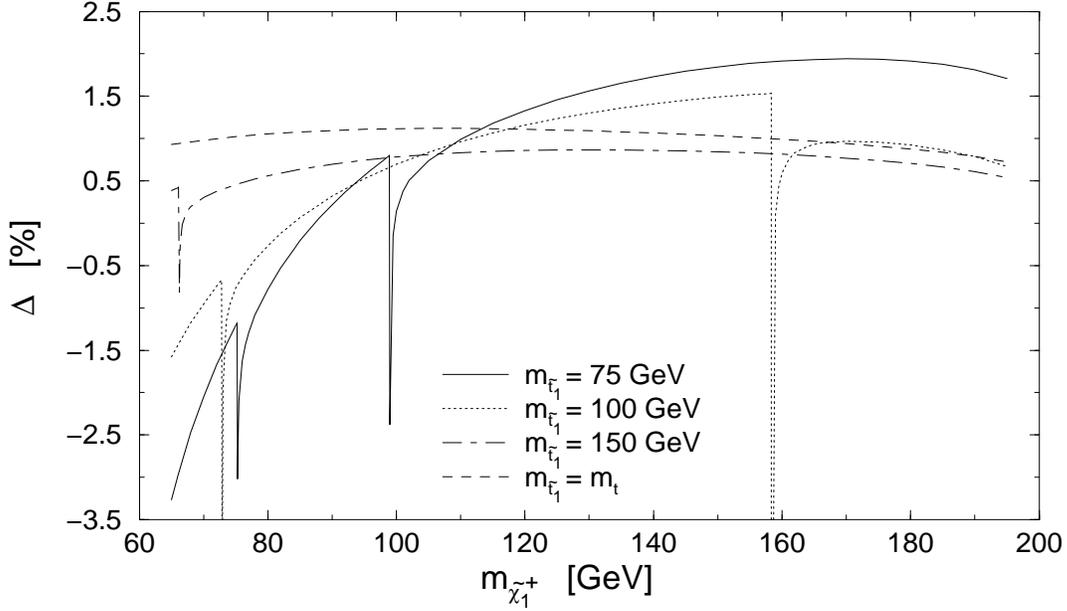,width=15cm}}
\caption[]{$e^+e^- \to t\bar{t}$. The $m_{\tilde{\chi}_1^+}$ dependence 
  of $\Delta$ for $\sqrt{s}=500$ GeV, $\Tb=1.6$, $M_A=250$ GeV, 
  $\mu=-200$ GeV and $A=m_{\tilde{t}_1}$.}
\label{mchar1.bat}
\end{figure}

For a more detailed discussion of the weak corrections, the dependence
of $\Delta$ on the lighter chargino mass $m_{\tilde{\chi}^+_1}$ is
shown in Fig. \ref{mchar1.bat}. Again the singularities correspond to 
thresholds in the top wave function renormalization. Given the 
experimental restrictions on $m_{\tilde{\chi}^+_1}$ and 
$m_{\tilde{t}_1}$, the values of $\Delta$ are typically within a few
percent.

\begin{figure}[htbp]
\centerline{\psfig{bbllx=20pt,bblly=90pt,bburx=600pt,bbury=420pt,%
figure=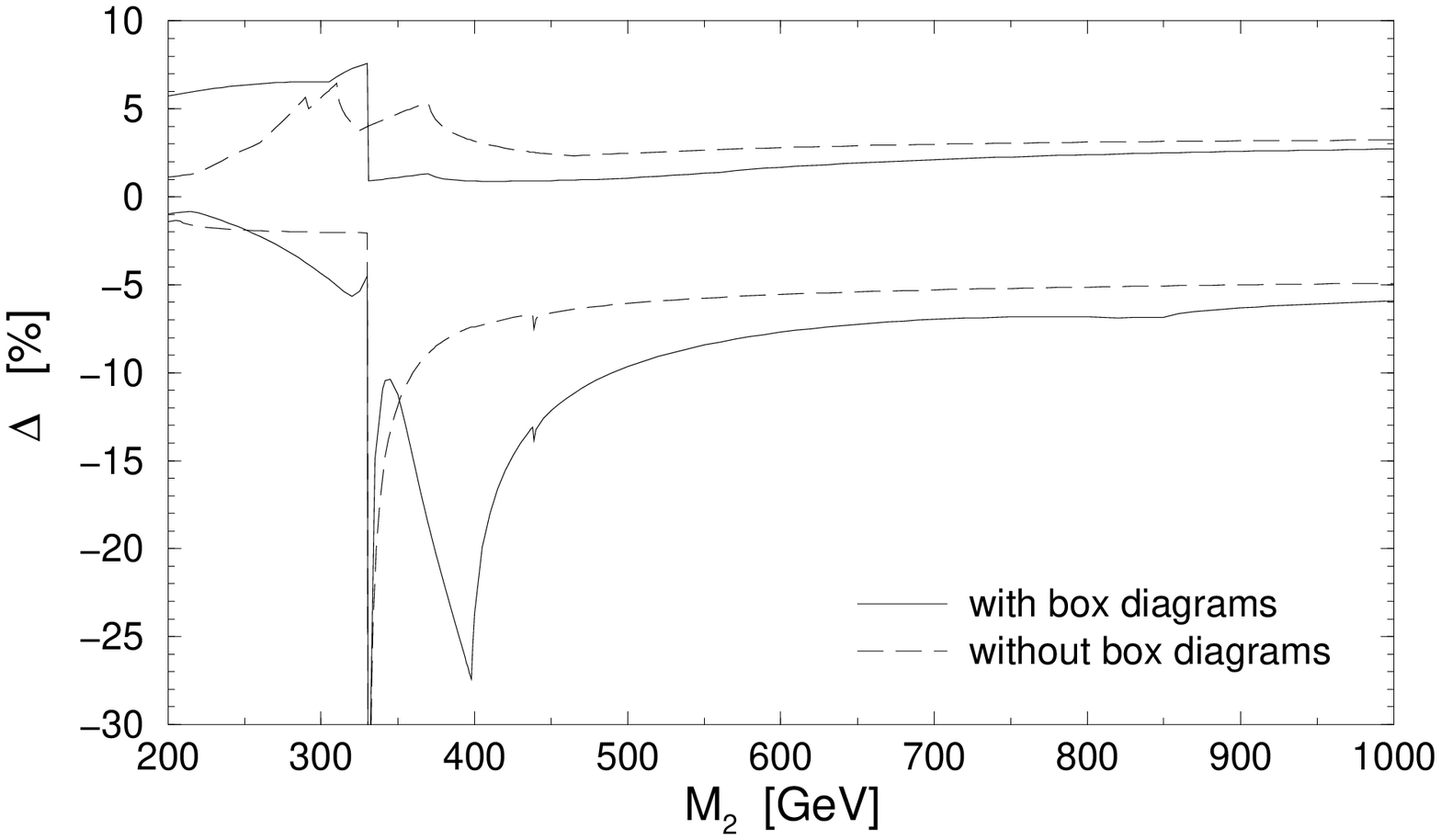,width=15cm}}
\caption[]{$e^+e^- \to t\bar{t}$. Predicted range of $\Delta$ for 
  $\sqrt{s}=500$ GeV, $\Tb=1.6$, $M_A=500$ GeV, 
  $100\GeV\leq m_{\tilde{t}_1}=A\leq 500\GeV$ and 
  $-500\GeV\leq\mu\leq 500\GeV$ (but $-100\GeV<\mu<100\GeV$ is 
  excluded). The predicted range is between the two upper and the two 
  lower lines.}
\label{all.bat}
\end{figure}

The result of a SUSY parameter scan for low $\Tb$ is presented in 
Fig. \ref{all.bat}, where $A=m_{\tilde{t}_1}$ is varied between 100 
and 500 GeV, $\mu$ between $-500$ and 500 GeV 
(but $-100\GeV<\mu<100\GeV$ is excluded) in steps of 50 GeV, 
and $M_2$ between 200 and 1000 GeV in steps of 10 GeV.  
The influence of the box diagrams is shown separately (solid line). 
The dip around $M_2=331\GeV$ (solid and dashed line) is a threshold 
singularity at $m_t=m_{\tilde{\chi}_1^+}+m_{\tilde{b}_1}$.
The dip around $M_2=398\GeV$ (dashed line) is a threshold effect 
corresponding to $m_{\tilde{\chi}^+_1}+m_{\tilde{\chi}^+_2}=\sqrt{s}$ 
from the crossed chargino box, because the $f\tilde{f}' \tilde{\chi}^+$ 
couplings are (for this parameters) up to 10 times larger then the 
$Z \tilde{\chi}^+_1 \tilde{\chi}^+_2$ coupling in the corresponding 
self-energy and vertex diagrams. 
(Therefore this threshold is invisible in the dashed line.)
The large values are due to nearby singularity effects at 
$m_t=m_{\tilde{\chi}_1^+}+m_{\tilde{b}_{1,2}}$ ($\Delta \sim -7\%$)
and production threshold effects at 
$m_{\tilde{\chi}_4^0}+m_{\tilde{t}_{1,2}}=\sqrt{s}$ ($\Delta \sim 6\%$).

\begin{figure}[htbp]
\centerline{\psfig{bbllx=20pt,bblly=90pt,bburx=600pt,bbury=420pt,%
figure=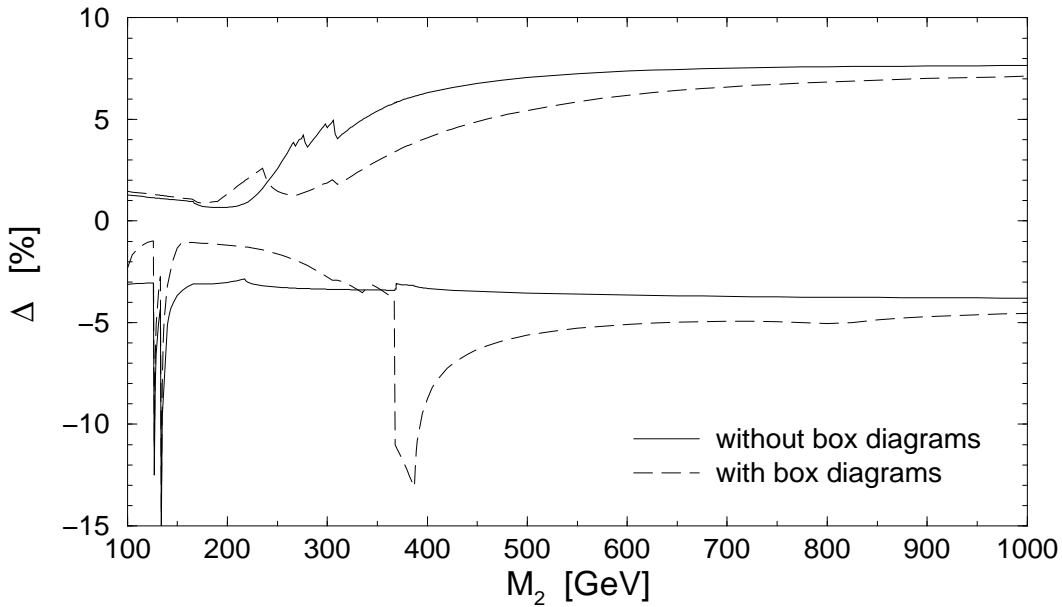,width=15cm}}
\caption[]{$e^+e^- \to t\bar{t}$. Predicted range of $\Delta$ for 
  $\sqrt{s}=500$ GeV, $\Tb=40$, $M_A=500$ GeV, 
  $100\GeV\leq m_{\tilde{t}_1}=A\leq 500\GeV$ and 
  $-500\GeV\leq\mu\leq 500\GeV$ (but $-100\GeV<\mu<100\GeV$ is 
  excluded). The predicted range is between the two upper and the two 
  lower lines.}
\label{all40.bat}
\end{figure}

In the large $\Tb$ scenario (Fig. \ref{all40.bat}) with the same 
parameters as in Fig. \ref{all.bat} we find a smaller negative boundary 
and a larger positive boundary ($-5\% \lsim \Delta \lsim 7\%$) than in
Fig. \ref{all.bat}. Here again these values are a nearby singularity 
effect at $m_t=m_{\tilde{\chi}_1^+}+m_{\tilde{b}_2}$ 
($\Delta \sim -5\%$) and a nearby threshold effect at 
$m_{\tilde{\chi}_1^+}+m_{\tilde{\chi}_1^+}=\sqrt{s}$ ($\Delta \sim 7\%$).
The large dip (dashed line) is again a threshold effect corresponding 
to $m_{\tilde{\chi}^+_1}+m_{\tilde{\chi}^+_2}=\sqrt{s}$ from the
crossed chargino box.

\section{Summary}
\label{chap4}
\setcounter{equation}{0}
\setcounter{figure}{0}
\setcounter{table}{0}

In this paper we have calculated the electroweak and SUSY-QCD 1-loop
corrections to the process $e^+e^- \to f\bar{f}$ ($f=\mu,\tau,q$) in
the MSSM. 

Quantitative numerical studies of the size of the virtual
SUSY effects have been performed for lepton pair and light quark pair
production for LEP energies, and for top quark pair production at 
$500\GeV$. Thereby, the parameters of the MSSM have been varied over a
wide range in order to get an overall impression of the size of
virtual SUSY effects. 

For the light fermions, the virtual SUSY contributions in the cross
sections amount to a few percent in the LEP region; they could be 
detectable for close-by SUSY states.

In the cross section for top production they can be larger, but 
typically stay below $10\%$ (with exception of the peaks at some 
thresholds). In general, the effects of box diagrams are sizeable,
mainly the SM box graphs. For the top case, also the SUSY box diagrams 
yield significant contributions.

For the forward-backward and left-right asymmetries no significant
deviations between the MSSM and the SM occur.

\section*{Acknowledgments}

We want to thank A. Kraft and S. Heinemeyer for helpful discussions.

\vfill
\newpage
\clearpage

\begin{appendix}


\section{Masses and mixing states of the  MSSM}
\label{appa}
\setcounter{equation}{0}
\setcounter{figure}{0}
\setcounter{table}{0}

\subsection{The Higgs sector}
\label{Higgs-Sektor}

The neutral Higgs sector is conventionally  fixed by choosing a value
for $\Tb := \frac{v_2}{v_1}$ and for the mass $M_{A^0}$ of the 
$\cal CP$-odd neutral Higgs boson $A^0$.

Radiative corrections for the neutral Higgs masses are important and 
dominated by terms proportional to $m_t^4$ \cite{HHOYY}. We take them 
into account in the $m_t^4$-approximation following Ref. \cite{ds}.

There are only small radiative corrections for the charged Higgs 
masses and the following equation holds only for 
$\MA \sim {\cal O}(M_W)$ \cite{Diaz} (we already used here the values 
$N_C^f=3$, $N_G=3$ and $N_H=2$, which are denoting the number of
colors, families and Higgs doublets, respectively):
\BEA
m^2_{H^\pm,{\rm eff}} &=& \frac{3\, g^2_2}{32\pi^2 M^2_W} \left[ 
    \frac{2\,m^2_t\, m^2_b}{\SQb\CQb} - M^2_W
    \left(\frac{m^2_t}{\SQb}+\frac{m^2_b}{\CQb}\right)
    +\frac{2}{3} M^4_W \right] \ln\frac{M^2_S}{m^2_t}
    \nonumber \\
& & +\MA^2 + M^2_W 
    +\frac{15 g^2_1 M^2_W}{48\pi^2}\, \ln\frac{M^2_S}{M^2_W}~,
\label{mHplus}
\EEA
$M_S$ is a soft SUSY breaking parameter defined in Eq. (\ref{Msusy}) 
and $g_1$, $g_2$ denotes the $U(1)$, $SU(2)$ gauge coupling, 
respectively. 

\subsection{Sfermions}
\label{misch}

For the simplest case we assume equal soft SUSY breaking parameters
for all sfermions (motivated by {\it minimal} SUGRA)\footnote{Since 
we ignore mixing between sfermions of different generations, the 
($3 \times 3$) matrices $\bf M^2_{\tilde{Q}}$, $\bf M^2_{\tilde{U}}$, 
$\bf M^2_{\tilde{D}}$, $\bf M^2_{\tilde{L}}$, $\bf M^2_{\tilde{E}}$, 
$\bf A_U$, $\bf A_D$ and $\bf A_L$ are diagonal. Furthermore we chose
all soft breaking parameters to be real.}:
\BE
{\bf M^2_{\tilde{Q}}} = {\bf M^2_{\tilde{U}}} = {\bf M^2_{\tilde{D}}} = 
{\bf M^2_{\tilde{L}}} = {\bf M^2_{\tilde{E}}} = M^2_S\,\id~, \qquad
{\bf A_U} = A_u\, \id~, \qquad {\bf A_L} = {\bf A_D} = A_d\, \id~.
\label{Msusy}
\EE
Then, the sfermion mass matrix is given by 
\BE
\renewcommand{\arraystretch}{1.5}
\ML M_S^2 + M_Z^2\CZb(I_3^f-Q_f\sw^2) + m_f^2 \quad & 
    m_f (A_{\{u;d\}} - \mu \{\cot\beta;\tan\beta\}) \\
    m_f (A_{\{u;d\}} - \mu \{\cot\beta;\tan\beta\}) & 
    M_S^2 + M_Z^2 \CZb Q_f\sw^2 + m_f^2 \VR~,
\label{Massenmatrix}
\EE
where $\{\cot\beta;\tan\beta\}$ refer to the corresponding 
$\{u;d\}$-type fermions. 

In the case of mixed sfermions the left- and right-handed sfermions
must be replaced by mass eigenstates of the sfermion mass matrix
(\ref{Massenmatrix}) \cite{hhg}:
\BEA
\tilde{f_1} &=& \tilde{f_L}\hspace{1mm}{\cos \tilde{\theta}_f} 
                + \tilde{f_R}\hspace{1mm}{\sin \tilde{\theta}_f}~,  
                \nonumber \\ 
\tilde{f_2} &=& \tilde{f_R}\hspace{1mm}{\cos \tilde{\theta}_f}
                - \tilde{f_L}\hspace{1mm}{\sin \tilde{\theta}_f}~, 
\EEA
where $\tilde{\theta}_f$ denotes the mixing angle.

\subsection{Charginos}
\label{Charginos}

The chargino mass eigenstates are obtained by diagonalizing the 
chargino mass matrix \cite{habkan,gunhab}
\BE
{\bf X} = \ML M_2 & \sqrt2\, M_W\, \Sb \\[1ex] 
          \sqrt2\, M_W\, \Cb & \mu \MR~,
\EE
containing the $SU(2)$ gaugino mass parameter $M_2$ with the help of
two unitary $(2 \times 2)$ matrices {\bf U} and {\bf V}:
\BE
{\bf U^*\,X\,V}^{-1} =
{\bf diag}(m_{\tilde{\chi}^+_1},m_{\tilde{\chi}^+_2})~. 
\EE
The squared entries in the diagonalized chargino mass matrix are obeying 
\BE
m^2_{\tilde{\chi}_{1,2}^+} = \frac{1}{2}\, \bigg\{ 
    M_2^2 + \mu^2 + 2M_W^2 \mp \Big[ (M_2^2-\mu^2)^2
    + 4M_W^4 (\CQZb + M_2^2+\mu^2+2\,\mu\, M_2\, \SZb) 
    \Big]^{\frac{1}{2}} \bigg\}~.
\label{Charmasse}
\EE

\subsection{Neutralinos}
\label{Neutralinos}

The four neutralino mass eigenstates are obtained from diagonalizing 
the ($4 \times 4$) neutralino mass matrix \cite{habkan,gunhab}
\BE
\renewcommand{\arraystretch}{1.2}
{\bf Y} = \MLv M_1 & 0 & -M_Z\sw\Cb & M_Z\sw\Sb \\ 0 & 
          M_2 & M_Z\cw\Cb & -M_Z\cw\Sb \\
          -M_Z\sw\Cb & M_Z\cw\Cb & 0 & -\mu \\ M_Z\sw\Sb & 
          -M_Z\cw\Sb & -\mu & 0 \MR~,
\label{Y}
\EE
with the help of a unitary matrix {\bf N}:
\BE
{\bf N^*\,Y\,N}^{-1} = 
{\bf diag}(m_{\tilde{\chi}^0_1},\ldots,m_{\tilde{\chi}^0_4})~.
\EE
The matrix {\bf Y} contains the $U(1)$ gaugino mass parameter $M_1$ as a
further input quantity. As conventionally done, we assume the SUSY-GUT
constraints, with the gluino mass $m_{\tilde{g}} \equiv |M_3|$
\BE
M_1 = \frac{5}{3} \tan^2\theta_w\, M_2~, \qquad
M_3 = \frac{\alpha_s}{\alpha}\, \sw^2\, M_2~,
\label{M-GUT}
\EE
if not stated differently.
 
Note that for our calculation, the matrices $\bf U$, $\bf V$ and $\bf N$
have been chosen to be real.

\newpage
\clearpage


\section{Self-energies}
\label{appb}
\setcounter{equation}{0}
\setcounter{figure}{0}
\setcounter{table}{0}

This appendix contains the analytical expressions for the 1-loop
self-energies of vector bosons and fermions.
We make use of the following notations: 
The isospin partner of  $f$ $(\tilde{f})$ is denoted by $f'$
$(\tilde{f'})$. Fermions with $I_3=\pm 1/2$ are denoted $f^\pm$, and
sfermions by  $\tilde{f}^\pm$, analogously.
The abbreviation 2HD denotes the SM sector plus the Two Higgs Doublet 
sector of the MSSM.

The Feynman rules of the MSSM are given in Refs. 
\cite{habkan,gunhab,hhg}. Note that the covariant derivative given
there is not compatible with the one used in this paper:
\BE
D_{\mu} = \dmu - i \frac{1}{2}\,g_2\,\sigma_l\,W_\mu^l + 
          i\frac{1}{2}\,g_1\,Y\,B_\mu~.
\EE
So we had to change the Feynman rules by substituting $\sin\theta_w$ 
(defined in Eq. (\ref{Theta})) with $-\sin\theta_w$ and in addition we
had to provide every Higgs field with a factor $(-1)$.

\subsection{Vector boson self-energies}
\label{appb1}

\begin{figure}[hb]
\centerline{\psfig{figure=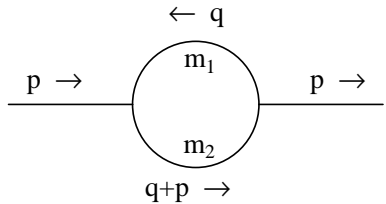,width=5cm}}
\caption[]{Masses and momenta for the vector boson self-energies.}
\label{fig:Vekimpkon}
\end{figure}

The vector boson self-energies are separated in 2HD-, sfermion-, and 
chargino/neutralino contributions. The momenta and internal masses  
are illustrated in Figure \ref{fig:Vekimpkon}. 
As a short hand notation we introduce the functions $F_1$, $F_2$
(which enter the fermion, chargino and neutralino loop contribution
to the vector boson self-energies):
\BEA
F_1(p^2,m_1,m_2,a,b,a',b') &=& 8 \bigg\{ (aa'+bb') 
      \Big[ -2 B_{22}(p^2,m_1,m_2) + A_0(m_2) 
    + m^2_1 B_0(p^2,m_1,m_2) \nonumber \\
& & + p^2 B_1(p^2,m_1,m_2) \Big] 
    - (ab'+ba') m_1 m_2 B_0(p^2,m_1,m_2) \bigg\}~, \nonumber \\
F_2(p^2,m_1,m_2) &=& 10 B_{22}(p^2,m_1,m_2)
    + \Big[ 4p^2+m^2_1+m^2_2 \Big] B_0(p^2,m_1,m_2) 
    + A_0(m_1) \nonumber \\ 
& & + A_0(m_2) - 2 \Big[ m^2_1+m^2_2-\frac{p^2}{3} \Big]~,
\EEA
with the 1- and 2-point functions $A_0$, $B_0$, $B_1$ and $B_{22}$
from Ref. \cite{pasvelt}, in the convention of Refs. 
\cite{hollik95,Oldenborgh}.

\vspace{5mm}

\vorn The self-energies are obtained as the sum of the terms depicted
below with their analytical expression.

\newpage
\clearpage

\subsubsection{Photon self-energy}

\begin{tabular}{p{4.2cm}@{\hspace{0.5cm}\hspace{0.5cm}}p{4.2cm}@{\hspace{0.5cm}\hspace{0.5cm}}p{4.2cm}}
\centerline{\raisebox{-0.7cm}{\psfig{bbllx=107pt,bblly=242pt,%
bburx=505pt,bbury=521pt,figure=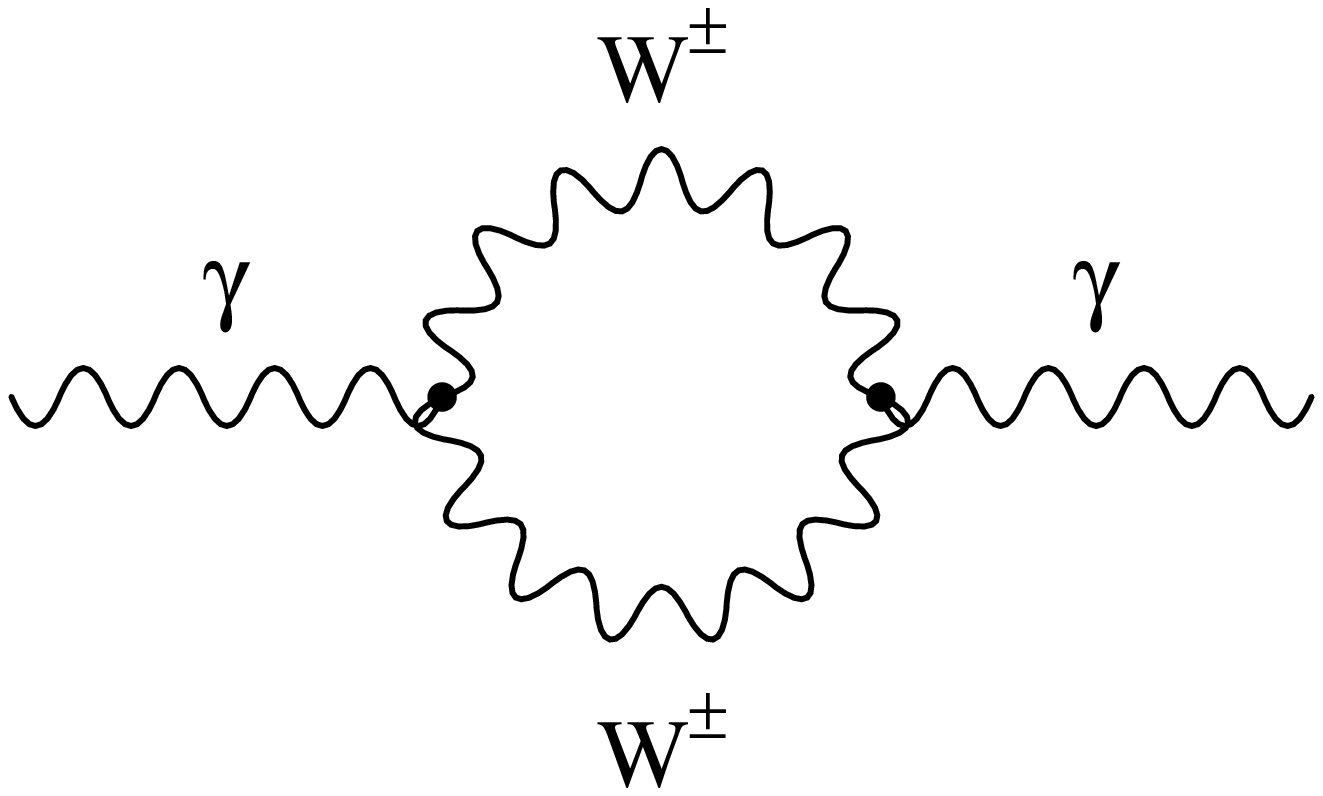,width=4cm}}} &
\centerline{\raisebox{-0.7cm}{\psfig{bbllx=107pt,bblly=242pt,%
bburx=505pt,bbury=530pt,figure=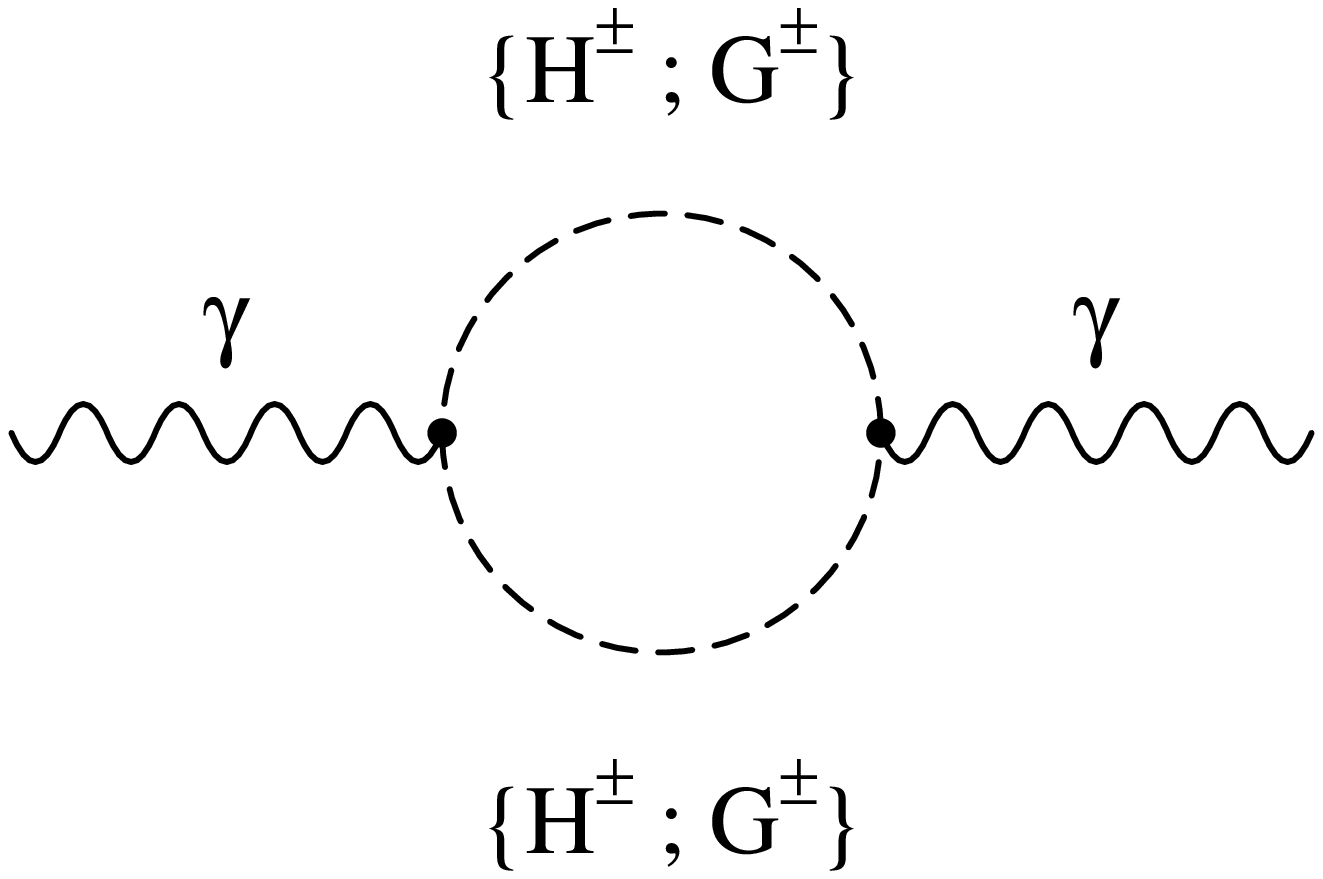,width=4cm}}} &
\centerline{\raisebox{-0.7cm}{\psfig{bbllx=107pt,bblly=242pt,%
bburx=505pt,bbury=521pt,figure=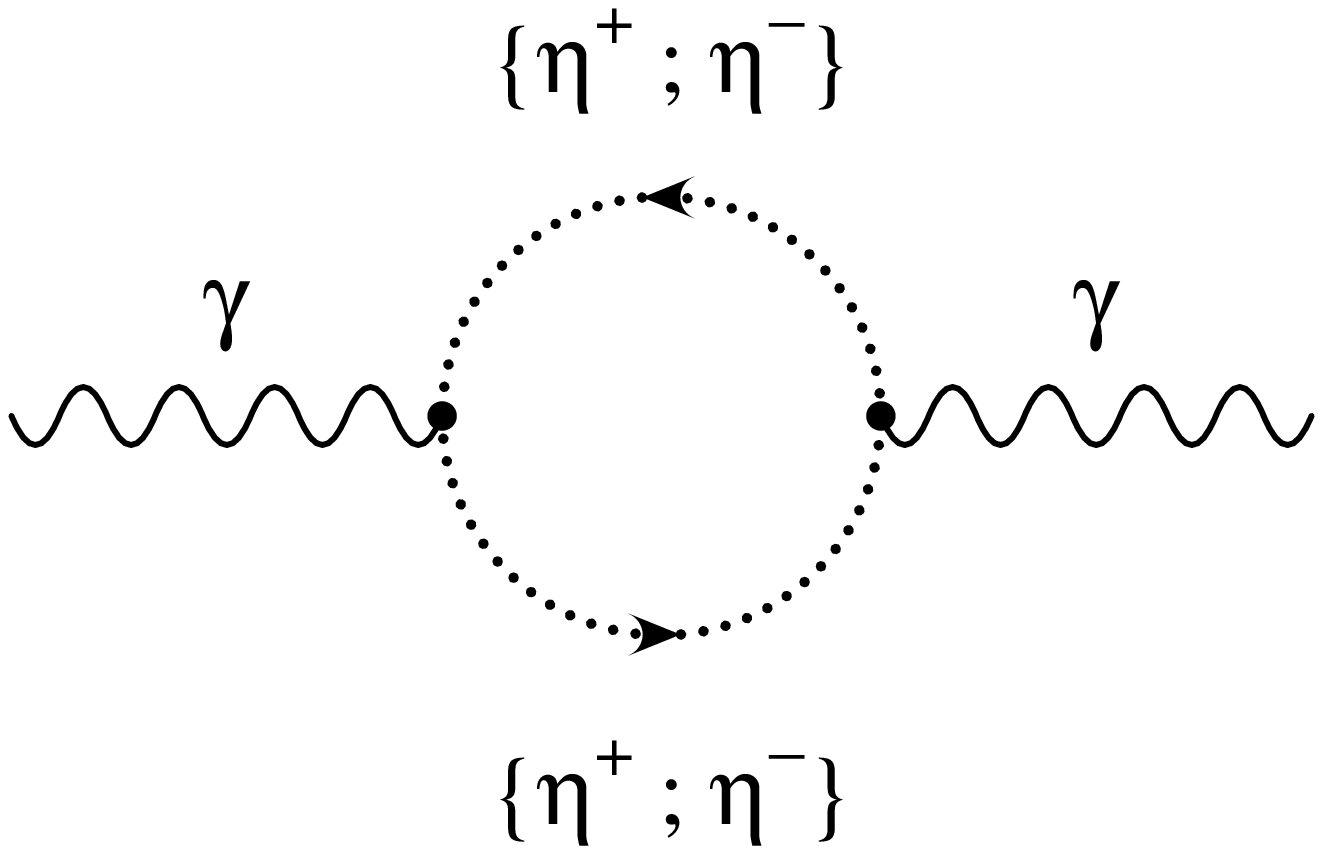,width=4cm}}} \\
\centerline{\psfig{bbllx=107pt,bblly=384pt,bburx=505pt,bbury=599pt,%
figure=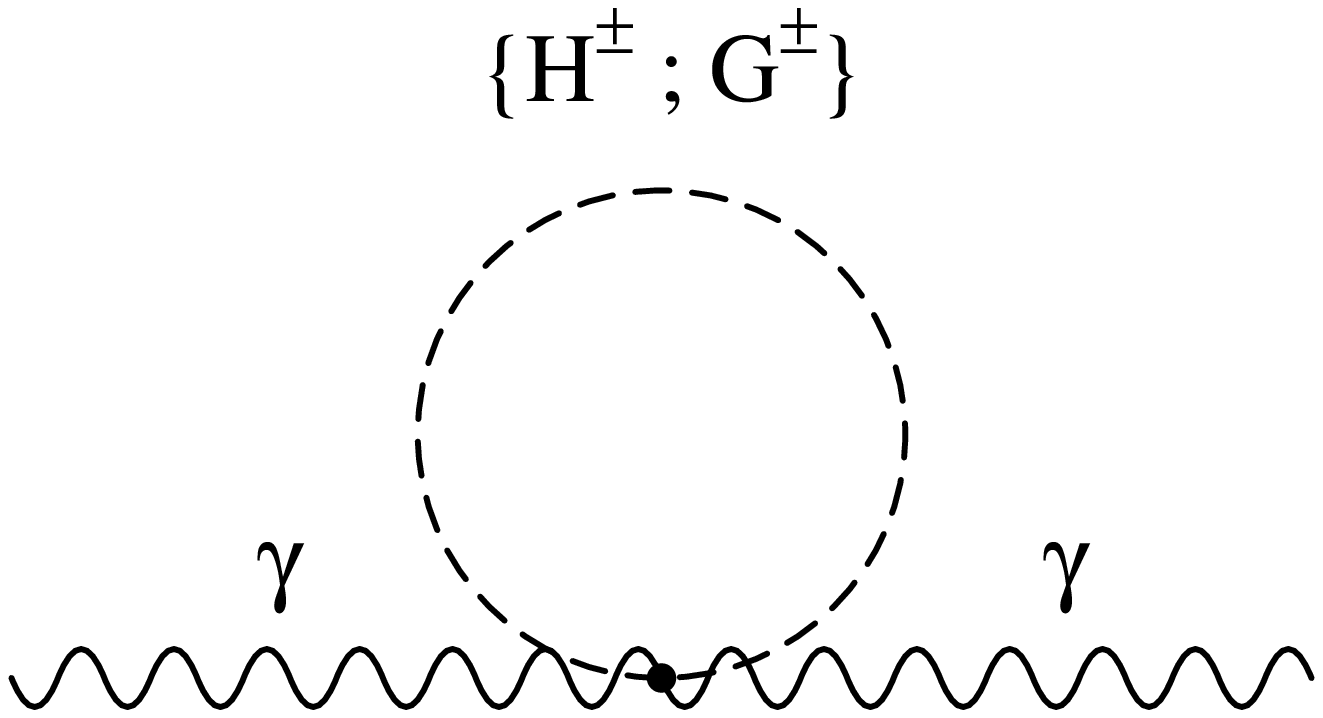,width=4cm}} &
\centerline{\psfig{bbllx=107pt,bblly=384pt,bburx=505pt,bbury=607pt,%
figure=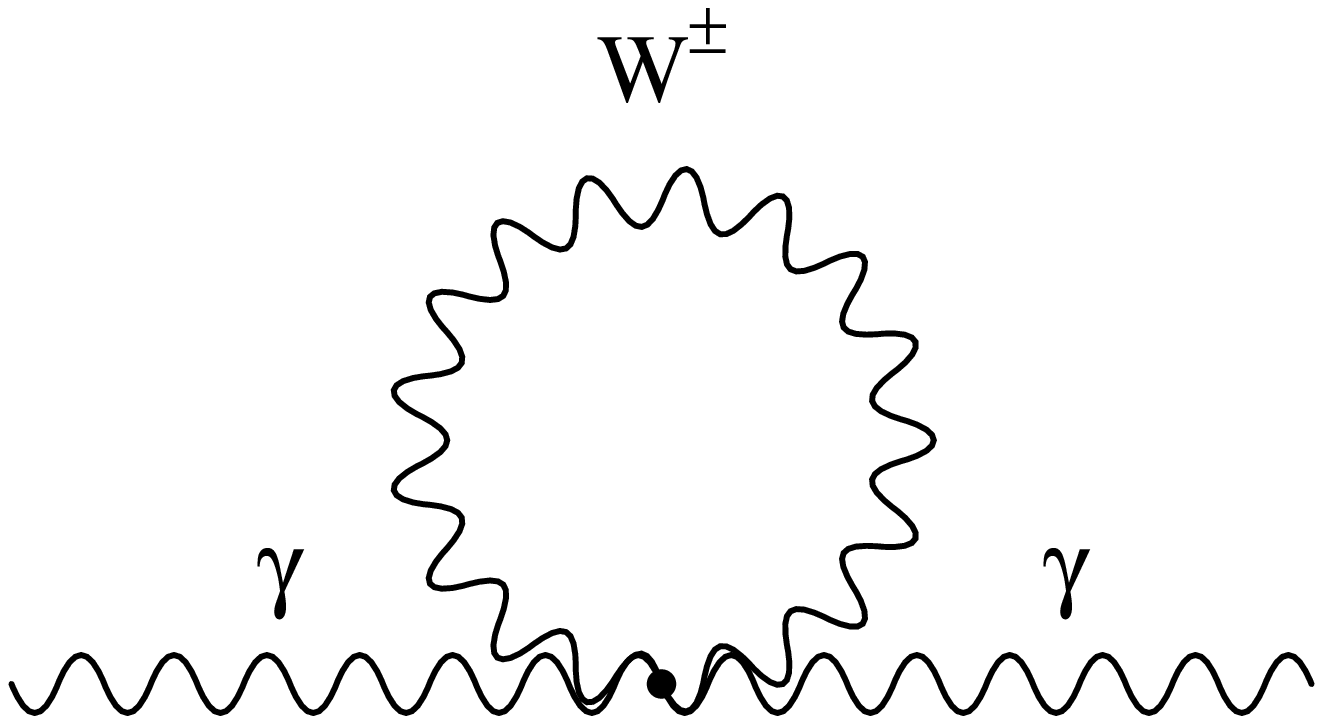,width=4cm}} &
\centerline{\raisebox{-0.7cm}{\psfig{bbllx=107pt,bblly=242pt,%
bburx=505pt,bbury=538pt,figure=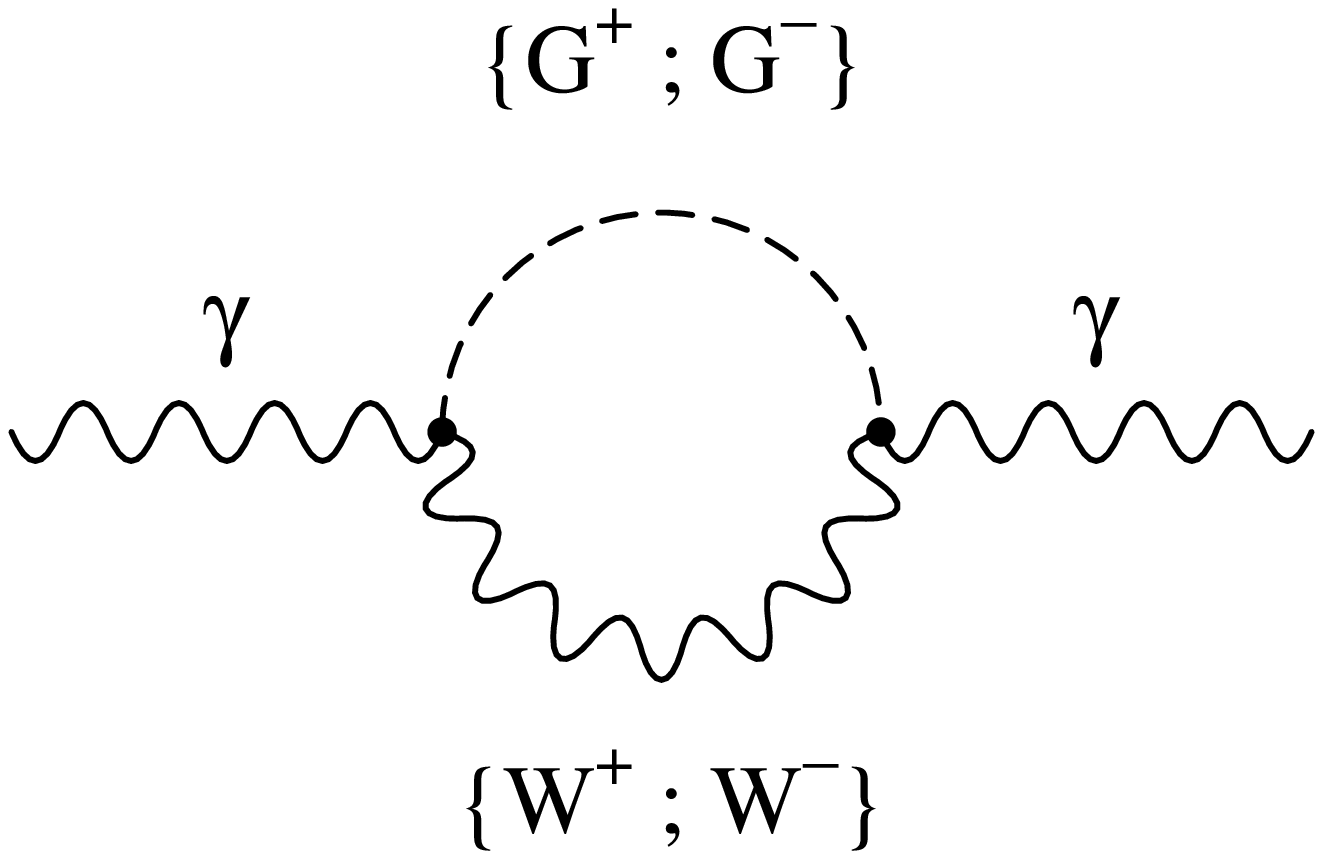,width=4cm}}} \\
\centerline{\raisebox{-0.7cm}{\psfig{bbllx=107pt,bblly=242pt,%
bburx=505pt,bbury=521pt,figure=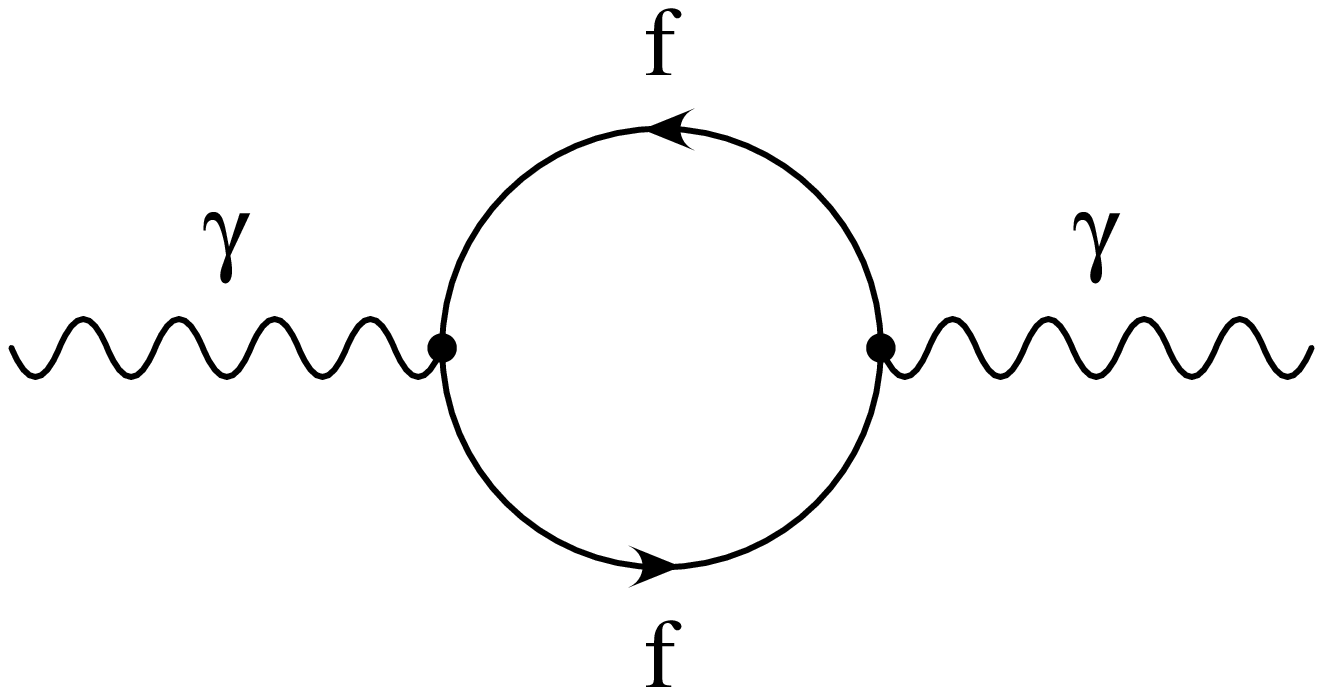,width=4cm}}}
\end{tabular}
\BEA
\Sigma^{\gamma\gamma}_{2HD}(p^2) &=& 
    -\frac{\alpha}{4\pi}\, \bigg\{ 
    4 B_{22}(p^2,m_{H^{\pm}},m_{H^{\pm}}) + 2 B_{22}(p^2,M_W,M_W) 
    \nonumber \\
& & -2 M_W^2 B_0(p^2,M_W,M_W) -2 A_0(m_{H^{\pm}}) -8 A_0(M_W) 
    +4 M_W^2 \nonumber \\
& & +F_2(p^2,M_W,M_W) +\sum_{f} N_C^f Q^2_f 
    F_1(p^2,m_f,m_f,\frac{1}{2},\frac{1}{2},\frac{1}{2},\frac{1}{2}) 
    \bigg\}~,
\EEA

\vspace{6mm}

\begin{tabular}{p{4.2cm}@{\hspace{0.5cm}\hspace{0.5cm}}p{4.2cm}@{\hspace{0.5cm}\hspace{0.5cm}}p{4.2cm}}
\centerline{\raisebox{-0.7cm}{\psfig{bbllx=107pt,bblly=242pt,%
bburx=505pt,bbury=530pt,figure=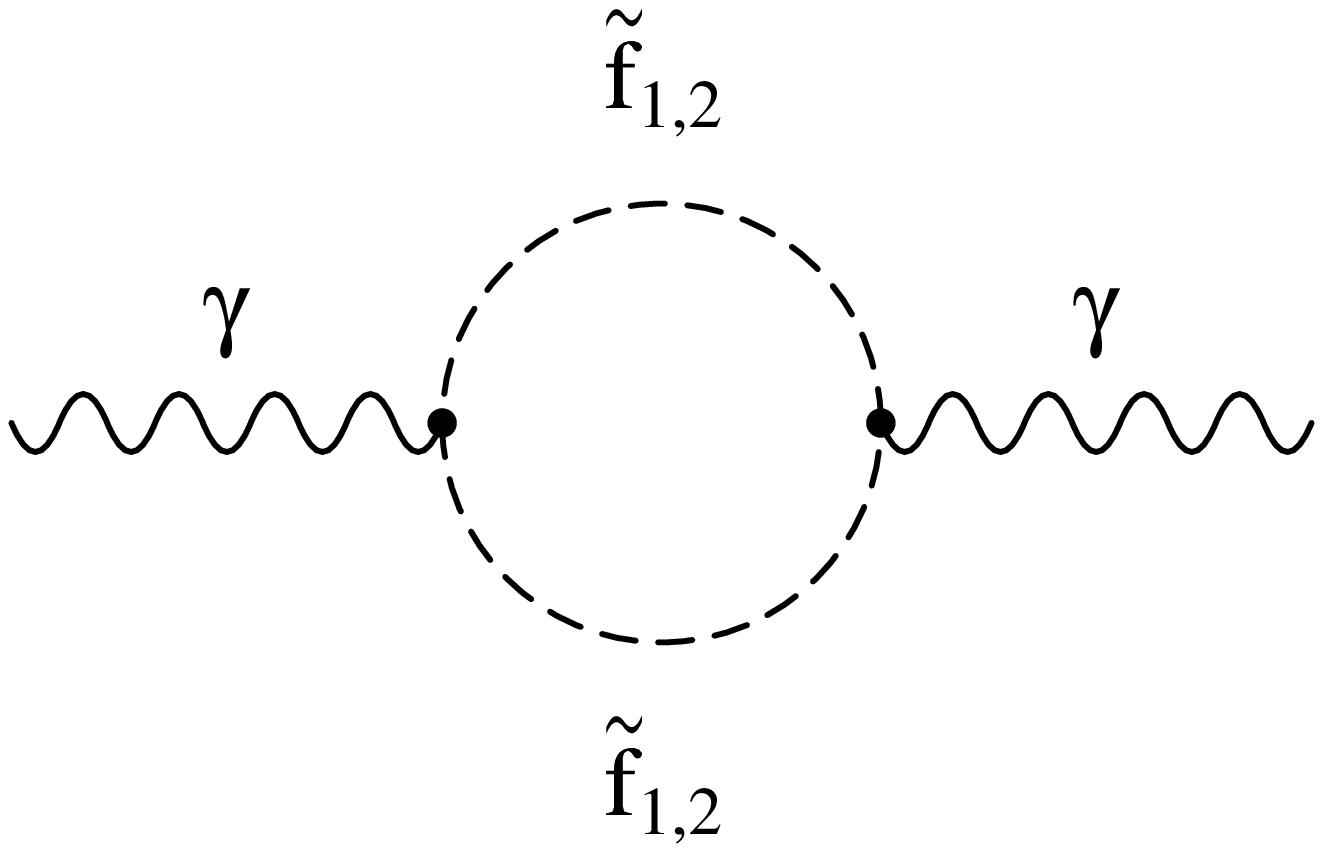,width=4cm}}} &
\centerline{\psfig{bbllx=107pt,bblly=384pt,%
bburx=505pt,bbury=599pt,figure=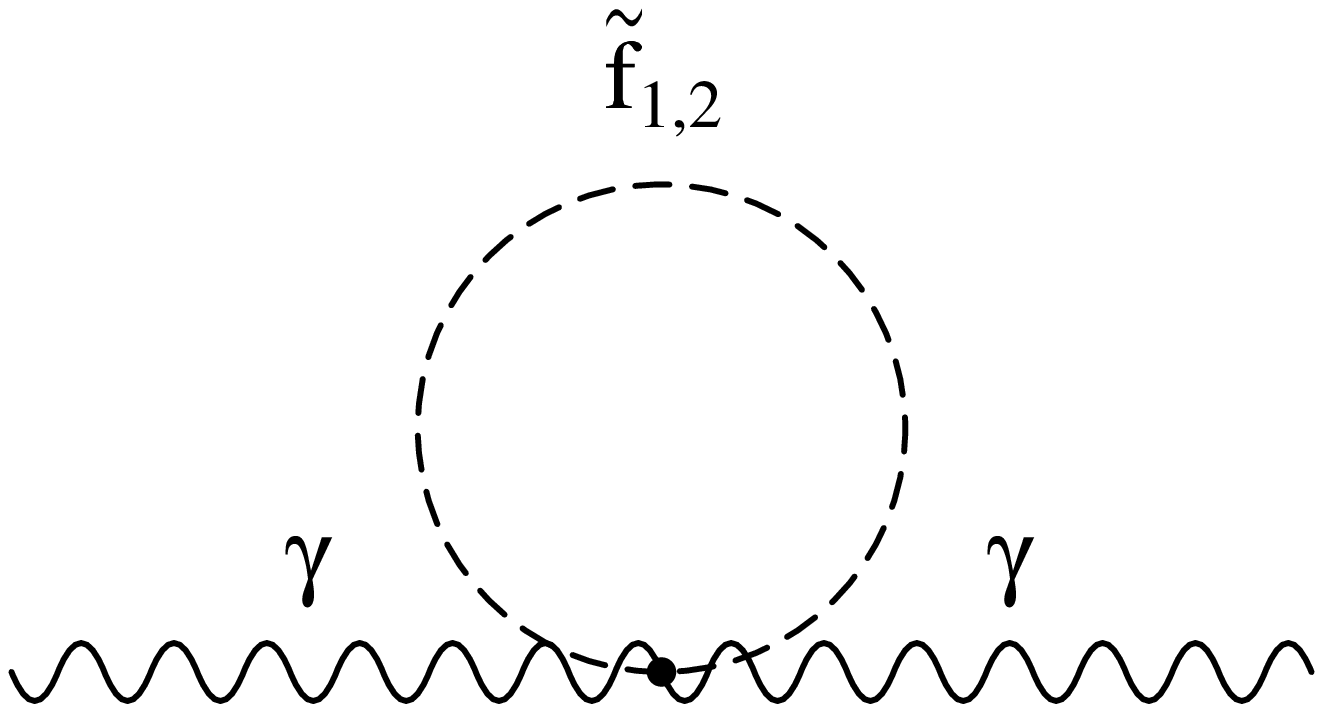,width=4cm}} &
\centerline{\raisebox{-0.7cm}{\psfig{bbllx=107pt,bblly=242pt,%
bburx=505pt,bbury=530pt,figure=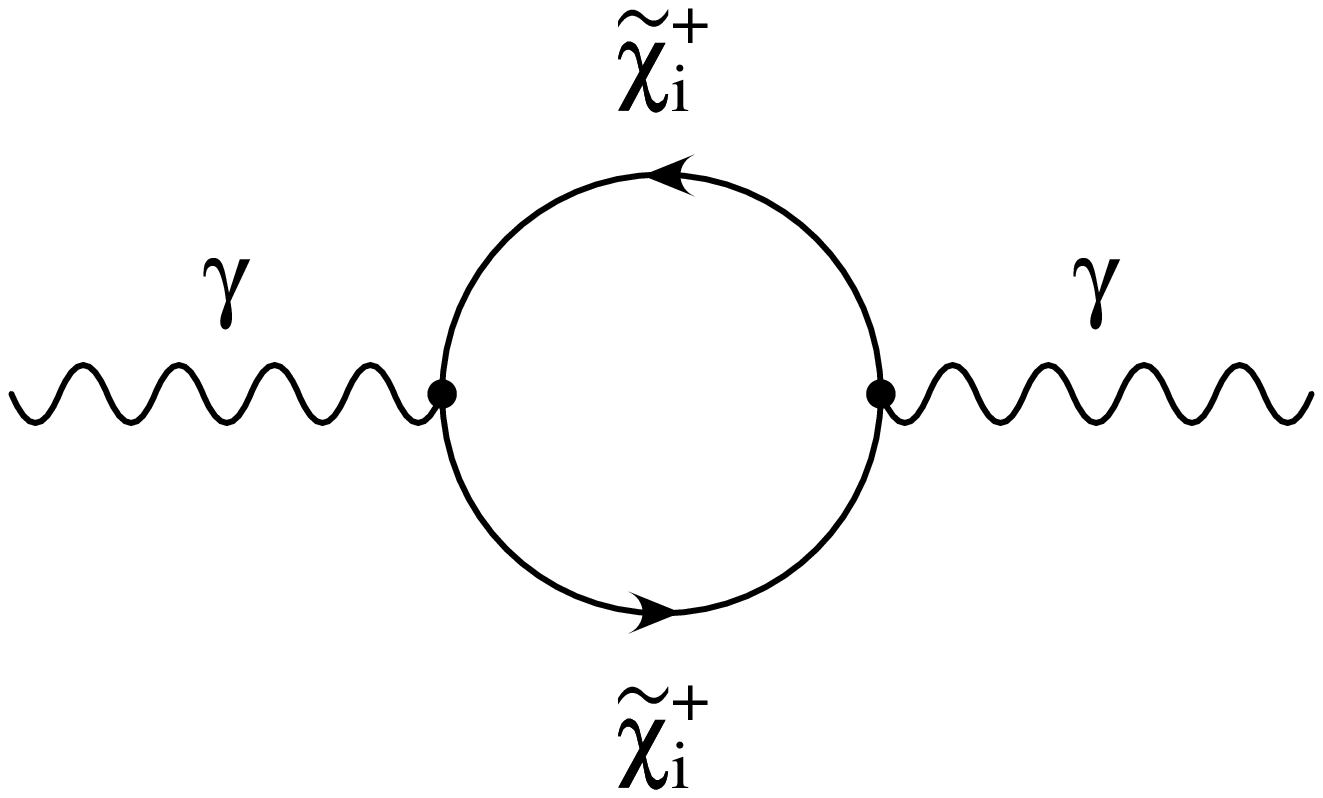,width=4cm}}}
\end{tabular}
\BEA
\Sigma^{\gamma\gamma}_{\tilde{f}}(p^2) &=& 
    -\frac{\alpha}{4\pi}\, 2\, \sum_{f}\, N_C^f\,
    Q^2_f\, \bigg\{ 2 B_{22}(p^2,m_{\tilde{f}_1},m_{\tilde{f}_1}) 
    + 2 B_{22}(p^2,m_{\tilde{f}_2},m_{\tilde{f}_2}) \nonumber \\
& & - A_0(m_{\tilde{f}_1}) - A_0(m_{\tilde{f}_2}) \bigg\} \nonumber \\
\Sigma^{\gamma\gamma}_{\tilde{\chi}}(p^2) &=& 
     -\frac{\alpha}{4\pi}\, \sum_{i=1}^2\, 
     F_1(p^2,m_{\tilde{\chi}^+_i},m_{\tilde{\chi}^+_i},
     \frac{1}{2},\frac{1}{2},\frac{1}{2},\frac{1}{2})~.
\EEA

\newpage

\subsubsection{$Z$ self-energy}

\begin{tabular}{p{4.2cm}@{\hspace{0.5cm}\hspace{0.5cm}}p{4.2cm}@{\hspace{0.5cm}\hspace{0.5cm}}p{4.2cm}}
\centerline{\raisebox{-0.7cm}{\psfig{bbllx=107pt,bblly=242pt,%
bburx=505pt,bbury=521pt,figure=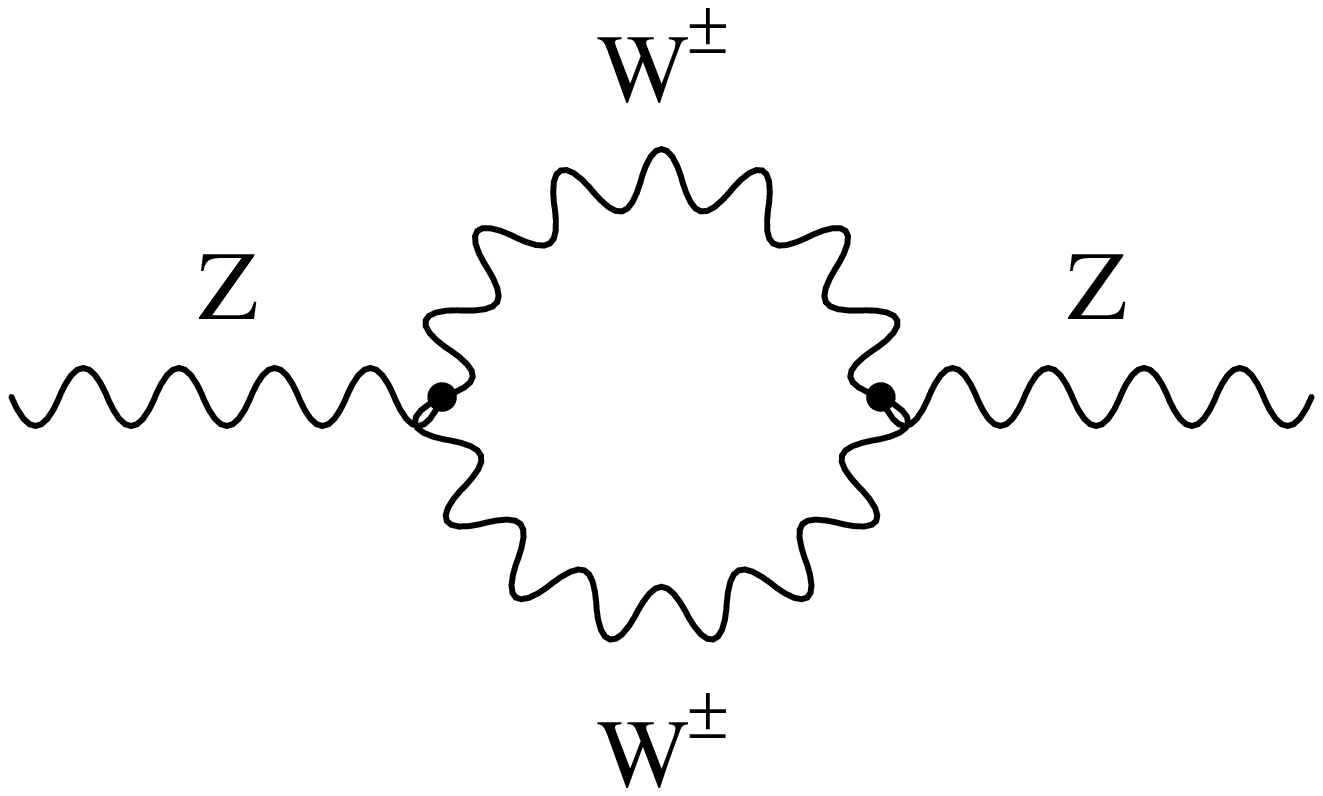,width=4cm}}} &
\centerline{\raisebox{-0.7cm}{\psfig{bbllx=107pt,bblly=242pt,%
bburx=505pt,bbury=530pt,figure=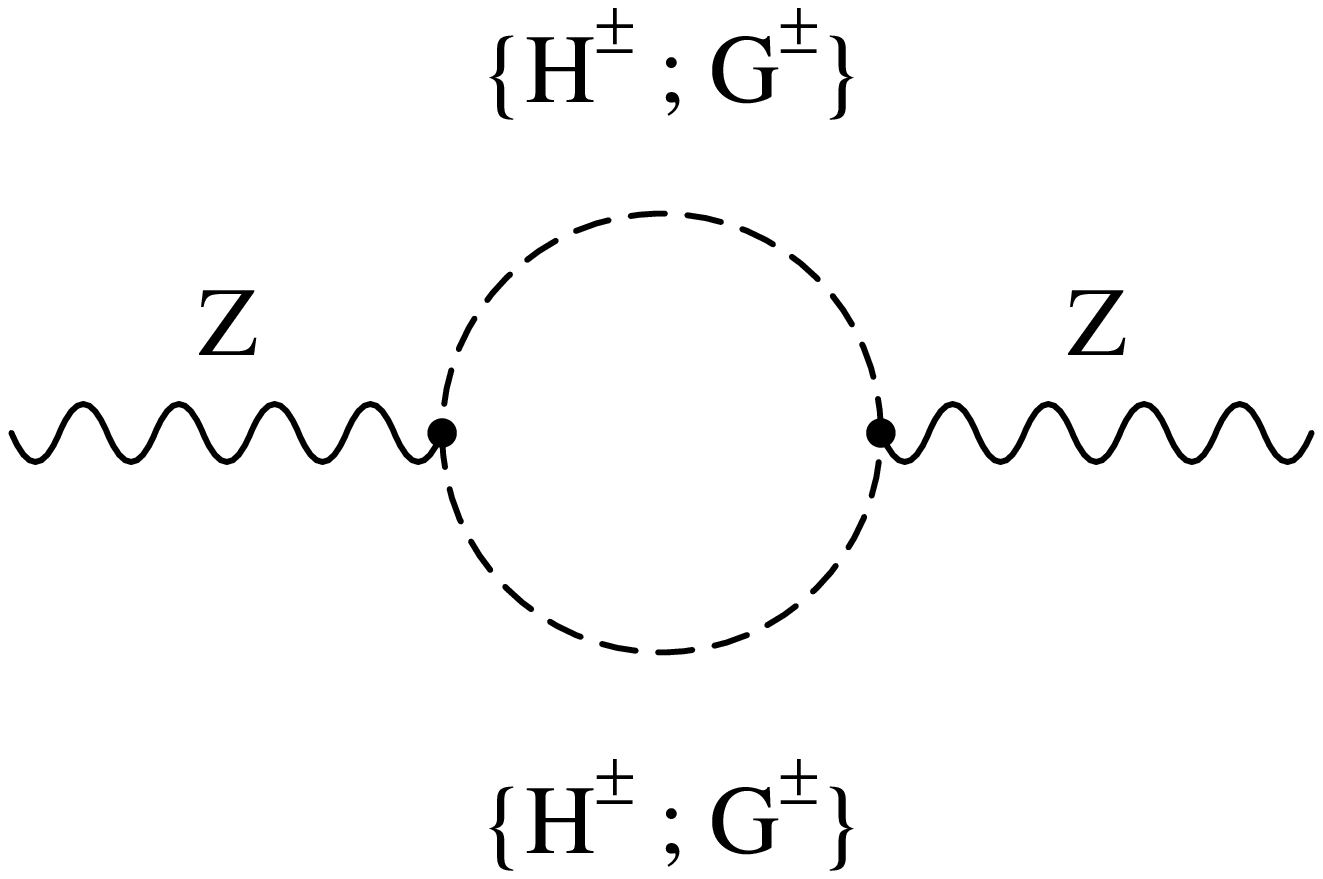,width=4cm}}} &
\centerline{\raisebox{-0.7cm}{\psfig{bbllx=107pt,bblly=242pt,%
bburx=505pt,bbury=521pt,figure=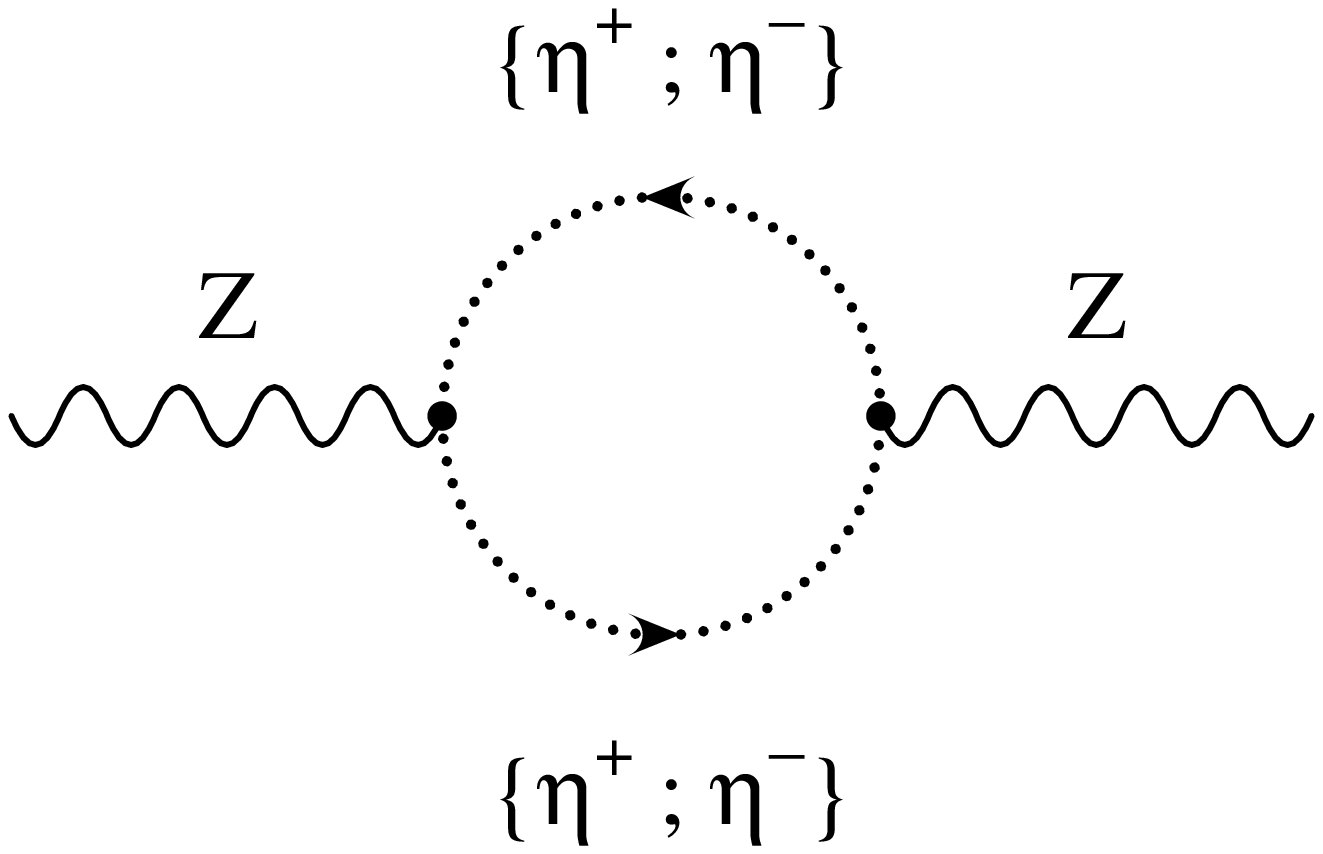,width=4cm}}} \\
\centerline{\psfig{bbllx=107pt,bblly=384pt,bburx=505pt,bbury=599pt,%
figure=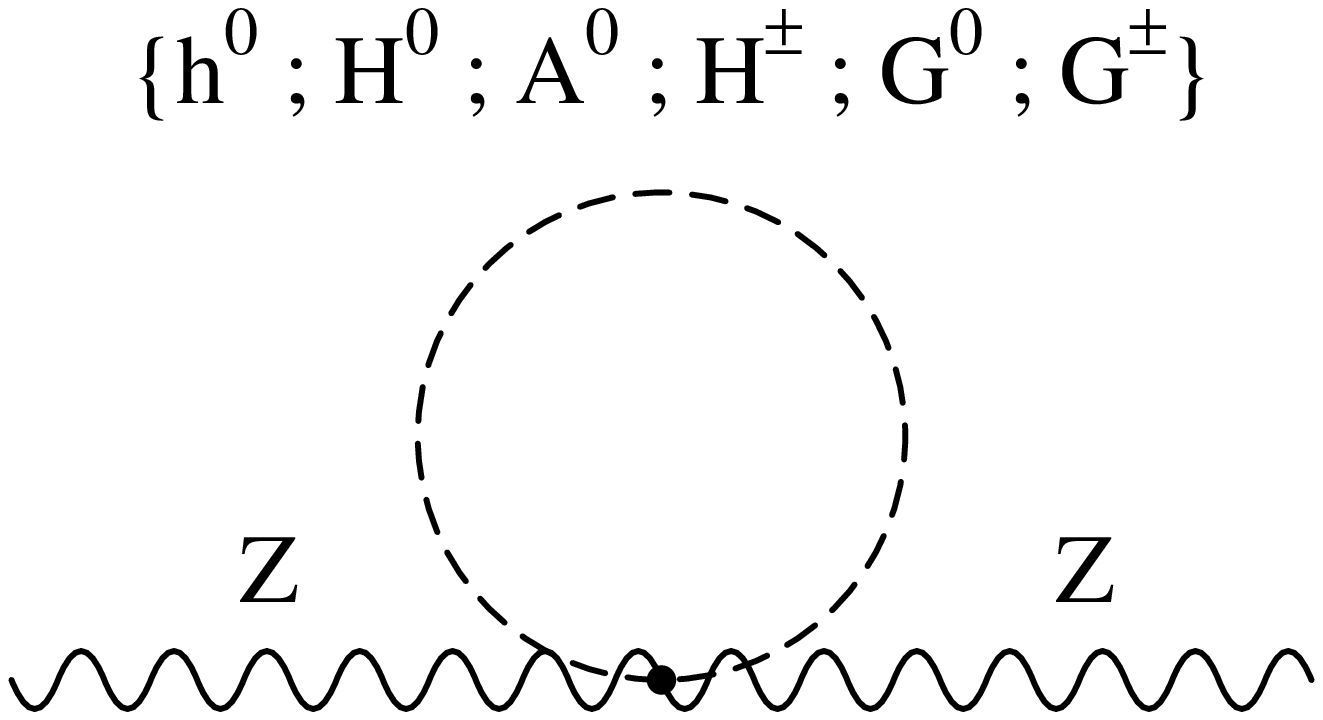,width=4cm}} &
\centerline{\psfig{bbllx=107pt,bblly=384pt,bburx=505pt,bbury=607pt,%
figure=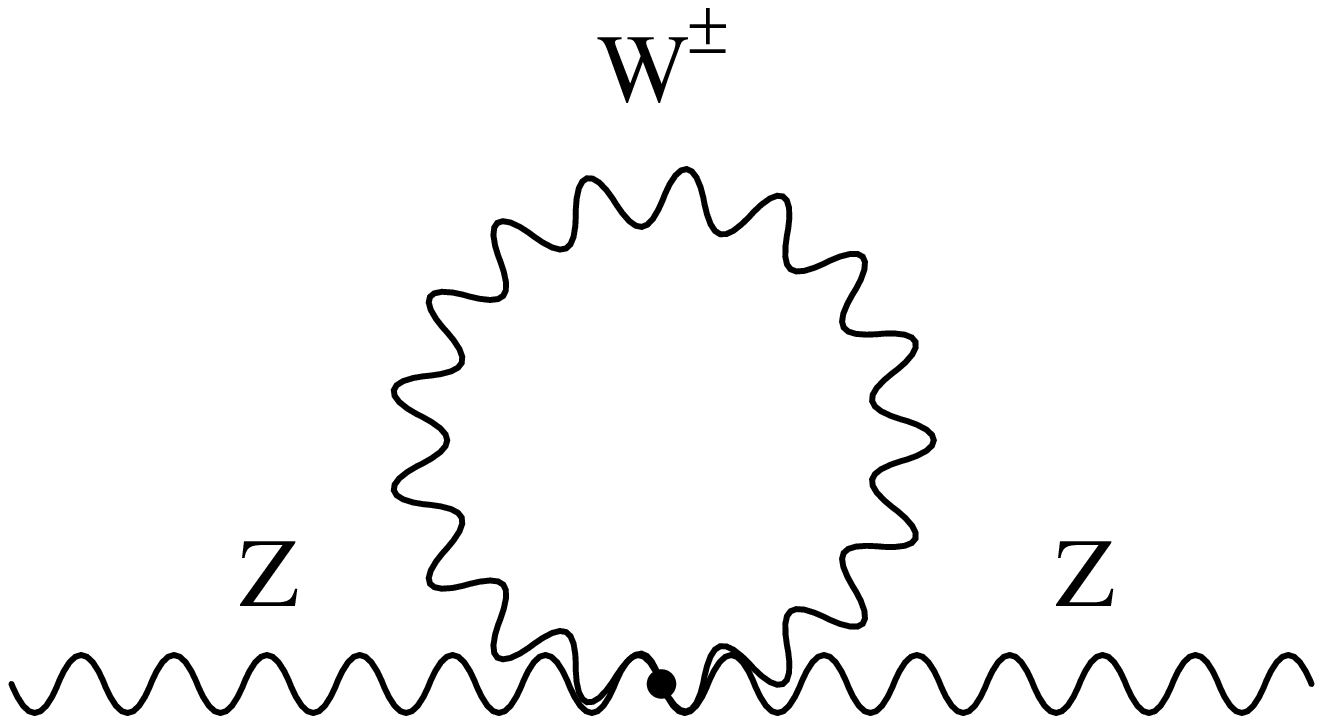,width=4cm}} &
\centerline{\raisebox{-0.7cm}{\psfig{bbllx=107pt,bblly=242pt,%
bburx=505pt,bbury=538pt,figure=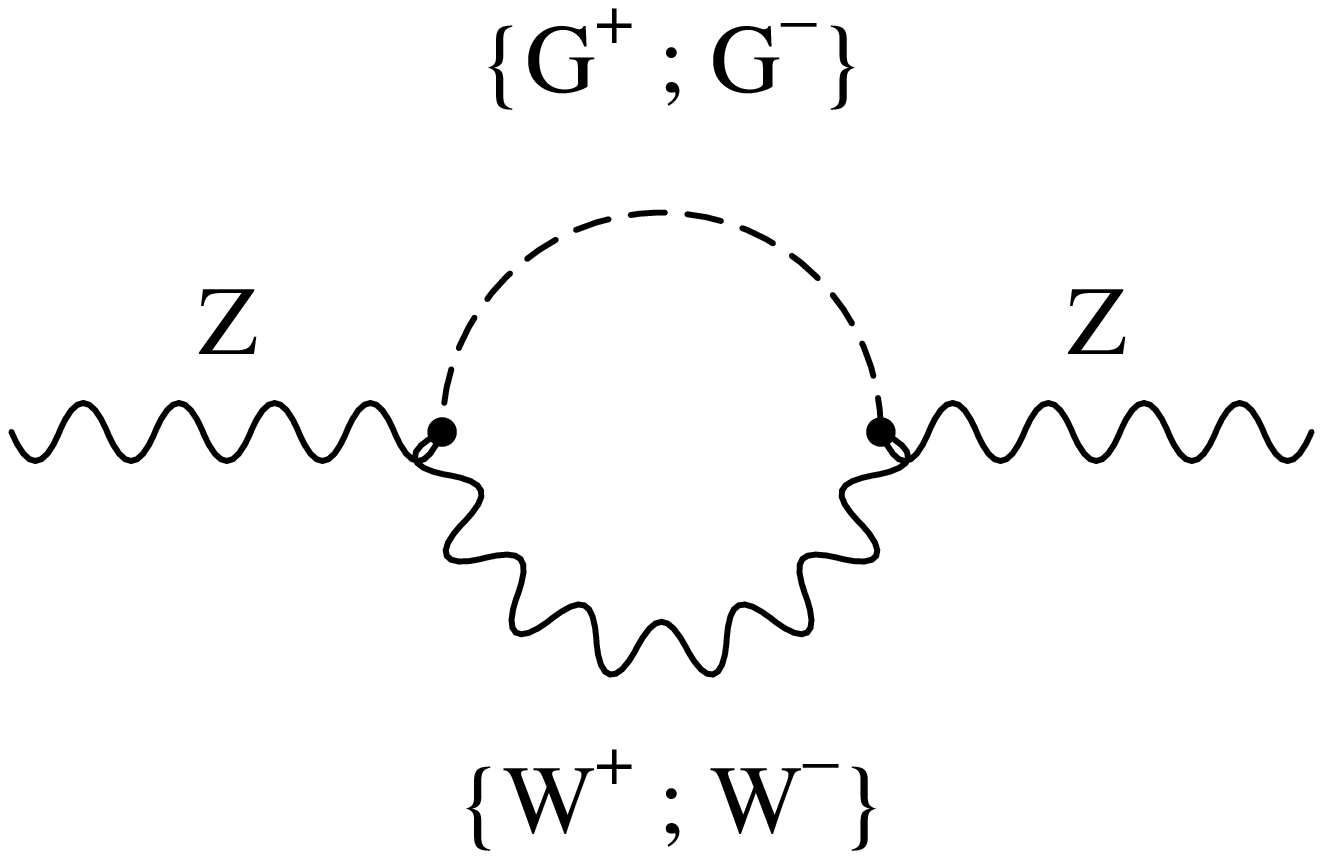,width=4cm}}} \\
\centerline{\raisebox{-0.7cm}{\psfig{bbllx=107pt,bblly=242pt,%
bburx=505pt,bbury=521pt,figure=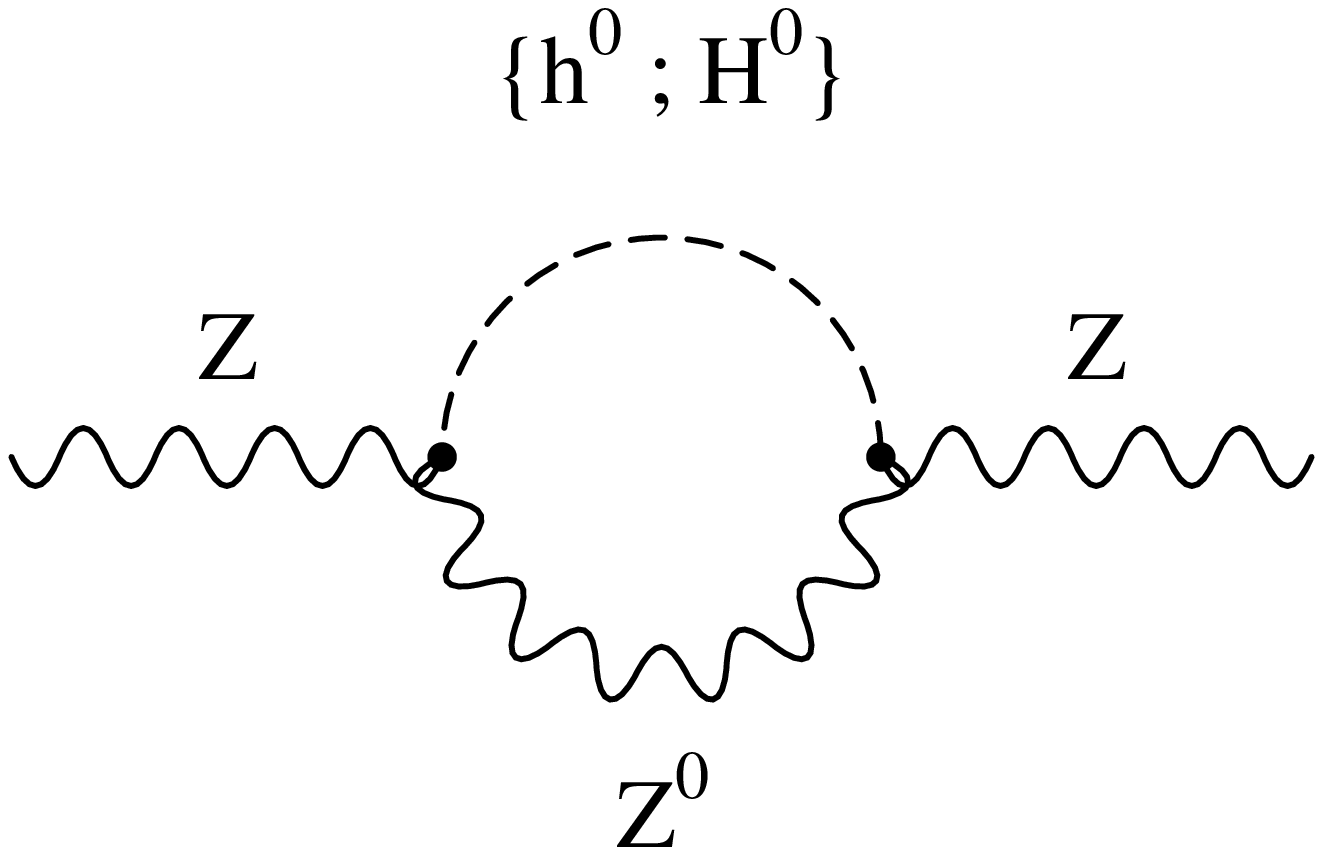,width=4cm}}} &
\centerline{\raisebox{-0.7cm}{\psfig{bbllx=107pt,bblly=242pt,%
bburx=505pt,bbury=521pt,figure=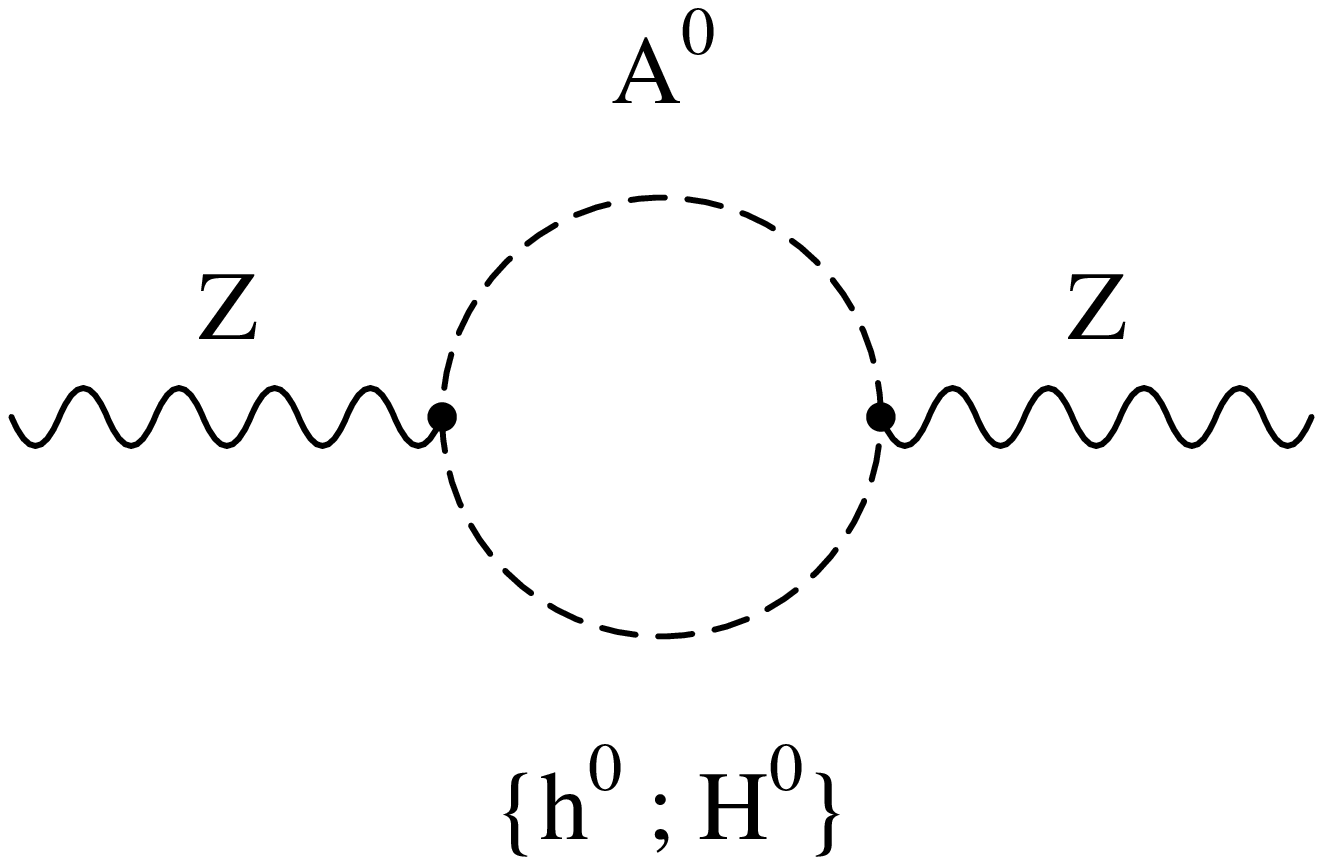,width=4cm}}} & 
\centerline{\raisebox{-0.7cm}{\psfig{bbllx=107pt,bblly=242pt,%
bburx=505pt,bbury=521pt,figure=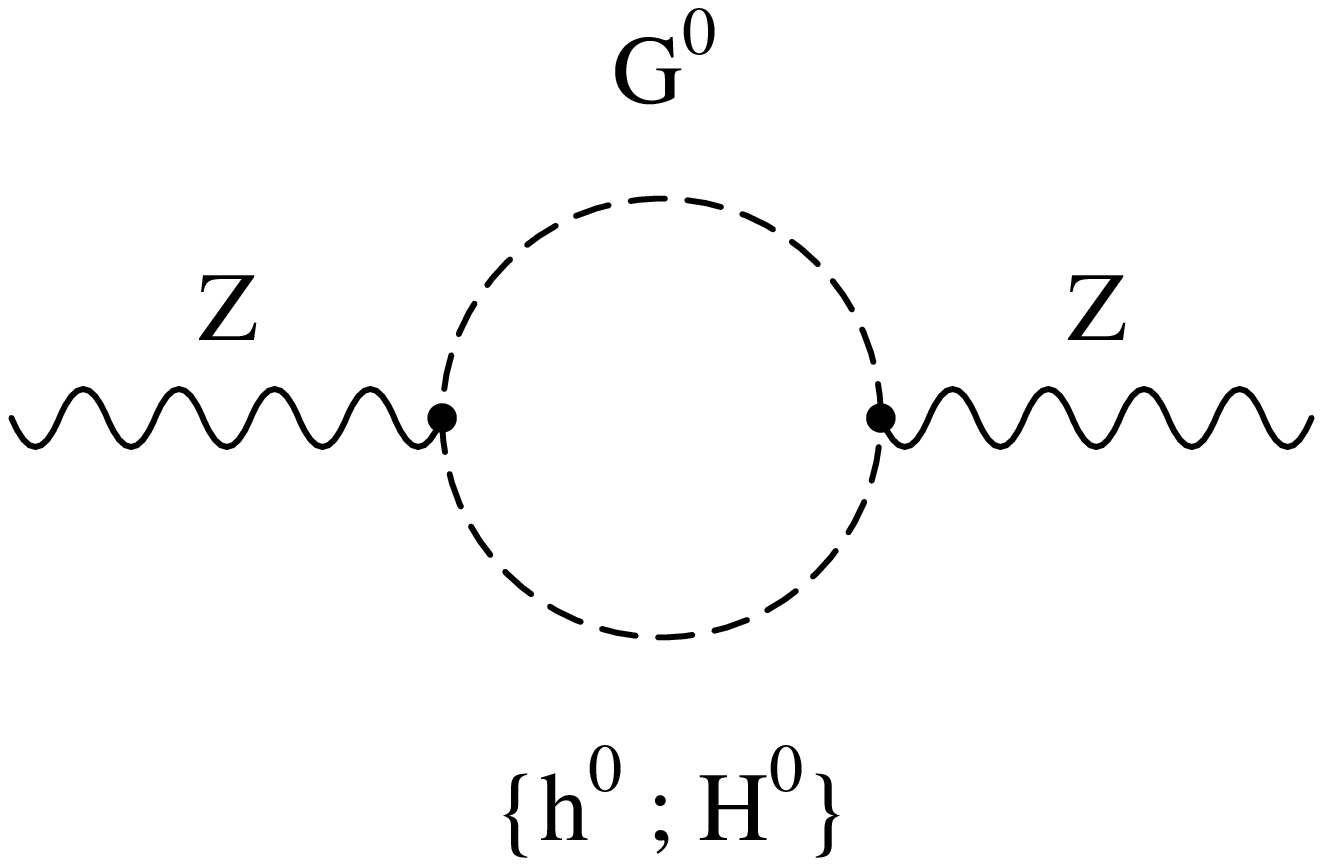,width=4cm}}} \\
\centerline{\raisebox{-0.7cm}{\psfig{bbllx=107pt,bblly=242pt,%
bburx=505pt,bbury=521pt,figure=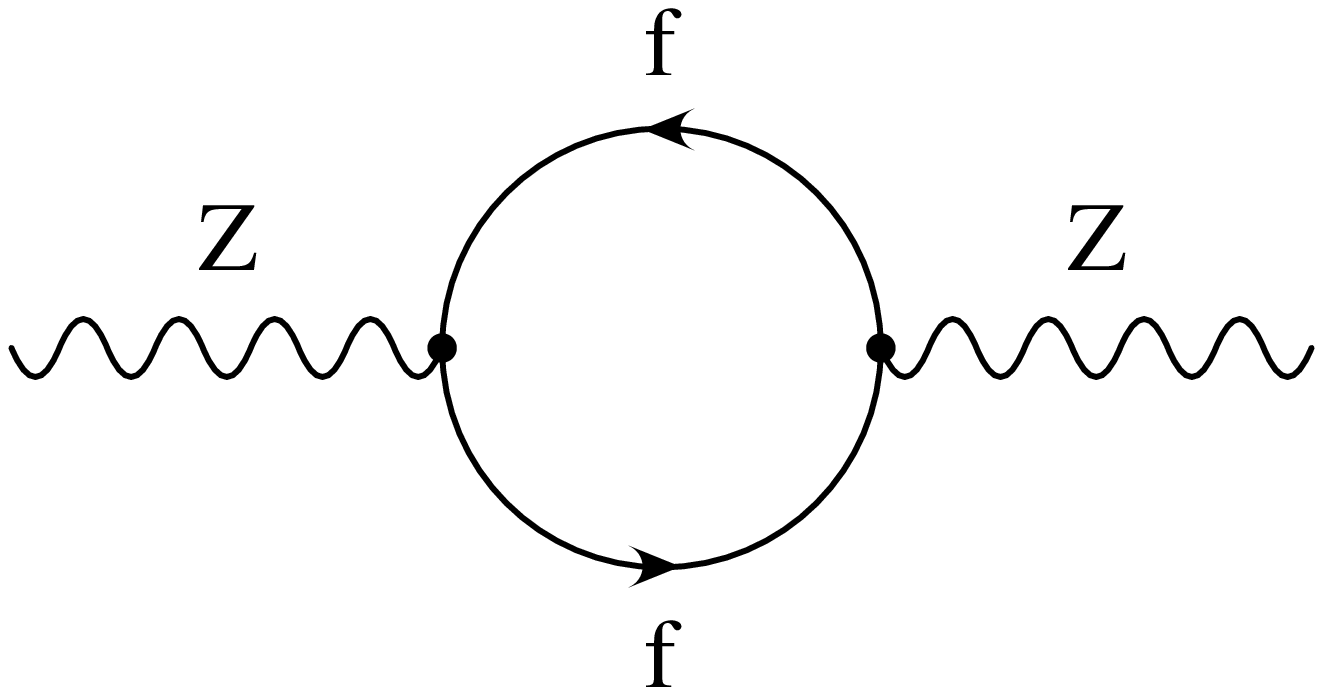,width=4cm}}} 
\end{tabular}
\BEA
\Sigma^{ZZ}_{2HD}(p^2) &=& \frac{\alpha}{4\pi}\,
    \frac{1}{\sw^2\cw^2}\, \bigg\{ -\SQba \Big[ B_{22}(p^2,\MA,m_{H^0})
    + B_{22}(p^2,M_Z,m_{h^0}) \Big] \nonumber \\
& & - \CQba \Big[ B_{22}(p^2,\MA,m_{h^0})
    + B_{22}(p^2,M_Z,m_{H^0}) \Big] \nonumber \\
& & - \cos^2 2\theta_w \Big[ B_{22}(p^2,m_{H^{\pm}},m_{H^{\pm}})
    + B_{22}(p^2,M_W,M_W) \Big] \nonumber \\
& & + 2\cw^4 B_{22}(p^2,M_W,M_W)
    + \frac{1}{4} \Big[ A_0(m_{h^0})+A_0(m_{H^0})+A_0(\MA)
    + A_0(M_Z) \Big] \nonumber \\
& & + \frac{\cos^2 2\theta_w}{2} 
      \Big[ A_0(M_W) + A_0(m_{H^{\pm}}) \Big]  + M_Z^2 \Big[ 
      2\sw^4 \cw^2 B_0(p^2,M_W,M_W) \nonumber \\ 
& & + \SQba B_0(p^2,m_{h^0},M_Z) 
    + \CQba B_0(p^2,m_{H^0},M_Z) \Big] \nonumber \\
& & + \cw^4 \Big[ 6 A_0(M_W)-4 M_W^2 \Big] - \cw^4 F_2(p^2,M_W,M_W) 
      \nonumber \\
& & - \sw^2 \cw^2\, \sum_{f}\, N_C^f\, F_1(p^2,m_f,m_f,
    \frac{v_f+a_f}{2},\frac{v_f-a_f}{2},\frac{v_f+a_f}{2},
    \frac{v_f-a_f}{2})  \bigg\}~,
\EEA

\newpage

\vspace*{2mm}

\begin{tabular}{p{4.2cm}@{\hspace{0.5cm}\hspace{0.5cm}}p{4.2cm}@{\hspace{0.5cm}\hspace{0.5cm}}p{4.2cm}}
\centerline{\raisebox{-0.7cm}{\psfig{bbllx=107pt,bblly=242pt,%
bburx=505pt,bbury=530pt,figure=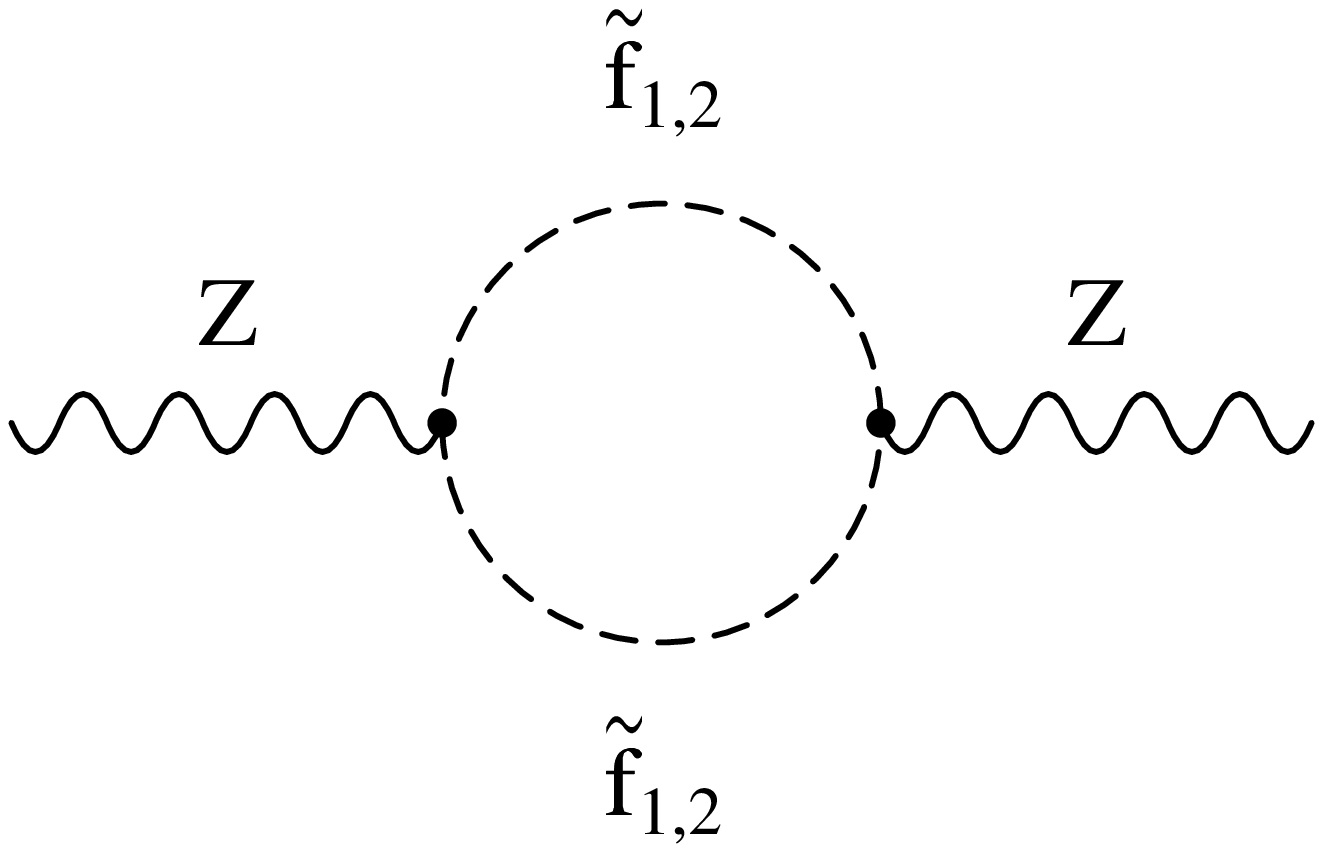,width=4cm}}} &
\centerline{\raisebox{-0.7cm}{\psfig{bbllx=107pt,bblly=242pt,%
bburx=505pt,bbury=530pt,figure=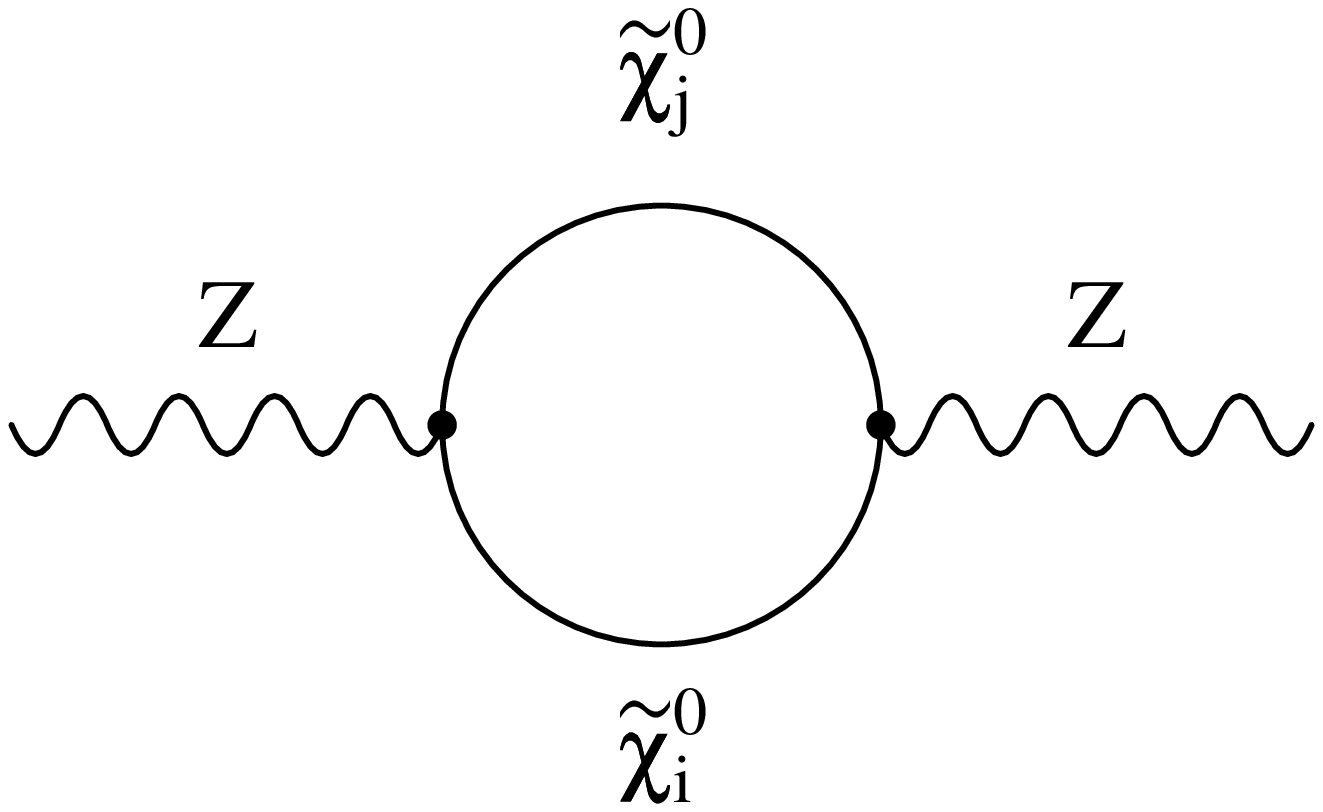,width=4cm}}} &
\centerline{\raisebox{-0.7cm}{\psfig{bbllx=107pt,bblly=242pt,%
bburx=505pt,bbury=530pt,figure=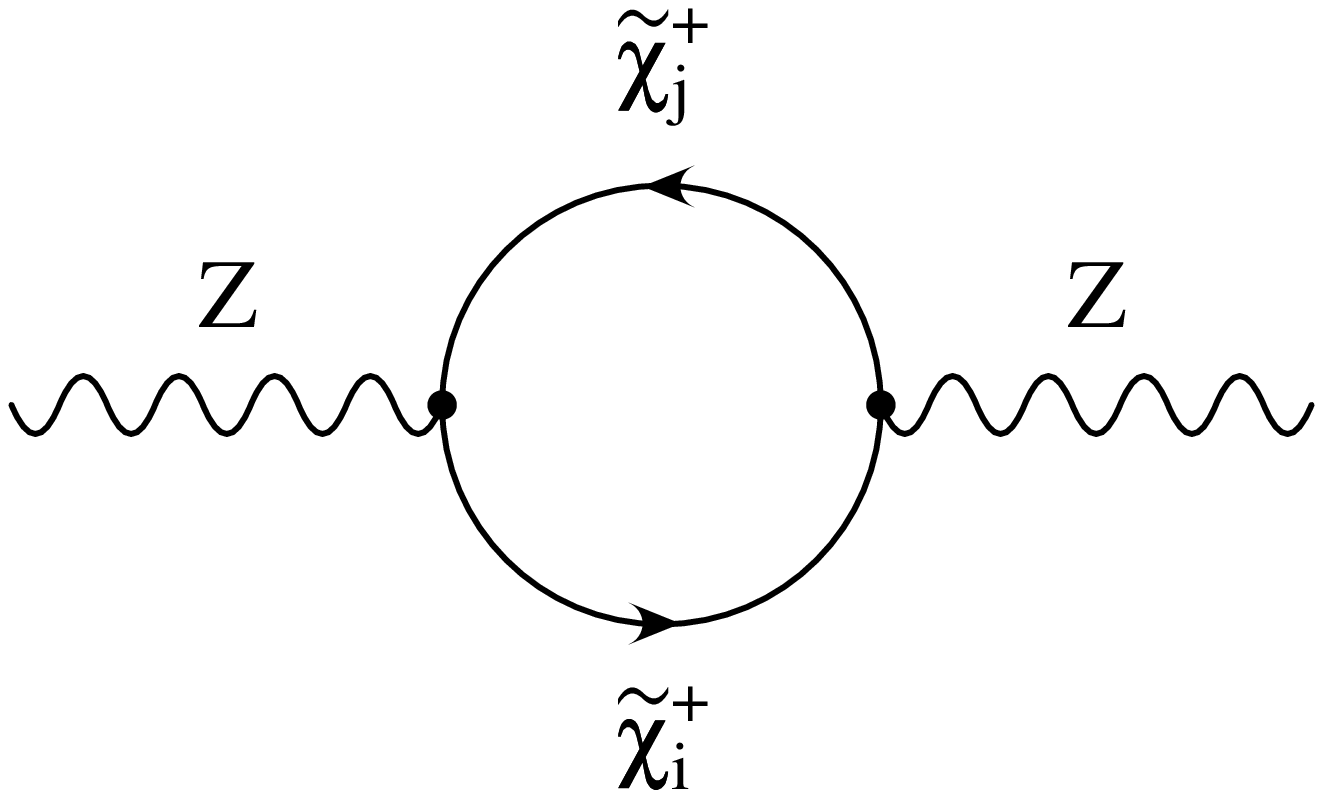,width=4cm}}} \\
\centerline{\psfig{bbllx=107pt,bblly=384pt,%
bburx=505pt,bbury=599pt,figure=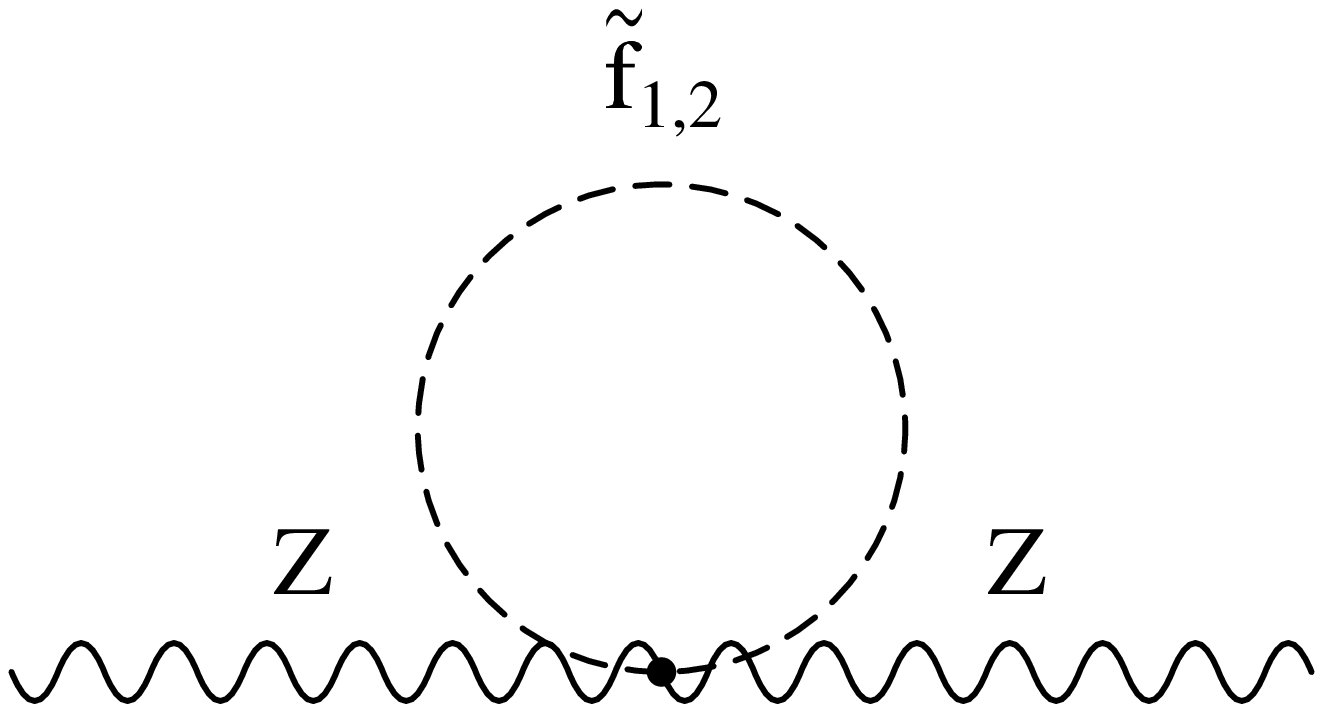,width=4cm}}
\end{tabular}
\BEA
\Sigma^{ZZ}_{\tilde{f}}(p^2) &=& -\frac{\alpha}{4\pi}\, 
    \sum_{f}\, N_C^f\, \bigg\{ \frac{\CtT^2 \StT^2}{\sw^2 \cw^2} 
    \Big[ B_{22}(p^2,m_{\tilde{f}_1},m_{\tilde{f}_2})
    + B_{22}(p^2,m_{\tilde{f}_2},m_{\tilde{f}_1}) \Big] \nonumber \\
& & + \frac{4(I^f_3 \CtT^2-Q_f \sw^2)^2}{\sw^2 \cw^2}
      B_{22}(p^2,m_{\tilde{f}_1},m_{\tilde{f}_1})
    + \frac{4(I^f_3 \StT^2-Q_f \sw^2)^2}{\sw^2 \cw^2}
      B_{22}(p^2,m_{\tilde{f}_2},m_{\tilde{f}_2}) \nonumber \\
& & -2\frac{(I^f_3-Q_f \sw^2)^2 \CtT^2+Q_f^2\sw^4\StT^2}{\sw^2 \cw^2}
     A_0(m_{\tilde{f}_1}) 
    -2\frac{(I^f_3-Q_f \sw^2)^2 \StT^2+Q_f^2\sw^4\CtT^2}{\sw^2 \cw^2}
     A_0(m_{\tilde{f}_2}) \bigg\} \nonumber \\
\Sigma^{ZZ}_{\tilde{\chi}}(p^2) &=& -\frac{\alpha}{4\pi}\,
    \frac{1}{4\sw^2\cw^2}\, \bigg\{
    \frac{1}{2}\, \sum_{i,j=1}^4 
    F_1(p^2,m_{\tilde{\chi}^0_j},m_{\tilde{\chi}^0_i},
    O''^L_{ij},O''^R_{ij},O''^L_{ji},O''^R_{ji})
    \nonumber \\
& & +\sum_{i,j=1}^2 F_1(p^2,m_{\tilde{\chi}^+_j},
    m_{\tilde{\chi}^+_i},O'^L_{ij},O'^R_{ij},O'^L_{ji},O'^R_{ji})
    \bigg\}~,    
\EEA

\vspace{5mm}

\vorn with the vector- and axial vector couplings
\BE
v_f = \frac{I_3^f-2\sw^2 Q_f}{2\sw\cw}~, \qquad
a_f = \frac{I_3^f}{2\sw\cw}~,
\label{vfaf}
\EE
and $\CtT:=\cos \tilde{\theta}_f$, $\StT:=\sin \tilde{\theta}_f$.
The coefficients of the chargino- and neutralino couplings are given by
the expressions
\BEA
O'^L_{ij} &=& -V_{i1} V^*_{j1} - \frac{1}{2} V_{i2} V^*_{j2} 
              + \delta_{ij} \sw^2~, \nonumber \\
O'^R_{ij} &=& -U^*_{i1} U_{j1} - \frac{1}{2} U^*_{i2} U_{j2} 
              + \delta_{ij} \sw^2~, \nonumber \\
O''^L_{ij} &=& - \frac{1}{2} N_{i3} N^*_{j3} 
               + \frac{1}{2} N_{i4} N^*_{j4}~, \nonumber \\
O''^R_{ij} &=& - O''^{L*}_{ij}~,
\label{Os}
\EEA
with the matrices $U_{ij}$, $V_{ij}$ for the charginos and $N_{ij}$ for 
the neutralinos, see Appendices \ref{Charginos}, \ref{Neutralinos}.

\newpage

\subsubsection{$W$ self-energy}

\begin{tabular}{p{4.2cm}@{\hspace{0.5cm}\hspace{0.5cm}}p{4.2cm}@{\hspace{0.5cm}\hspace{0.5cm}}p{4.2cm}}
\centerline{\raisebox{-0.7cm}{\psfig{bbllx=107pt,bblly=242pt,%
bburx=505pt,bbury=521pt,figure=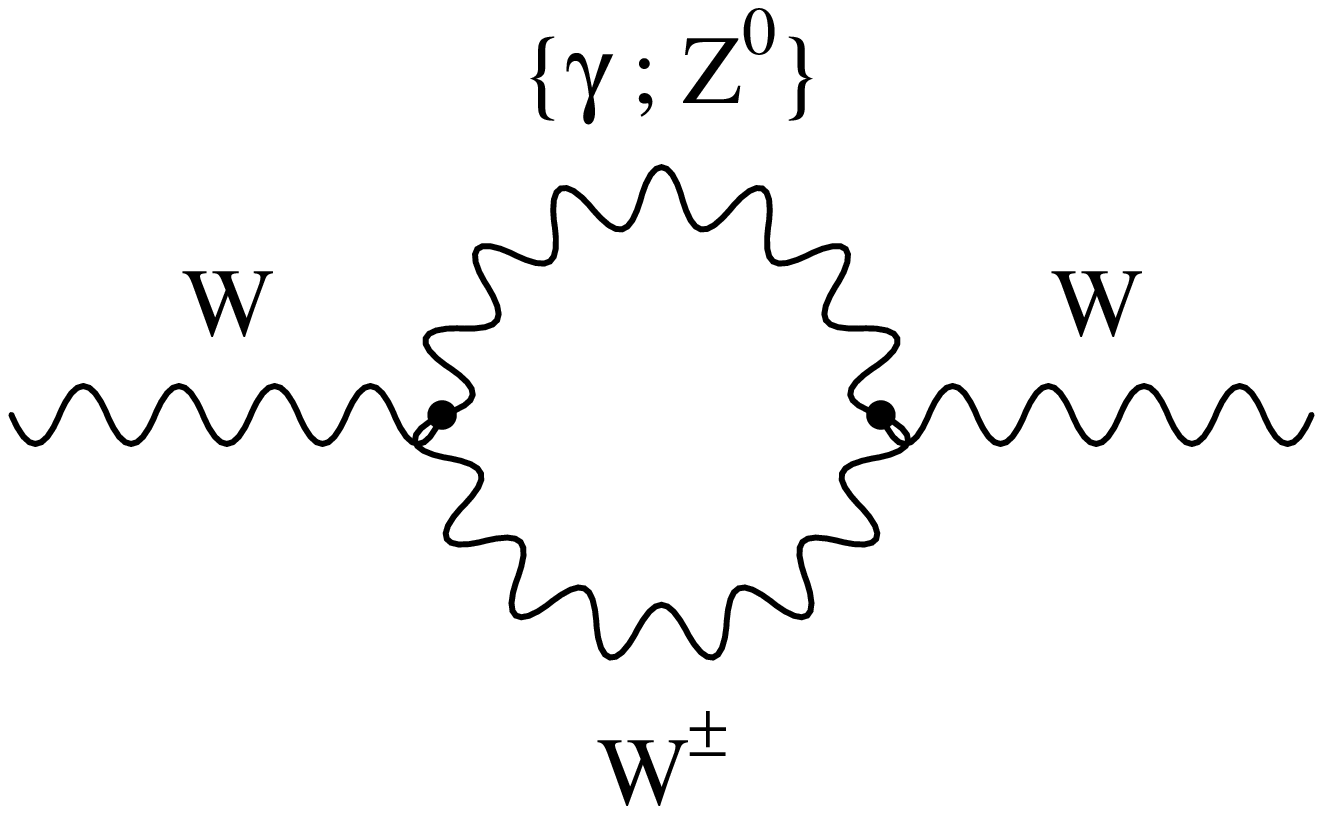,width=4cm}}} &
\centerline{\raisebox{-0.7cm}{\psfig{bbllx=107pt,bblly=242pt,%
bburx=505pt,bbury=530pt,figure=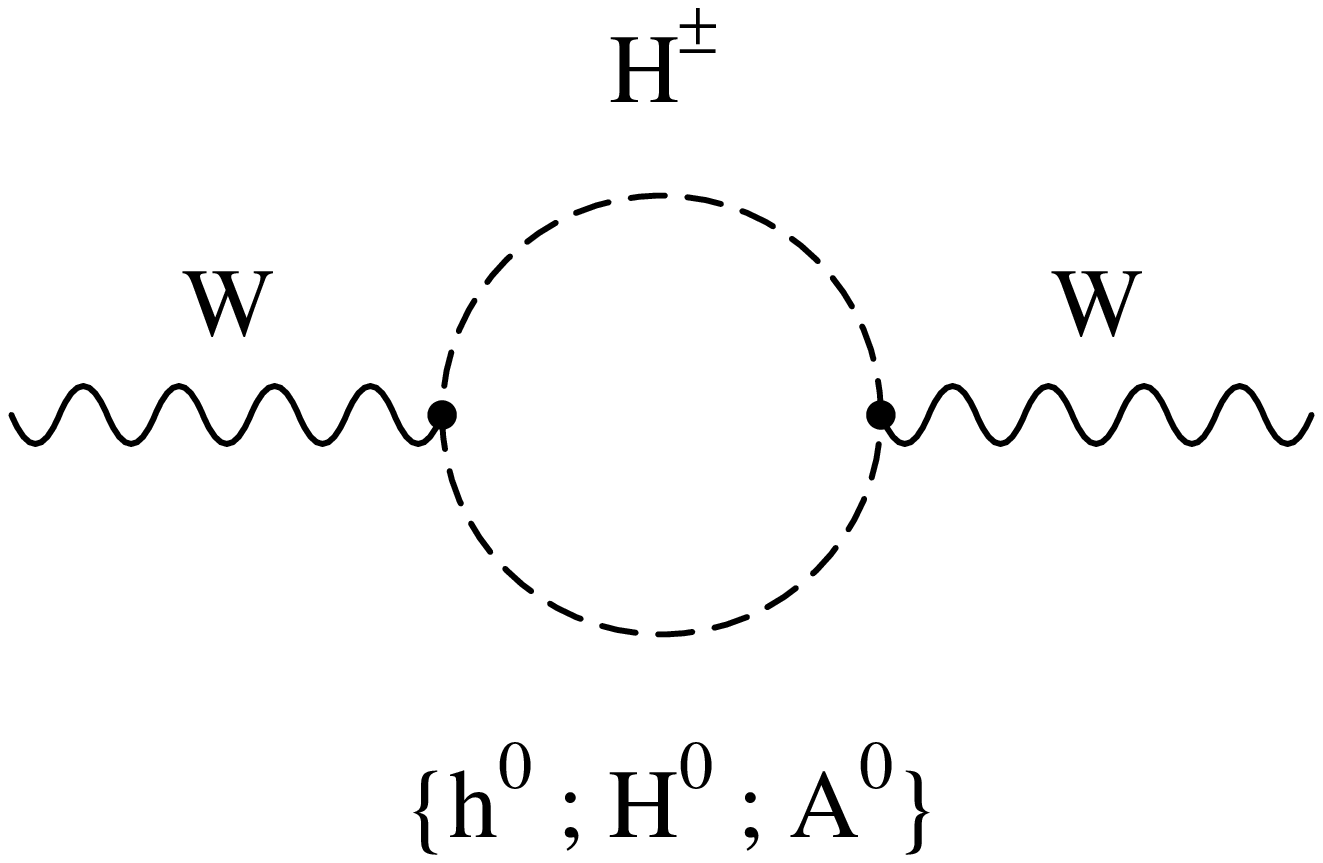,width=4cm}}} &
\centerline{\raisebox{-0.7cm}{\psfig{bbllx=107pt,bblly=242pt,%
bburx=505pt,bbury=521pt,figure=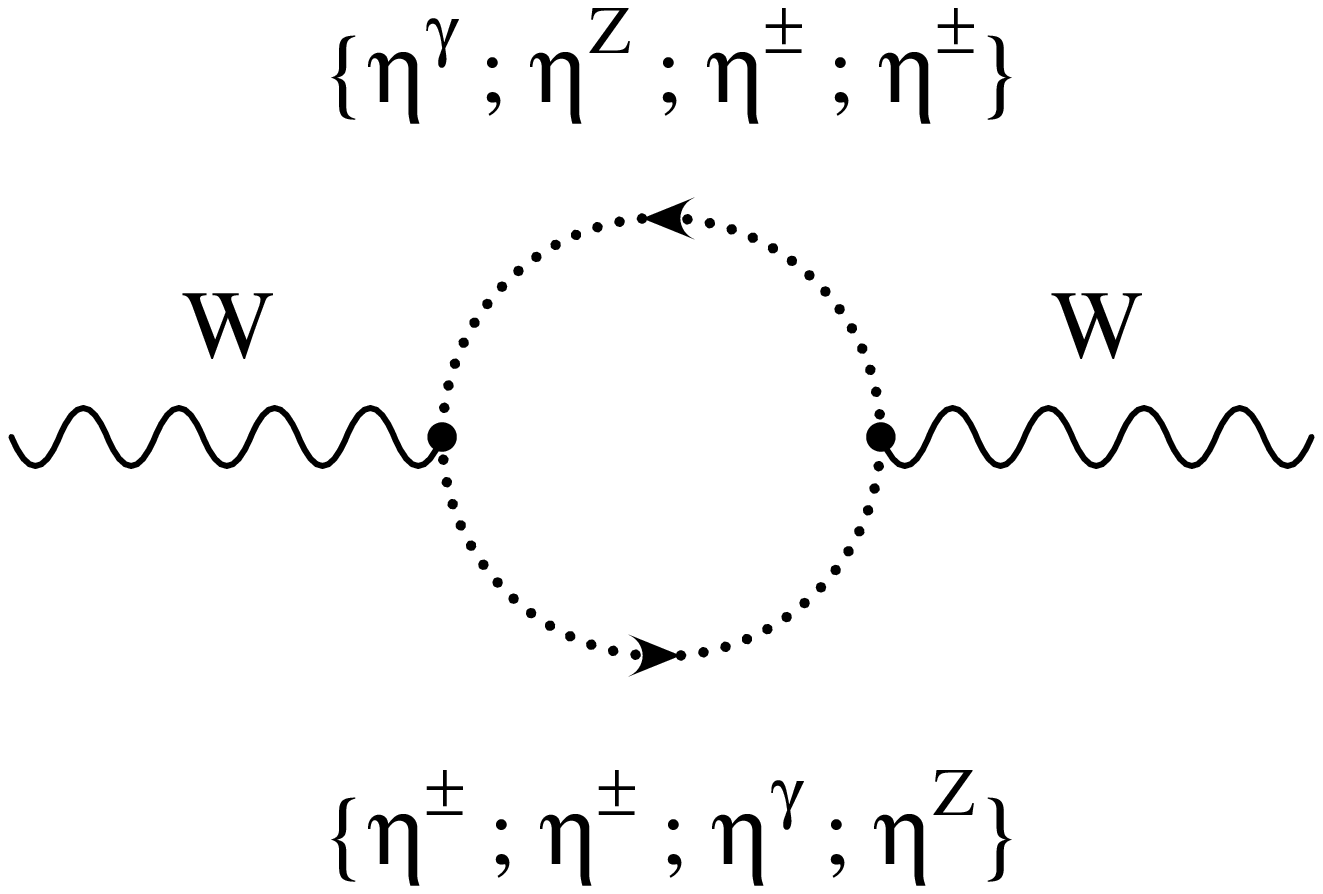,width=4cm}}} \\
\centerline{\psfig{bbllx=107pt,bblly=384pt,bburx=505pt,bbury=599pt,%
figure=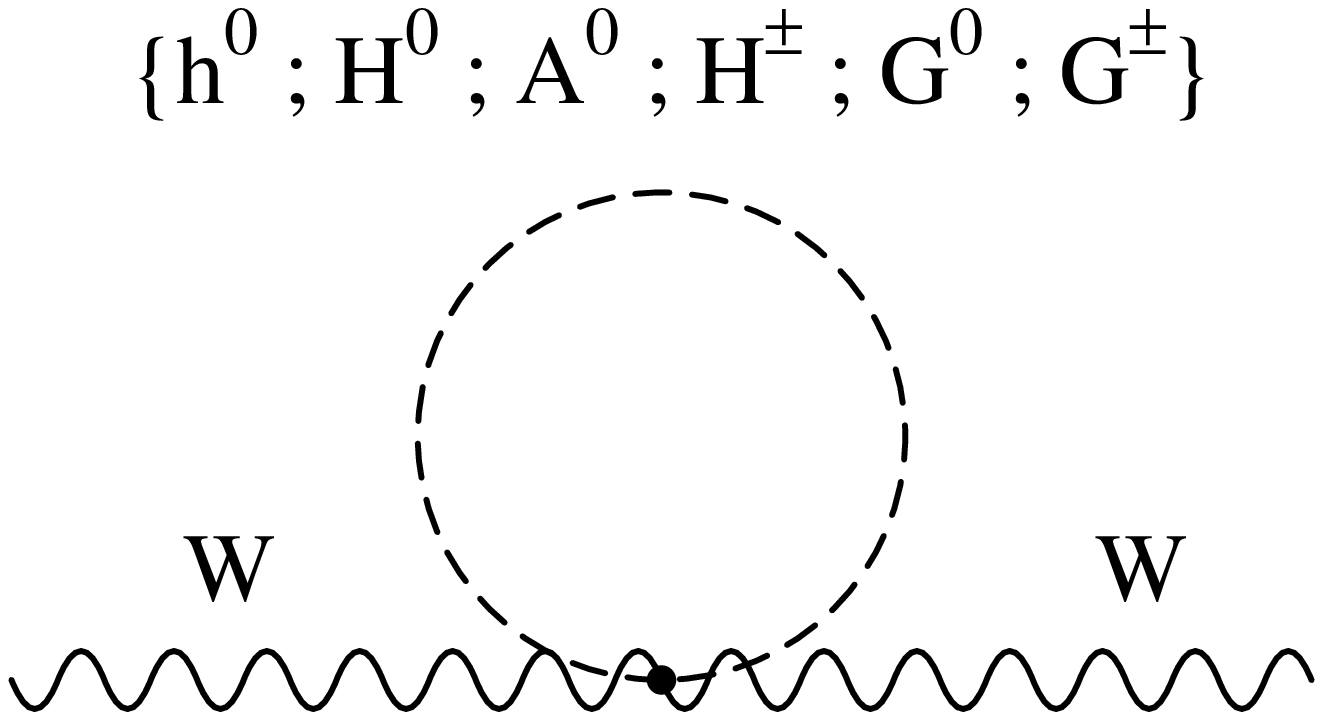,width=4cm}} &
\centerline{\psfig{bbllx=107pt,bblly=384pt,bburx=505pt,bbury=607pt,%
figure=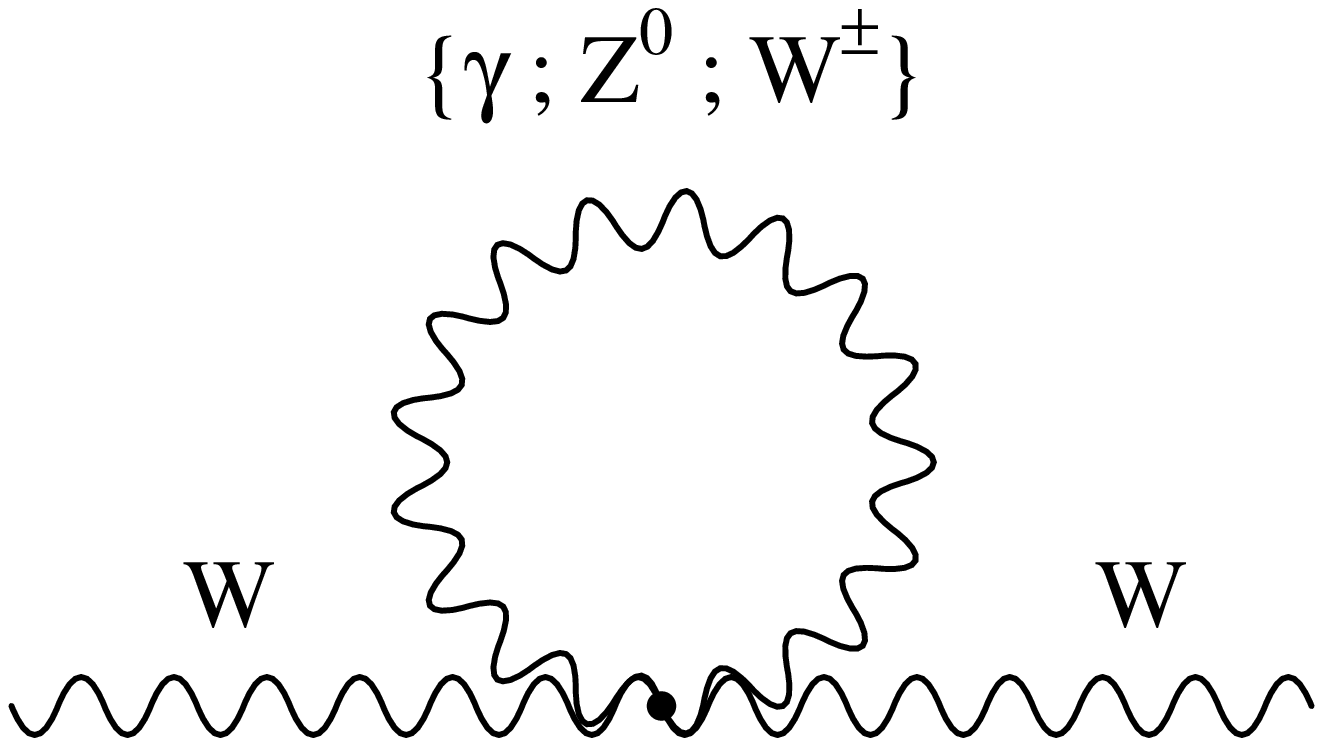,width=4cm}} &
\centerline{\raisebox{-0.7cm}{\psfig{bbllx=107pt,bblly=242pt,%
bburx=505pt,bbury=538pt,figure=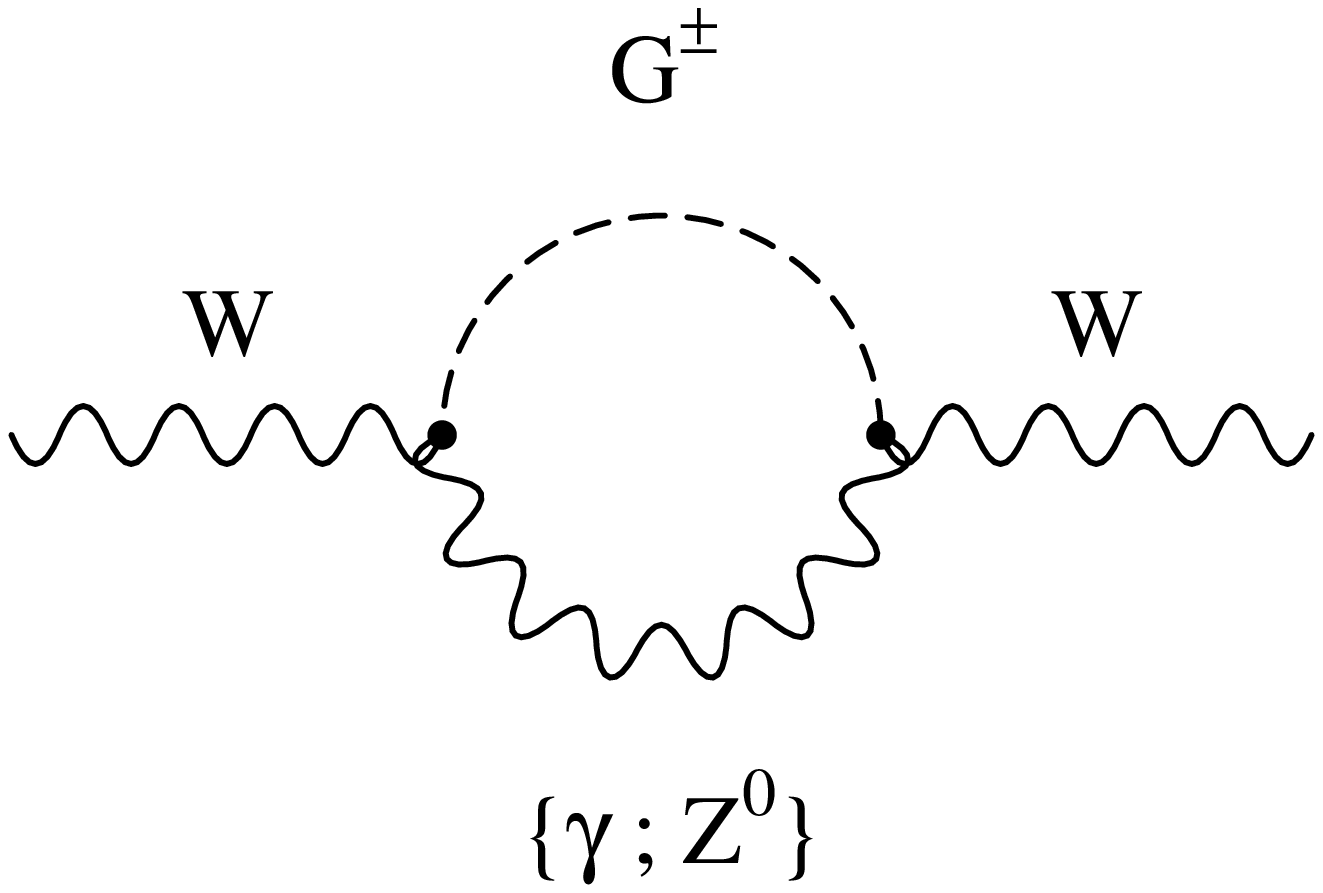,width=4cm}}} \\
\centerline{\raisebox{-0.7cm}{\psfig{bbllx=107pt,bblly=242pt,%
bburx=505pt,bbury=521pt,figure=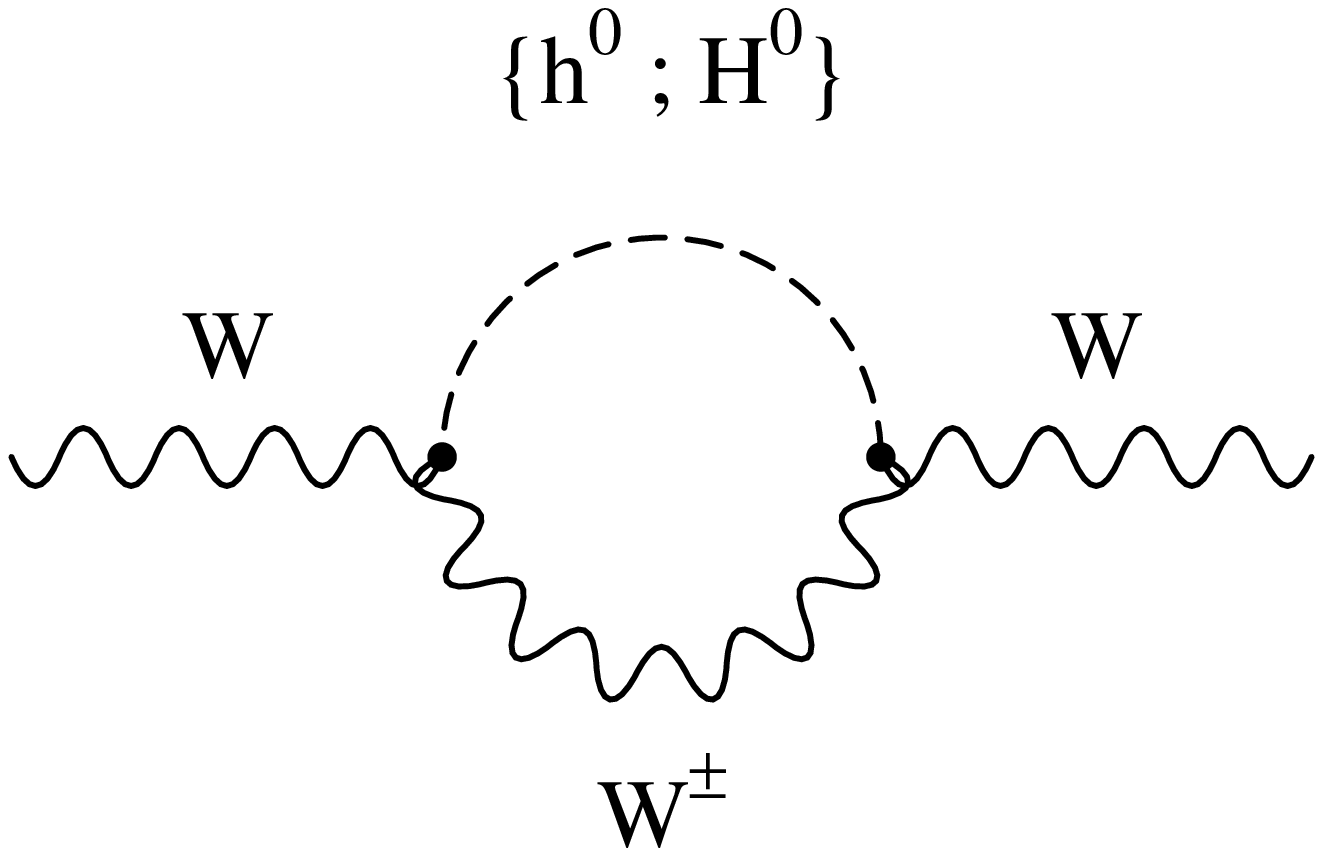,width=4cm}}} &
\centerline{\raisebox{-0.7cm}{\psfig{bbllx=107pt,bblly=242pt,%
bburx=505pt,bbury=521pt,figure=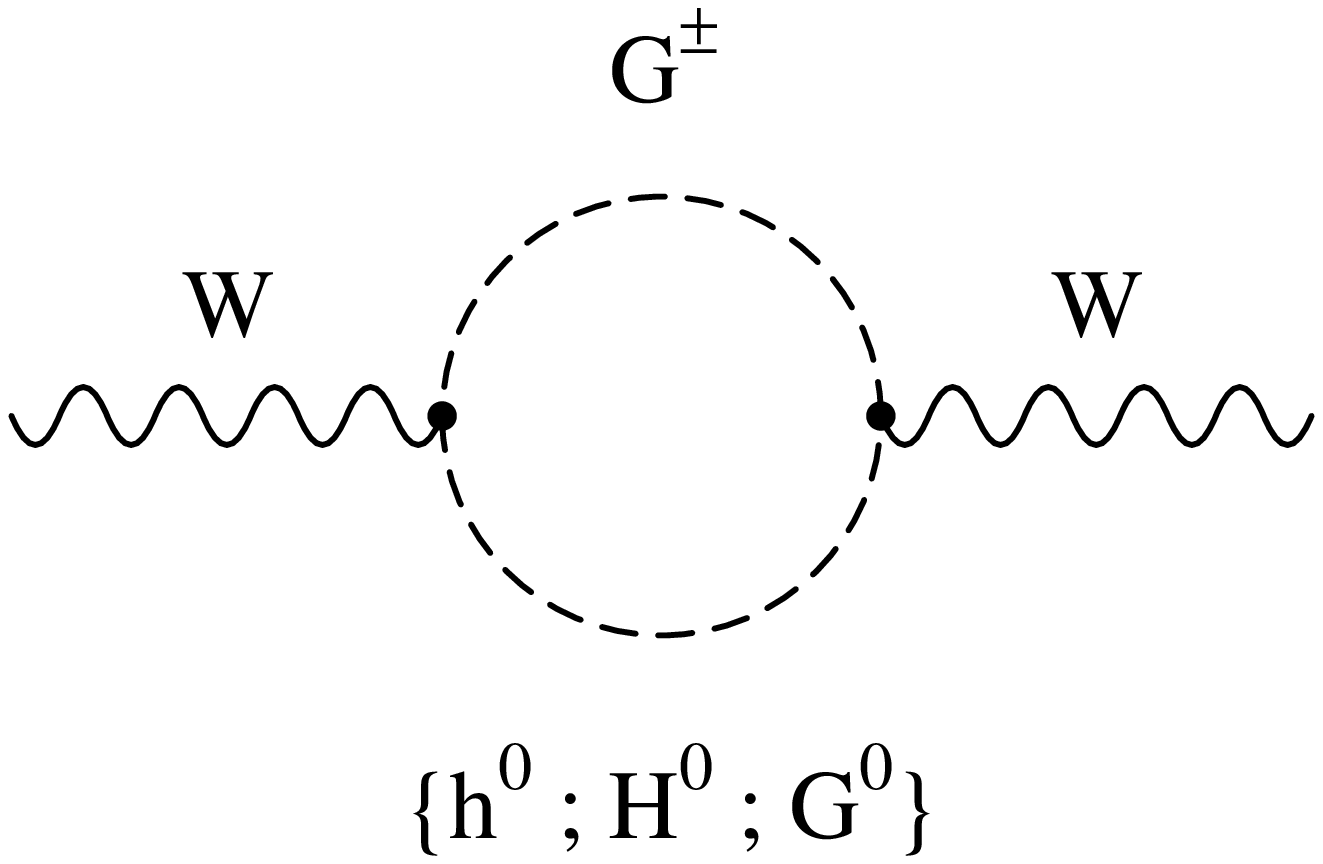,width=4cm}}} &
\centerline{\raisebox{-0.7cm}{\psfig{bbllx=107pt,bblly=242pt,%
bburx=505pt,bbury=521pt,figure=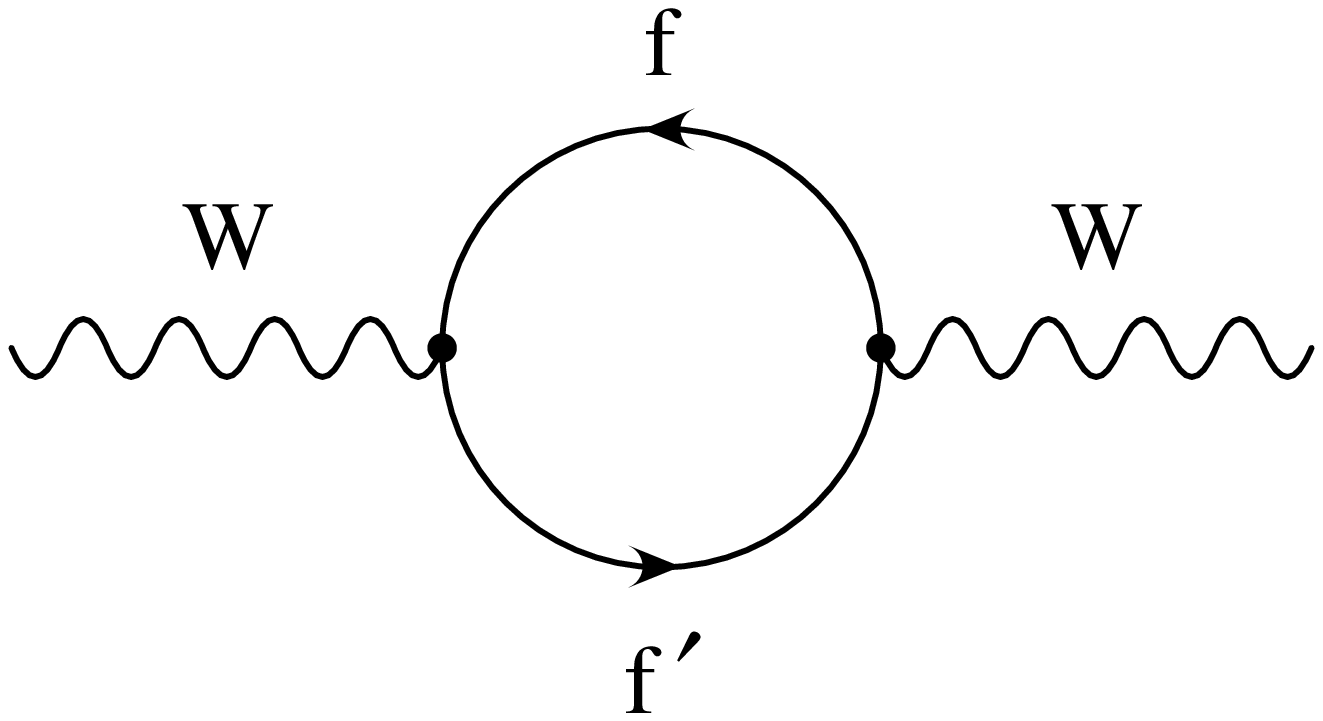,width=4cm}}} 
\end{tabular}

\BEA
\Sigma^{WW}_{2HD}(p^2) &=& \frac{\alpha}{4\pi}\, 
    \frac{1}{\sw^2}\, \bigg\{ -\SQba \Big[B_{22}(p^2,m_{H^{\pm}},m_{H^0})
    + B_{22}(p^2,M_W,m_{h^0}) \Big] \nonumber \\
& & - \CQba \Big[ B_{22}(p^2,m_{H^{\pm}},m_{h^0})
    + B_{22}(p^2,M_W,m_{H^0}) \Big] \nonumber \\
& & - B_{22}(p^2,M_W,M_Z) - B_{22}(p^2,m_{H^{\pm}},\MA) 
      \nonumber \\
& & + 2\sw^2 B_{22}(p^2,0,M_W) + 2\cw^2 B_{22}(p^2,M_W,M_Z)
      \nonumber \\ 
& & + \frac{1}{4} \Big[ A_0(m_{h^0})+A_0(m_{H^0})+A_0(\MA)
    + A_0(M_Z)+2A_0(M_W)+2A_0(m_{H^{\pm}}) \Big] \nonumber \\
& & + M_W^2 \Big[  \SQba B_0(p^2,m_{h^0},M_W) 
    + \CQba B_0(p^2,m_{H^0},M_W) \nonumber \\ 
& & + \sw^2 B_0(p^2,M_W,0) + \frac{\sw^4}{\cw^2}
      B_0(p^2,M_W,M_Z) \Big] \nonumber \\
& & + 3 A_0(M_W) - 2 M_W^2 +\cw^2 \Big[3 A_0(M_Z)-2 M_Z^2 \Big] 
      \nonumber \\ 
& & - \cw^2 F_2(p^2,M_Z,M_W) - \sw^2 F_2(p^2,0,M_W) \nonumber \\
& & - \sum_{doublets} \frac{N_C^f}{2}\, F_1(p^2,m_{f^+},m_{f^-},
      \frac{1}{2},0,\frac{1}{2},0) \bigg\}~,
\EEA

\newpage

\vspace*{2mm}

\begin{tabular}{p{4.2cm}@{\hspace{0.5cm}\hspace{0.5cm}}p{4.2cm}@{\hspace{0.5cm}\hspace{0.5cm}}p{4.2cm}}
\centerline{\raisebox{-0.7cm}{\psfig{bbllx=107pt,bblly=242pt,%
bburx=505pt,bbury=530pt,figure=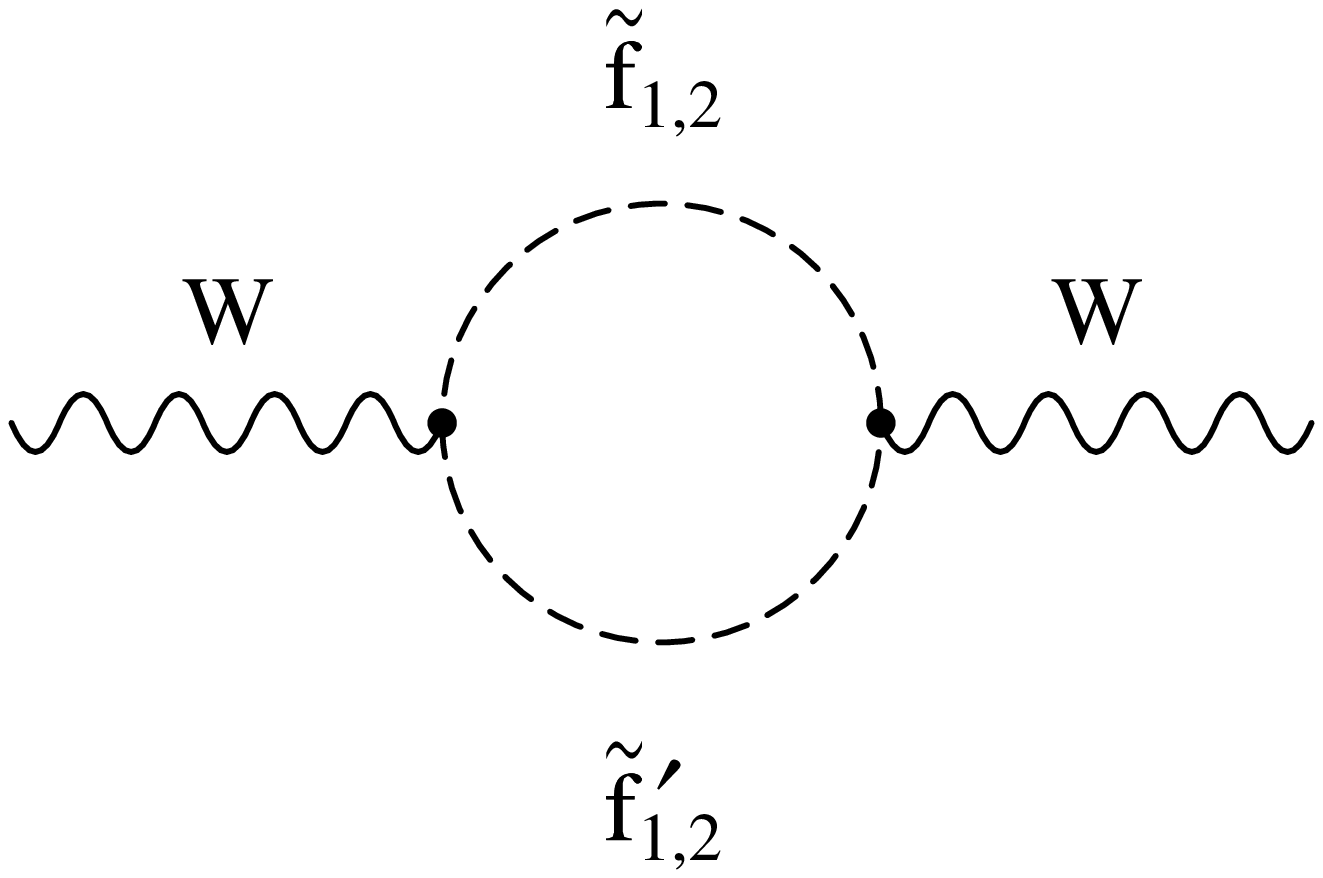,width=4cm}}} &
\centerline{\psfig{bbllx=107pt,bblly=384pt,%
bburx=505pt,bbury=599pt,figure=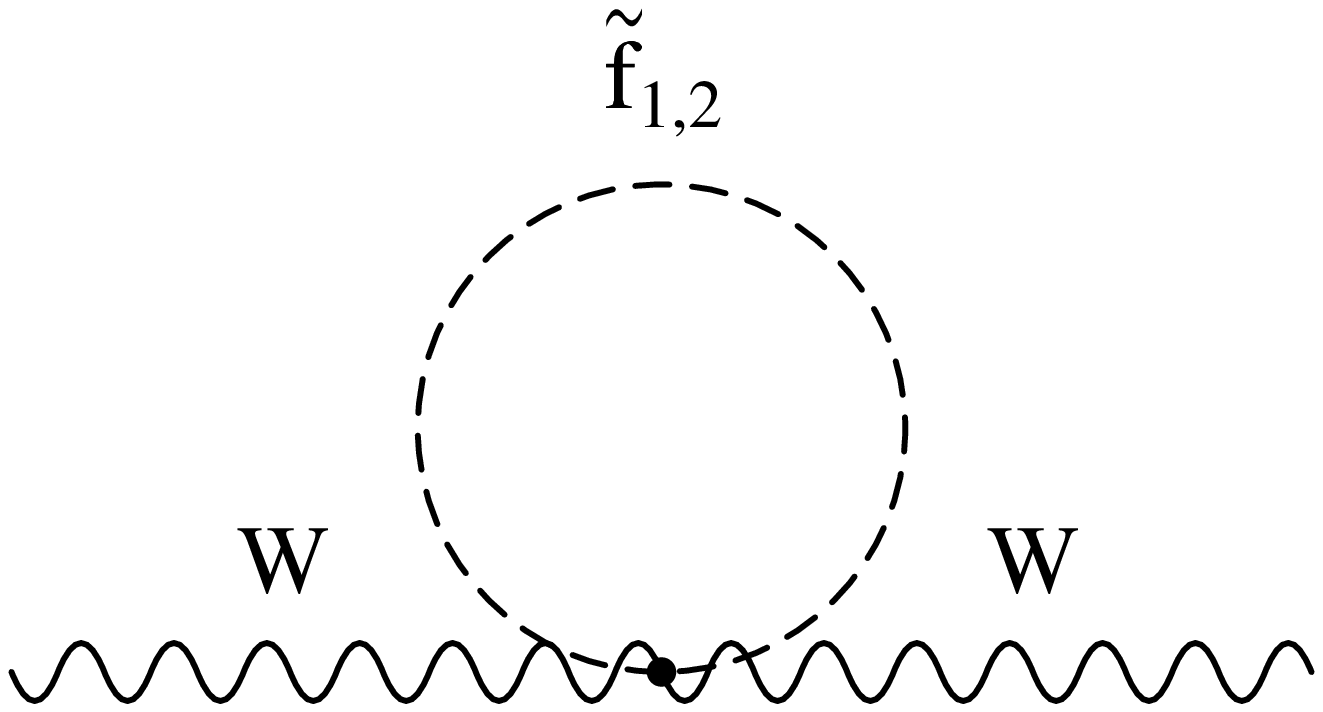,width=4cm}} &
\centerline{\raisebox{-0.7cm}{\psfig{bbllx=107pt,bblly=242pt,%
bburx=505pt,bbury=530pt,figure=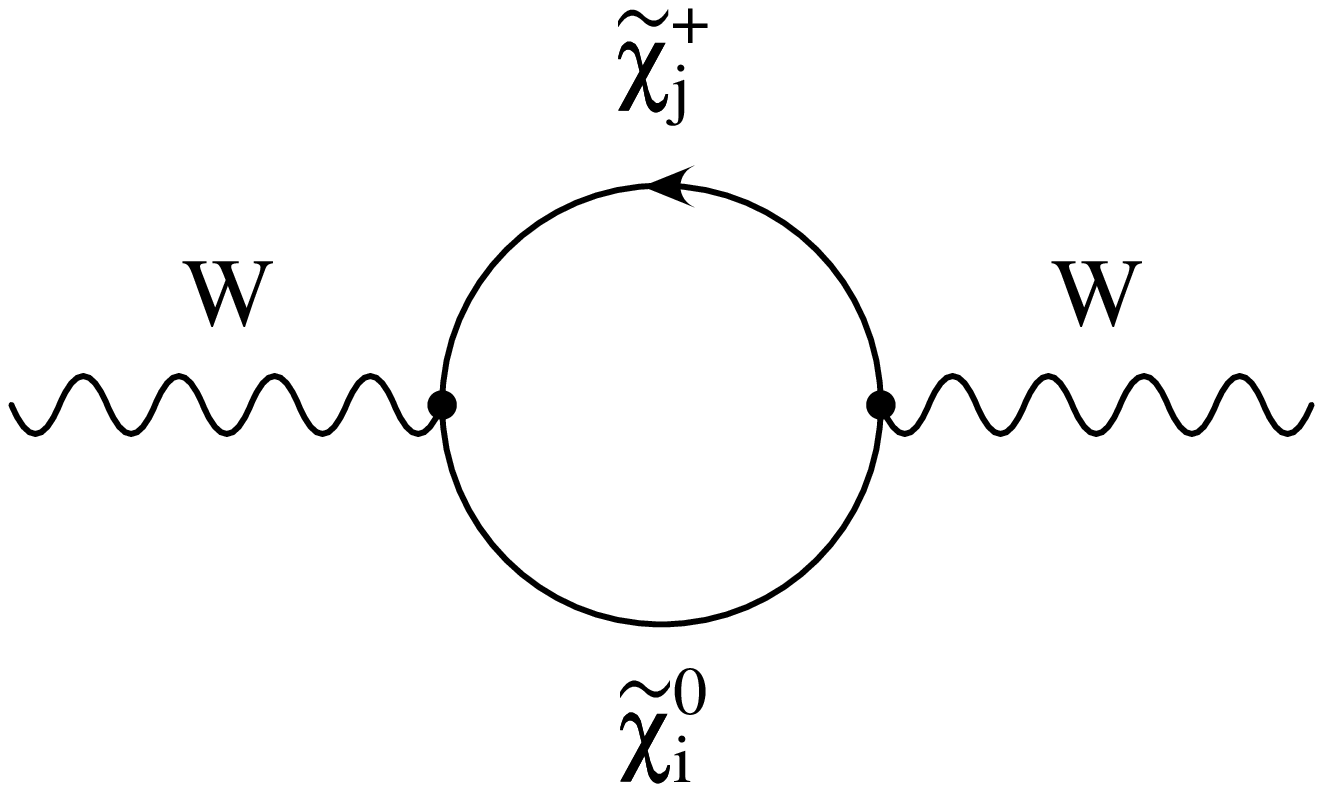,width=4cm}}}
\end{tabular}

\BEA
\Sigma^{WW}_{\tilde{f}}(p^2) &=& -\frac{\alpha}{4\pi} 
    \frac{1}{\sw^2} \sum_{doublets} N_C^f\, \bigg\{
    2 c_{\tilde{\theta}^+}^2 c_{\tilde{\theta}^-}^2  
    B_{22}(p^2,m_{\tilde{f}^+_1},m_{\tilde{f}^-_1}) \nonumber \\
& & +2 c_{\tilde{\theta}^+}^2 s_{\tilde{\theta}^-}^2  
    B_{22}(p^2,m_{\tilde{f}^+_1},m_{\tilde{f}^-_2}) \nonumber \\
& & +2 s_{\tilde{\theta}^+}^2 c_{\tilde{\theta}^-}^2  
    B_{22}(p^2,m_{\tilde{f}^+_2},m_{\tilde{f}^-_1}) 
    +2 s_{\tilde{\theta}^+}^2 s_{\tilde{\theta}^-}^2  
    B_{22}(p^2,m_{\tilde{f}^+_2},m_{\tilde{f}^-_2}) \bigg\} 
    \nonumber \\
& & +\frac{\alpha}{4\pi} \frac{1}{2\,\sw^2} 
    \sum_{f} N_C^f\, \bigg\{ c_{\tilde{\theta}}^2\,
    A_0(m_{\tilde{f}_1}) + s_{\tilde{\theta}}^2\,
    A_0(m_{\tilde{f}_2}) \bigg\} \nonumber \\
\Sigma^{WW}_{\tilde{\chi}}(p^2) &=& -\frac{\alpha}{4\pi}\, 
   \frac{1}{4\sw^2}\, \sum_{i=1}^4\, \sum_{j=1}^2\,  
   F_1(p^2,m_{\tilde{\chi}^+_j},m_{\tilde{\chi}^0_i},
   O^L_{ij},O^R_{ij},O^{L*}_{ij},O^{R*}_{ij})~,
\EEA

\vspace{5mm}

\vorn with
\BEA
O^L_{ij} &=& -\frac{1}{\sqrt{2}}\, N_{i4} V^*_{j2} + N_{i2} V^*_{j1}~, 
             \nonumber \\
O^R_{ij} &=& +\frac{1}{\sqrt{2}}\, N^*_{i3} U_{j2} + N^*_{i2} U_{j1}~,
\EEA
containing the matrices $U_{ij}$, $V_{ij}$, $N_{ij}$ for the chargino
and neutralino sector, see Appendices \ref{Charginos}, \ref{Neutralinos}.

\newpage

\subsubsection{Photon-$Z$ mixing}

\begin{tabular}{p{4.2cm}@{\hspace{0.5cm}\hspace{0.5cm}}p{4.2cm}@{\hspace{0.5cm}\hspace{0.5cm}}p{4.2cm}}
\centerline{\raisebox{-0.7cm}{\psfig{bbllx=107pt,bblly=242pt,%
bburx=505pt,bbury=521pt,figure=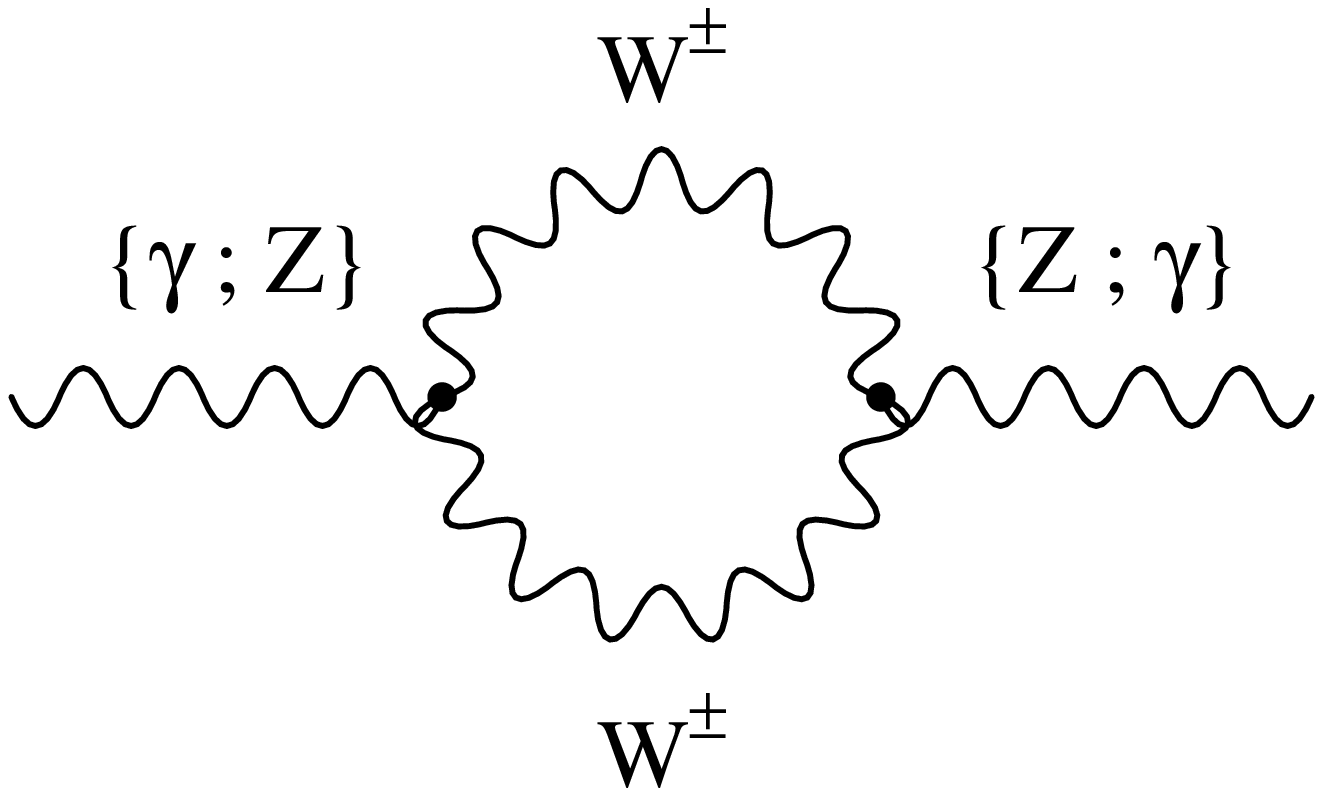,width=4cm}}} &
\centerline{\raisebox{-0.7cm}{\psfig{bbllx=107pt,bblly=242pt,%
bburx=505pt,bbury=530pt,figure=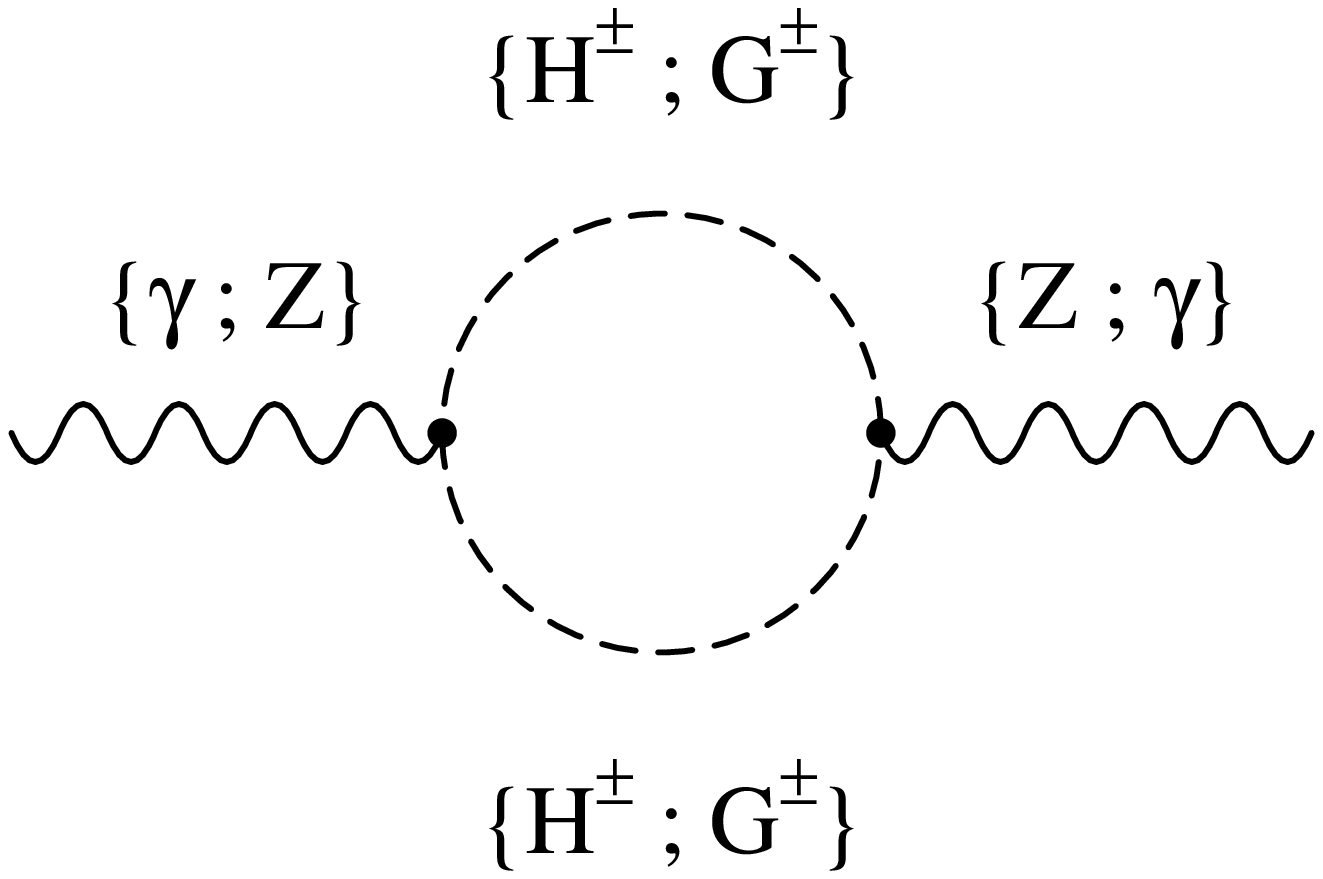,width=4cm}}} &
\centerline{\raisebox{-0.7cm}{\psfig{bbllx=107pt,bblly=242pt,%
bburx=505pt,bbury=521pt,figure=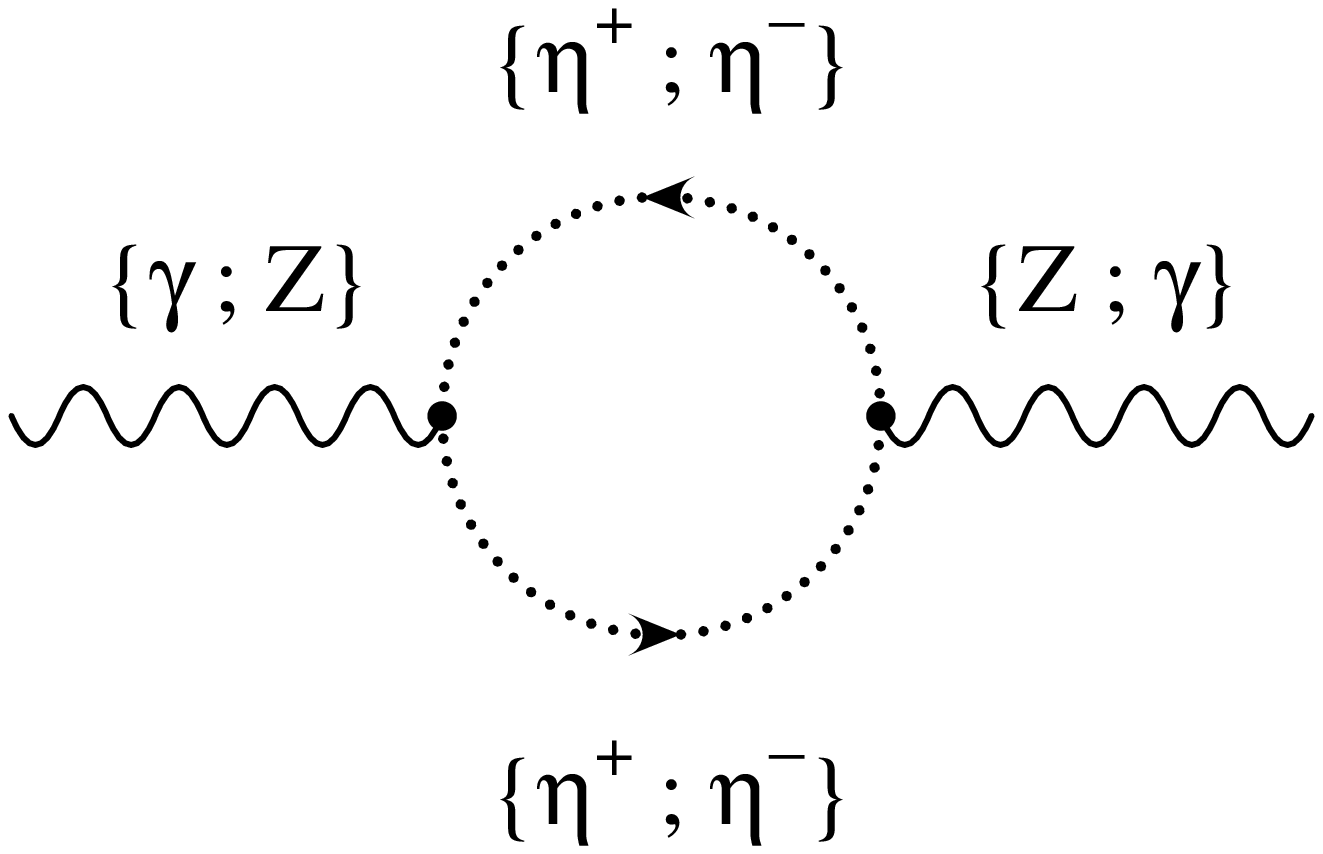,width=4cm}}} \\
\centerline{\psfig{bbllx=107pt,bblly=384pt,bburx=505pt,bbury=599pt,%
figure=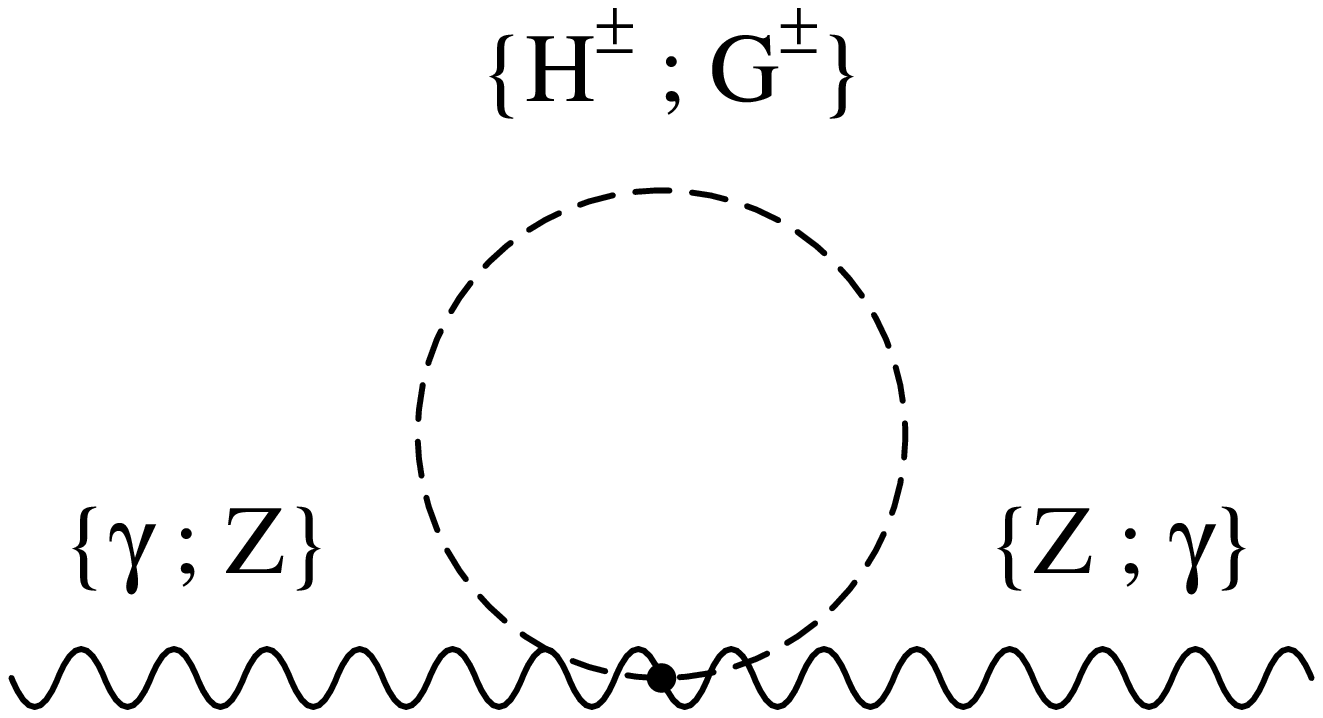,width=4cm}} &
\centerline{\psfig{bbllx=107pt,bblly=384pt,bburx=505pt,bbury=607pt,%
figure=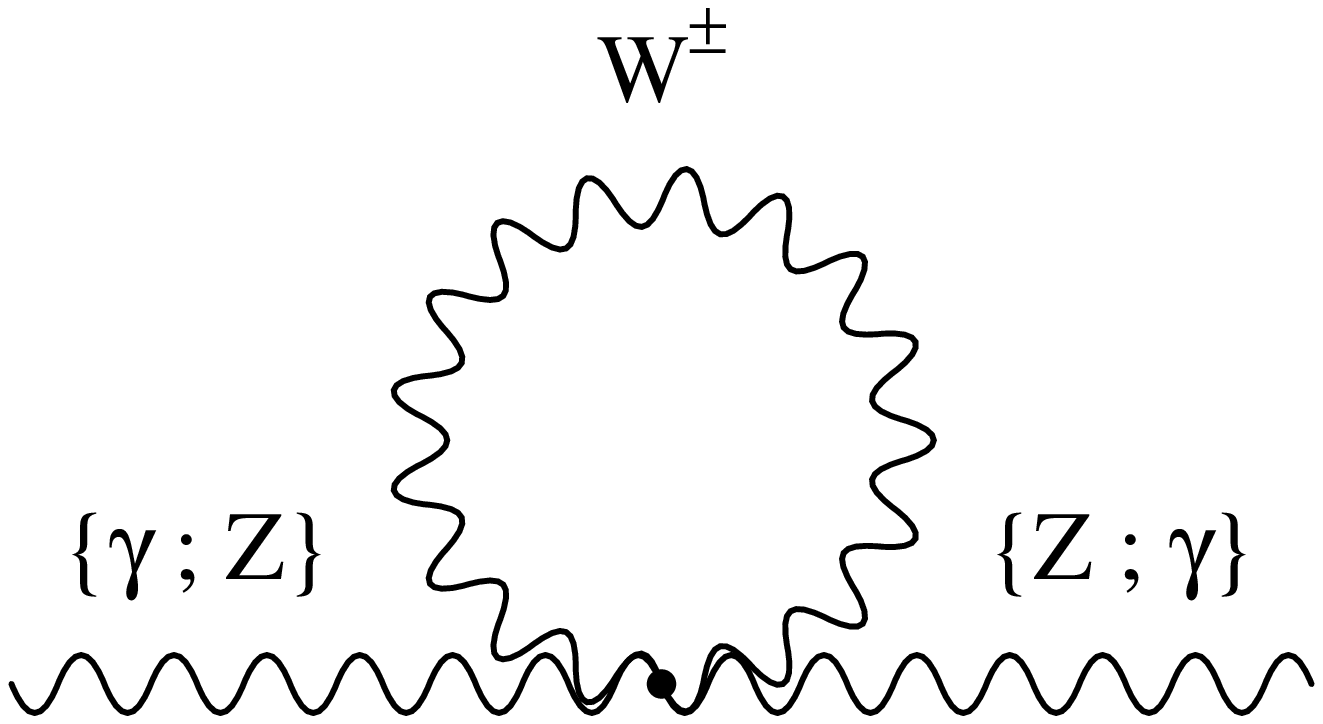,width=4cm}} &
\centerline{\raisebox{-0.7cm}{\psfig{bbllx=107pt,bblly=242pt,%
bburx=505pt,bbury=538pt,figure=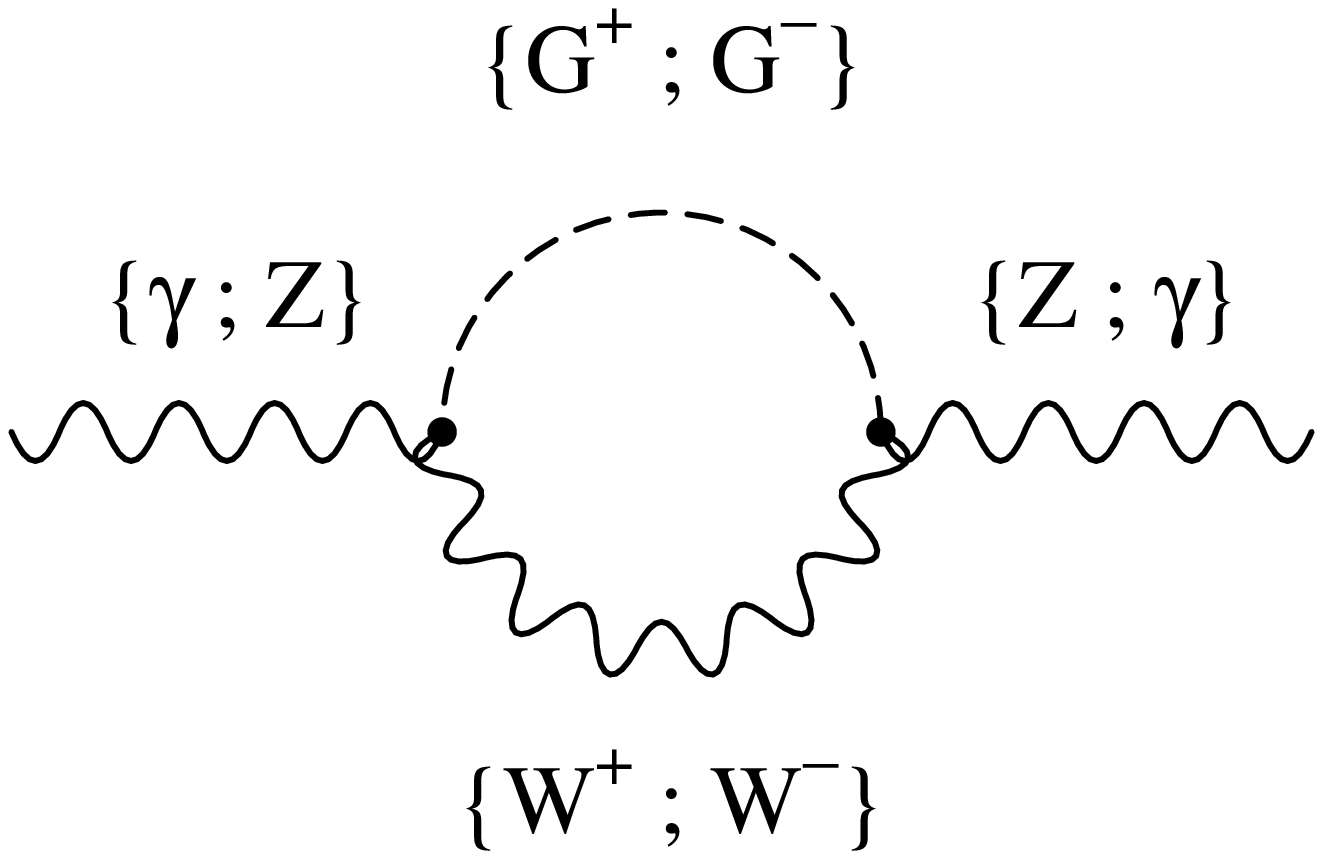,width=4cm}}} \\
\centerline{\raisebox{-0.7cm}{\psfig{bbllx=107pt,bblly=242pt,%
bburx=505pt,bbury=521pt,figure=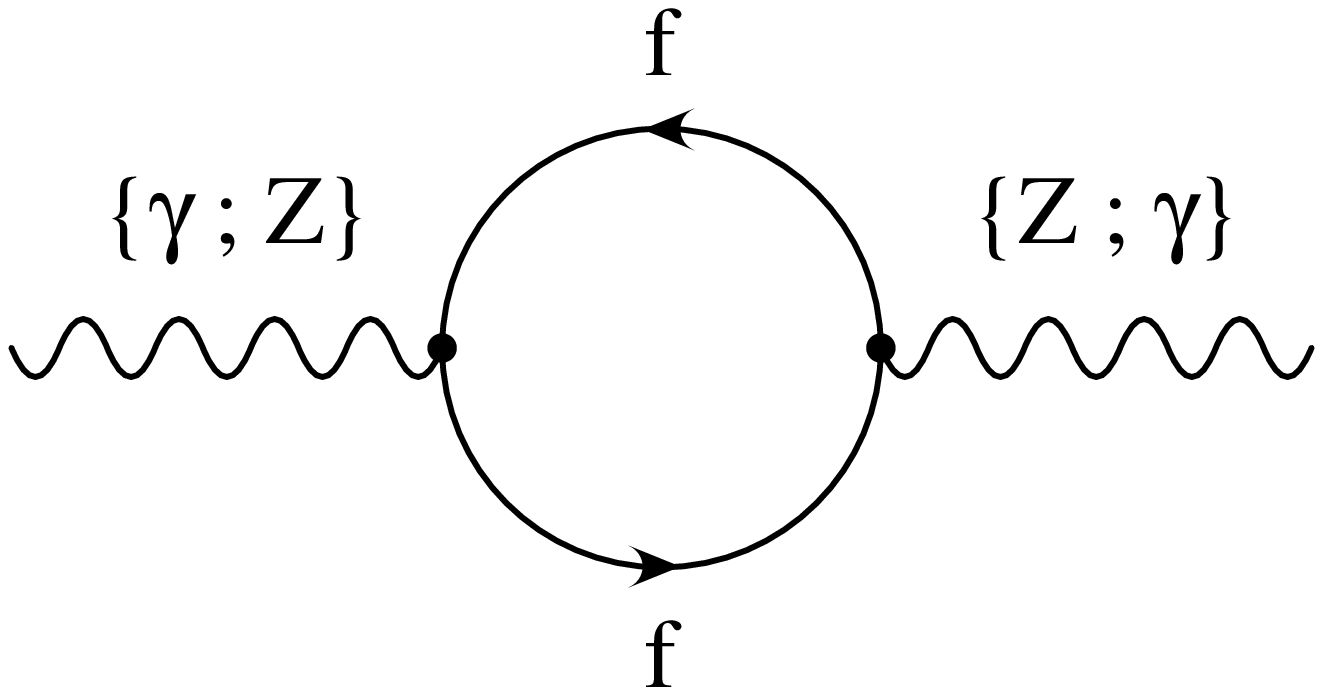,width=4cm}}}
\end{tabular}

\BEA
\Sigma^{\gamma Z}_{2HD}(p^2) &=& -\frac{\alpha}{4\pi}\,
    \bigg\{ -\frac{2\cos 2\theta_w}{\sw\cw} \Big[ 
    B_{22}(p^2,m_{H^{\pm}},m_{H^{\pm}})
    + B_{22}(p^2,M_W,M_W) \Big] \nonumber \\
& & +\frac{2\cw}{\sw} B_{22}(p^2,M_W,M_W)
    -2\sw\cw M_Z^2 B_{0}(p^2,M_W,M_W)  \nonumber \\
& & +\frac{\cos 2\theta_w}{\sw\cw} \Big[ A_0(m_{H^{\pm}}) + A_0(M_W) 
     \Big] 
    +\frac{\cw}{\sw} \Big[ 6 A_0(M_W) - 4 M_W^2) \Big] \nonumber \\ 
& & -\frac{\cw}{\sw} F_2(p^2,M_W,M_W) \nonumber \\
& & -\sum_{f}\, N_C^f\, Q_f\, F_1(p^2,m_f,m_f,\frac{1}{2},
     \frac{1}{2},\frac{v_f-a_f}{2},\frac{v_f+a_f}{2}) \bigg\}~,
\EEA

\newpage

\begin{tabular}{p{4.2cm}@{\hspace{0.5cm}\hspace{0.5cm}}p{4.2cm}@{\hspace{0.5cm}\hspace{0.5cm}}p{4.2cm}}
\centerline{\raisebox{-0.7cm}{\psfig{bbllx=107pt,bblly=242pt,%
bburx=505pt,bbury=530pt,figure=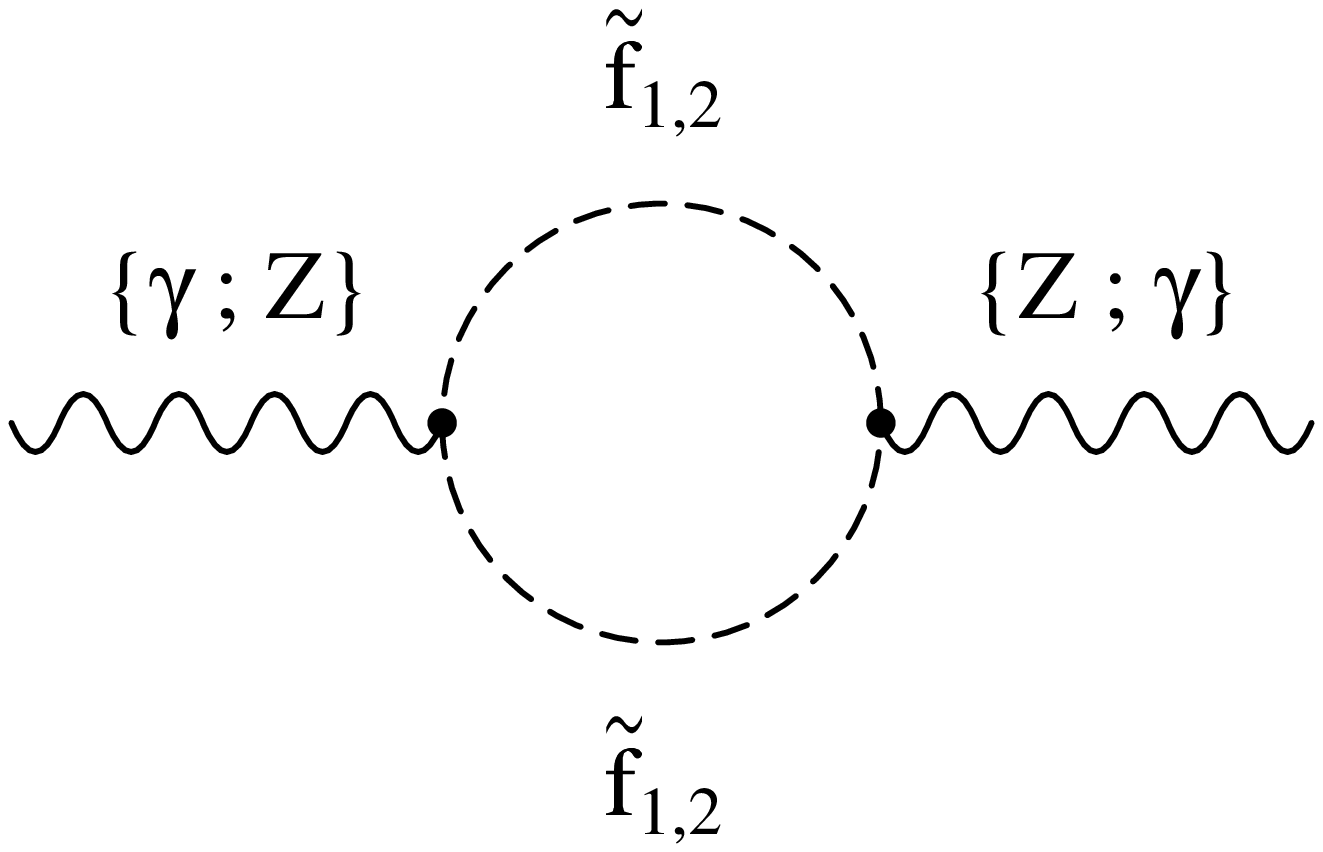,width=4cm}}} &
\centerline{\psfig{bbllx=107pt,bblly=384pt,%
bburx=505pt,bbury=599pt,figure=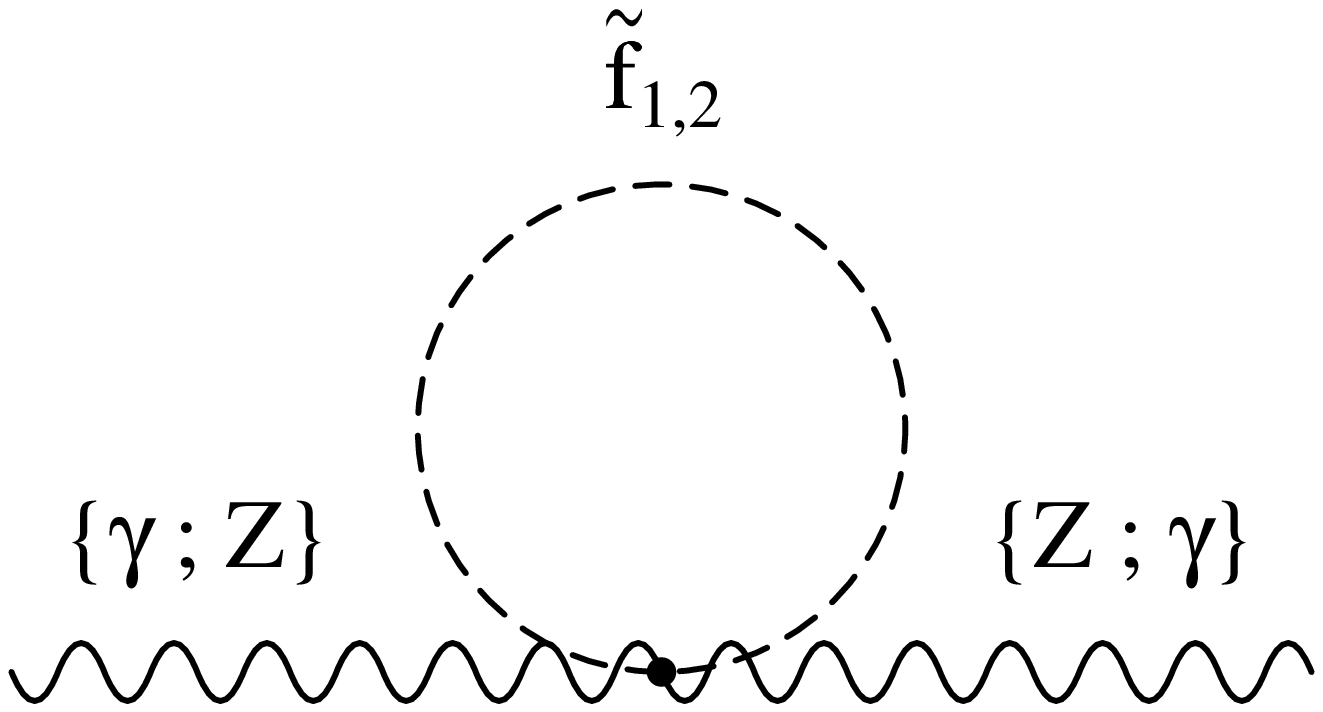,width=4cm}} &
\centerline{\raisebox{-0.7cm}{\psfig{bbllx=107pt,bblly=242pt,%
bburx=505pt,bbury=530pt,figure=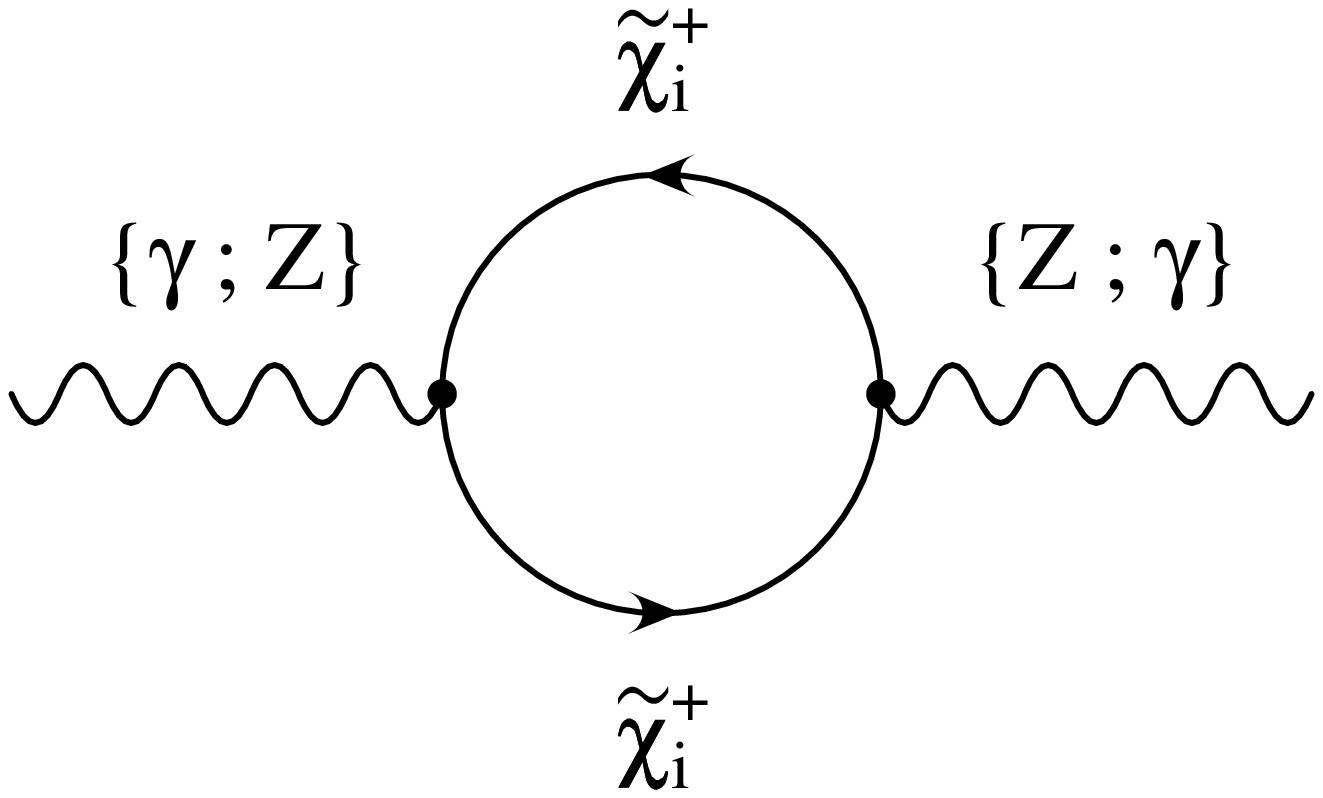,width=4cm}}}
\end{tabular}

\BEA
\Sigma^{\gamma Z}_{\tilde{f}}(p^2) &=& 
    \frac{\alpha}{4\pi}\, 2\, \sum_{f}\, N_C^f\, Q_f 
    \bigg\{ \frac{I^f_3 \CtT^2-Q_f \sw^2}{\sw\cw} \Big[
    2 B_{22}(p^2,m_{\tilde{f}_1},m_{\tilde{f}_1}) 
    - A_0(m_{\tilde{f}_1}) \Big] \nonumber \\
& & +\frac{I^f_3 \StT^2-Q_f \sw^2}{\sw\cw} \Big[
    2 B_{22}(p^2,m_{\tilde{f}_2},m_{\tilde{f}_2})
    - A_0(m_{\tilde{f}_2}) \Big]  \bigg\} \nonumber \\
\Sigma^{\gamma Z}_{\tilde{\chi}}(p^2) &=& 
    -\frac{\alpha}{4\pi} \frac{1}{2\sw\cw} 
    \sum_{i=1}^2 F_1(p^2,m_{\tilde{\chi}^+_i},
    m_{\tilde{\chi}^+_i},\frac{1}{2},\frac{1}{2},O'^L_{ii},O'^R_{ii})~.
\EEA

\vspace{1cm}

\subsection{Fermion self-energies}
\label{appb2}

\begin{figure}[h]
\centerline{\psfig{figure=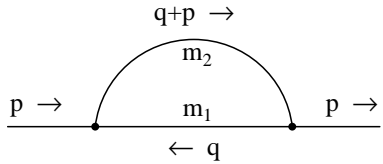,width=5cm}}
\caption[]{Masses and momenta for the fermion self-energies.}
\label{fig:Fimpkon}
\end{figure}

The fermion self-energy can be expressed in terms of the scalar 
functions $\Sigma^f_{\{S,V,A\}}$ as coefficients for the scalar-, 
vector-, and axial vector part:
\BE
\Sigma^f(p) = \pslash \Sigma_V^f(p^2)+\pslash \gamma_5 \Sigma_A^f(p^2)
              + m_f \Sigma_S^f(p^2)~.
\EE
The contributions are separated in a (weak) 2HD part, a (weak) SUSY 
part and a gluino part. 
The momenta and internal masses are illustrated in figure 
\ref{fig:Fimpkon}.
For computing the fermion-number-violating vertices the algorithm of
Ref. \cite{DEHK} is utilized. The arrows within the lines denote the 
charge flow and long, thin arrows denote the fermion flow. 
 
\vspace{8mm}

\newpage

\vorn We introduce the following definitions and shorthand notations:
\BEA
X^{+\{d;u\}}_{i1} &:=& -\frac{1}{\sw} 
                      \{V_{i1};U_{i1}\}\cos \tilde{\theta}_{f'}
                      +\frac{m_{f'}}{\sqrt{2} M_W\sw} \: \Big\{
                      \frac{V_{i2}}{\Sb};\frac{U_{i2}}{\Cb} \Big\}
                      \sin \tilde{\theta}_{f'}~, \nonumber \\
X^{+\{d;u\}}_{i2} &:=& +\frac{1}{\sw}
                      \{V_{i1};U_{i1}\}\sin \tilde{\theta}_{f'}
                      +\frac{m_{f'}}{\sqrt{2} M_W\sw} \: \Big\{
                      \frac{V_{i2}}{\Sb};\frac{U_{i2}}{\Cb} \Big\}
                      \cos \tilde{\theta}_{f'}~, \nonumber \\
Y^{+\{d;u\}}_{i1} &:=& +\frac{m_f}{\sqrt{2} M_W\sw} \: \Big\{
                      \frac{U_{i2}}{\Cb};\frac{V_{i2}}{\Sb} \Big\} 
                      \cos \tilde{\theta}_{f'}~, \nonumber \\
Y^{+\{d;u\}}_{i2} &:=& -\frac{m_f}{\sqrt{2} M_W\sw} \: \Big\{
                      \frac{U_{i2}}{\Cb};\frac{V_{i2}}{\Sb} \Big\}
                      \sin \tilde{\theta}_{f'}~, \nonumber \\
X^{0\{d;u\}}_{i1} &:=& -\sqrt{2} \Big[ Q_f N'_{i1} \mp 
                       \frac{\frac{1}{2} \pm Q_f\sw^2}{\sw\cw} N'_{i2} 
                       \Big] \cos \tilde{\theta}_f 
                       -\frac{m_f}{\sqrt{2} M_W\sw} 
                       \Big\{\frac{N_{i3}}{\Cb};\frac{N_{i4}}{\Sb}\Big\} 
                       \sin \tilde{\theta}_f~, \nonumber \\
X^{0\{d;u\}}_{i2} &:=& +\sqrt{2} \Big[ Q_f N'_{i1} \mp 
                       \frac{\frac{1}{2} \pm Q_f\sw^2}{\sw\cw} N'_{i2} 
                       \Big] \sin \tilde{\theta}_f 
                       -\frac{m_f}{\sqrt{2} M_W\sw}
                       \Big\{\frac{N_{i3}}{\Cb};\frac{N_{i4}}{\Sb}\Big\}
                       \cos \tilde{\theta}_f~, \nonumber \\
Y^{0\{d;u\}}_{i1} &:=& -\frac{m_f}{\sqrt{2} M_W\sw} \Big\{ 
                       \frac{N_{i3}}{\Cb};\frac{N_{i4}}{\Sb} \Big\}
                       \cos \tilde{\theta}_f + \sqrt{2}
                       \Big[ Q_f N'_{i1} - \frac{Q_f\sw}{\cw} N'_{i2} 
                       \Big] \sin \tilde{\theta}_f~,\nonumber \\ 
Y^{0\{d;u\}}_{i2} &:=& +\frac{m_f}{\sqrt{2} M_W\sw} \Big\{
                       \frac{N_{i3}}{\Cb};\frac{N_{i4}}{\Sb} \Big\}
                       \sin \tilde{\theta}_f + \sqrt{2} 
                       \Big[ Q_f N'_{i1} - \frac{Q_f\sw}{\cw} N'_{i2} 
                       \Big] \cos \tilde{\theta}_f~,
\label{viai}
\EEA
with
\BEA
N'_{i1} &:=& N_{i1} \cw + N_{i2} \sw~,  \nonumber \\ 
N'_{i2} &:=& N_{i2} \cw - N_{i1} \sw~. 
\EEA
and the matrices $U_{ij}$, $V_{ij}$ for the charginos and $N_{ij}$ for 
the neutralinos, see Appendices \ref{Charginos}, \ref{Neutralinos}.

The color factor $C_F$ is defined in Eq. (\ref{cf}), and the factors 
$g^f_H$ of the Yukawa couplings can be found in Table \ref{tab:ver}.

\vspace{8mm}

\begin{table}[h]
\begin{center}
$\renewcommand{\arraystretch}{1.6}
\tabcolsep3mm
\begin{array}{|c||*{5}{c|}}
\hline
\quad g^f_H\quad & H=\mbox{SM-Higgs} & \quad H=h^0\quad & 
\quad H=H^0\quad & \quad H=A^0\quad & \quad H=H^{\pm}\quad \\
\hline\hline 
f=u\mbox{-type} & 1 & \D \frac{\Ca}{\Sb} & \D \frac{\Sa}{\Sb} & 
\cot\beta & \cot\beta  \\
\hline 
f=d\mbox{-type} & 1 & \D -\frac{\Sa}{\Cb} & \D \frac{\Ca}{\Cb} & 
\tan\beta & \tan\beta  \\
\hline
\end{array}$
\end{center}
\caption[]{\label{tab:ver} Yukawa coupling coefficients.}
\end{table}

\newpage

\vspace*{5mm}

\begin{tabular}{p{4.2cm}@{\hspace{0.5cm}\hspace{0.5cm}}p{4.2cm}@{\hspace{0.5cm}\hspace{0.5cm}}p{4.2cm}}
\centerline{\psfig{bbllx=118pt,bblly=349pt,bburx=494pt,bbury=560pt,%
figure=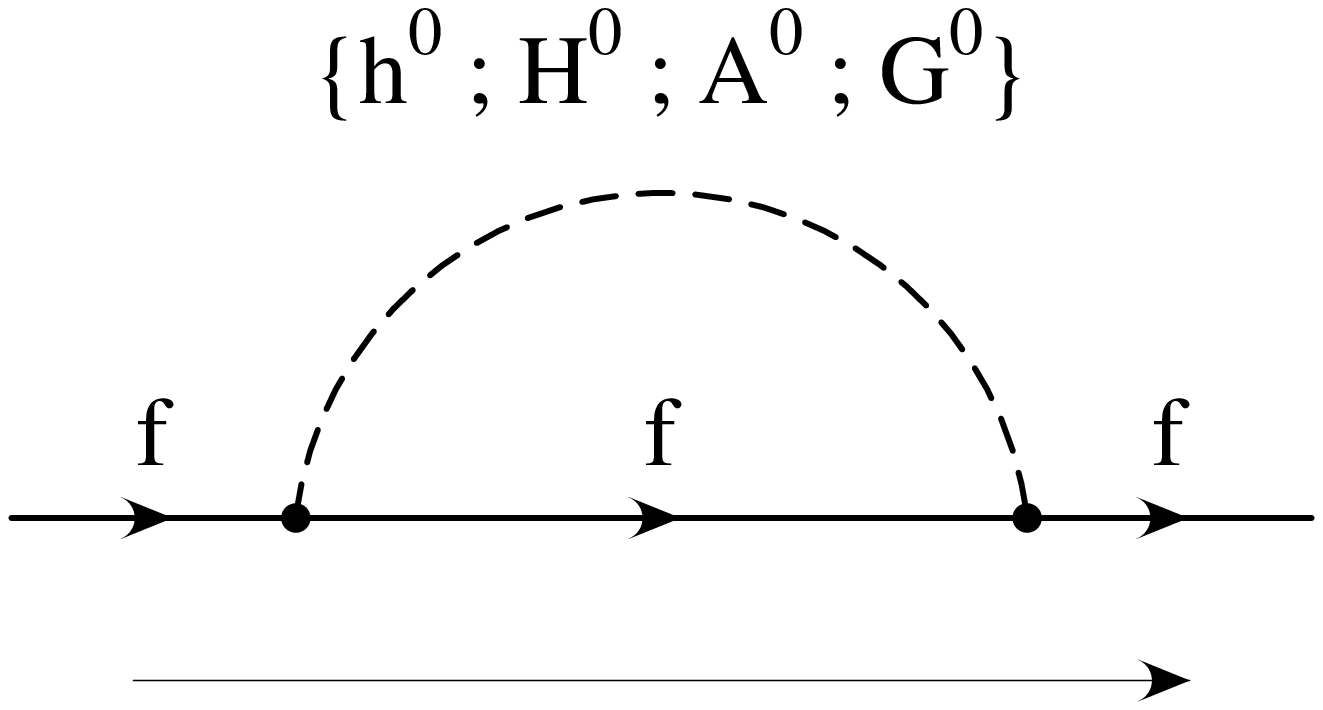,width=4cm}} &
\centerline{\psfig{bbllx=118pt,bblly=349pt,bburx=494pt,bbury=560pt,%
figure=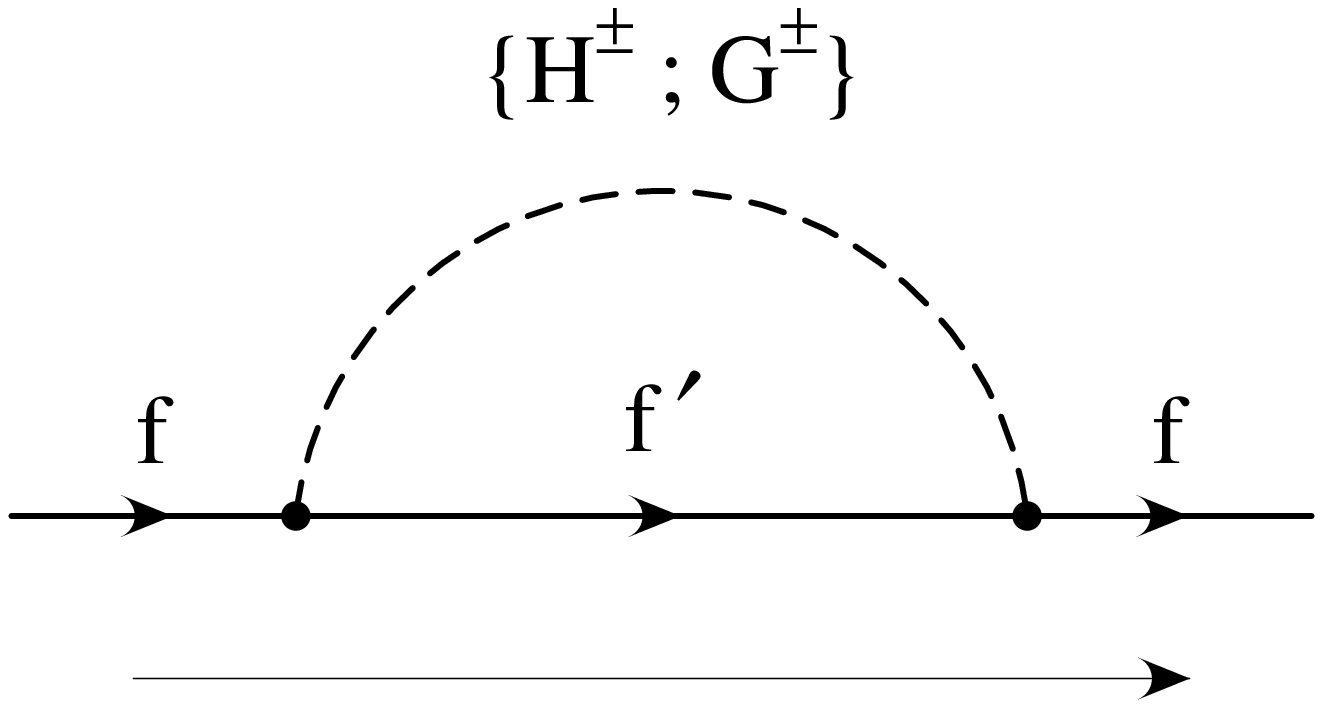,width=4cm}} &
\centerline{\psfig{bbllx=118pt,bblly=349pt,bburx=494pt,bbury=560pt,%
figure=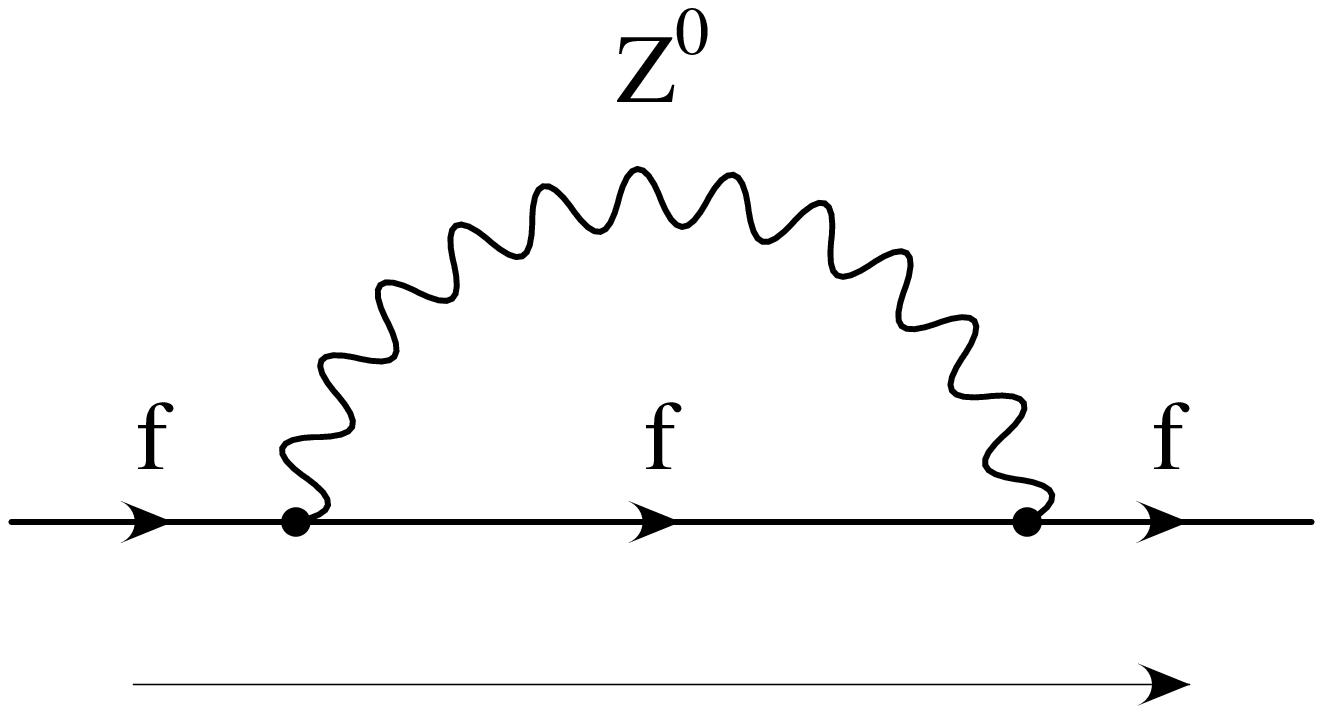,width=4cm}} \\
\centerline{\psfig{bbllx=118pt,bblly=349pt,bburx=494pt,bbury=560pt,%
figure=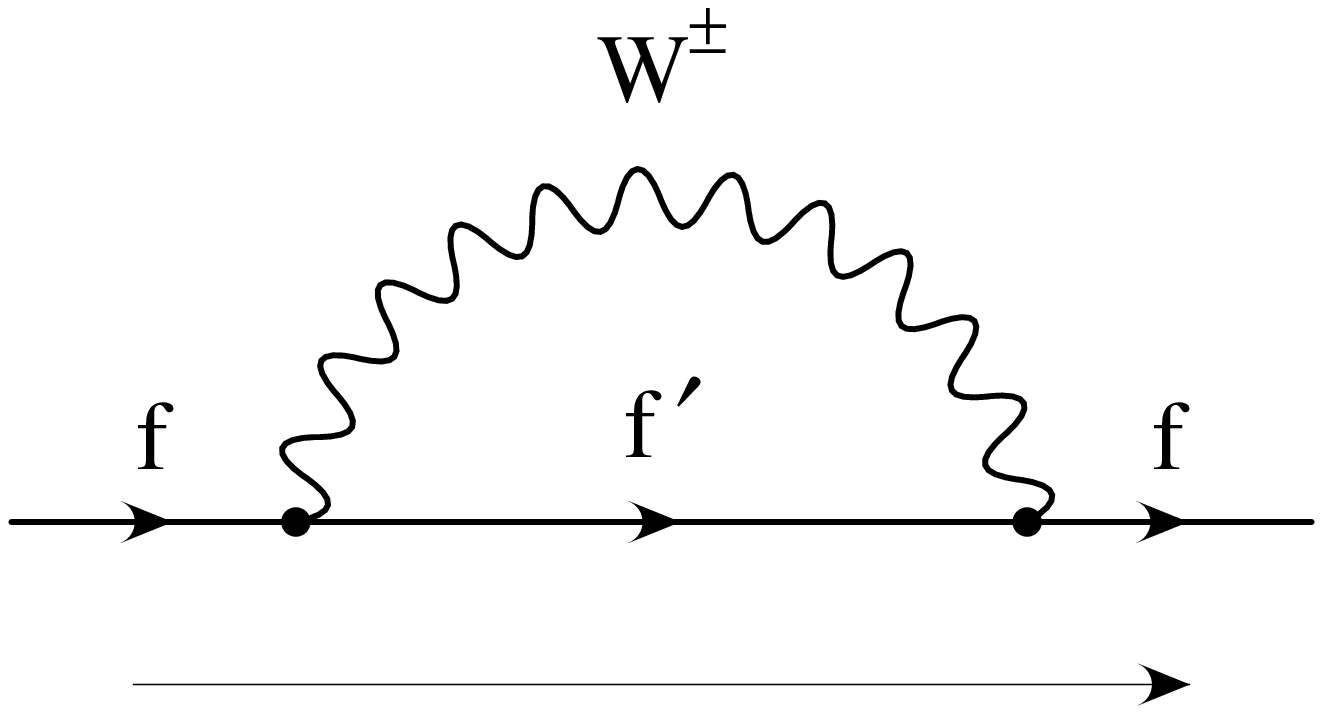,width=4cm}} &
\centerline{\psfig{bbllx=118pt,bblly=349pt,bburx=494pt,bbury=560pt,%
figure=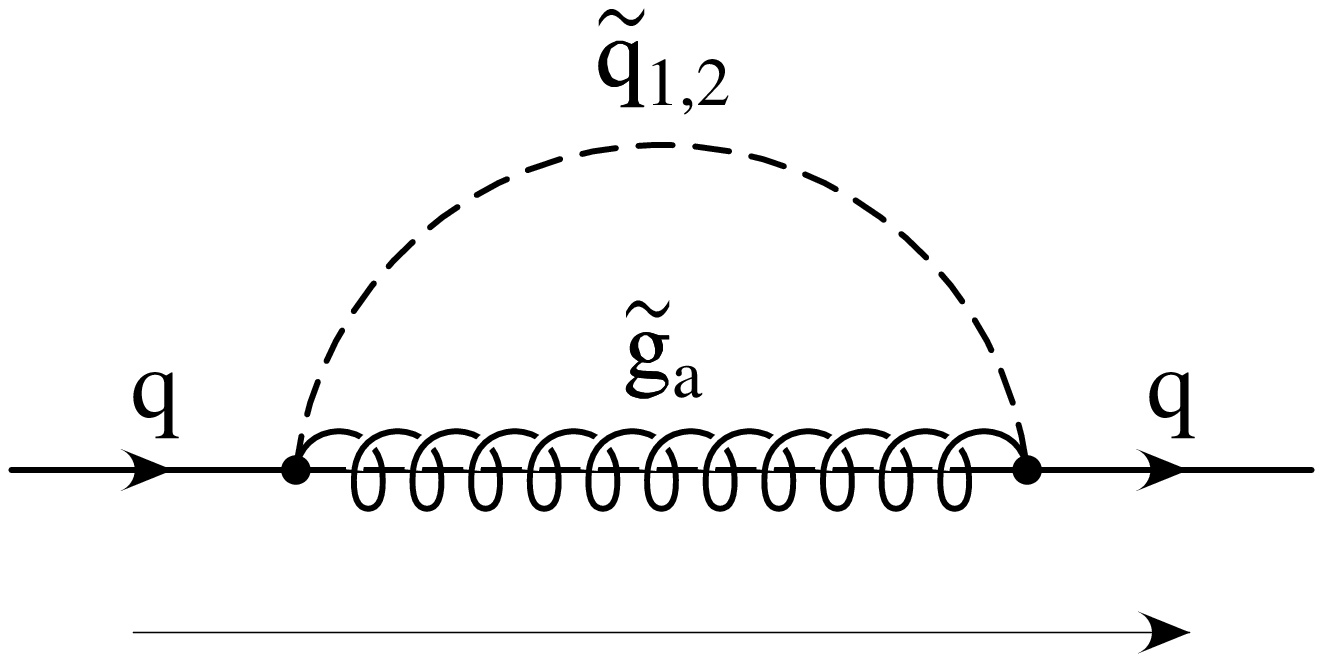,width=4cm}} &
\centerline{\psfig{bbllx=118pt,bblly=349pt,bburx=494pt,bbury=560pt,%
figure=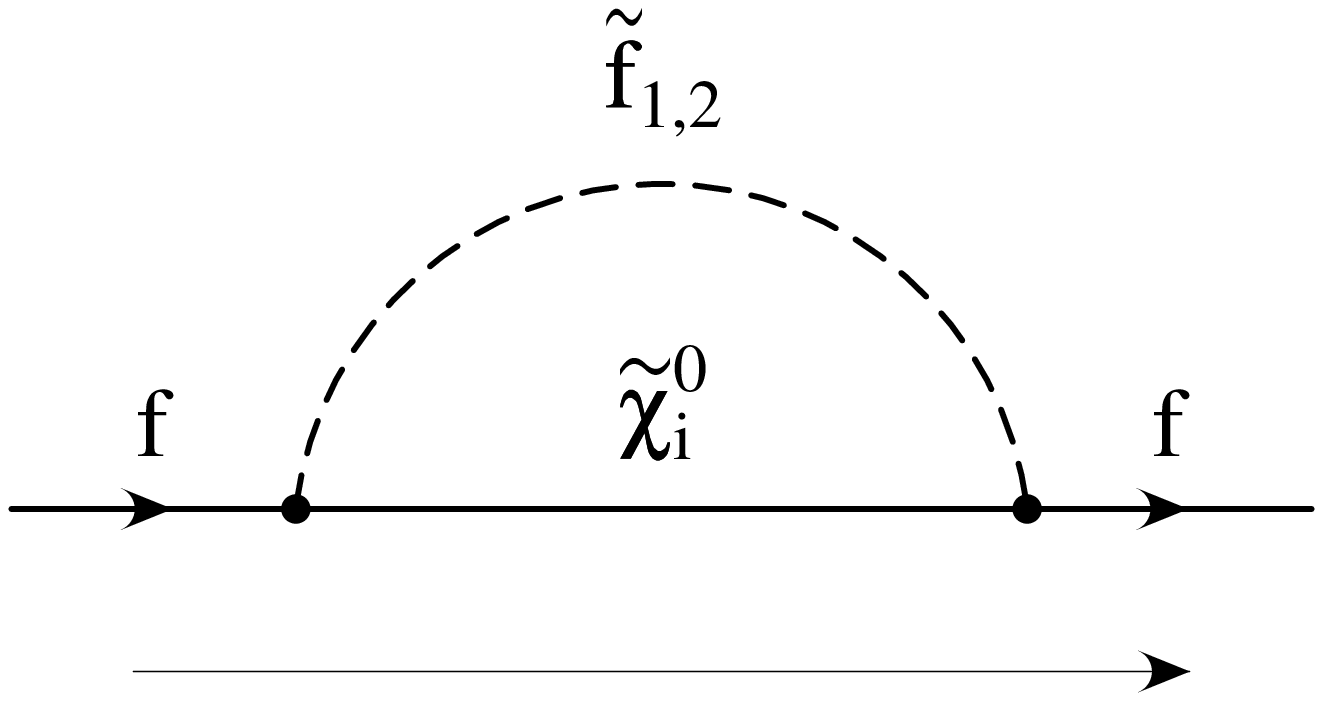,width=4cm}} \\
\centerline{\psfig{bbllx=118pt,bblly=349pt,bburx=494pt,bbury=560pt,%
figure=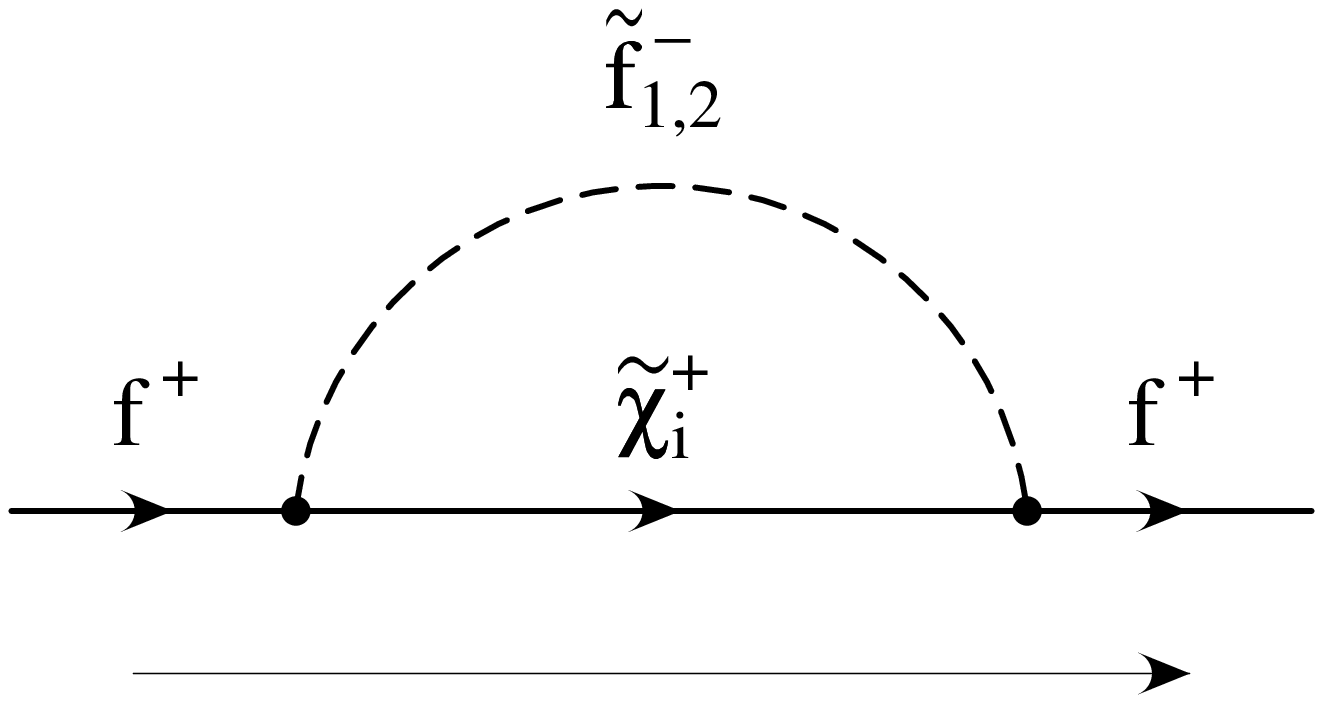,width=4cm}} & 
\centerline{\psfig{bbllx=118pt,bblly=349pt,bburx=494pt,bbury=560pt,%
figure=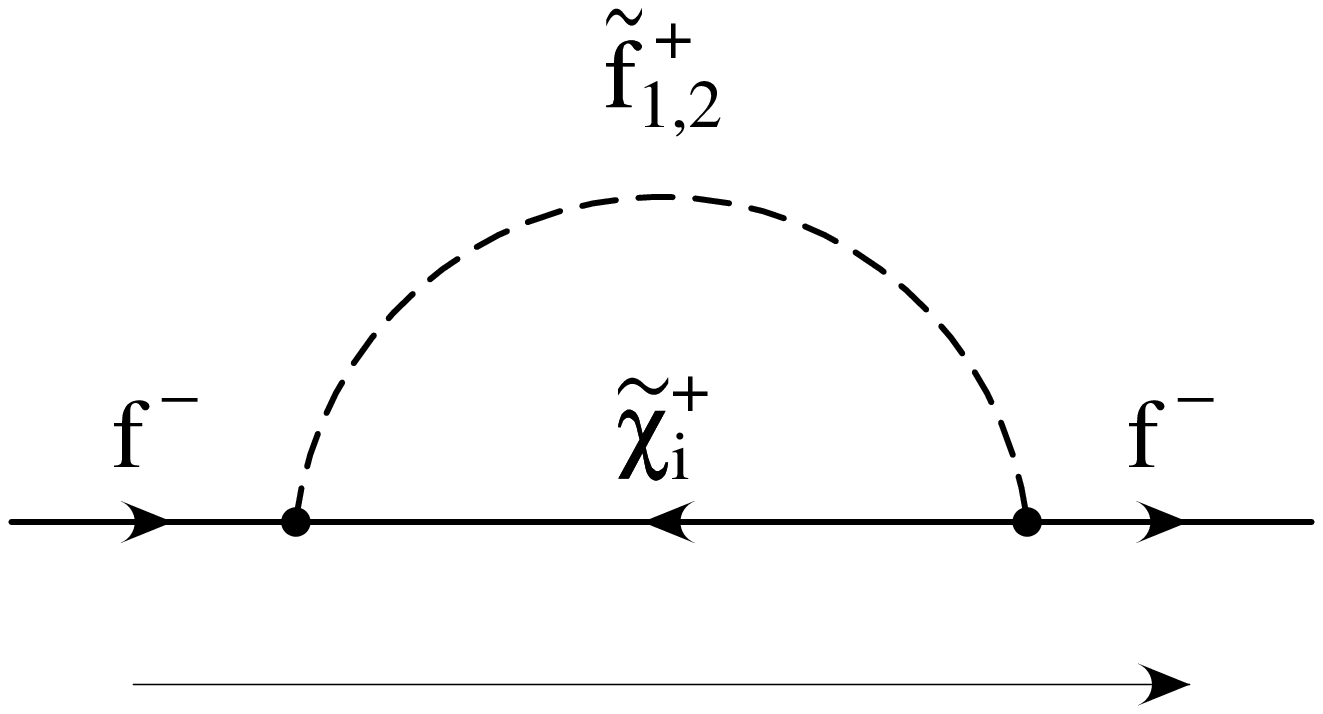,width=4cm}} & 
\end{tabular}

\vspace{1cm}

\subsubsection*{The scalar part $\Sigma_S^f(p^2)$:}

\BEA
\Sigma^f_{S,2HD}(p^2) &=& -\frac{\alpha}{4\pi} 
    \bigg\{ (v_f^2-a_f^2) \Big[ 4 B_0(p^2,m_f,M_Z) - 2 \Big] \nonumber \\
& & +\frac{m^2_{f'}}{2\sw^2 M_W^2} \Big[ B_0(p^2,m_{f'},M_W) 
    - B_0(p^2,m_{f'},m_{H^{\pm}}) \Big] \nonumber \\
& & +\frac{m^2_f}{4\sw^2 M_W^2} \Big[ B_0(p^2,m_f,M_Z) 
    -(g^f_{h^0})^2 B_0(p^2,m_f,m_{h^0}) \nonumber \\
& & -(g^f_{H^0})^2 B_0(p^2,m_f,m_{H^0})
    +(g^f_{A^0})^2 B_0(p^2,m_f,\MA) \Big] \bigg\}~, \nonumber \\
\Sigma^f_{S,gluino}(p^2) &=& -\frac{\alpha_s(M^2_Z)}{4\pi}\, 
    \frac{m_{\tilde{g}}}{m_f} \, (2 \, C_F \, \StT \CtT) \Big[
    B_0(p^2,m_{\tilde{g}},m_{\tilde{q}_1})
    -B_0(p^2,m_{\tilde{g}},m_{\tilde{q}_2}) \Big]~, \nonumber \\
\Sigma^f_{S,SUSY}(p^2) &=& \frac{\alpha}{4\pi} 
      \bigg\{ \sum_{i=1}^4 \frac{m_{\tilde{\chi}^0_i}}{m_f} 
      \Big[ \re (X^{0f}_{i1} Y^{0f}_{i1})\, 
      B_0(p^2,m_{\tilde{\chi}^0_i},m_{\tilde{f}_1})
    + \re (X^{0f}_{i2} Y^{0f}_{i2})\, 
      B_0(p^2,m_{\tilde{\chi}^0_i},m_{\tilde{f}_2}) \Big] \nonumber \\
& & + \sum_{i=1}^2 \frac{m_{\tilde{\chi}^+_i}}{m_f} \Big[
      \re (X^{+f}_{i1} Y^{+f}_{i1})\, 
      B_0(p^2,m_{\tilde{\chi}^+_i},m_{\tilde{f}'_1}) \nonumber \\
& & + \, \re (X^{+f}_{i2} Y^{+f}_{i2})\, 
       B_0(p^2,m_{\tilde{\chi}^+_i},m_{\tilde{f}'_2}) \Big] \bigg\}~.
\EEA

\newpage

\subsubsection*{The vector part $\Sigma_V^f(p^2)$:}

\BEA
\Sigma^f_{V,2HD}(p^2) &=& -\frac{\alpha}{4\pi} 
    \bigg\{ (v_f^2+a_f^2) \Big[ 2 B_1(p^2,m_f,M_Z) + 1 \Big] 
    +\frac{1}{4\sw^2} \Big[ 2 B_1(p^2,m_{f'},M_W) + 1 \Big] \nonumber \\
& & +\frac{m^2_f}{4\sw^2 M_W^2} \Big[ (g^f_{h^0})^2 B_1(p^2,m_f,m_{h^0}) 
    +(g^f_{H^0})^2 B_1(p^2,m_f,m_{H^0}) \nonumber \\
& & +(g^f_{A^0})^2 B_1(p^2,m_f,\MA) + B_1(p^2,m_f,M_Z) 
    \Big] \nonumber \\
& & +\frac{1}{4\sw^2 M_W^2} 
    \Big[ (m^2_f+m^2_{f'}) B_1(p^2,m_{f'},M_W) \nonumber \\ 
& & + (m^2_f (g^f_{H^{\pm}})^2 + m^2_{f'} (g^{f'}_{H^{\pm}})^2)
    B_1(p^2,m_{f'},m_{H^{\pm}}) \Big]
    \bigg\}~, \nonumber \\
\Sigma^f_{V,gluino}(p^2) &=& -\frac{\alpha_s(M^2_Z)}{4\pi}\, C_F \, 
         \Big[ B_1(p^2,m_{\tilde{g}},m_{\tilde{q}_1})
         + B_1(p^2,m_{\tilde{g}},m_{\tilde{q}_2}) \Big]~,  \nonumber \\
\Sigma^f_{V,SUSY}(p^2) &=& -\frac{\alpha}{4\pi} 
    \Bigg\{ \sum_{i=1}^2\bigg(\frac{1}{2}
    \Big[ |X^{+f}_{i1}|^2 + |Y^{+f}_{i1}|^2 \Big] 
    B_1(p^2,m_{\tilde{\chi}^+_i},m_{\tilde{f}'_1}) \nonumber \\
& & +\frac{1}{2} \Big[ |X^{+f}_{i2}|^2 + |Y^{+f}_{i2}|^2 \Big] 
    B_1(p^2,m_{\tilde{\chi}^+_i},m_{\tilde{f}'_2})\bigg) \nonumber \\
& & +\sum_{i=1}^4 \bigg( \frac{1}{2} 
    \Big[ |X^{0f}_{i1}|^2 + |Y^{0f}_{i1}|^2 \Big] 
    B_1(p^2,m_{\tilde{\chi}^0_i},m_{\tilde{f}_1}) \nonumber \\
& & +\frac{1}{2} \Big[ |X^{0f}_{i2}|^2 + |Y^{0f}_{i2}|^2 \Big] 
    B_1(p^2,m_{\tilde{\chi}^0_i},m_{\tilde{f}_2}) \bigg)
    \Bigg\}~.
\EEA

\subsubsection*{The axial vector part $\Sigma_A^f(p^2)$:}

\BEA
\Sigma^f_{A,2HD}(p^2) &=& -\frac{\alpha}{4\pi} 
    \bigg\{ -2 v_f a_f \Big[ 2 B_1(p^2,m_f,M_Z) + 1 \Big] 
    -\frac{1}{4\sw^2} \Big[ 2 B_1(p^2,m_{f'},M_W) + 1 \Big] \nonumber \\
& & +\frac{1}{4\sw^2 M_W^2} 
    \Big[ (m^2_f-m^2_{f'}) B_1(p^2,m_{f'},M_W) \nonumber \\ 
& & +(m^2_f (g^f_{H^{\pm}})^2 - m^2_{f'} (g^{f'}_{H^{\pm}})^2)
    B_1(p^2,m_{f'},m_{H^{\pm}}) \Big]
    \bigg\}~, \nonumber \\
\Sigma^f_{A,gluino}(p^2) &=& -\frac{\alpha_s(M^2_Z)}{4\pi}\, C_F \, 
    \bigg\{\, \CtT^2 \, \Big[ B_1(p^2,m_{\tilde{g}},m_{\tilde{q}_2})
         - B_1(p^2,m_{\tilde{g}},m_{\tilde{q}_1}) \Big] \nonumber \\ 
& & +\, \StT^2 \, \Big[ B_1(p^2,m_{\tilde{g}},m_{\tilde{q}_1})
    - B_1(p^2,m_{\tilde{g}},m_{\tilde{q}_2}) \Big] \bigg\}~, \nonumber \\
\Sigma^f_{A,SUSY}(p^2) &=& \frac{\alpha}{4\pi} \Bigg\{
    \sum_{i=1}^2\bigg(\frac{1}{2} \Big[ 
    |X^{+f}_{i1}|^2 - |Y^{+f}_{i1}|^2\Big] 
    B_1(p^2,m_{\tilde{\chi}^+_i},m_{\tilde{f}'_1}) \nonumber \\
& & +\frac{1}{2} \Big[ |X^{+f}_{i2}|^2 - |Y^{+f}_{i2}|^2 \Big] 
    B_1(p^2,m_{\tilde{\chi}^+_i},m_{\tilde{f}'_2})\bigg) \nonumber \\
& & +\sum_{i=1}^4 \bigg( \frac{1}{2} 
    \Big[ |X^{0f}_{i1}|^2 - |Y^{0f}_{i1}|^2 \Big] 
    B_1(p^2,m_{\tilde{\chi}^0_i},m_{\tilde{f}_1}) \nonumber \\
& & +\frac{1}{2} \Big[ |X^{0f}_{i2}|^2 - |Y^{0f}_{i2}|^2 \Big] 
    B_1(p^2,m_{\tilde{\chi}^0_i},m_{\tilde{f}_2}) \bigg)
    \Bigg\}~.
\EEA


\section{Vertex corrections}
\label{appc}
\setcounter{equation}{0}
\setcounter{figure}{0}
\setcounter{table}{0}

This appendix contains all 3-point vertex diagrams for 
$\gamma f\bar{f}$ and $Z f\bar{f}$ within the MSSM at the 1-loop level 
(without virtual photons and gluons). The diagrams with Higgs bosons are 
negligible for $f=e$.

In addition to Appendix \ref{appb} we define the following shorthand 
notations for the couplings:
\BE
\begin{array}{*{5}{c}}
\D g_{Z,R} = -\frac{\sw}{\cw}\, Q_f & , &
\D g_{Z,L} = \frac{I_3^f-\sw^2 Q_f}{\sw\cw} & , &
\D g_C = \frac{1}{\sqrt{2}\sw}\, \frac{m_f}{M_W} \\[3ex]
\D g_{Z,R}^\prime = -\frac{\sw}{\cw}\, Q_{f^\prime} & , &
\D g_{Z,L}^\prime = \frac{I_3^{f^\prime}-\sw^2Q_{f^\prime}}{\sw\cw} & , &
\D g_C^\prime = \frac{1}{\sqrt{2}\sw}\, \frac{m_{f^\prime}}{M_W} \\[3ex]
\D g_N = -\frac{1}{2\sw}\, \frac{m_f}{M_W}~.
\end{array}
\label{gezet}
\EE
3-point functions $C$ are discussed in Refs. 
\cite{pasvelt,thooftvelt,Denner}, but we use the convention of 
Refs. \cite{hollik95,Oldenborgh}.
The arguments for the $C$-functions are $C(p_1,p_2,m_1,m_2,m_3)$.
For the electron vertex corrections we have $p_1=p$, $p_2=-p-\bar{p}$, 
$p_3=\bar{p}$, and for the outgoing fermions we have $p_1=-\bar{k}$, 
$p_2=k+\bar{k}$, $p_3=-k$.

\begin{figure}[h]
\centerline{{\psfig{figure=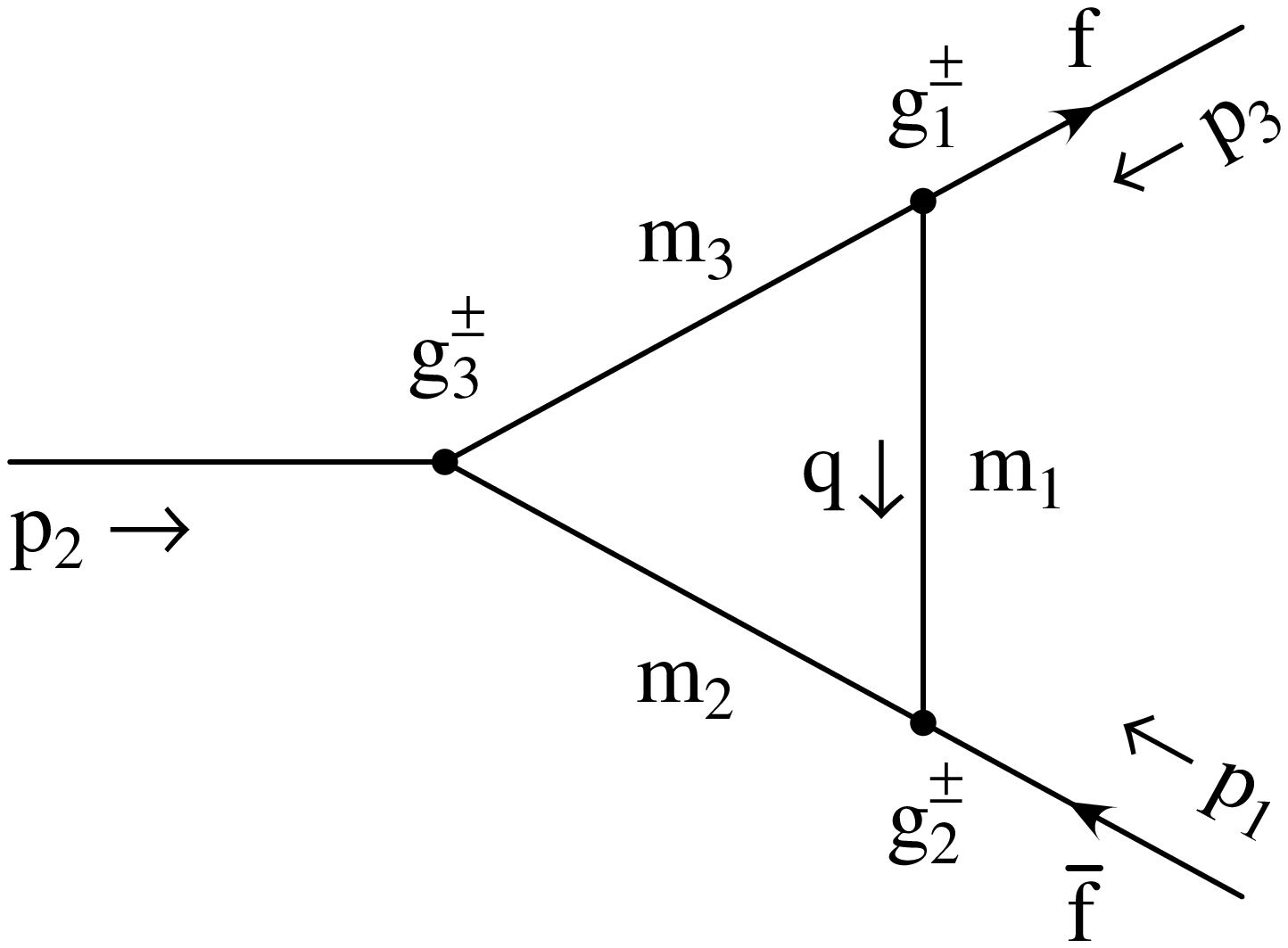,width=5cm}}} 
\caption[]{The vertex topology.}
\label{fig:Vertop}
\end{figure}

The conventions on momenta and internal masses are illustrated in Figure 
\ref{fig:Vertop} (the arrows within the lines denote the charge flow
and long, thin arrows denote the fermion flow if necessary).

\vspace{8mm}

\vspace{8mm}

Next we give the generic expressions for the form factors of each
class of vertex diagrams. The contributing particles and their
couplings are listed in the attached tables.

\newpage

\setcounter{dummy}{1}
\subsection{Class-\Roman{dummy} diagram}
\vspace{5mm}
\centerline{\psfig{figure=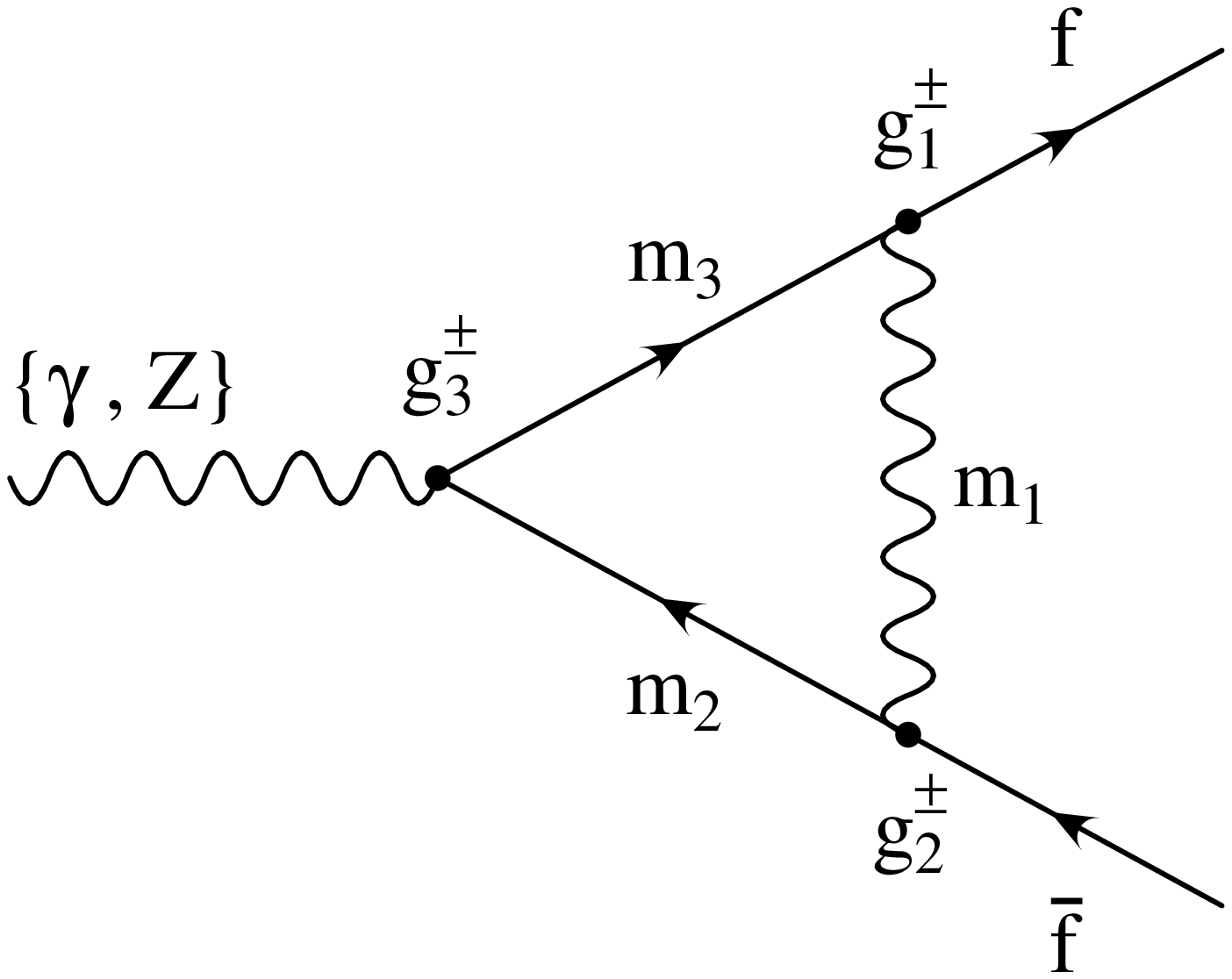,width=4.5cm}}
\vspace{5mm}
\BEA
A_I^\pm &=& \Big[ 4\, C_{24} - 2 + 2 m_f^2 (2\, C_0+3\, C_{11}+C_{21})
            -2 s(C_0+C_{11}+C_{23}-C_{22}) \Big] g_1^\pm g_2^\pm 
             g_3^\pm \nonumber\\
        & & +2 m_f^2 (C_0+C_{11})g_1^\mp g_2^\mp g_3^\mp
            -2 m_{\hat{f}}^2\, C_0\, g_1^\pm g_2^\pm g_3^\mp~,
            \nonumber\\
B_I^\pm &=& 2 \bigg\{ m_{\hat{f}}(C_0+C_{11})g_1^\mp g_2^\pm 
            (g_3^\pm + g_3^\mp)
            -m_f \Big[ (C_0+C_{11}+C_{12}+C_{23})g_1^\pm g_2^\pm g_3^\pm
            \nonumber \\
        & & +(C_0+2\, C_{11}-C_{12}+C_{21}-C_{23})g_1^\mp g_2^\mp g_3^\mp
            \Big] \bigg\}~.
\label{vert1}
\EEA
\vspace{1cm}

\renewcommand{\arraystretch}{1.8}
\[
\arraycolsep4mm
\begin{array}{|c@{}c@{}c||*{6}{c|}}\hline
\multicolumn{9}{|c|}{\rule[-2mm]{0pt}{6mm}
\mbox{\bf Masses and coupling constants for the $\gamma f\bar{f}$ 
  vertex I}} \\
  \hline
  m_1\; & m_2\; & m_3 & \quad g_1^+\quad & \quad g_1^-\quad & 
  \quad g_2^+\quad & \quad g_2^-\quad & \quad g_3^+\quad & 
  \quad g_3^-\quad \\
  \hline\hline
  \rm Z & \rm f & \rm f & 
  g_{Z,R} & g_{Z,L} & g_{Z,R} & g_{Z,L} & -Q_f & -Q_f \\
  \hline
  \rm W & \rm {f'} & \rm {f'} & 0 & \D \frac{1}{\sqrt{2}\, \sw} &
  0 & \D \frac{1}{\sqrt{2}\, \sw} & -Q_{f'} & -Q_{f'} \\
  \hline
\end{array} 
\]
\vspace{1cm}
\[\arraycolsep4mm
\begin{array}{|c@{}c@{}c||*{6}{c|}}\hline
\multicolumn{9}{|c|}{\rule[-2mm]{0pt}{6mm}
\mbox{\bf Masses and coupling constants for the $Z f\bar{f}$ vertex I}} \\
  \hline
  m_1\; & m_2\; & m_3 & \quad g_1^+\quad & \quad g_1^-\quad & 
  \quad g_2^+\quad & \quad g_2^-\quad & \quad g_3^+\quad & 
  \quad g_3^-\quad \\
  \hline\hline
  \rm Z & \rm f & \rm f & 
  g_{Z,R} & g_{Z,L} & g_{Z,R} & g_{Z,L} & g_{Z,R} & g_{Z,L} \\
  \hline
  \rm W & \rm {f'} & \rm {f'} & 0 & \D \frac{1}{\sqrt{2}\, \sw} &
  0 &\D \frac{1}{\sqrt{2}\, \sw} & g'_{Z,R} & g'_{Z,L} \\
  \hline
\end{array}
\]

\newpage

\setcounter{dummy}{2}
\subsection{Class-\Roman{dummy} diagram}
\vspace{5mm}
\centerline{\psfig{figure=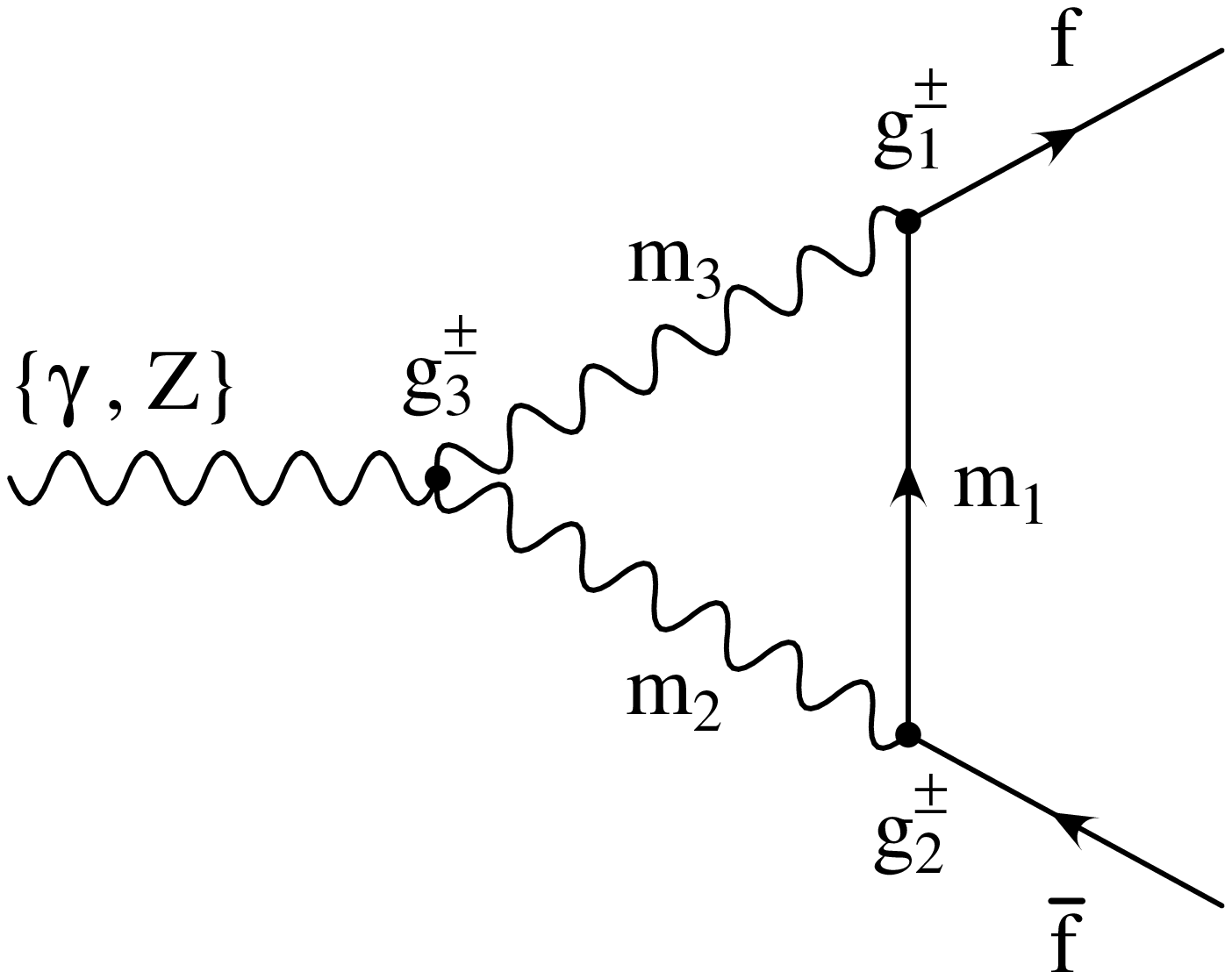,width=4.5cm}}
\vspace{5mm}
\BEA
A_{I\!I}^+ &=& 3 m_f^2\, C_{11}\, g_1^- g_2^- g_3^-~, \nonumber\\
A_{I\!I}^- &=& \Big[ 12\, C_{24} - 2 + 2m_f^2\, C_{21} 
               -2s (C_{23}-C_{22}) + (5m_f^2-2s)C_{11} \Big] 
               g_1^- g_2^- g_3^-~, \nonumber\\
B_{I\!I}^+ &=& m_f (2 C_{21}-2 C_{23}-C_{12}) 
               g_1^- g_2^- g_3^-~,\nonumber\\
B_{I\!I}^- &=& m_f (2\, C_{23}-C_{11}+C_{12})g_1^- g_2^- g_3^-~.
\label{vert2}
\EEA
\vspace{1cm}
\[
\arraycolsep5mm
\begin{array}{|c@{}c@{}c||*{5}{c|}}\hline
\multicolumn{8}{|c|}{\rule[-2mm]{0pt}{6mm}
\mbox{\bf Masses and coupling constants for the $\gamma f\bar{f}$ 
  vertex II}} \\
  \hline
  m_1\; & m_2\; & m_3 & \quad g_1^+\quad & \quad g_1^-\quad &
  \quad g_2^+\quad & \quad g_2^-\quad & \; g_3^+ = g_3^- \; \\
  \hline\hline
  \rm {f'} & \rm W & \rm W & 0 &\D \frac{1}{\sqrt{2}\, \sw} & 
  0 &\D \frac{1}{\sqrt{2}\, \sw} & -2I_3^f \\
  \hline
\end{array}
\]
\vspace{1cm}
\[
\arraycolsep5mm
\begin{array}{|c@{}c@{}c||*{5}{c|}}\hline
\multicolumn{8}{|c|}{\rule[-2mm]{0pt}{6mm}
\mbox{\bf Masses and coupling constants for the $Z f\bar{f}$ 
  vertex II}} \\
  \hline
  m_1\; & m_2\; & m_3 & \quad g_1^+\quad & \quad g_1^-\quad &
  \quad g_2^+\quad & \quad g_2^-\quad & \; g_3^+ = g_3^- \; \\
  \hline\hline
  \rm {f'} & \rm W & \rm W &\D 0 &\D \frac{1}{\sqrt{2}\, \sw} & 
  0 &\D \frac{1}{\sqrt{2}\, \sw} &\D 2I_3^f\, \frac{\cw}{\sw} \\
  \hline
\end{array}
\]

\newpage

\setcounter{dummy}{3}
\subsection{Class-\Roman{dummy} diagram}
\vspace{5mm}
\begin{center}
\begin{tabular}{ccc}
\raisebox{-22mm}{\psfig{bbllx=34pt,bblly=160pt,bburx=494pt,%
bbury=587pt,figure=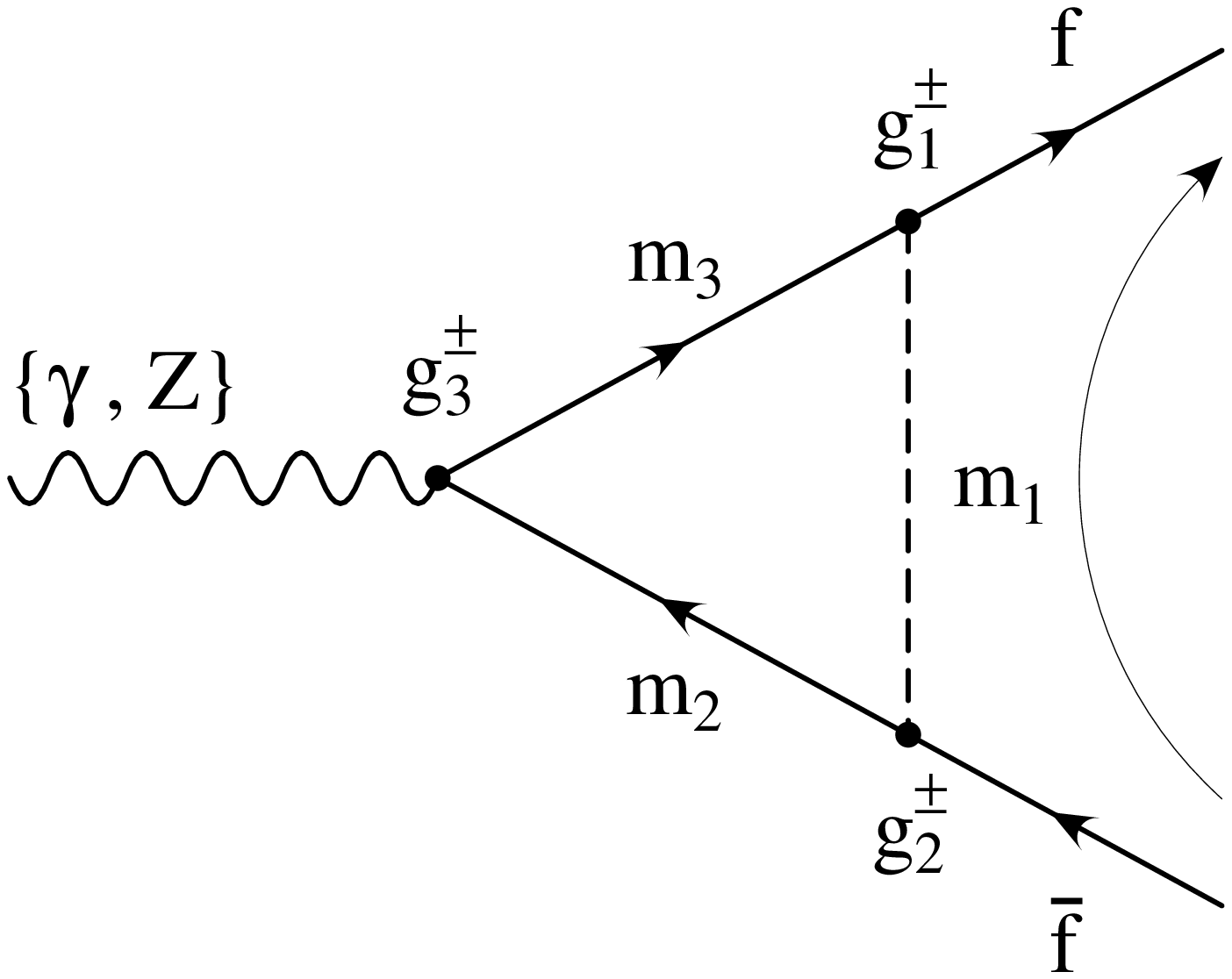,width=4.5cm}} & \hspace{1cm} &
\raisebox{-22mm}{\psfig{bbllx=34pt,bblly=160pt,bburx=494pt,%
bbury=587pt,figure=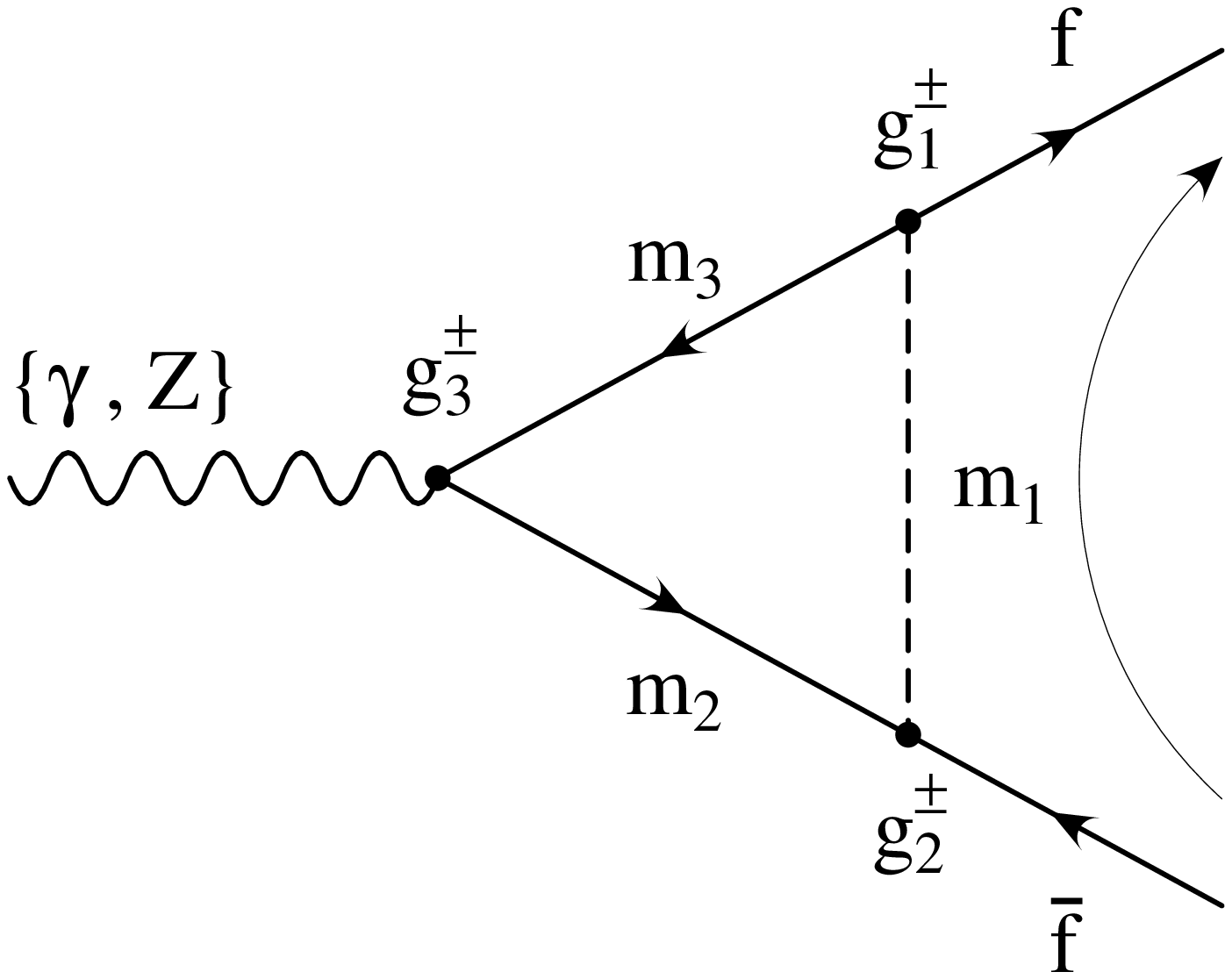,width=4.5cm}}
\end{tabular}
\end{center}
\vspace{5mm}
\vorn In class-\rm{\Roman{dummy}}, the dashed line indicates either a 
Higgs particle or a sfermion.
\BEA
A_{V}^\pm &=& \Big[ 2\, C_{24} - \frac{1}{2}
              +m_f^2 (C_{11}+C_{21}) - s (C_{23}-C_{22}) \Big]
               g_1^\mp g_2^\pm g_3^\mp \nonumber\\
          & & -m_2 m_3\, C_0\, g_1^\mp g_2^\pm g_3^\pm
              -m_f^2 (C_0+C_{11}) g_1^\pm g_2^\mp g_3^\pm \nonumber\\
          & & -m_f m_2 \Big[ (C_0+C_{12}) g_1^\pm g_2^\pm g_3^\pm
              -(C_{11}-C_{12}) g_1^\mp g_2^\mp g_3^\mp \Big]~,
              \nonumber \\
          & & -m_f m_3 \Big[(C_0+C_{11}-C_{12})g_1^\mp g_2^\mp g_3^\pm
              -C_{12}\, g_1^\pm g_2^\pm g_3^\mp \Big] \nonumber \\
B_{V}^\pm &=& -\bigg\{ \Big[ m_2 (C_{11}-C_{12}) g_3^\pm 
              + m_3\, C_{12}\, g_3^\mp \Big] g_1^\pm g_2^\pm \nonumber\\
          & & +m_f \Big[ (C_{11}-C_{12}+C_{21}-C_{23})
              g_1^\pm g_2^\mp g_3^\pm + (C_{12}+C_{23})
              g_1^\mp g_2^\pm g_3^\mp \Big] \bigg\}~.
\label{vert5}
\EEA

The tables contain the extra notations
\BEA 
O^f_a &=& \left\{ \begin{array}{rl} 
          O'^R_{ji}\quad & \mbox{for $f$ = $u$-type} \\
         -O'^L_{ij}\quad & \mbox{for $f$ = $d$-type} \end{array} 
          \right.~, \\ 
O^f_b &=& \left\{ \begin{array}{rl} 
          O'^L_{ji}\quad & \mbox{for $f$ = $u$-type} \\
         -O'^R_{ij}\quad & \mbox{for $f$ = $d$-type} \end{array} 
          \right.~. 
\EEA
$O'^{L,R}_{ij}$ and $O''^{L,R}_{ij}$ are defined in Eq. (\ref{Os})
and $X^{0f}_{kl}$, $X^{+f}_{kl}$,  $Y^{0f}_{kl}$ and $Y^{+f}_{kl}$ in 
Eqs. (\ref{viai}).

\newpage

\[
\arraycolsep3mm
\begin{array}{|c@{}c@{}c||*{6}{c|}}\hline
\multicolumn{9}{|c|}{\rule[-2mm]{0pt}{6mm}
\mbox{\bf Masses and coupling constants for the $\gamma f\bar{f}$ 
  vertex \rm{\Roman{dummy}} }} \\
  \hline
  m_1\; & m_2\; & m_3 & \quad g_1^+ \quad & \quad g_1^-\quad &
  \quad g_2^+\quad & \quad g_2^-\quad & \qquad g_3^+\qquad &
  \qquad g_3^-\qquad \\
  \hline\hline
  \rm H^0 & \rm f & \rm f & g_N\, g^f_{H^0} & g_N\, g^f_{H^0} &
  g_N\, g^f_{H^0} & g_N\, g^f_{H^0} & -Q_f & -Q_f \\
  \hline
  \rm h^0 & \rm f & \rm f & g_N\, g^f_{h^0} & g_N\, g^f_{h^0} &
  g_N\, g^f_{h^0} & g_N\, g^f_{h^0} & -Q_f & -Q_f \\
  \hline
  \rm A^0 & \rm f & \rm f & -i\, g_N\, g^f_{A^0} &
  i\, g_N\, g^f_{A^0} & -i\, g_N\, g^f_{A^0} &
  i\, g_N\, g^f_{A^0} & -Q_f & -Q_f \\
  \hline
  \rm G^0 & \rm f & \rm f & -i\, g_N\, 2I_3^f &
  i\, g_N\, 2I_3^f & -i\, g_N\, 2I_3^f &
  i\, g_N\, 2I_3^f & -Q_f & -Q_f \\
  \hline
  \rm H^\pm & \rm {f'} & \rm {f'} & g'_C\, g^{f'}_{H^\pm} &
  g_C\, g^f_{H^\pm} & g_C\, g^f_{H^\pm} & g'_C\, g^{f'}_{H^\pm} & 
  -Q_{f'} & -Q_{f'} \\
  \hline
  \rm G^\pm & \rm {f'} & \rm {f'} & g'_C\, 2I_3^{f'} &
  g_C\, 2I_3^f &\D g_C\, 2I_3^f & g'_C\, 2I_3^{f'} & -Q_{f'} & -Q_{f'} \\ 
  \hline
  \rm \tilde{f}'_1 & \rm \tilde{\chi}^+_i & \rm \tilde{\chi}^+_i &
  X^{+f}_{i1} & (Y^{+f}_{i1})^* & Y^{+f}_{i1} & (X^{+f}_{i1})^* & 
  -2I_3^f & -2I_3^f \\ 
  \hline
  \rm \tilde{f}'_2 & \rm \tilde{\chi}^+_i & \rm \tilde{\chi}^+_i &
  X^{+f}_{i2} & (Y^{+f}_{i2})^* & Y^{+f}_{i2} & (X^{+f}_{i2})^* & 
  -2I_3^f & -2I_3^f \\ 
  \hline
\end{array}
\]
\vspace{1cm}
\[
\arraycolsep3mm
\begin{array}{|c@{}c@{}c||*{6}{c|}}\hline
\multicolumn{9}{|c|}{\rule[-2mm]{0pt}{6mm}
\mbox{\bf Masses and coupling constants for the $Z f\bar{f}$ 
  vertex \rm{\Roman{dummy}} }} \\
  \hline
  m_1\; & m_2\; & m_3 & \quad g_1^+ \quad & \quad g_1^ -\quad &
  \quad g_2^+ \quad & \quad g_2^- \quad & \qquad g_3^+ \qquad &
  \qquad g_3^- \qquad \\
  \hline\hline
  \rm H^0 & \rm f & \rm f & g_N\, g^f_{H^0} & g_N\, g^f_{H^0} &
  g_N\, g^f_{H^0} & g_N\, g^f_{H^0} & g_{Z,R} &\D g_{Z,L} \\
  \hline
  \rm h^0 & \rm f & \rm f & g_N\, g^f_{h^0} & g_N\, g^f_{h^0} &
  g_N\, g^f_{h^0} & g_N\, g^f_{h^0} & g_{Z,R} & g_{Z,L} \\
  \hline
  \rm A^0 & \rm f & \rm f &\D -i\, g_N\, g^f_{A^0} &
  i\, g_N\, g^f_{A^0} & -i\, g_N\, g^f_{A^0} &
  \D i\, g_N\, g^f_{A^0} & g_{Z,R} &\D g_{Z,L} \\
  \hline
  \rm G^0 & \rm f & \rm f & -i\, g_N\, 2I_3^f & i\, g_N\, 2I_3^f & 
  -i\, g_N\, 2I_3^f & i\, g_N\, 2I_3^f & g_{Z,R} &\D g_{Z,L} \\
  \hline
  \rm H^\pm & \rm {f'} & \rm {f'} & g'_C\, g^{f'}_{H^\pm} &
  g_C\, g^f_{H^\pm} & g_C\, g^f_{H^\pm} &
  g'_C\, g^{f'}_{H^\pm} & g'_{Z,R} &\D g'_{Z,L} \\
  \hline
  \rm G^\pm & \rm {f'} & \rm {f'} & g'_C\, 2I_3^{f'} & g_C\, 2I_3^f & 
  g_C\, 2I_3^f & g'_C\, 2I_3^{f'} & g'_{Z,R} & g'_{Z,L} \\ 
  \hline
  \rm \tilde{f}_1 & \rm \tilde{\chi}^0_i & \rm \tilde{\chi}^0_j &
  X^{0f}_{j1} & (Y^{0f}_{j1})^* & Y^{0f}_{i1} & (X^{0f}_{i1})^* &
  \D -\frac{1}{\sw\, \cw}\, O''^R_{ji} & 
  \D -\frac{1}{\sw\, \cw}\, O''^L_{ji} \\ 
  \hline
  \rm \tilde{f}_2 & \rm \tilde{\chi}^0_i & \rm \tilde{\chi}^0_j &
  X^{0f}_{j2} & (Y^{0f}_{j2})^* & Y^{0f}_{i2} & (X^{0f}_{i2})^* &
  \D -\frac{1}{\sw\, \cw}\, O''^R_{ji} & 
  \D -\frac{1}{\sw\, \cw}\, O''^L_{ji} \\
  \hline
  \rm \tilde{f}'_1 & \rm \tilde{\chi}^+_i & \rm \tilde{\chi}^+_j &
  X^{+f}_{j1} & (Y^{+f}_{j1})^* & Y^{+f}_{i1} & (X^{+f}_{i1})^* &
  \D -\frac{1}{\sw\, \cw}\, O^f_a & \D -\frac{1}{\sw\, \cw}\, O^f_b \\ 
  \hline
  \rm \tilde{f}'_2 & \rm \tilde{\chi}^+_i & \rm \tilde{\chi}^+_j &
  X^{+f}_{j2} & (Y^{+f}_{j2})^* & Y^{+f}_{i2} & (X^{+f}_{i2})^* &
  \D -\frac{1}{\sw\, \cw}\, O^f_a & \D -\frac{1}{\sw\, \cw}\, O^f_b \\
  \hline
\end{array}
\]

\newpage

\setcounter{dummy}{4}
\subsection{Class-\Roman{dummy} diagram}
\vspace{5mm}
\begin{tabular}{ccc}
\raisebox{-25mm}{\psfig{bbllx=34pt,bblly=160pt,bburx=494pt,%
bbury=587pt,figure=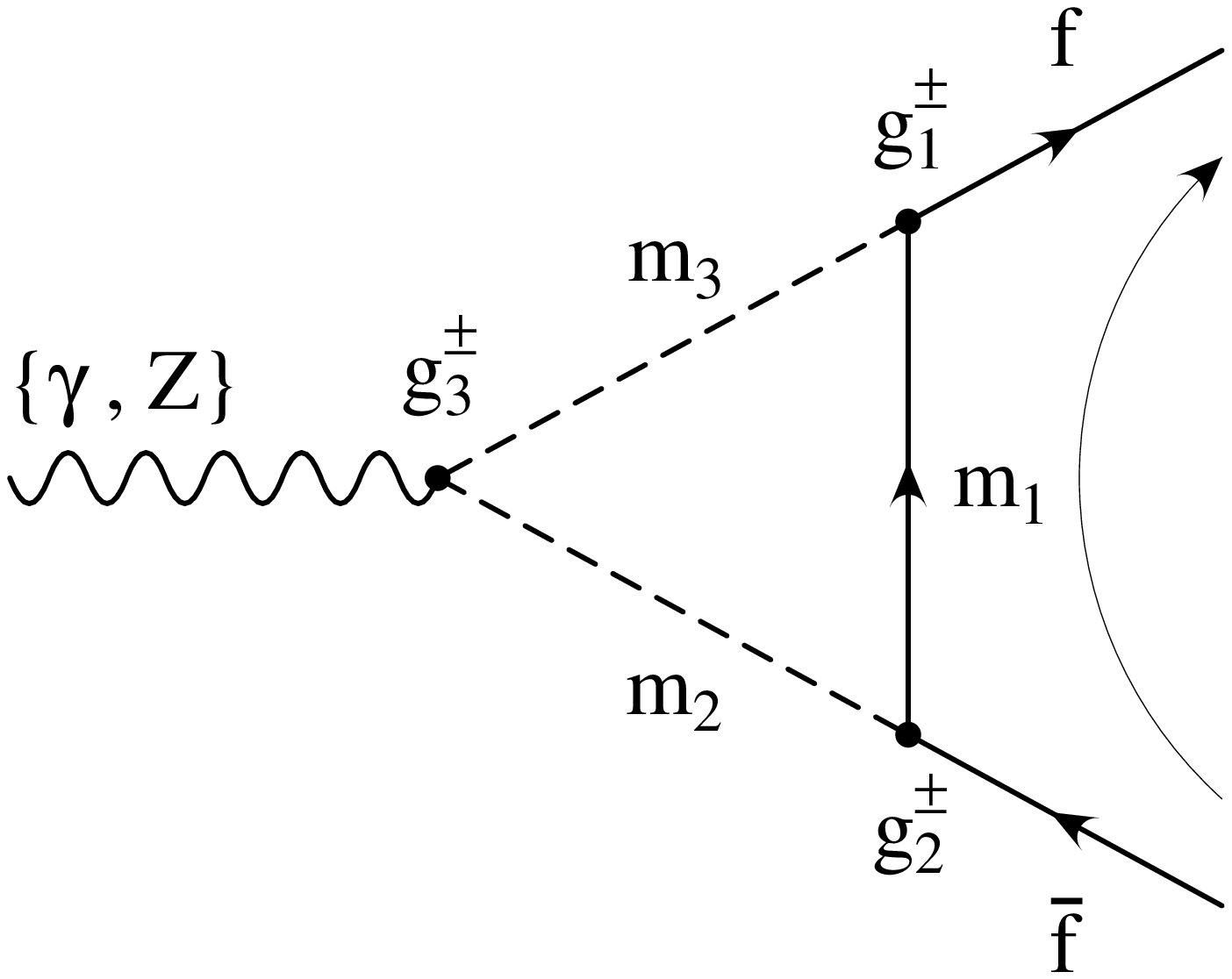,width=4.5cm}} &
\raisebox{-25mm}{\psfig{bbllx=34pt,bblly=160pt,bburx=494pt,%
bbury=587pt,figure=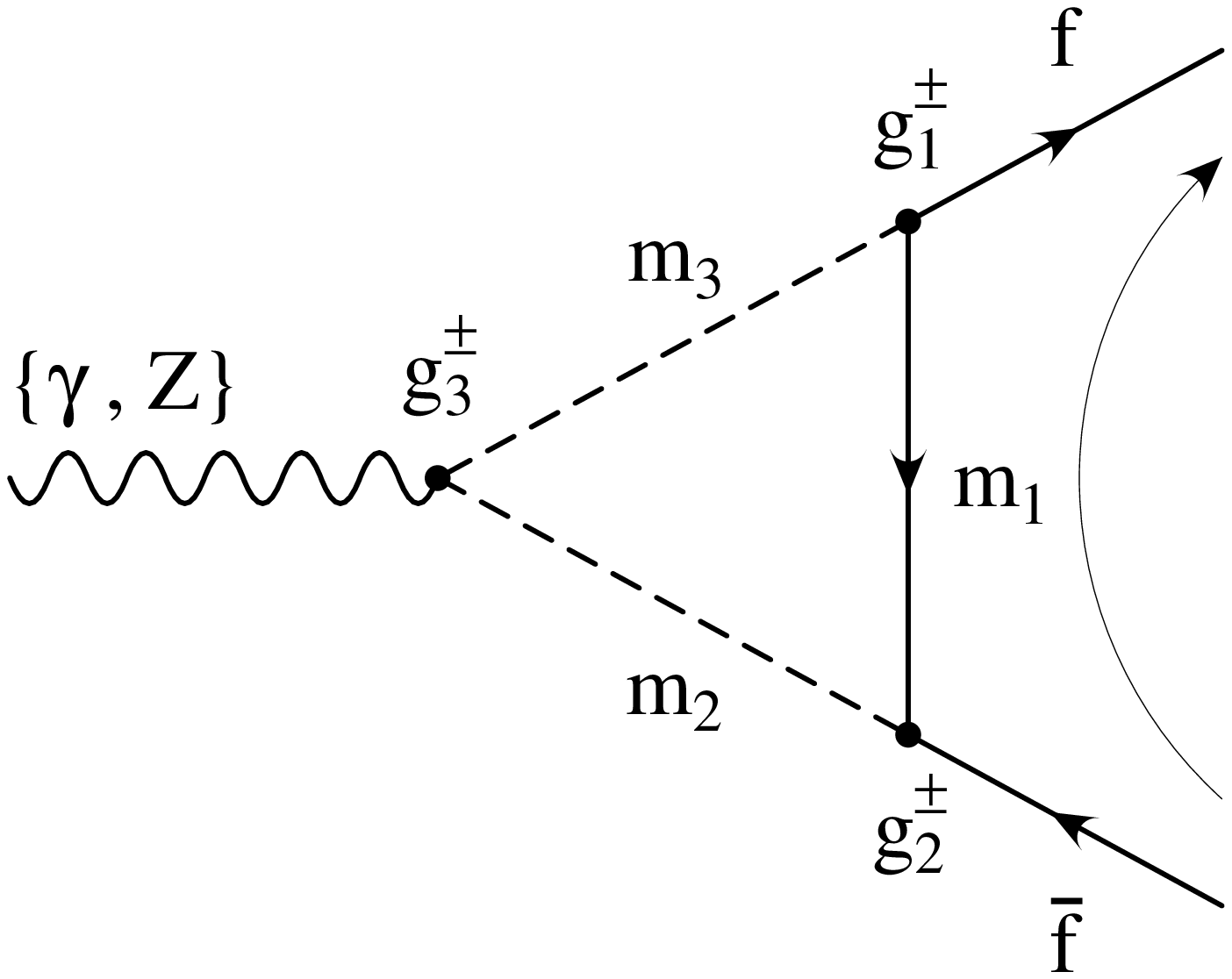,width=4.5cm}} &
\raisebox{-25mm}{\psfig{bbllx=34pt,bblly=160pt,bburx=494pt,%
bbury=587pt,figure=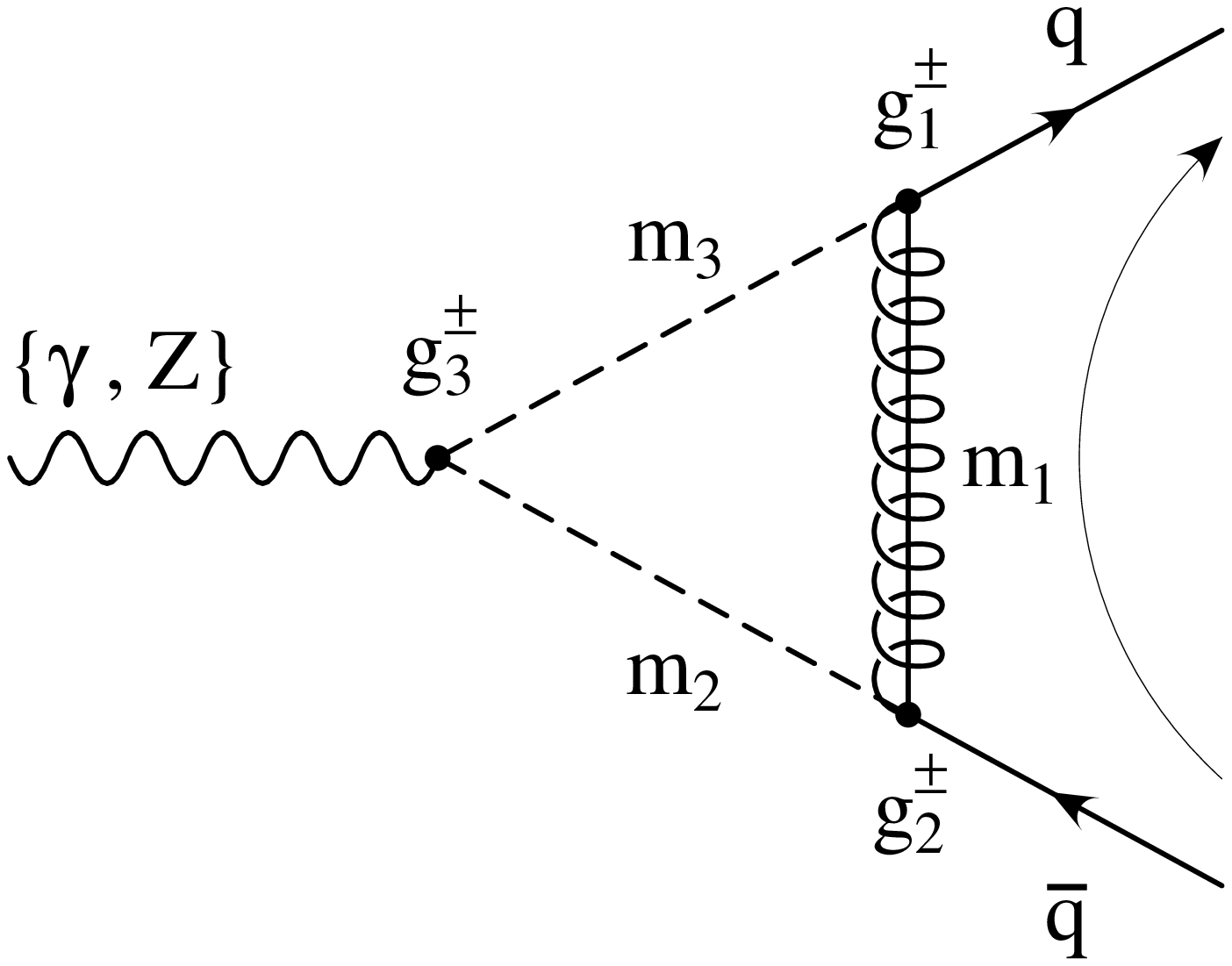,width=4.5cm}}
\end{tabular}

\vspace{5mm}

\vorn The signature for the dashed lines is as in class-V.
In addition there is the gluino diagram.
\BEA
A_{V\!I}^\pm &=& 2\, C_{24}\, g_1^\mp g_2^\pm g_3^\pm~,  \nonumber\\
B_{V\!I}^\pm &=& g_3^\pm \bigg\{ m_f \Big[ (C_{12}+C_{23}) g_1^\mp g_2^\pm
                 + (C_{11}-C_{12}+C_{21}-C_{23}) g_1^\pm g_2^\mp \Big]
                 \nonumber \\ 
             & & -m_1 (C_0+C_{11}) g_1^\pm g_2^\pm \bigg\}~.
\label{vert6}
\EEA

The following shorthand notations are used:
\BEA
g_{Z,1} &=& \frac{I^f_3\, \CtT^2 - Q_f\, \sw^2}{\sw\,\cw}~, \nonumber \\
g_{Z,2} &=& \frac{I^f_3\, \StT^2 - Q_f\, \sw^2}{\sw\,\cw}~, \nonumber \\
g_Z     &=& -\frac{I^f_3\, \StT\, \CtT}{\sw\,\cw}~, \nonumber \\
g'_{Z,1} &=& \frac{I^{f'}_3\, \CtTp^2 - Q_{f'}\, \sw^2}{\sw\,\cw}~, 
             \nonumber \\
g'_{Z,2} &=& \frac{I^{f'}_3\, \StTp^2 - Q_{f'}\, \sw^2}{\sw\,\cw}~,
             \nonumber \\
g'_Z     &=& -\frac{I^{f'}_3\, \StTp\, \CtTp}{\sw\,\cw}~,
\EEA
with $\CtT:=\cos \tilde{\theta}_f$, $\StT:=\sin \tilde{\theta}_f$ and
$\CtTp:=\cos \tilde{\theta}_{f'}$, $\StTp:=\sin \tilde{\theta}_{f'}$.

\newpage

\[
\arraycolsep2mm
\begin{array}{|c@{}c@{}c||*{5}{c|}}\hline
\multicolumn{8}{|c|}{\rule[-2mm]{0pt}{6mm}
\mbox{\bf Masses and coupling constants for the $\gamma f\bar{f}$ 
  vertex \rm{\Roman{dummy}} }} \\
  \hline
  m_1\; & m_2\; & m_3 & \quad g_1^+ \quad & \quad g_1^-\quad &
  \quad g_2^+\quad & \quad g_2^-\quad & \quad g_3^+ = g_3^- \quad \\
  \hline\hline
  \rm {f'} & \rm H^\pm & \rm H^\pm  & g'_C\, g^{f'}_{H^\pm} &
  g_C\, g^f_{H^\pm} & g_C\, g^f_{H^\pm} & g'_C\, g^{f'}_{H^\pm} & 
  -2I_3^f \\
  \hline
  \rm {f'} & \rm G^\pm & \rm G^\pm & g'_C\, 2I_3^{f'} &
  g_C\, 2I_3^f &\D g_C\, 2I_3^f & g'_C\, 2I_3^{f'} & -2I_3^f \\ 
  \hline
  \rm \tilde{\chi}^0_i & \rm \tilde{f}_1 & \rm \tilde{f}_1 &
  X^{0f}_{i1} & (Y^{0f}_{i1})^* & Y^{0f}_{i1} & (X^{0f}_{i1})^* & 
  -Q_f \\ 
  \hline
  \rm \tilde{\chi}^0_i & \rm \tilde{f}_2 & \rm \tilde{f}_2 &
  X^{0f}_{i2} & (Y^{0f}_{i2})^* & Y^{0f}_{i2} & (X^{0f}_{i2})^* & 
  -Q_f \\
  \hline
  \rm \tilde{\chi}^+_i & \rm \tilde{f}'_1 & \rm \tilde{f}'_1 &
  X^{+f}_{i1} & (Y^{+f}_{i1})^* & Y^{+f}_{i1} & (X^{+f}_{i1})^* & 
  -Q_{f'} \\ 
  \hline
  \rm \tilde{\chi}^+_i & \rm \tilde{f}'_2 & \rm \tilde{f}'_2 &
  X^{+f}_{i2} & (Y^{+f}_{i2})^* & Y^{+f}_{i2} & (X^{+f}_{i2})^* & 
  -Q_{f'} \\
  \hline
  \rm \tilde{g}_a & \rm \tilde{q}_1 & \rm \tilde{q}_1 &
  \D -\sqrt{2}\, \frac{g_3}{e}\, \CtT\, T^a_{jk}\, & 
  \D  \sqrt{2}\, \frac{g_3}{e}\, \StT\, T^a_{jk}\, & 
  \D  \sqrt{2}\, \frac{g_3}{e}\, \StT\, T^a_{ij}\, & 
  \D -\sqrt{2}\, \frac{g_3}{e}\, \CtT\, T^a_{ij}\, & -Q_q \\ 
  \hline
  \rm \tilde{g}_a & \rm \tilde{q}_2 & \rm \tilde{q}_2 &
  \D \sqrt{2}\, \frac{g_3}{e}\, \StT\, T^a_{jk}\, & 
  \D \sqrt{2}\, \frac{g_3}{e}\, \CtT\, T^a_{jk}\, & 
  \D \sqrt{2}\, \frac{g_3}{e}\, \CtT\, T^a_{ij}\, & 
  \D \sqrt{2}\, \frac{g_3}{e}\, \StT\, T^a_{ij}\, & -Q_q \\ 
  \hline
\end{array}
\]

\vspace{1.5cm} 

The $T^a_{ij}\; (a=1,2,\ldots,8\; ; \;i,j=1,2,3)$ are the matrix
elements of the $SU(3)_C$ generators, with the property
\BE
\sum_{a,j}\, T^a_{ij}\, T^a_{jk} = 
\frac{1}{4}\, \sum_{a,j}\, \lambda^a_{ij}\, \lambda^a_{jk} =
\frac{1}{4}\, \sum_{j}\, (-\frac{2}{3}\, \delta_{ij}\, \delta_{jk} 
+ 2\, \delta_{ik}\, \delta_{jj})\, \stackrel{i=k}{=}\,
 \frac{4}{3} =: C_F~.
\label{cf} 
\EE

$X^{0f}_{kl}$, $X^{+f}_{kl}$,  $Y^{0f}_{kl}$ and $Y^{+f}_{kl}$ are 
defined in Eq. (\ref{viai}). The factor $1/e$ in the gluino couplings 
only appears because of the definition of the matrix elements 
(\ref{inver}) and (\ref{outver}).

\newpage

\[
\arraycolsep2mm
\begin{array}{|c@{}c@{}c||*{5}{c|}}\hline
\multicolumn{8}{|c|}{\rule[-2mm]{0pt}{6mm}
\mbox{\bf Masses and coupling constants for the $Z f\bar{f}$ 
  vertex \rm{\Roman{dummy}} }} \\
  \hline
  m_1\; & m_2\; & m_3 & \quad g_1^+ \quad & \quad g_1^-\quad & 
  \quad g_2^+\quad & \quad g_2^-\quad & \; g_3^+ = g_3^- \; \\
  \hline\hline
  \rm f & \rm H^0 & \rm A^0 & -i\, g_N\, g^f_{A^0} & 
  i\, g_N\, g^f_{A^0} & g_N\, g^f_{H^0} & g_N\, g^f_{H^0} &
  \D -i\, \frac{\Sba}{2\sw\cw}  \\
  \hline
  \rm f & \rm H^0 & \rm G^0 & -i\, g_N\, 2I_3^f &
  i\, g_N\, 2I_3^f & g_N\, g^f_{H^0} & g_N\, g^f_{H^0} &
  \D i\, \frac{\Cba}{2\sw\cw}  \\
  \hline
  \rm f & \rm h^0 & \rm A^0 & -i\, g_N\, g^f_{A^0} &
  i\, g_N\, g^f_{A^0} & g_N\, g^f_{h^0} & g_N\, g^f_{h^0} &
  \D i\, \frac{\Cba}{2\sw\cw}  \\
  \hline
  \rm f & \rm h^0 & \rm G^0 & -i\, g_N\, 2I_3^f &
  i\, g_N\, 2I_3^f & g_N\, g^f_{h^0} & g_N\, g^f_{h^0} &
  \D i\, \frac{\Sba}{2\sw\cw}  \\
  \hline
  \rm f & \rm A^0 & \rm H^0 & g_N\, g^f_{H^0} &
  g_N\, g^f_{H^0} & -i\, g_N\, g^f_{A^0} & i\, g_N\, g^f_{A^0} &
  \D i\, \frac{\Sba}{2\sw\cw}  \\
  \hline
  \rm f & \rm G^0 & \rm H^0 & g_N\, g^f_{H^0} &
  g_N\, g^f_{H^0} & -i\, g_N\, 2I_3^f & i\, g_N\, 2I_3^f &
  \D -i\, \frac{\Cba}{2\sw\cw}  \\
  \hline
  \rm f & \rm A^0 & \rm h^0 & g_N\, g^f_{h^0} &
  g_N\, g^f_{h^0} & -i\, g_N\, g^f_{A^0} & i\, g_N\, g^f_{A^0} &
  \D -i\, \frac{\Cba}{2\sw\cw}  \\
  \hline
  \rm f & \rm G^0 & \rm h^0 & g_N\, g^f_{h^0} &
  g_N\, g^f_{h^0} & -i\, g_N\, 2I_3^f & i\, g_N\, 2I_3^f &
  \D -i\, \frac{\Sba}{2\sw\cw}  \\
  \hline
  \rm {f'} & \rm H^\pm & \rm H^\pm & g'_C\, g^{f'}_{H^\pm} &
  g_C\, g^f_{H^\pm} & g_C\, g^f_{H^\pm} & g'_C\, g^{f'}_{H^\pm} &
  \D -2I_3^f\, \frac{\sw^2-\cw^2}{2\sw\cw} \\
  \hline
  \rm {f'} & \rm G^\pm & \rm G^\pm & g'_C\, 2I_3^{f'} &
  g_C\, 2I_3^f & g_C\, 2I_3^f & g'_C\, 2I_3^{f'} &
  \D -2I_3^f\, \frac{\sw^2-\cw^2}{2\sw\cw} \\ 
  \hline
  \rm \tilde{\chi}^0_i & \rm \tilde{f}_1 & \rm \tilde{f}_1 &
  X^{0f}_{i1} & (Y^{0f}_{i1})^* & Y^{0f}_{i1} & (X^{0f}_{i1})^* & 
  g_{Z,1} \\ 
  \hline
  \rm \tilde{\chi}^0_i & \rm \tilde{f}_1 & \rm \tilde{f}_2 &
  X^{0f}_{i2} & (Y^{0f}_{i2})^* & Y^{0f}_{i1} & (X^{0f}_{i1})^* & g_Z \\
  \hline
  \rm \tilde{\chi}^0_i & \rm \tilde{f}_2 & \rm \tilde{f}_1 &
  X^{0f}_{i1} & (Y^{0f}_{i1})^* & Y^{0f}_{i2} & (X^{0f}_{i2})^* & g_Z \\
  \hline
  \rm \tilde{\chi}^0_i & \rm \tilde{f}_2 & \rm \tilde{f}_2 &
  X^{0f}_{i2} & (Y^{0f}_{i2})^* & Y^{0f}_{i2} & (X^{0f}_{i2})^* & 
  g_{Z,2} \\
  \hline
  \rm \tilde{\chi}^+_i & \rm \tilde{f}'_1 & \rm \tilde{f}'_1 &
  X^{+f}_{i1} & (Y^{+f}_{i1})^* & Y^{+f}_{i1} & (X^{+f}_{i1})^* & 
  g'_{Z,1} \\ 
  \hline
  \rm \tilde{\chi}^+_i & \rm \tilde{f}'_1 & \rm \tilde{f}'_2 &
  X^{+f}_{i2} & (Y^{+f}_{i2})^* & Y^{+f}_{i1} & (X^{+f}_{i1})^* & 
  g'_Z \\
  \hline
  \rm \tilde{\chi}^+_i & \rm \tilde{f}'_2 & \rm \tilde{f}'_1 &
  X^{+f}_{i1} & (Y^{+f}_{i1})^* & Y^{+f}_{i2} & (X^{+f}_{i2})^* & 
  g'_Z \\
  \hline
  \rm \tilde{\chi}^+_i & \rm \tilde{f}'_2 & \rm \tilde{f}'_2 &
  X^{+f}_{i2} & (Y^{+f}_{i2})^* & Y^{+f}_{i2} & (X^{+f}_{i2})^* & 
  g'_{Z,2} \\
  \hline
  \rm \tilde{g}_a & \rm \tilde{q}_1 & \rm \tilde{q}_1 &
  \D -\sqrt{2}\, \frac{g_3}{e}\, \CtT\, T^a_{jk}\, & 
  \D  \sqrt{2}\, \frac{g_3}{e}\, \StT\, T^a_{jk}\, & 
  \D  \sqrt{2}\, \frac{g_3}{e}\, \StT\, T^a_{ij}\, & 
  \D -\sqrt{2}\, \frac{g_3}{e}\, \CtT\, T^a_{ij}\, & g_{Z,1} \\ 
  \hline
  \rm \tilde{g}_a & \rm \tilde{q}_1 & \rm \tilde{q}_2 &
  \D  \sqrt{2}\, \frac{g_3}{e}\, \StT\, T^a_{jk}\, & 
  \D  \sqrt{2}\, \frac{g_3}{e}\, \CtT\, T^a_{jk}\, & 
  \D  \sqrt{2}\, \frac{g_3}{e}\, \StT\, T^a_{ij}\, & 
  \D -\sqrt{2}\, \frac{g_3}{e}\, \CtT\, T^a_{ij}\, & g_Z \\ 
  \hline
  \rm \tilde{g}_a & \rm \tilde{q}_2 & \rm \tilde{q}_1 &
  \D -\sqrt{2}\, \frac{g_3}{e}\, \CtT\, T^a_{jk}\, & 
  \D  \sqrt{2}\, \frac{g_3}{e}\, \StT\, T^a_{jk}\, & 
  \D  \sqrt{2}\, \frac{g_3}{e}\, \CtT\, T^a_{ij}\, & 
  \D  \sqrt{2}\, \frac{g_3}{e}\, \StT\, T^a_{ij}\, & g_Z \\ 
  \hline
  \rm \tilde{g}_a & \rm \tilde{q}_2 & \rm \tilde{q}_2 &
  \D \sqrt{2}\, \frac{g_3}{e}\, \StT\, T^a_{jk}\, & 
  \D \sqrt{2}\, \frac{g_3}{e}\, \CtT\, T^a_{jk}\, & 
  \D \sqrt{2}\, \frac{g_3}{e}\, \CtT\, T^a_{ij}\, & 
  \D \sqrt{2}\, \frac{g_3}{e}\, \StT\, T^a_{ij}\, & g_{Z,2} \\ 
  \hline
\end{array}
\]

\newpage

\setcounter{dummy}{5}
\subsection{Class-\Roman{dummy} diagram}
\vspace{5mm}
\centerline{\psfig{figure=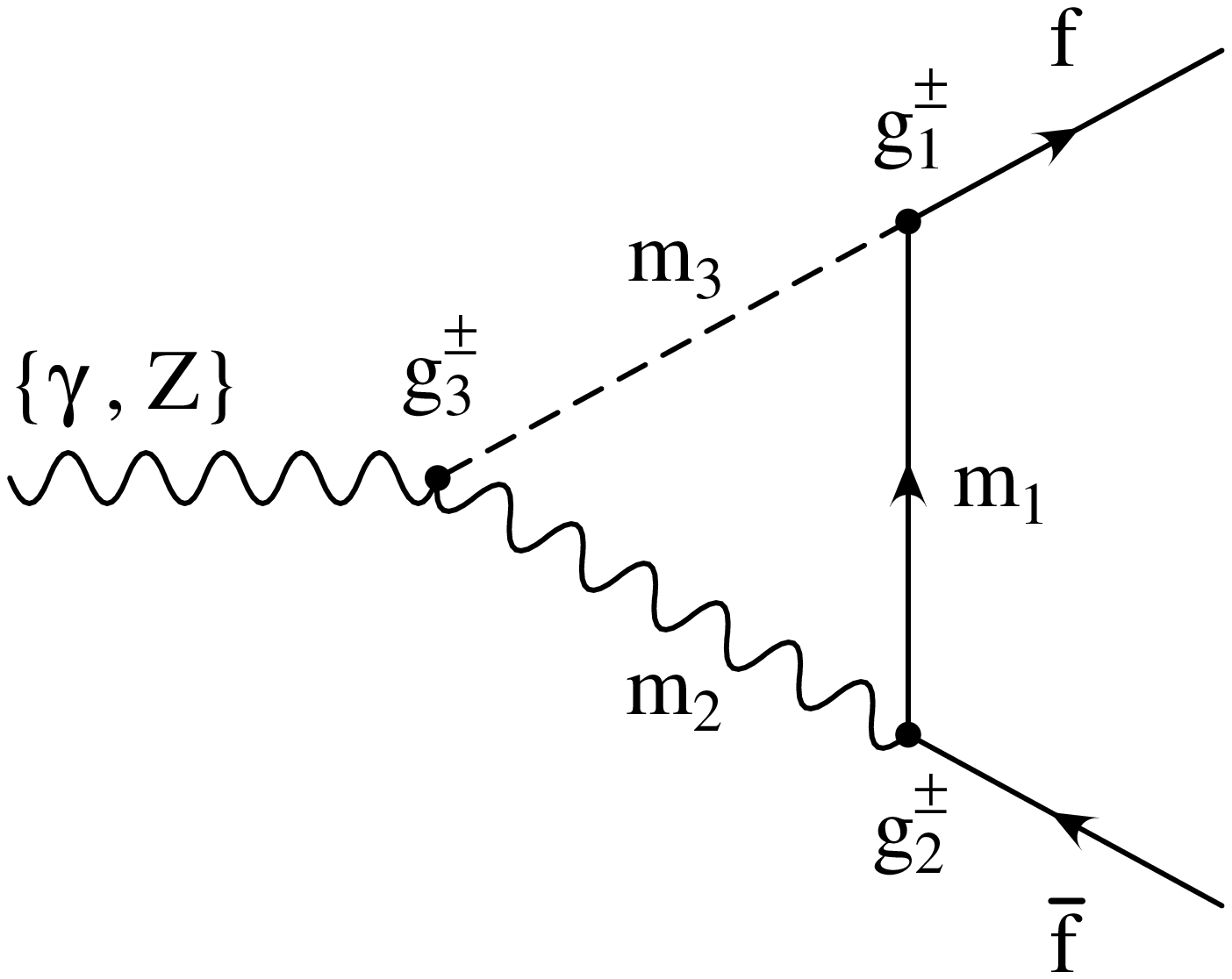,width=4.5cm}}
\vspace{5mm}
\BEA
A_{I\!I\!I}^\pm &=& g_3^\pm \bigg\{ 
                    \Big[ m_f(C_{11}-C_{12})g_1^\mp g_2^\mp 
                    - C_{12}\, g_1^\pm g_2^\pm \Big]
                    +m_1\, C_0\, g_1^\mp g_2^\pm \bigg\}~,
                    \nonumber\\
B_{I\!I\!I}^\pm &=& (C_{12}-C_{11}) g_1^\pm g_2^\pm g_3^\pm~.
\label{vert3}
\EEA
\vspace{1cm}
\[
\arraycolsep5mm
\begin{array}{|c@{}c@{}c||*{5}{c|}}\hline
\multicolumn{8}{|c|}{\rule[-2mm]{0pt}{6mm}
\mbox{\bf Masses and coupling constants for the $\gamma f\bar{f}$ 
  vertex \rm{\Roman{dummy}} }} \\
  \hline
  m_1\; & m_2\; & m_3 & \quad g_1^+\quad & \quad g_1^-\quad & 
  \quad g_2^+\quad & \quad g_2^-\quad & \qquad g_3^+ = g_3^- \qquad \\
  \hline\hline
  \rm {f'} & \rm W & \rm G^\pm & 2I_3^{f'}\, g'_C & 2I_3^f\, g_C &
  0  &\D \frac{1}{\sqrt{2}\, \sw} & -M_W  \\
  \hline
\end{array}
\]
\vspace{1cm}
\[
\arraycolsep5mm
\begin{array}{|c@{}c@{}c||*{5}{c|}}\hline
\multicolumn{8}{|c|}{\rule[-2mm]{0pt}{6mm}
\mbox{\bf Masses and coupling constants for the $Z f\bar{f}$ 
  vertex \rm{\Roman{dummy}} }} \\
  \hline
  m_1\; & m_2\; & m_3 & \quad g_1^+\quad & \quad g_1^-\quad &
  \quad g_2^+\quad & \quad g_2^-\quad & \qquad g_3^+ = g_3^- \qquad \\
  \hline\hline
  \rm f & \rm Z & \rm H^0 & g_N \, g^f_{H^0} & g_N \, g^f_{H^0} &
  g_{Z,R}  & g_{Z,L} &\D M_W \, \frac{\Cba}{\sw\, \cw^2} \\
  \hline
  \rm f & \rm Z & \rm h^0 & g_N \, g^f_{h^0} & g_N \, g^f_{h^0} &
  g_{Z,R}  & g_{Z,L} &\D M_W \, \frac{\Sba}{\sw\, \cw^2} \\
  \hline
  \rm {f'} & \rm W & \rm G^\pm & 2I_3^{f'}\, g'_C & 2I_3^f\, g_C &
  0  &\D \frac{1}{\sqrt{2}\, \sw} &\D -M_W \, \frac{\sw}{\cw} \\
  \hline
\end{array}
\]

\newpage

\setcounter{dummy}{6}
\subsection{Class-\Roman{dummy} diagram}
\vspace{5mm}
\centerline{\psfig{figure=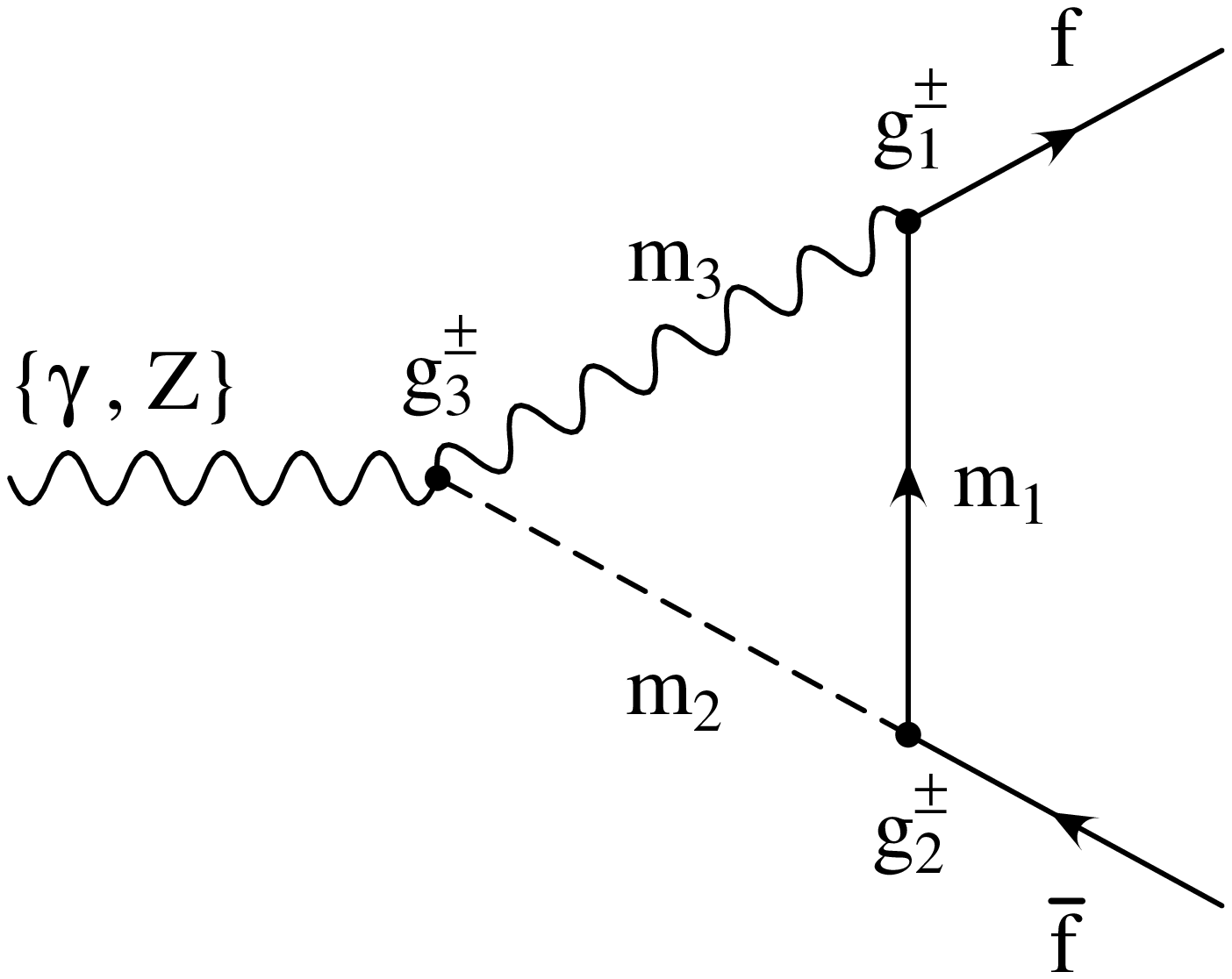,width=4.5cm}}
\vspace{5mm}
\BEA
A_{I\!V}^\pm &=& g_3^\pm \bigg\{ \Big[ m_f(C_{12}-C_{11}) g_1^\mp g_2^\pm 
                 + C_{12}\, g_1^\pm g_2^\mp \Big]
                 +m_1\, C_0\, g_1^\pm g_2^\pm \bigg\}~,
                 \nonumber\\
B_{I\!V}^\pm &=& -C_{12}\, g_1^\pm g_2^\mp g_3^\pm~.
\label{vert4}
\EEA
\vspace{1cm}
\[
\arraycolsep5mm
\begin{array}{|c@{}c@{}c||*{5}{c|}}\hline
\multicolumn{8}{|c|}{\rule[-2mm]{0pt}{6mm}
\mbox{\bf Masses and coupling constants for the $\gamma f\bar{f}$ 
  vertex \rm{\Roman{dummy}} }} \\
  \hline
  m_1\; & m_2\; & m_3 & \quad g_1^+\quad & \quad g_1^-\quad &
  \quad g_2^+\quad & \quad g_2^-\quad & 
  \qquad g_3^+ = g_3^- \qquad \\
  \hline\hline
  \rm {f'} & \rm G^\pm & \rm W & 0  &\D \frac{1}{\sqrt{2}\, \sw} &
  2I_3^f\, g_C &\D 2I_3^{f'}\, g'_C & -M_W  \\
  \hline
\end{array}
\]
\vspace{1cm}
\[
\arraycolsep5mm
\begin{array}{|c@{}c@{}c||*{5}{c|}}\hline
\multicolumn{8}{|c|}{\rule[-2mm]{0pt}{6mm}
\mbox{\bf Masses and coupling constants for the $Z f\bar{f}$ 
  vertex \rm{\Roman{dummy}} }} \\ 
  \hline
  m_1\; & m_2\; & m_3 & \quad g_1^+\quad & \quad g_1^-\quad &
  \quad g_2^+\quad & \quad g_2^-\quad & 
  \qquad g_3^+ = g_3^- \qquad \\
  \hline\hline
  \rm f & \rm H^0 & \rm Z & g_{Z,R}  & g_{Z,L} & g_N \, g^f_{H^0} & 
  g_N \, g^f_{H^0} & \D M_W \, \frac{\Cba}{\sw\, \cw^2} \\
  \hline
  \rm f & \rm h^0 & \rm Z & g_{Z,R}  & g_{Z,L} & g_N \, g^f_{h^0} & 
  g_N \, g^f_{h^0} & \D M_W \, \frac{\Sba}{\sw\, \cw^2} \\
  \hline
  \rm {f'} & \rm G^\pm & \rm W & 0 &\D \frac{1}{\sqrt{2}\, \sw} &
  2I_3^f\, g_C & 2I_3^{f'}\, g'_C &\D -M_W \, \frac{\sw}{\cw} \\
  \hline
\end{array}
\]

\newpage


\section{Box diagrams}
\label{appd}
\setcounter{equation}{0}
\setcounter{figure}{0}
\setcounter{table}{0}

Here we list the input for the box diagram contributions in the MSSM,
as specified in Appendix~\ref{appa}, with exception of the QED graphs
involving virtual photons. Because of $m_e \ll M_W$, Higgs exchange 
diagrams are negligible.

Masses and couplings of the involved particles are put together in
Table \ref{tabdVb} for the direct and in Table \ref{tabcVb} for the 
crossed vector boson box diagrams, and for the direct and crossed SUSY
box diagrams in Tables \ref{tabdNb} to \ref{tabcCb}.

\begin{figure}[h]
\centerline{\psfig{figure=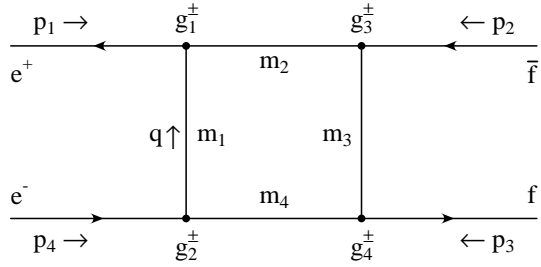,width=7cm}}
\caption[]{The topology of the direct box diagrams.}
\label{fig:Boxtopd}
\end{figure}

\begin{figure}[h]
\centerline{\psfig{figure=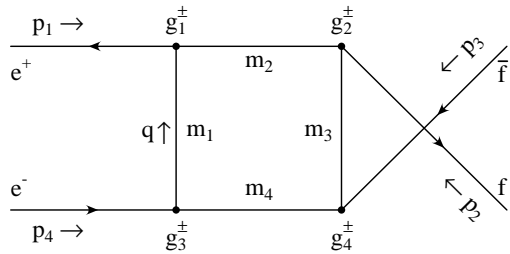,width=7cm}}
\caption[]{The topology of the crossed box diagrams.}
\label{fig:Boxtopc}
\end{figure}

Conventions on momenta and internal masses are illustrated in Figure 
\ref{fig:Boxtopd} and in Figure \ref{fig:Boxtopc}. The arrows within
the lines denote the charge flow, and long thin arrows denote the 
fermion flow if necessary.

\vspace{3mm}

\vspace{3mm}

\vorn For the direct box diagram the arguments of the 
$C_0$-function\footnote{The arguments of the $C_0$-function results
  here from the reduction of the tensor integral $D_{\mu\nu}$.}
are $C(-p_2,p_1+p_4,m_2,m_3,m_4)$ and the arguments of the $D$-functions
are $D(p_1,p_2,p_3,m_1,m_2,m_3,m_4)$, with $p_1=\bar{p}$, $p_2=-\bar{k}$, 
$p_3=-k$ and  $p_4=p$.

\vspace{3mm}

\vspace{3mm}

\vorn For the crossed box diagram the arguments of the $C_0$-function
are $C(-p_2,p_1+p_4,m_2,m_3,m_4)$ and the arguments of the $D$-functions
are $D(p_1,p_2,p_3,m_1,m_2,m_3,m_4)$, with $p_1=\bar{p}$, $p_2=-k$,
$p_3=-\bar{k}$ and $p_4=p$.

The couplings $g_{Z,R}$ and $g_{Z,L}$ are defined in Eqs. (\ref{gezet}), 
and the SUSY couplings $X^{0f}_{kl}$, $X^{+f}_{kl}$, $Y^{0f}_{kl}$ and 
$Y^{+f}_{kl}$ in Eqs. (\ref{viai}).

\newpage

\subsection{The direct vector boson box diagram}

\vspace{4mm}
\centerline{\psfig{figure=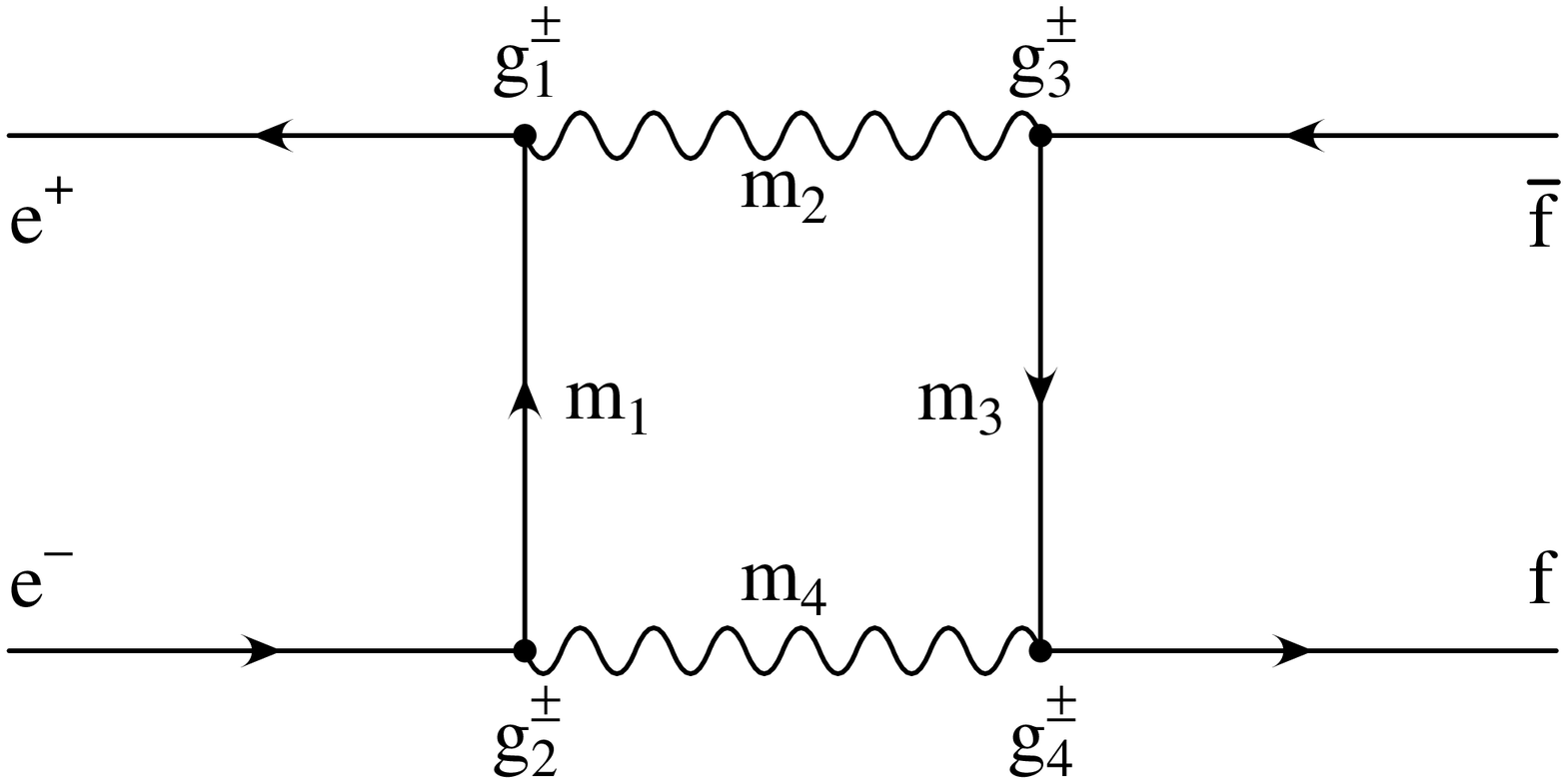,width=7cm}}

\begin{table}[h]
\begin{center}
$\renewcommand{\arraystretch}{1.8}
\arraycolsep2mm
\begin{array}{|c@{}c@{}c@{}c||*{8}{c|}}\hline
\multicolumn{12}{|c|}{\rule[-2ex]{0mm}{6ex}
\mbox{\bf Masses and coupling constants for the direct 
  vector boson box diagram}} \\
  \hline
  m_1\; & m_2\; & m_3\; & m_4 & 
  \quad g_1^+\quad & \quad g_1^-\quad & \quad g_2^+\quad & 
  \quad g_2^-\quad & \quad g_3^+\quad & \quad g_3^-\quad & 
  \quad g_4^+\quad & \quad g_4^-\quad  \\
  \hline\hline
  \rm e & \rm Z & \rm f & \rm Z &
  g^e_{Z,R} & g^e_{Z,L} & g^e_{Z,R} & g^e_{Z,L} &
  g^f_{Z,R} & g^f_{Z,L} & g^f_{Z,R} & g^f_{Z,L}  \\
  \hline
  \nu_e & \rm W & \rm f' & \rm W & 0 &\D \frac{1}{\sqrt{2}\, \sw} & 0 & 
  \D \frac{1}{\sqrt{2}\, \sw} & 0 &
  \D \frac{\frac{1}{2}-I_3^f}{\sqrt{2}\, \sw} & 0 &
  \D \frac{\frac{1}{2}-I_3^f}{\sqrt{2}\, \sw} \\
  \hline
\end{array}$
\end{center}
\caption[]{\label{tabdVb}Entries in the direct vector boson box diagram.}
\end{table}

\vspace{5mm}

\subsection{The crossed vector boson box diagram}

\vspace{4mm}
\centerline{\psfig{figure=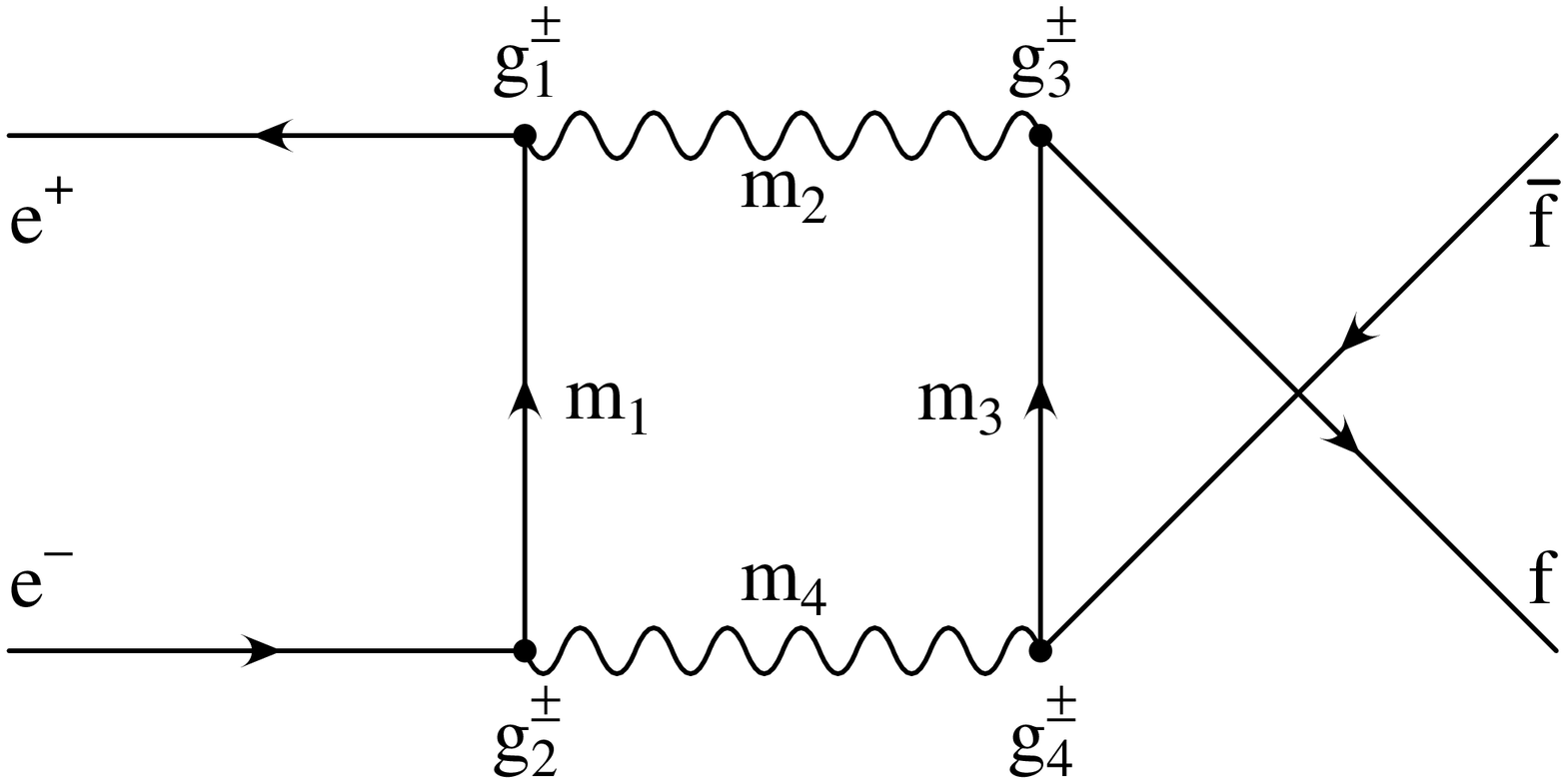,width=7cm}}

\begin{table}[h]
\begin{center}
$\renewcommand{\arraystretch}{1.8}
\arraycolsep2mm
\begin{array}{|c@{}c@{}c@{}c||*{8}{c|}}\hline
\multicolumn{12}{|c|}{\rule[-2ex]{0mm}{6ex}
\mbox{\bf Masses and coupling constants for the crossed vector boson 
  box diagram}} \\
  \hline
  m_1\; & m_2\; & m_3\; & m_4 & 
  \quad g_1^+\quad & \quad g_1^-\quad & \quad g_2^+\quad & 
  \quad g_2^-\quad & \quad g_3^+\quad & \quad g_3^-\quad & 
  \quad g_4^+\quad & \quad g_4^-\quad \\
  \hline\hline
  \rm e & \rm Z & \rm f & \rm Z &
  g^e_{Z,R} & g^e_{Z,L} & g^e_{Z,R} & g^e_{Z,L} &
  g^f_{Z,R} & g^f_{Z,L} & g^f_{Z,R} & g^f_{Z,L}  \\
  \hline
  \nu_e & \rm W & \rm f' & \rm W & 0 &\D \frac{1}{\sqrt{2}\, \sw} & 0 & 
  \D \frac{1}{\sqrt{2}\, \sw} & 0 &
  \D \frac{\frac{1}{2}+I_3^f}{\sqrt{2}\, \sw} & 0 & 
  \D \frac{\frac{1}{2}+I_3^f}{\sqrt{2}\, \sw} \\
  \hline
\end{array}$
\end{center}
\caption[]{\label{tabcVb}Entries in the crossed vector boson box diagram.}
\end{table}

\newpage

\subsection{The direct neutralino box diagram}

\centerline{\psfig{figure=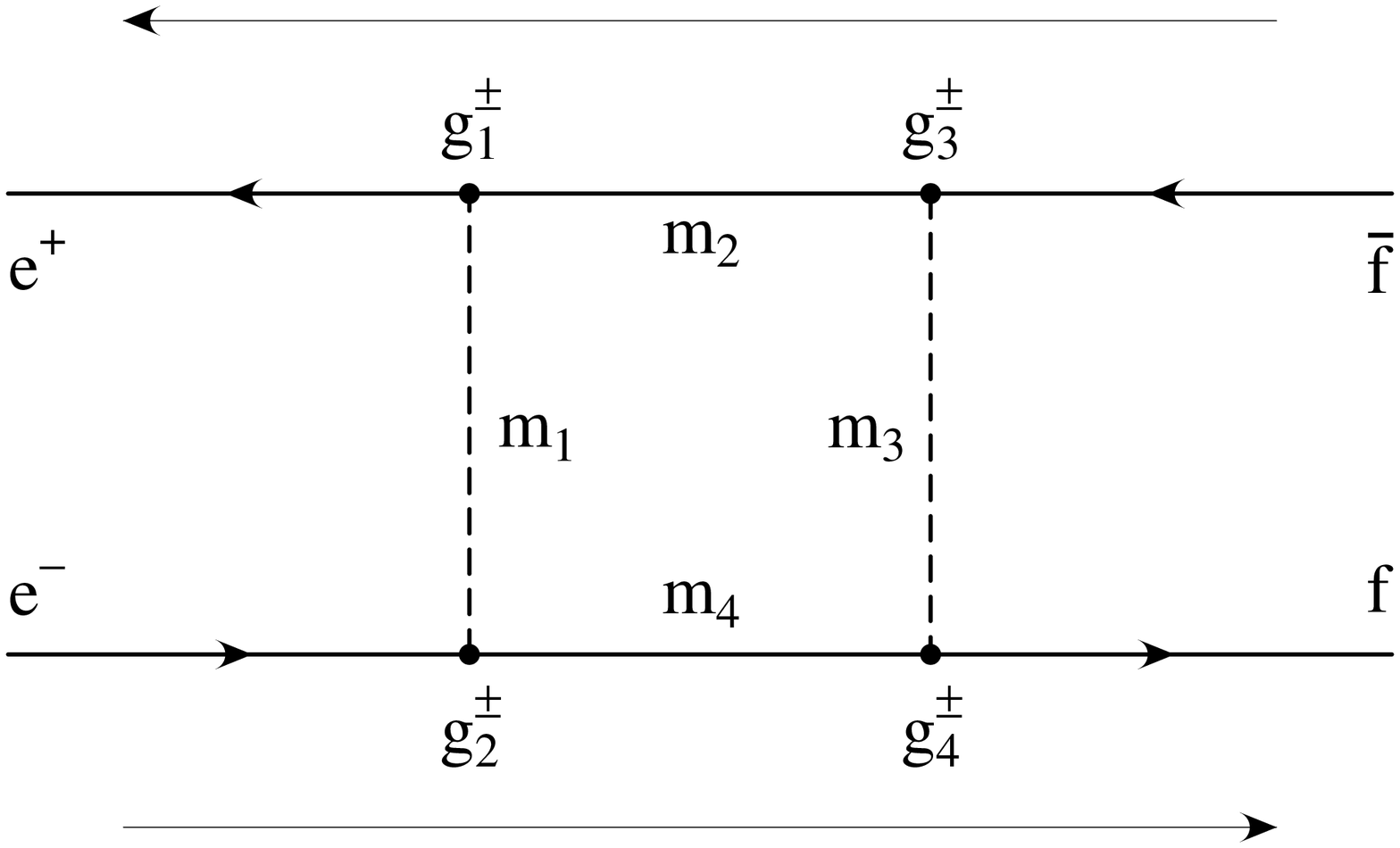,width=7cm}}

\renewcommand{\arraystretch}{1.3}

\begin{table}[h]
\begin{center}
$\arraycolsep2mm
\begin{array}{|c@{}c@{}c@{}c||*{8}{c|}}\hline
\multicolumn{12}{|c|}{\rule[-2ex]{0mm}{6ex}
\mbox{\bf Masses and coupling constants for the direct neutralino 
  box diagram}} \\
  \hline
  m_1\; & m_2\; & m_3\; & m_4 & 
  \quad g_1^+\quad & \quad g_1^-\quad & \quad g_2^+\quad & 
  \quad g_2^-\quad & \quad g_3^+\quad & \quad g_3^-\quad & 
  \quad g_4^+\quad & \quad g_4^-\quad \\
  \hline\hline
  \tilde{e}_1 & \tilde{\chi}^0_j & \tilde{f}_1 & \tilde{\chi}^0_i &
  X^{0e}_{j1} & (Y^{0e}_{j1})^* & Y^{0e}_{i1} & (X^{0e}_{i1})^* &
  Y^{0f}_{j1} & (X^{0f}_{j1})^* & X^{0f}_{i1} & (Y^{0f}_{i1})^* \\
  \hline
  \tilde{e}_1 & \tilde{\chi}^0_j & \tilde{f}_2 & \tilde{\chi}^0_i &
  X^{0e}_{j1} & (Y^{0e}_{j1})^* & Y^{0e}_{i1} & (X^{0e}_{i1})^* &
  Y^{0f}_{j2} & (X^{0f}_{j2})^* & X^{0f}_{i2} & (Y^{0f}_{i2})^* \\
  \hline
  \tilde{e}_2 & \tilde{\chi}^0_j & \tilde{f}_1 & \tilde{\chi}^0_i &
  X^{0e}_{j2} & (Y^{0e}_{j2})^* & Y^{0e}_{i2} & (X^{0e}_{i2})^* &
  Y^{0f}_{j1} & (X^{0f}_{j1})^* & X^{0f}_{i1} & (Y^{0f}_{i1})^* \\
  \hline
  \tilde{e}_2 & \tilde{\chi}^0_j & \tilde{f}_2 & \tilde{\chi}^0_i &
  X^{0e}_{j2} & (Y^{0e}_{j2})^* & Y^{0e}_{i2} & (X^{0e}_{i2})^* &
  Y^{0f}_{j2} & (X^{0f}_{j2})^* & X^{0f}_{i2} & (Y^{0f}_{i2})^* \\
  \hline
\end{array}$
\end{center}
\caption[]{\label{tabdNb}Entries in the direct neutralino box diagram.}
\end{table}

\subsection{The crossed neutralino box diagram}

\centerline{\psfig{figure=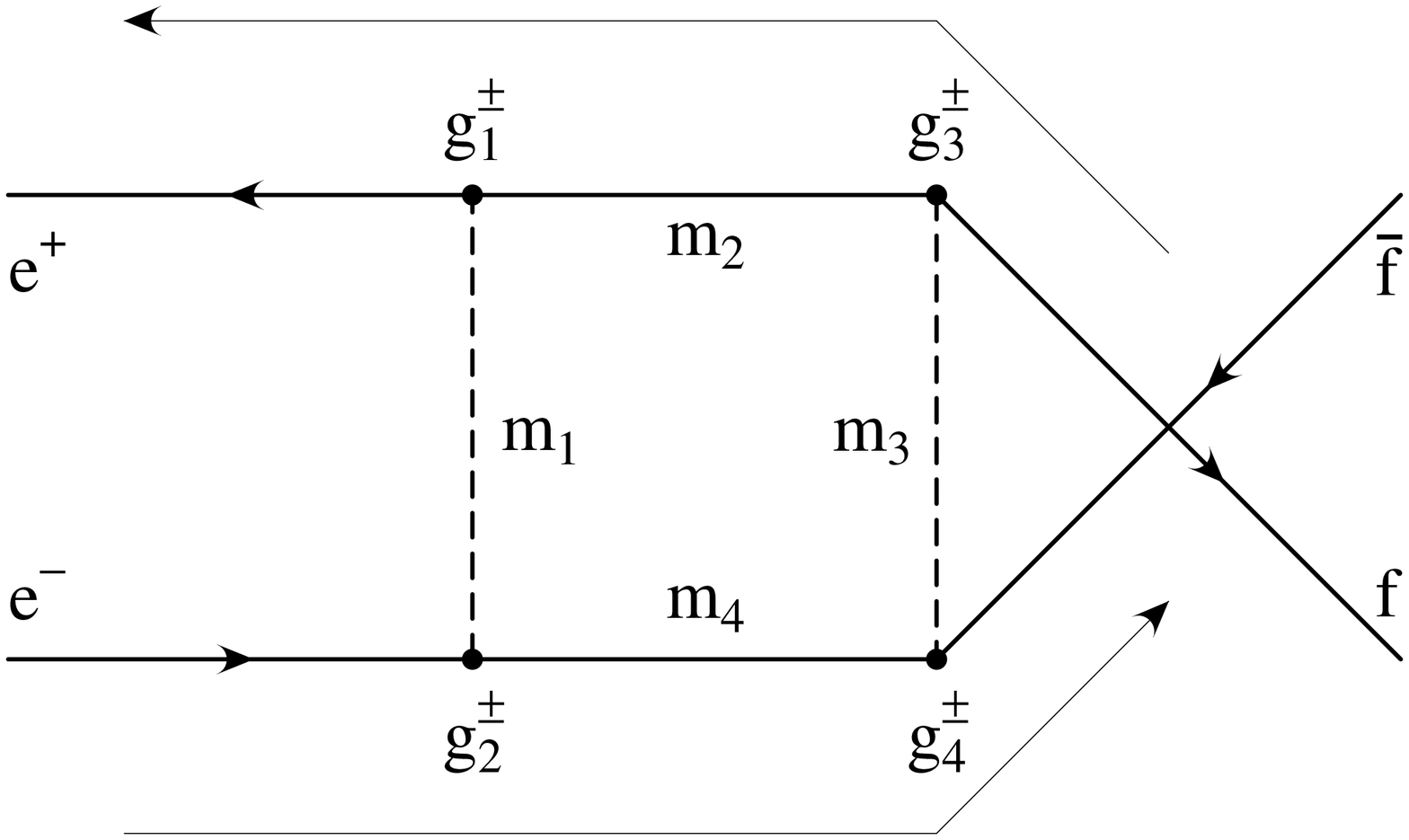,width=7cm}}

\begin{table}[h]
\begin{center}
$\arraycolsep2mm
\begin{array}{|c@{}c@{}c@{}c||*{8}{c|}}\hline
\multicolumn{12}{|c|}{\rule[-2ex]{0mm}{6ex}
\mbox{\bf Masses and coupling constants for the crossed neutralino 
  box diagram}} \\
  \hline
  m_1\; & m_2\; & m_3\; & m_4 & 
  \quad g_1^+\quad & \quad g_1^-\quad & \quad g_2^+\quad & 
  \quad g_2^-\quad & \quad g_3^+\quad & \quad g_3^-\quad & 
  \quad g_4^+\quad & \quad g_4^-\quad \\
  \hline\hline
  \tilde{e}_1 & \tilde{\chi}^0_j & \tilde{f}_1 & \tilde{\chi}^0_i &
  X^{0e}_{j1} & (Y^{0e}_{j1})^* & Y^{0e}_{i1} & (X^{0e}_{i1})^* &
  X^{0f}_{j1} & (Y^{0f}_{j1})^* & Y^{0f}_{i1} & (X^{0f}_{i1})^* \\
  \hline
  \tilde{e}_1 & \tilde{\chi}^0_j & \tilde{f}_2 & \tilde{\chi}^0_i &
  X^{0e}_{j1} & (Y^{0e}_{j1})^* & Y^{0e}_{i1} & (X^{0e}_{i1})^* &
  X^{0f}_{j2} & (Y^{0f}_{j2})^* & Y^{0f}_{i2} & (X^{0f}_{i2})^* \\
  \hline
  \tilde{e}_2 & \tilde{\chi}^0_j & \tilde{f}_1 & \tilde{\chi}^0_i &
  X^{0e}_{j2} & (Y^{0e}_{j2})^* & Y^{0e}_{i2} & (X^{0e}_{i2})^* &
  X^{0f}_{j1} & (Y^{0f}_{j1})^* & Y^{0f}_{i1} & (X^{0f}_{i1})^* \\
  \hline
  \tilde{e}_2 & \tilde{\chi}^0_j & \tilde{f}_2 & \tilde{\chi}^0_i &
  X^{0e}_{j2} & (Y^{0e}_{j2})^* & Y^{0e}_{i2} & (X^{0e}_{i2})^* &
  X^{0f}_{j2} & (Y^{0f}_{j2})^* & Y^{0f}_{i2} & (X^{0f}_{i2})^* \\
  \hline
\end{array}$
\end{center}
\caption[]{\label{tabcNb}Entries in the crossed neutralino box diagram.}
\end{table}

\newpage

\subsection{The direct chargino box diagram}

\vspace{6mm}
\centerline{\psfig{figure=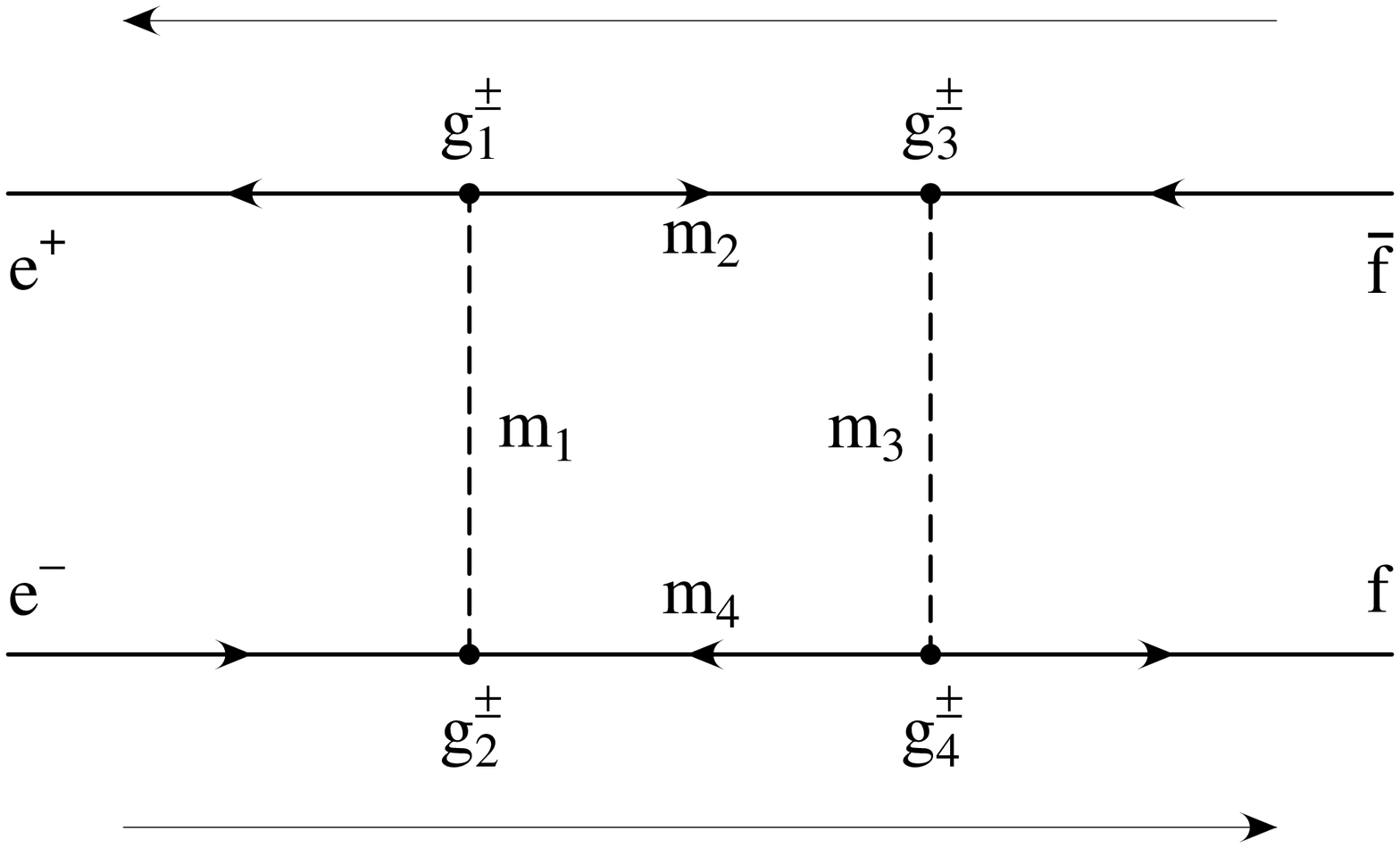,width=7cm}}
\vspace{6mm}

\renewcommand{\arraystretch}{1.3}

\begin{table}[h]
\begin{center}
$\arraycolsep2mm
\begin{array}{|c@{}c@{}c@{}c||*{8}{c|}}\hline
\multicolumn{12}{|c|}{\rule[-2ex]{0mm}{6ex}
\mbox{\bf Masses and coupling constants for the direct chargino
  box diagram}} \\
  \hline
  m_1\; & m_2\; & m_3\; & m_4 & 
  \quad g_1^+\quad & \quad g_1^-\quad & \quad g_2^+\quad & 
  \quad g_2^-\quad & \quad g_3^+\quad & \quad g_3^-\quad & 
  \quad g_4^+\quad & \quad g_4^-\quad \\
  \hline\hline
  \tilde{\nu}_e & \tilde{\chi}^+_j & \tilde{f}'_1 & \tilde{\chi}^+_i &
  X^{+e}_{j1} & (Y^{+e}_{j1})^* & Y^{+e}_{i1} & (X^{+e}_{i1})^* &
  Y^{+f}_{j1} & (X^{+f}_{j1})^* & X^{+f}_{i1} & (Y^{+f}_{i1})^* \\
  \hline
  \tilde{\nu}_e & \tilde{\chi}^+_j & \tilde{f}'_2 & \tilde{\chi}^+_i &
  X^{+e}_{j1} & (Y^{+e}_{j1})^* & Y^{+e}_{i1} & (X^{+e}_{i1})^* &
  Y^{+f}_{j2} & (X^{+f}_{j2})^* & X^{+f}_{i2} & (Y^{+f}_{i2})^* \\
  \hline
\end{array}$
\end{center}
\caption[]{\label{tabdCb}Entries in the direct chargino box diagram.}
\end{table}

\vspace{5mm}

\subsection{The crossed chargino box diagram}

\vspace{6mm}
\centerline{\psfig{figure=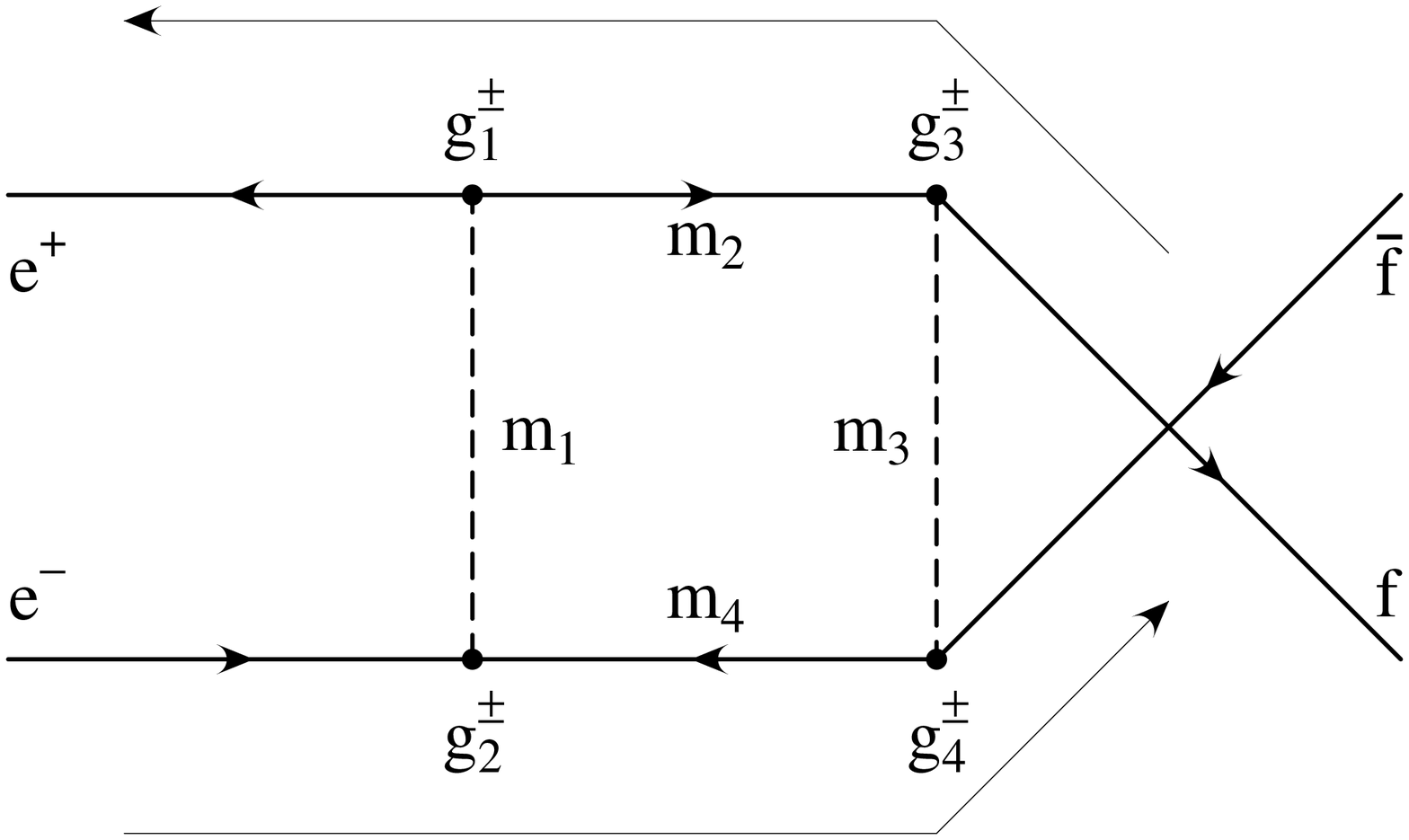,width=7cm}}
\vspace{6mm}

\begin{table}[h]
\begin{center}
$\arraycolsep2mm
\begin{array}{|c@{}c@{}c@{}c||*{8}{c|}}\hline
\multicolumn{12}{|c|}{\rule[-2ex]{0mm}{6ex}
\mbox{\bf Masses and coupling constants for the crossed chargino 
  box diagram}} \\
  \hline
  m_1\; & m_2\; & m_3\; & m_4 & 
  \quad g_1^+\quad & \quad g_1^-\quad & \quad g_2^+\quad & 
  \quad g_2^-\quad & \quad g_3^+\quad & \quad g_3^-\quad & 
  \quad g_4^+\quad & \quad g_4^-\quad \\
  \hline\hline
  \tilde{\nu}_e & \tilde{\chi}^+_j & \tilde{f}'_1 & \tilde{\chi}^+_i &
  X^{+e}_{j1} & (Y^{+e}_{j1})^* & Y^{+e}_{i1} & (X^{+e}_{i1})^* &
  X^{+f}_{j1} & (Y^{+f}_{j1})^* & Y^{+f}_{i1} & (X^{+f}_{i1})^* \\
  \hline
  \tilde{\nu}_e & \tilde{\chi}^+_j & \tilde{f}'_2 & \tilde{\chi}^+_i &
  X^{+e}_{j1} & (Y^{+e}_{j1})^* & Y^{+e}_{i1} & (X^{+e}_{i1})^* &
  X^{+f}_{j2} & (Y^{+f}_{j2})^* & Y^{+f}_{i2} & (X^{+f}_{i2})^* \\
  \hline
\end{array}$
\end{center}
\caption[]{\label{tabcCb}Entries in the crossed chargino box diagram.}
\end{table}

\newpage

\end{appendix}


\end{document}